\documentclass[12pt]{book}
\usepackage{amsmath,graphicx} 
\usepackage{multirow} 
\usepackage{amsfonts}
\usepackage{amssymb}
\usepackage{fancyhdr}
\fancypagestyle{plain}{\fancyhead{} 
}
\def\hybrid{\topmargin -20pt    \oddsidemargin 0pt
        \headheight 0pt \headsep 15pt
        \textwidth 6.25in       
        \textheight 8.6 in       
        \marginparwidth .875in
        \parskip 5pt plus 1pt   \jot = 1.5ex}
\hybrid

\numberwithin{equation}{chapter}
\numberwithin{table}{section}

\newcommand{\beq}{\begin{equation}}
\newcommand{\eeq}{\end{equation}}
\newcommand{\bea}{\begin{eqnarray}}
\newcommand{\eea}{\end{eqnarray}}
\newcommand{\ba}{\begin{array}}
\newcommand{\ea}{\end{array}}
\newcommand{\bt}{\begin{tabular}}
\newcommand{\et}{\end{tabular}}
\newcommand{\bc}{\begin{center}}
\newcommand{\ec}{\end{center}}

\newcommand{\Ox}{\Omega}

\newcommand{\cC}{\mathcal{C}}
\newcommand{\cD}{\mathcal{D}}
\newcommand{\cL}{\mathcal{L}}
\newcommand{\cS}{\mathcal{S}}
\newcommand{\cK}{\mathcal{K}}
\newcommand{\cN}{\mathcal{N}}
\newcommand{\cW}{\mathcal{W}}
\newcommand{\cG}{\mathcal{G}}
\newcommand{\cA}{\mathcal{A}}

\newcommand{\cB}{\mathcal{B}}
\newcommand{\cF}{\mathcal{F}}

\newcommand{\cV}{\mathcal{V}}
 
\newcommand{\KK}{\mathcal{K}}
\newcommand{\MM}{\mathcal{M}}
\newcommand{\cM}{\mathcal M}

\newcommand{\cO}{\mathcal{O}}
\newcommand{\OO}{\mathcal{O}}

\newcommand{\IM}{\text{Im}\, \mathcal{M}}

\newcommand{\RM}{\text{Re}\, \mathcal{M}}
\newcommand{\I}{\text{Im}}
\newcommand{\R}{\text{Re}}

\newcommand{\Kcs}{K^{\text{cs}}}

\newcommand{\pev}{{\varphi^{ev}}}
\newcommand{\pevb}{{\bar \varphi^{ev}}}
\newcommand{\podd}{{\varphi^{odd}}}
\newcommand{\poddb}{{\bar \varphi^{odd}}}
\newcommand{\fuh}{{\mathcal{\hat U}}}
\newcommand{\fu}{{\mathcal{U}}}
\newcommand{\fe}{{\mathcal{E}}}
\newcommand{\feh}{{\mathcal{\hat E}}}
\newcommand{\fa}{{\mathcal{ A}}}

\newcommand{\bi}{{\bar \imath}}
\newcommand{\ib}{{\bar\imath }}
\newcommand{\jb}{{\bar \jmath }}
\newcommand{\bj}{{\bar\jmath}}

\newcommand{\Kh}{{\hat{K}}}
\newcommand{\Lh}{{\hat{L}}}
\newcommand{\Ah}{{\hat{A}}}
\newcommand{\Bh}{{\hat{B}}}

\newcommand{\Mh}{{\hat{M}}}
\newcommand{\Nh}{{\hat{N}}}
\newcommand{\kh}{{\hat{k}}}
\newcommand{\ah}{{\hat{a}}}
\newcommand{\bh}{{\hat{b}}}
\newcommand{\ch}{{\hat{c}}}

\newcommand{\Mext}{{\mathbb{M}^{3,1}}}
\newcommand{\Mint}{{Y}}

\DeclareMathOperator{\vol}{vol}

\newcommand{\dd}{d}

\newcommand{\bbZ}{\mathbb{Z}}
\newcommand{\bbR}{\mathbb{R}}
\newcommand{\bbC}{\mathbb{C}}

\newcommand{\nn}{\nonumber}
\newcommand{\RE}{\textrm{Re} \,}

\newcommand{\cref}{{\bf [check ref]}}

\newcommand{\N}{\Theta}

\newcommand{\Jc}{J_{\rm c}}
\newcommand{\Omegac}{\Omega_{\rm c}}
\newcommand{\cc}{c}
\newcommand{\CC}{C}

\newcommand{\Em}{\varphi}          
\newcommand{\WV}{\mathcal{W}}      
\newcommand{\FD}{F}

\begin{document}

\begin{titlepage}
\begin{center}

\hfill hep-th/0507153\\
\hfill ZMP-HH/05-16\\
\vskip 1.5cm

{\large \bf The effective action of type II Calabi-Yau orientifolds}
\footnote{%
This article is based on the Ph.D.~thesis of the author.}\\

\vskip 1cm

{\bf Thomas W.\ Grimm }\footnote{%
From September 1, 2005: Department of Physics, University of Wisconsin, Madison WI 53706, USA }  \\
\vskip 0.5cm

{\em II. Institut f{\"u}r Theoretische Physik\\
Universit{\"a}t Hamburg, Luruper Chaussee 149\\
 D-22761 Hamburg, Germany}\\
\vskip 5pt
and
\vskip 5pt
{\em Zentrum f\"ur Mathematische Physik \\
Universit\"at Hamburg, 
Bundesstrasse 55\\
D-20146 Hamburg, Germany}\\

\vskip 10pt

 {\tt  thomas.grimm@desy.de} \\

\end{center}

\vskip .5cm

\begin{center} {\bf ABSTRACT } \end{center}

\noindent

This article first reviews the calculation of 
the $N = 1$ effective action for generic type IIA and type IIB Calabi-Yau orientifolds 
in the presence of background fluxes by using a Kaluza-Klein reduction. The K\"ahler potential, 
the gauge kinetic functions and the flux-induced superpotential are 
determined in terms of geometrical data of the Calabi-Yau orientifold and the background fluxes. 
As a new result, it is shown that the chiral description directly relates to Hitchin's generalized geometry 
encoded by special odd and even forms on a threefold, whereas a dual formulation with several 
linear multiplets makes contact to the underlying $N=2$ special geometry.  
In type IIB setups, the flux-potentials can be expressed 
in terms of superpotentials, D-terms and, generically, a 
massive linear multiplet. The type IIA superpotential depends on all geometric 
moduli of the theory. It is reviewed, how type IIA orientifolds arise as a special 
limit of M-theory compactified on specific $G_2$ manifolds by matching the effective actions. In a similar 
spirit type IIB orientifolds are shown to descend from F-theory 
on a specific class of Calabi-Yau fourfolds.
In addition, mirror symmetry for Calabi-Yau orientifolds
is briefly discussed and it is shown that the $N = 1$ chiral coordinates linearize the appropriate 
instanton actions.

\vfill

\end{titlepage}

\vspace*{10cm}
\begin{center}
   {\bf \large Acknowledgments}
\end{center}
This article is based on my Ph.D.~thesis. 
First of all I would like to express my deep gratitude to my supervisor Prof.~Jan Louis 
for his continuous support, expert advises and encouragement.  
The collaboration with Mariana Gra\~na, Hans Jockers, Frederic Schuller and Mattias Wohlfarth
was very enjoyable and fruitful. I esspecially like to thank my office
mates Iman Benmachiche, Olaf Hohm, Hans Jockers, Andrei Micu and Anke Knauf for providing a very 
delighting athmosphere and the numerous discussions about physiscs 
and beyond. I am also indebted to David Cerde\~no, Vincente Cort\'es, Frederik Denef, Sergei Gukov, Henning Samtleben, 
Sakura Sch\"afer-Nameki, Shamit Kachru, Boris~K\"ors, Paolo Merlatti, Thorsten Pr\"ustel, 
Waldemar Schulgin, Silvia Vaul\`a and 
Martin Weidner for various discussions and correspondence. I am grateful to my lovely girlfriend,
for supporting me through the last years. 

\vspace{.2cm} 
This work is supported by the DFG -- The German Science Foundation, 
the DAAD~--~the German Academic Exchange Service, and the European RTN Program MRTN-CT-2004-503369.

\thispagestyle{plain}

\tableofcontents

\renewcommand{\thetable}{\arabic{chapter}.\arabic{table}}

\chapter{Introduction}

The Standard Model of particle physics extended by massive neutrinos 
has been tested to a very high precision and is believed 
to correctly describe the known elementary particles and their interactions.
Experimentally, the only missing ingredient is the scalar Higgs particle, which gives 
masses to the leptons and quarks, once it acquires a vacuum expectation 
value. The Standard Model provides a realistic model of a renormalizable
gauge theory. Despite its impressive success there are also various theoretical 
drawbacks, such as the large number of free parameters, the hierarchy and naturalness problem as
well as the missing unification with gravity. These indicate that it cannot be viewed 
as a fundamental theory, but rather should arise as an effective description.

A natural extension of the Standard Model is provided by supersymmetry, which serves as a 
fundamental symmetry between bosons and fermions. Supersymmetry predicts a superpartner 
for all known particles and thus basically doubles the particle content of the theory.
However, none of the superpartners was ever detected in an accelerator experiment, which 
implies that supersymmetry is appearing in its broken phase. The supersymmetric Standard 
Model solves some of the problems of the Standard Model \cite{revSusy}. 
Even in its (softly) broken phase it forbids large quantum corrections to scalar masses.
This allows the Higgs mass to remain to be of order the weak scale also in a theory 
with a higher mass scale. Furthermore, assuming the supersymmetric Standard Model to be valid
up to very high scales, the renormalization group flow predicts a unification of all
three gauge-couplings. This supports the idea of an underlying theory 
relevant beyond the Standard Model scales. However, it remains to unify 
these extensions with gravity.

On the other hand, we know that General Relativity  
links the geometry of spacetime with the distribution of the matter densities. Einsteins 
theory is very different in nature. It is a classical theory which is 
hard to quantize due to its ultra-violet divergences (see however \cite{LQG}). 
This fact constraints its range of validity to 
phenomena, where quantum effects are of negligible importance. However, there is no 
experimental evidence which contradicts large scale predictions based on General 
Relativity.

Facing these facts General Relativity and the Standard Model seems to be incompatible, 
in the sense that neither of them allows to naturally adapt the other. This becomes 
important in regimes where both theories have to be applied in order to describe the 
correct physics. Early time cosmology or physics of black holes are only 
two regimes where the interplay of quantum and gravitational effects become important. 
To nevertheless approach this theoretically interesting questions one might 
hope for a fundamental quantum theory combining the Standard Model and 
General Relativity. Until now one does not know what this unifying theory 
is, but one has at least one possible candidate. This theory is 
known as String Theory, which was studied intensively from various directions in
the last thirty years. A comprehensive introduction to the 
subject can be found in \cite{GSWbook,JPbook,Zwiebach}.

Perturbative String Theory is a quantum theory of one-dimensional extended objects which 
replace the ordinary point particles. These fundamental strings can appear in various vibrational 
modes which at low energies are identified with different particles. The characteristic
length of the string is $\sqrt{\alpha'}$, where $\alpha'$ is the Regge slope.
Hence, the extended nature of the strings only becomes apparent close to the string 
scale $1/\sqrt{\alpha'}$. The string spectrum naturally includes a mode corresponding to the graviton. 
This implies that Sting Theory indeed includes gravity and as we will further discuss below 
reduces to Einsteins theory at low energies. It most likely provides a renormalizable 
quantum theory of gravity around a given background. It avoids the ultra-violet divergences of graviton 
scattering amplitudes in field theory by smearing out the location of the interactions.

The extended nature of the fundamental strings poses strong consistency constraints on the 
theory. Non-tachyonic String Theories (Superstring Theories) require space-time supersymmetry
and predict a ten-dimensional space-time at weak coupling. 
Altogether there are only five consistent String Theories, which are called 
type IIA, type IIB, heterotic $SO(32)$ and $E_8 \times E_8$ and type I. These theories are connected
by various dualities and one may eventually hope to unify all of them into one fundamental theory \cite{Dual, JPbook}.

As striking a proper formulation of such a fundamental theory might be, much of its uniqueness and beauty could 
be spoiled in attempting to extract four-dimensional results. This is equally true for the five String Theories 
formulated in ten dimensions. One approach to reduce String Theory from ten to four space-time dimensions 
is compactification on a geometric background of the form $\Mext \times \Mint$. $\Mext$ is 
identified with our four-dimensional world, while $\Mint$ is chosen to be small and compact, 
such that these six additional dimensions are not visible in experiments. This however induces a 
high amount of ambiguity, since String Theory allows for various consistent choices of $\Mint$.
Eventually one would hope to find a String Field Theory formulated in ten dimensions, which 
resolves this ambiguity and dynamically chooses a certain background. 
However, such a theory is still lacking and one is forced to take a sideway 
to find and explore consistent string backgrounds. 

For a given background, the ten-dimensional theory can then be reduced to four dimensions by a 
Kaluza-Klein compactification \cite{KaluzaKlein} (for a review on Kaluza-Klein reduction see 
e.g.\ \cite{KK-review}). This amounts to expanding 
the fields into modes of $\Mint$ and results in a full tower of Kaluza-Klein modes for 
each of the string excitations. Additionally there are winding modes corresponding to 
strings winding around cycles in $\Mint$. 
Generically it is hard and phenomenologically not interesting 
to deal with these infinite towers of modes and an effective description is needed.

In order to extract an effective formulation one may first integrate out the 
massive string excitations with masses of order $1/\sqrt{\alpha'}$. This is possible 
due to the fact that the string scale $1/\sqrt{\alpha'}$ is usually set to be of order the 
Planck scale such that gravity couples with Newtonian strength. 
In the point-particle limit $\alpha' \rightarrow 0$ the effective theory 
describing the massless string modes is a supergravity theory (see e.g.~\cite{GSWbook,JPbook}). 
It can be constructed by calculating string scattering amplitudes for massless states.
One then infers an effective action for these fields encoding the same 
tree level scattering vertices. An example is the three-graviton scattering amplitude
in String Theory, which in an effective description can be equivalently obtained from the 
ten-dimensional Einstein-Hilbert term. 
Repeating the same reasoning for all other massless string modes 
yields a ten-dimensional supergravity theory for each of the five String Theories.

In a similar spirit one can also extract an effective Kaluza-Klein theory. For a 
compact internal manifold $\Mint$ the first massive Kaluza-Klein 
modes have a mass of order $1/R$, where $R$ is the `average radius' of $\Mint$. Hence, 
choosing $\Mint$ to be sufficiently small these modes become heavy and can be integrated 
out. On the other hand, $\Mint$ has to be large enough that winding modes of length $\sqrt{\alpha'}$ 
can be discarded. Together for $p$ being the characteristic momentum of the lower-dimensional fields 
an effective description of the massless modes is valid in the regime $1/p \gg R \gtrsim \sqrt{\alpha'}$.

The structure of the four-dimensional theory obtained by such a reduction highly depends
on the chosen internal manifold. The properties of $\Mint$ determine the amount of supersymmetry
and the gauge-groups of the lower-dimensional theory. Generically one insists that $\Mint$ 
preserves some of the ten-dimensional supersymmetries. This is due to the fact that string
theory on supersymmetric backgrounds is under much better control and various consistency 
conditions are automatically satisfied. It turns out that looking for a supersymmetric theory 
with a four-dimensional Minkowski background the internal manifold has to be a  
Calabi-Yau manifold \cite{Greene}. From a phenomenological point of view the resulting low-energy 
supergravity theories need to include gauged matter fields filling the spectrum of the desired 
gauge theory such as the supersymmetric Standard Model. However, parameters like the size and 
shape of the compact space appear as massless neutral scalar fields in four dimensions. 
They label the continuous degeneracy of consistent backgrounds 
$\Mint$ and are generically not driven to any particular value; 
they are moduli of the theory. In a Standard Model-like vacuum these moduli have to be massive, such 
that they are not dynamical in the low-energy effective action. Therefore one needs to identify 
a mechanism in String Theory which induces a potential for these scalars. As it is 
well-known for supergravity theories this potential can provide at the same time
a way to spontaneously break supersymmetry. 

To generate a moduli-dependent potential in a consisted String Theory setup is a non-trivial task 
and requires further refinements of the standard compactifications. Recently, much effort was
made to establish controllable mechanisms to stabilize moduli fields in type II 
String Theory. The three most popular approaches are the inclusion of background 
fluxes \cite{Strominger1}--\cite{TGL2}, instanton corrections \cite{Witten,HM,non-pert,Curio} and 
gaugino condensates \cite{gaugino,non-pert}.
This raised the hope to find examples of string vacua with all moduli being 
fixed \cite{KKLT,KachruK,TGL2,DSFGK}. Moreover, phenomenologically interesting scenarios 
for particle physics and cosmology can be constructed within these setups \cite{reviewPP,reviewcosmo}. 

In contrast to $E_8 \times E_8$ and $SO(32)$ heterotic String Theory and type I 
strings both Type II String Theory do not consist of non-Abelian gauge-groups 
in their original formulation. Thus most of the model building was first concentrated
on the heterotic String Theory as well as type I strings. 
This has changed after the event of the D-branes \cite{JP,JPbook,D-branes,AD}, 
which naturally induce non-trivial 
gauge theories. It turned out that compactifications with space-time filling D-branes 
combined with moduli potentials due to fluxes or non-perturbative
effects provide a rich arena for model building in particle physics as well as cosmology 
\cite{reviewPP,reviewcosmo}. One of the reasons is that consistent setups 
with D-branes and fluxes generically demand a generalization of the Kaluza-Klein Ansatz to so-called warped 
compactifications \cite{BB1,Verlinde,GSS,GKP}. Remarkably, these compactifications provide 
a String Theory realization of models with large hierarchies \cite{Verlinde,DRS,Mayr,GSS,GKP}
as they were first suggested in \cite{RS}.
 
One of the major motivation of this work it to analyze the low energy 
dynamics of the (bulk) supergravity moduli fields within a brane world setup 
with a non-vanishing potential. Hence, we will more carefully introduce the basic 
constituents in the following.

\section{Compactification and moduli stabilization}
\label{geom_red}

Sting Theory is consistently formulated in a ten-dimensional space-time.
In order in order to make contact with our four-dimensional observed world
one is forced to assign six of these dimensions to an invisible sector. 
This can be achieve by choosing these dimensions to be 
small and compact and not detectable in present experiments. 
Even though the additional dimensions are not observed directly,
they influence the resulting four-dimensional physics in a crucial way.
 
The idea of geometric compactification is rather old 
and goes back to the work of Kaluza and Klein in 1920 considering 
compactification of five-dimensional gravity on a circle \cite{KaluzaKlein}. They
aimed at combining gravity with $U(1)$ gauge theory in a higher-dimensional 
theory. Through our motivations have changed, the techniques are very similar
and can be generalized to the reduction from ten to four dimensions.

In the Kaluza-Klein reduction one starts by specifying an Ansatz for the background
space-time \cite{KK-review}. Topologically it is assumed to be a manifold of the product structure 
\beq \label{product_Ansatz}
  \cM_{10} = \Mext \times \Mint\ ,
\eeq
where $\Mext$ represent the four observed non-compact dimensions and $\Mint$ correspond 
to the compact internal manifold. On this space one specifies a block-diagonal 
background metric
\beq \label{background_metric}
   ds^2 =  g^{(4)}_{\mu \nu}(x)\, dx^\mu dx^\nu + g^{(6)}_{mn}(y)\, dy^m dy^n  
\eeq  
where $g^{(4)}_{\mu \nu}$ is a four-dimensional Minkowski metric and $g^{(6)}_{ab}$ is 
the metric on the compact internal subspace. More generally, one can include 
a nontrivial warp factor $e^{2A(y)}$ depending on the internal coordinates $y$
into the Ansatz \eqref{background_metric}. This amounts to replacing 
$g^{(4)}_{\mu \nu}(x)$ with $e^{2A(y)} g^{(4)}_{\mu \nu}(x)$ which is 
the most general Ansatz for a Poincar\'e invariant four-dimensional metric \cite{deWitSD,Strominger1,BB1,Verlinde,GSS,GKP}.
The functional form of the warp factor is then determined by demanding the 
background Ansatz to be a solution of the supergravity theory. It becomes 
a non-trivial function in the presence of localized sources such as D-branes.
However, for simplicity we will restrict ourselves to the Ansatz \eqref{background_metric} 
in the following. 

The lower-dimensional theory is obtained by expanding all fields into modes of 
the internal manifold $\Mint$. As an illustrative example we discuss
the Kaluza-Klein reduction of a ten-dimensional scalar $\Phi(x,y)$ fulfilling
the ten-dimensional Laplace equation $\Delta_{10} \Phi = 0$ \cite{KK-review}. Using the Ansatz \eqref{background_metric}
the Laplace operator splits as $\Delta_{10}=\Delta_4 + \Delta_6$ and we may apply the
fact that $\Delta_6$ on a compact space has a discrete spectrum. The coefficients arising in 
the expansion of $\Phi(x,y)$ into eigenfunctions of $\Delta_6$ 
are fields depending only on the coordinates of $\Mext$. From a four dimensional
point of view the term $\Delta_6 \Phi$ thus appears as a mass term. 
One ends up with an infinite tower of massive states with masses
quantized in terms of $1/R$, where $R$ is the `radius' of $\Mint$ such that 
$\text{Vol}(\Mint)$ is of order $R^6$. Choosing the internal manifold to be small 
enough the massive Kaluza-Klein states become heavy and can be integrate 
out. The resulting effective theory encodes the dynamics of the four-dimensional fields
associated with the massless Kaluza-Klein modes satisfying
\beq \label{zero_modes}
   \Delta_6 \Phi(x,y)\ =\ 0\ .
\eeq
In chapter \ref{TypeII} we review how this procedure can be generalized to all other 
fields present in the ten-dimensional supergravity theories. This also includes the metric itself \cite{KK-review}. 
Equation \eqref{background_metric} specifies the ten-dimensional background metric and a gravity theory 
describes variations around this Ansatz.
In the non-compact dimensions these correspond to the four-dimensional graviton and 
the effective action reduces to the standard Einstein-Hilbert term for the metric.
The situation changes for the internal part of the metric. Massless fluctuations of $g_{mn}(y)$
around its background value, such as changes of the size and shape of $\Mint$, correspond to scalar and vector
fields in four-dimensions. 
As a result the four-dimensional 
theory consists of a huge set of scalar and vector fields arising 
as coefficients in the expansion of the ten-dimensional fields into zero modes of $\Mint$. 
In order that the four dimensional theory inherits some of the supersymmetries 
of the underlying ten-dimensional supergravity theory one restricts to background 
manifolds with structure group in $SU(3)$ such as Calabi-Yau manifolds or six-tori. 
This implies that the Kaluza-Klein modes reside in supermultiplets with dynamics encoded 
by a supergravity theory.

As already remarked above 
every compactification induces a set of massless neutral 
scalars called moduli. In Calabi-Yau compactifications
it typically consists of more then 100 scalar fields parameterizing the 
geometry of $\Mint$, which is clearly in conflict with the known particle spectrum. 
It is a long-standing problem to find a  
mechanism within String Theory to generate a potential for these fields. 
Finding such a potential will fix
their values in a vacuum and make them sufficiently massive such 
that they can be discarded from the observable spectrum. 
Above we already listed the three most popular possibilities to generate such a
potential: background fluxes, instanton corrections
and gaugino condensation. Let us now focus our attention to the first mechanism,
since fluxes will play a major role in this work.

To include background fluxes amounts to allowing for non-trivial vacuum expectation value of certain field
strengths \cite{Strominger1}--\cite{TGL2}. Take as an example a tensor field $B_2$. If its field strength 
$H_3 = dB_2$ admits a background flux $H_3^{\rm flux} = \big<dB_2 \big>$, the kinetic term of 
$B_2$ yields a contribution \cite{GKP} 
\beq \label{flux_pot}
   \int_{\cM_{10}} H_3^{\rm flux} \wedge * H_3^{\rm flux}\ ,
\eeq
which via the Hodge-$*$ couples to the metric and its deformations. Insisting on four-dimensional 
Poincar\'e invariance of the background, non-trivial fluxes can only be induced on internal three-cycles 
$\gamma$. 
The terms \eqref{flux_pot} induce a non-trivial 
potential for the deformations of the internal metric $g_{mn}(y)$ which generically 
stabilizes the corresponding moduli fields at a scale $m_{\rm flux} \sim \alpha'/R^3$ \cite{GKP,DWG}.

There are at least 
two further important points to remark. Firstly, note that in general one is not completely free to 
choose the fluxes, but rather has to obey certain consistency conditions. 
Fluxes generically induce a charge which has to be canceled on a compact space. 
Hence, the setup needs to be enriched by objects carrying a negative charge \cite{GKP}. 
Secondly, it is usually the case that fluxes do not stabilize all moduli of the theory. 
In order to induce a potential for the remaining fields, one needs to include 
non-perturbative effects such as instantons and gaugino condensates. 
Various recent work \cite{non-pert} is intended to get some deeper insight
into the nature of these corrections.


\section{Brane World Scenarios}
\label{braneworlds}

In the middle of the 90's, the discovery of the D-brane opened 
a new perspective for String Theory \cite{JP}. On the one hand, D-branes where
required to fill the conjectured web of string dualities \cite{Dual,JPbook}. 
Their appearance supports the hope for a more fundamental underlying theory 
unifying all the known String Theories. Moreover, they led to the conjecture of various 
new connections between String Theories and supersymmetric gauge theories, 
such as the celebrated AdS/CFT correspondence \cite{reviewAdSCFT}. 
From a direct phenomenological point of view, they opened a whole new arena 
for model building \cite{reviewPP}, since they come equipped with a gauge theory.

More precisely, D-branes are extended objects defined as subspaces of the ten-dimensional 
space-time on which open strings can end \cite{JP,JPbook,D-branes,AD}. Open strings with both ends on the same D-brane 
correspond to an $U(1)$ gauge field in the low energy effective action. This gauge group gets enhanced 
to $U(\cN)$ when putting a stack of $\cN$ D-branes on top of each other. At lowest order this 
induces a Yang-Mills gauge theory in the low-energy effective action. 
This fact allows to construct phenomenologically attractive
models from space-time filling D-branes consistently included in a compactification
of type II String Theory \cite{reviewPP}. The basic idea is that the Standard Model, 
or rather its supersymmetric extensions, is realized on a stack of space-time filling D-branes. 
The matter fields arise from dynamical excitations of the brane around its background configuration. 
This is similar to the situation in standard compactifications discussed in the beginning of the 
previous section, where moduli fields parameterize fluctuations of the background 
metric on $\Mint$. The crucial difference is that fluctuations of the D-branes are charged under 
the corresponding gauge group and can yield chiral fermions in topologically non-trivial 
configuration \cite{reviewPP}. 

In addition to the applications in Particle Physics, D-branes can serve as essential
ingredients to construct cosmological models. Their non-perturbative nature can be 
used to circumvent the no-go theorem excluding the possibility of de Sitter vacua 
in String Theory \cite{KKLT,BKQ,reviewcosmo}. 
Furthermore, similar to the fundamental string, D-branes are dynamical objects, which can 
move through the ten-dimensional ambient space. In certain circumstances this 
dynamical behavior was conjectured to be linked to a cosmological evolution \cite{reviewcosmo}.  

There are basically three steps to extract phenomenological data from brane world 
scenarios. Firstly, one has to actually construct consistent examples 
yielding the desired gauge groups, field content and amount of supersymmetry. 
Secondly, to determine the dynamics of the theory one needs to evaluate the 
low energy effective action of the brane excitations and the gauge neutral 
bulk moduli. This can then be combined with the approach 
to generate potentials by a flux-background and non-perturbative effects.

The resulting theory may exhibit various phenomenologically interesting features. 
As briefly discussed in section \ref{geom_red} it can yield moduli stabilization in the vacuum.
Moreover, if the vacuum breaks supersymmetry this generically results in a set of soft supersymmetry 
breaking terms for the charged matter fields on the D-branes (see ref. \cite{KL,BIM} for 
a generic string inspired supergravity analysis). 
These can be computed from the effective low energy action
as it has been carried out in refs.\ \cite{Soft1,GGJL,Soft2}.
On the other hand, anti-branes (or brane fluxes) 
can be used to generate a positive cosmological constant \cite{KKLT,BKQ}. 

Even though this general approach sounds promising, it is extremely hard to 
address all these issues at once. Hence, one is usually forced to 
either concentrate on specific models or on one or the other ingredient 
to develop techniques and to extract general results. 

\begin{figure}[h]
\begin{center}
\includegraphics[height=5cm]{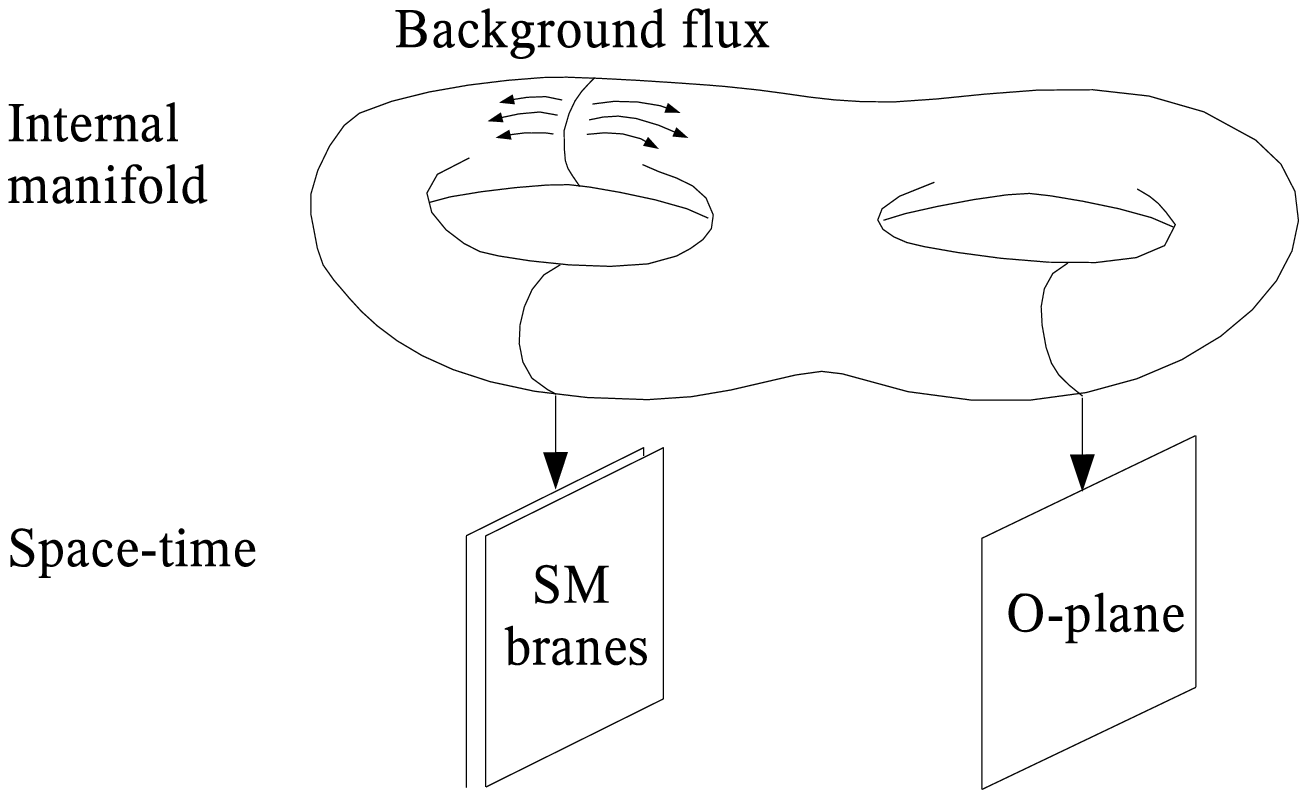}
\caption{\textit{Brane-world scenario on $\Mext \times \Mint$ 
         with space-time filling D-branes, orientifold planes and background fluxes.}}
\label{braneworld_fig}
\end{center}
\end{figure}

As an example, one can already check if space-time filling D-branes and fluxes 
alone can be consistently included in a compactification. 
Namely, since D-branes are charged under certain fields of the 
bulk supergravity theory they contribute a source term in the Bianchi identities 
of these fields \cite{JP,JPbook,D-branes,AD}. This is similarly true for non-trivial 
background fluxes. One can next apply the Gau\ss~law for the compact internal space such that 
consistency requires internal sources to cancel. In this respect D-branes 
are the higher dimensional analog of say positively charged particles. 
Putting such a particle in a compact space, the field lines have to end somewhere 
and we have to require for negative sources. In String Theory these negative sources are 
either appropriately chosen anti-D-branes or `orientifold planes' \cite{JP,AD}. 
Even though it is  possible to construct consistent 
scenarios with D-branes and anti-D-branes only, one may further insist to keep a 
$D=4$ supergravity theory. This is mainly due 
to the fact that these models are under much better control
and are not plagued by instabilities.
This favors the inclusion of appropriate orientifold planes, since there 
negative tension cancels the run-away potentials for the moduli induced by D-branes.
In figure \ref{braneworld_fig} we schematically picture some ingredients of a
brane-world model.

Orientifold planes arise in String Theories constructed 
form type II strings by modding out world-sheet parity plus a geometric 
symmetry $\sigma$ of $\Mext \times \Mint$ \cite{JP,AD}. On the level of the full String Theory 
this implies that non-orientable string world-sheets, such as the Klein bottle 
or the M\"obius strip, are allowed. 
Focusing on the effective action orientifolds break part or all of 
the supersymmetry of the low-energy theory. By imposing 
appropriate conditions on the  orientifold projection and the included D-branes 
the setup can be adjusted to preserve exactly half of the original supersymmetry. 

{}From a phenomenological point of view spontaneously broken $\cN=1$
theories are of particular interest.
Starting from type II String Theories in ten space-time dimensions, 
one can compactify on Calabi-Yau
threefolds to obtain  $\cN=2$ theories in four dimensions.
This $\cN=2$ is further broken to $\cN=1$ if in addition
background D-branes and orientifold planes are present \cite{Ori,GKP,AAHV,BBKL,BH}.
The presence of background fluxes or other
effects generating a potential results in a 
spontaneously broken  $\cN=1$ theory \cite{Bachas}--\cite{TGL2}. 
To examine this setup on the level of the effective action is one of the
motivations for this work. Note that all these brane world scenarios are conjectured
to admit a higher dimensional origin in a more fundamental theory, which we briefly introduce next.
However, it is important to keep in mind that this unifying theory is much 
less understood then the five String Theories.

\subsection{From dualities to M- and F-theory}

\begin{figure}[h]
\begin{center}
\includegraphics[height=5cm]{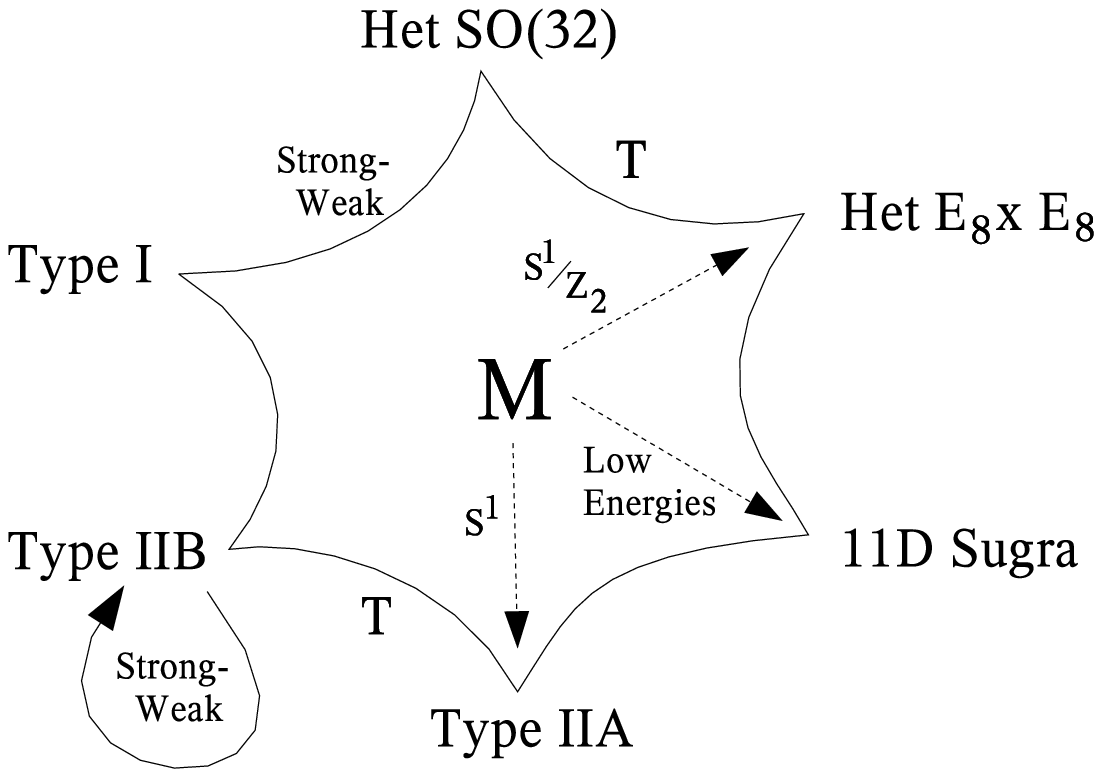}
\caption{\textit{The duality web of String Theories.}}
\label{web}
\end{center}
\end{figure}

At the first glance in seems as if we have to choose one or
the other String Theory in which we aim to construct a specific
model. However, it turns out that many of these choices are 
actually equivalent and linked by various dualities \cite{Dual,JPbook}. The full set of 
dualities forms a interlocking web between all five String
Theories (see figure \ref{web}).

As an example type IIA compactified on a circle of radius $R$
is shown to be equivalent to type IIB on a circle of radius $1/R$ \cite{JPbook,revT-dual}. 
This duality is termed T-duality and relates two String Theories 
at weak string coupling \cite{revT-dual}. There are also strong/weak dualities
such as S-duality, which is a symmetry of the type IIB String
Theory \cite{Dual}. Both of these dualities can be generalized and applied 
to standard Calabi-Yau compactifications as well as 
brane-world scenarios. 

A prominent example is mirror symmetry which can be interpreted as performing
several T-dualities \cite{SYZ}. 
It associates to each Calabi-Yau manifold $Y$ a corresponding mirror Calabi-Yau $\tilde Y$ \cite{Mirror}. 
Within the framework of String Theory it can be argued that 
type IIA compactified on $Y$ is fully equivalent to type IIB strings on 
$\tilde Y$. From a mathematical point of view mirror symmetry 
exchanges the odd cohomologies of $Y$ with the even cohomologies of $\tilde Y$
and vice versa. Even stronger it suggests that the moduli spaces of the 
two Calabi-Yau manifolds are identified. Remarkably, in specific 
examples this allows to calculate stringy corrections to the theory on $Y$
from geometrical data of $\tilde Y$. Mirror symmetry can be generalized
to setups with D-branes \cite{Aspinwall} and eventually should identify type IIA and type IIB 
brane world scenarios. This raises various non-trivial questions such 
as in which way mirror symmetry applies to flux compactifications \cite{Vafa_NCY,GLMW}.

Let us also introduce S-duality in slightly more detail \cite{Dual,JPbook}. Type 
IIB String Theory contains in addition to the fundamental string 
also a D-string (D1-brane). 
It can now be argued that the theory where the fundamental string 
is at low coupling $g_s$, and hence the D-string is very heavy, is 
dual to a theory at $1/g_s$ with the role of both strings exchanged. 
Carefully identifying the fields, S-duality is also shown to be a 
symmetry of the corresponding type IIB low-energy effective action.
This strong/weak duality is actually part of a larger symmetry group 
$Sl(2,\bbZ)$. It has been suggested in \cite{Vafa} that this duality 
group admits a geometric interpretation in terms of two additional 
toroidal dimensions. This twelve dimensional construction was named
F-theory. The additional two dimensions are necessarily a compact torus,
which however in compactifications can be non-trivially fibered over 
the compactification manifold. This naturally applies to type IIB 
brane-world scenarios, which generically admit backgrounds corresponding 
these non-trivial compactifications \cite{Sen,DRS,GKP}.

The existence of these various dualities suggests that the ten-dimensional 
String Theories are actually just different limits of a more fundamental 
theory \cite{Dual} as pictured in figure \ref{web}. 
This mysterious theory unifying all five String Theories was named 
M-theory. In general, not much is known about its actual
formulation and the required structures 
are far less understood then the one for String Theory. However,
there are certain regimes in which one believes to find some
hints of its existence. This also includes the existence 
D-branes, which fit into this picture as they occur from 
higher-dimensional objects termed M-branes. There also is  
a unique supergravity theory in eleven dimensions \cite{CJS}, 
which is interpreted to be the low-energy limit of M-theory. 
In the final chapter of this article it will be this 
low-energy theory which allows us lift the orientifold
compactifications to M-theory.

\subsection{Topics and outline of this article}

After this brief general introduction let us now turn to the actual topics
of this article. As just discussed, an essential step to extract 
phenomenological properties of string vacua with 
(spontaneously broken) $N=1$ supersymmetry in brane world 
scenarios is to determine the low energy effective 
action. In this work we focus 
on type IIA and IIB String Theory compactified on generic Calabi-Yau 
orientifolds and determine their low energy effective action in 
terms of geometrical data of the Calabi-Yau orientifold and 
the background fluxes. We include D-branes 
for consistency, but freeze their matter fields (and moduli) concentrating 
on the couplings of the bulk moduli. We also provide a detailed discussion 
of the resulting $N=1$ moduli space in the chiral and the dual linear 
multiplet description and check mirror symmetry in the large volume--large complex 
structure limit. Moreover, we show at the level of the effective actions 
that Calabi-Yau orientifolds with fluxes admit a natural embedding 
into F- and M-theory compactifications. 

In \textit{chapter \ref{TypeII}} we first briefly review standard Calabi-Yau 
compactifications of type IIA and type IIB supergravity and discuss the resulting 
$N=2$ supergravity action.  In doing so we focus 
on the geometry of the moduli space $\cM^{\rm SK} \times \cM^{\rm Q}$ spanned by 
the scalars of the $N=2$ supergravity theory. Supersymmetry constrains 
it to locally admit this product form, where $\cM^{\rm SK}$ is a
special K\"ahler manifold and $\cM^{\rm Q}$ is a quaternionic manifold.  
Furthermore, we introduce $N=2$ mirror symmetry on the level of 
the effective action and present a somewhat non-standard construction 
of the mirror map between the IIA and IIB quaternionic moduli spaces
reproducing the results of \cite{BGHL}. 

In \textit{chapter \ref{effective_actO}} we immediately turn to the 
compactification of type II theories on Calabi-Yau orientifolds. 
We start with a more detailed introduction to setups with 
D-branes and orientifold planes and comment on consistency and 
supersymmetry conditions. As already mentioned in section \ref{braneworlds} 
orientifold planes are essential ingredients to obtain supersymmetric theories in 
brane-world compactifications. They arise in String Theories modded
out by a geometrical symmetry $\sigma$ of $\Mext \times \Mint$
in addition to the world-sheet parity operation. We demand 
$\Mint$ to be a generic Calabi-Yau manifold admitting an 
isometric involutive symmetry $\sigma$. It turns out that in order to preserve
$N=1$ supersymmetry $\sigma$ has to be a holomorphic 
map in type IIB and an anti-holomorphic map in type IIA 
compactifications. Taking into account further properties of $\sigma$ 
one finds three supersymmetric setups \cite{AAHV,BH}: (1) IIB orientifolds with 
$O3/O7$ planes, (2) IIB orientifolds with $O5/O9$ planes and 
(3) IIA orientifolds with $O6$ planes. 

The spectrum of these
theories was first determined in \cite{BH}.
However, the effective action was only computed 
for special cases of type IIB Calabi-Yau orientifolds with $O3/O7$ 
planes \cite{GKP,BBHL}. In \cite{TGL1} we generalized these results and 
also included an analysis of $O5/O9$ setups. 
For type IIA brane-world scenarios the calculation of the low energy 
supergravity theory was mainly concerned with orbifolds of six-tori \cite{CP,reviewPP} for 
which conformal field theory techniques can be applied. Complementary, 
the dynamics of the bulk theory can extracted for general type IIA Calabi-Yau orientifolds by 
using a Kaluza-Klein reduction as shown in our publication \cite{TGL2}.
In chapter \ref{effective_actO} we review the first parts of refs. \cite{TGL1,TGL2} 
and determine the $N=1$ effective action of all three setups. We extract
the K\"ahler potential and the gauge-kinetic couplings by first assuming that 
no background fluxes are present. The $N=1$ moduli space is shown to 
be a local product $\tilde \cM^{\rm SK} \times \tilde \cM^{\rm Q}$, where 
$\tilde \cM^{\rm SK}$ is a special K\"ahler manifold inside $\cM^{\rm SK}$
and $\tilde \cM^{\rm Q}$ is a K\"ahler manifold inside the quaternionic 
manifold $\cM^{\rm Q}$. 

We end chapter \ref{effective_actO} with a discussion of
mirror symmetry for Calabi-Yau orientifolds and determine 
the necessary conditions
on the involutive symmetries of the mirror IIA and IIB orientifold theories. 
By specifying two types of 
special coordinates on the IIA side, we are able to identify the large complex 
structure limit of IIA orientifolds with the large volume limits of IIB orientifolds
with $O3/O7$ and $O5/O9$ planes. 

In \textit{chapter \ref{lin_geom_of_M}} we present a more detailed analysis of
the $N=1$ moduli space geometry of Calabi-Yau orientifold compactifications \cite{TGL1,TGL2}.
The special K\"ahler manifold $\tilde \cM^{\rm SK}$ inherits its geometrical 
structure directly from $N=2$, such that we focus our attention
to the K\"ahler manifold $\tilde \cM^Q$ inside the quaternionic space.  
We show that the definition of the K\"ahler coordinates as well as
certain no-scale type conditions can be more easily understood 
in terms of the `dual' formulation where some chiral multiplets of the Calabi-Yau 
orientifold are replaced by linear multiplets. A linear multiplet consists of 
a real scalar and an anti-symmetric two-tensor as bosonic fields. In the massless case 
this two-tensor is dual to a second real scalar and one is led back to the chiral description. 
In order to do set the 
stage for the orientifold analysis we first 
review $N=1$ supergravity with several linear multiplets following \cite{BGG}.
The transformation into linear multiplets corresponds to a Legendre transformation of 
the K\"ahler potentials and coordinates. In the dual picture 
the characteristic functions for type IIB orientifolds take a particularly 
simple form. Moreover, in type IIA orientifolds the Legendre transform is 
essential to make contact with the underlying $N=2$ special geometry.
As a byproduct we determine an entire new class of no-scale
K\"ahler potentials which in the chiral formulation
can only be given implicitly as the solution of some constraint equation.
These new insights will enable us to give an direct construction of the K\"ahler 
manifold $\tilde \cM^{\rm Q}$ in analogy to the moduli space of supersymmetric Lagrangian submanifolds \cite{Hitchin2}. 
Moreover, this sets the stage to generalize the reduction to orientifolds of certain
non-Calabi-Yau manifolds introduced in \cite{HitchinGCM,Gualtieri}.

In \textit{chapter \ref{fluxesAB}} we redo the Kaluza-Klein compactification by additionally allowing
for non-trivial background fluxes. For $O3/O7$ orientifolds this amounts to 
a generalization of the analysis presented in \cite{GKP,BBHL} and
confirms that the Gukov-Vafa-Witten superpotential \cite{GVW} encodes 
the potential due to background fluxes. However, we show that 
for orientifolds with $O5/O9$ planes background fluxes generically 
result in a non-trivial superpotential, $D$-terms as well as
a direct mass term for a linear multiplet.
Following this observation, supergravity theories with massive linear
multiplets coupled to vector and chiral multiplets where further analyzed 
in \cite{mass_tensors}. Surprisingly, in type IIA orientifolds with background fluxes 
the superpotential depends on all (bulk) moduli fields of the theory. 
In \cite{KachruK} an equivalent observation was made for the underlying $N=2$
theory. This suggests that all geometric moduli can be stabilized in a supersymmetric 
vacuum \cite{KachruK,TGL2}. In ref.~\cite{DeWGKT} this 
was shown to be possible at large volume and small string coupling (see also \cite{VZ}).  

The IIA superpotential is expected to receive non-perturbative corrections
from world-sheet as well as D-brane instantons. In the final section of 
chapter \ref{fluxesAB} we derive that for supersymmetric type IIA and type IIB instantons 
the respective actions are linear in the chiral coordinates and therefore can result in holomorphic
corrections to the superpotential.

In \textit{chapter \ref{M-F-embedding}} we embed type IIB and type IIA 
orientifolds into F- and M-theory compactifications.  
Orientifolds with $O3/O7$-planes can be obtained as a limit of 
F-theory compactified on elliptically fibered Calabi-Yau fourfolds \cite{Sen}.
We check this correspondence on the level of the effective action
by first compactifying M-theory on a specific Calabi-Yau fourfold and comparing
the result with the effective action of $O3/O7$ orientifolds compactified 
on a circle to $D=3$. The low energy effective action of 
M-theory compactifications on Calabi-Yau four-folds 
was determined in \cite{HL,BHS} and we use their results in
a slightly reformulated version. Moreover, it turns out 
that this duality is best understood in the dual pictures
where three-dimensional vector multiplets are kept in the 
spectrum and the K\"ahler potential is an explicit function of the 
moduli. We determine simple solutions to the 
fourfold consistency conditions for which we find perfect matching between the 
orientifold and M-theory compactifications. This correspondence can be lifted
to $D=4$ where M-theory  on an elliptically fibered Calabi-Yau fourfold descents 
to an F-theory compactification. 

We end this chapter by also discussing the embedding of type IIA
orientifolds into a specific class of $G_2$ compactifications of M-theory 
as suggested in \cite{KMcG}. 
Restricting the general results of \cite{PT,HM,Hitchin1,GPap,BW} to a specific 
$G_2$ manifold and neglecting the contributions arising from the singularities
we show agreement between the low energy effective
actions \cite{TGL2}. In \cite{TGL2} we discovered that only parts of the 
orientifold flux superpotential decent from fluxes in an M-theory 
compactifications on manifolds with $G_2$ holonomy.  
However, as we will argue one of the missing terms is
generated on $G_2$ structure manifolds with non-trivial
fibrations. However, the higher-dimensional origin of the term 
involving the mass parameter of massive type IIA supergravity 
remains mysterious.

This article is mainly based on the publications 
\cite{TGL1} and \cite{TGL2} of the author. 
However, we also present various new results. 
Namely, it turns out to be possible to reformulate the results of 
\cite{TGL1,TGL2} in a very elegant and powerful way adapted to
Hitchin's analysis of special even and odd forms on six-manifolds \cite{HitchinGCM,Hitchin1}.
This allows for a better understanding of the N=1 moduli 
space inside the quaternionic manifold and 
suggests a generalization to non-Calabi-Yau orientifolds.
Moreover, we included a detailed analysis of 
the orientifold limit of the F-theory embedding of type IIB orientifolds. 
In addition we identify the higher-dimensional origin of a second flux term of the 
IIA orientifold superpotential being due to a non-trivial fibration of a 
$G_2$ structure manifold.

%
%

\chapter{Calabi-Yau compactifications of Type II theories}
\label{TypeII}

In this section we review compactifications of type IIA and type IIB 
supergravity on a Calabi-Yau manifold $Y$. These lead to $N=2$ supergravity
theories in four dimensions expressed in terms of the characteristic data 
of the Calabi-Yau space. We start our discussion with some mathematical preliminaries.
In section \ref{CY-mfds} we introduce Calabi-Yau manifolds and give a short 
description of their moduli spaces. In a next step we turn to compactifications
of IIA and IIB supergravity on Calabi-Yau manifolds in section \ref{revIIB} and \ref{revIIA}.
Finally, in section \ref{revMirror} we give a brief account of $N=2$ mirror symmetry 
applied at the level of the effective action. The mirror map for the quaternionic
moduli spaces will be constructed. 

\section{Calabi-Yau manifolds and their moduli space}
\label{CY-mfds}

String theory is consistently formulated in a ten-dimensional target space. 
In order to reduce to a four-dimensional observable world, we choose 
the background to be of the form $\cM_{10}=\Mext \times \Mint$ as 
already given in \eqref{product_Ansatz}.
Here $\Mint$ is a compact six-dimensional manifold, which, in principle, we are 
free to choose. Due to this Ansatz, the Lorentz group of $\cM_{10}$ decomposes 
as $SO(9,1)\rightarrow SO(3,1)\times SO(6)$, where $SO(6)$ is the generic
structure group of a sixfold. 
However, demanding $\Mint$ to preserve the minimal amount of supersymmetry
one has to pick a manifold with structure group $SU(3)$. 
They admit one globally defined spinor $\eta$, since the $SO(6)$ spinor 
representation $\bf{4}$ decomposes to $\bf{1}\oplus\bf{3}$.
Further demanding this spinor $\eta$ to be covariantly constant 
reduces the class of background manifolds to manifolds with 
$SU(3)$ holonomy \cite{GSWbook}. These spaces are called Calabi-Yau manifolds and   
are complex K\"ahler manifolds, which are in addition Ricci flat \cite{Huebsch}.

In terms of $\eta$ one can globally define a covariantly constant
two-from $J$ (the K\"ahler form) and a three-form $\Omega$ (the holomorphic three-form). 
For a fixed complex structure these 
fulfill the algebraic conditions
\beq
  J \wedge J \wedge J\ \propto\ \Omega \wedge \bar \Omega\ , \qquad J \wedge \Omega = 0\ . 
\eeq
where the proportionality factor depends on the normalization of $\Omega$ with respect to $J$.
Performing a Kaluza-Klein reduction on the background \eqref{product_Ansatz} 
the massless four-dimensional fields arise as the zero modes of 
the internal Laplacian \eqref{zero_modes} \cite{GSWbook,JPbook}. These zero modes are in one-to-one correspondence 
with harmonic forms on $Y$ and thus their multiplicity is counted by the dimension
of the non-trivial cohomologies of the Calabi-Yau manifold. The Calabi-Yau condition poses 
strong constraints on the Hodge decomposition of the cohomology groups.
The only non-vanishing cohomology groups are the even and odd cohomologies 
\bea \label{odd_even_cohom}
  H^{ev}& =& H^{(0,0)} \oplus H^{(1,1)} \oplus  H^{(2,2)} \oplus H^{(3,3)}\ , \\   
  H^{odd}& =& H^{(3,0)} \oplus H^{(2,1)} \oplus  H^{(1,2)} \oplus H^{(0,3)}\ . \nn
\eea  
Their dimensions $h^{(p,q)}=\dim H^{(p,q)}$ can be summarized in the Hodge diamond 
as follows
\beq \label{hodge_diamond}
  \begin{array}{c}
   h^{(0,0)}\\
   h^{(1,0)} \qquad h^{(0,1)}\\
    h^{(2,0)}\qquad h^{(1,1)} \qquad h^{(0,2)}\\
   h^{(3,0)}\qquad h^{(2,1)}\qquad h^{(1,2)} \qquad h^{(0,3)}\\
   h^{(3,1)}\qquad h^{(2,2)} \qquad h^{(1,3)}\\
   h^{(3,2)} \qquad h^{(2,3)}\\
   h^{(3,3)}
  \end{array}\ =\
  \begin{array}{c}
   1\\
   0 \qquad 0\\
   0\qquad h^{(1,1)} \qquad 0\\
   1\qquad h^{(2,1)}\qquad h^{(2,1)} \qquad 1\\
   0\qquad h^{(1,1)} \qquad 0\\
   0 \qquad 0\\
   1
  \end{array}\ .
\eeq
Let us introduce a basis for the different cohomology 
groups by always choosing the unique harmonic representative in 
each cohomology class. The basis of harmonic $(1,1)$-forms
we denote by $\omega_A$ with dual harmonic $(2,2)$-forms
$\tilde \omega^A $ which form a basis of $H^{(2,2)}(Y)$.
$(\alpha_{\hat K}, \beta^{\hat L})$ are harmonic three-forms
and form a real, symplectic basis on $H^{(3)}(Y)$. Together 
the non-trivial intersection numbers are summarized as
\beq \label{int-numbers1}
  \int_Y \omega_A \wedge \tilde \omega^B = \delta_A^B\ , \qquad 
  \int_Y \alpha_{\hat K} \wedge \beta^{\hat L} = \delta^{\hat L}_{\hat K}\ ,
\eeq
with all other intersections vanishing. Finally, we denote by $\vol(Y)$ the 
harmonic volume $(3,3)$-form of the Calabi-Yau space.  
In Table \ref{CYbasis} we summarize 
the non-trivial cohomology groups on $Y$ and denote their basis elements. 

\begin{table}[h]
\begin{center}
\begin{tabular}{| c | c | c |} \hline
   \rule[-0.3cm]{0cm}{0.8cm} cohomology group&
   dimension & basis
   \\ \hline
   \rule[-0.3cm]{0cm}{0.8cm} $H^{(1,1)}$  &
   $h^{(1,1)}$ & $\omega_A$
   \\ \hline
   \rule[-0.3cm]{0cm}{0.8cm} $H^{(2,2)}$  & $h^{(1,1)}$ & $\tilde \omega^A$
   \\ \hline
   \rule[-0.3cm]{0cm}{0.8cm} $H^{(3)}$  & $2h^{(2,1)}+2$ &
   $(\alpha_{\hat K},\beta^{\hat L})$ \\ \hline
\rule[-0.3cm]{0cm}{0.8cm} $H^{(2,1)}$  &
   $h^{(2,1)}$ &
   $\chi_K$
   \\ \hline
   \rule[-0.3cm]{0cm}{0.8cm} $H^{(3,3)}$  &
   $1$&
   $\vol$
   \\ \hline
\end{tabular}
\caption{\small \label{CYbasis}
\textit{Cohomology groups on $Y$ and their basis elements.}}
\end{center}
\end{table}

In sections \ref{revIIA} and  \ref{revIIB} we explain how these harmonics 
yield four-dimensional massless fields, when expanding the ten-dimensional 
supergravity forms. Furthermore, there are additional massless modes arising as
deformations of the metric $g_{i\jb}$.  
Considering variations $R_{mn}(g+\delta g)$ of the Ricci-tensor which 
respect the Ricci-flatness condition $R_{mn}=0$ forces $\delta g$ to satisfy a 
differential equation (the Lichnerowicz equation). 
Solutions to this equation can be identified in case of a Calabi-Yau manifold 
with the harmonic $(1,1)$- and $(2,1)$-forms, which parameterize K\"ahler structure and 
complex structure deformations of $Y$ \cite{Tian, CdO, Huebsch}.  The deformations of the 
K\"ahler form $J = i {g}_{i\bj}\, dy^i \wedge d\bar y^\jb$ 
give rise to $h^{(1,1)}$ real scalars $v^{A}$ and one expands \footnote{%
Globally only those deformations are allowed which keep the volume 
of $Y$ as well as its two- and four-cycles positive, i.e.\
$\int_Y J \wedge J \wedge J \ge 0$, $\int_{S_4} J \wedge J  \ge 0$ and $\int_{S_2} J \ge 0$.
These conditions are preserved under positive rescalings of the fields $v^A$, such that 
they span a $h^{(1,1)}-$dimensional cone.}
\beq\label{def-v}
  g_{i\bj} + \delta g_{i\bj} = -i\, J_{i\bj}  = -i\, v^{A}\, (\omega_{A})_{i\bj}  \ , \qquad A = 1, \ldots, h^{(1,1)}\ .
\eeq
These real deformations are complexified by the $h^{(1,1)}$ real scalars $b^A(x)$ 
arising in the expansion of the B-field present in both type II string theories. 
More precisely we introduce the complex fields
\beq \label{def-t_II}
t^A = b^A + i\, v^A\ ,
\eeq
which parameterize the $h^{(1,1)}-$dimensional complexified K\"ahler cone \cite{CdO}.

The second set of deformations are variations of the complex structure
of $Y$. They are parameterized by complex scalar fields $z^{K}$
and are in one-to-one correspondence with harmonic
$(1,2)$-forms 
\beq\label{cs}
  \delta{g}_{ij} =  \frac{i}{||\Omega||^2}\, \bar z^{K} 
  (\bar \chi_{K})_{i\ib\bj}\,
  \Omega^{\ib\bj}{}_j \ , \quad K=1,\ldots,h^{(1,2)}\ ,
\eeq
where $\Omega$ is the holomorphic (3,0)-form,
$\bar\chi_{K}$ denotes a basis of $H^{(1,2)}$ and we abbreviate
$||\Omega||^2\equiv \frac1{3!}\Omega_{ijk}\bar\Omega^{ijk}$.

Together the complex scalars $z^K$ and $t^A$ span the geometric moduli 
space of the Calabi-Yau manifold. It is shown to be locally a product 
\beq \label{geom-mod}
   \cM^{\rm cs} \times \cM^{\rm ks}\ ,
\eeq
where both factors are special K\"ahler manifolds of complex dimension 
$h^{(2,1)}$ and $h^{(1,1)}$ respectively. To make that more precise let us first 
discuss $\cM^{\rm cs}$. Its metric $G_{K \bar L}$ is given by \cite{Tian,Strominger2,CdO}
\beq \label{chi_barchi}
  G_{K \bar L} = -\frac{\int_Y \chi_K \wedge \bar \chi_{ L}}{\int_Y \Ox \wedge \bar \Ox} \ ,
\eeq
where $\chi_K$ is related to the
variation of the three-form
$\Omega$ via Kodaira's formula
\beq \label{Kod-form}
\chi_K(z,\bar z) = \partial_{z^K} \Omega(z)+ \Omega(z)\, \partial_{z^K}\Kcs   \ .
\eeq
With the help of \eqref{Kod-form} one shows that $G_{K \bar L}$ is a K\"ahler manifold, 
since we can locally find complex coordinates $z^K$ and a function
$K(z,\bar z)$ such that 
\beq\label{csmetric}
 G_{K \bar L} = {\partial}_{z^K}\partial_{\bar z^L}\  \Kcs\ , \qquad \Kcs = -\ln\Big[ i \int_Y \Ox \wedge \bar \Ox\Big] 
      = -\ln i\Big[\bar Z^\Kh\mathcal{F}_\Kh - Z^\Kh\bar{\mathcal{F}}_\Kh \Big]
\ ,
\eeq
where the holomorphic periods  $Z^\Kh, \mathcal{F}_\Kh$ are defined as
\beq \label{pre-z}
Z^\Kh(z) = \int_Y \Omega(z) \wedge \beta^\Kh\ , \qquad 
\cF_\Kh(z) = \int_Y \Omega(z) \wedge \alpha_\Kh\ , 
\eeq
or in other words $\Omega$ enjoys the expansion
\beq\label{Omegaexp}
  \Omega(z) = Z^\Kh(z)\, \alpha_\Kh - \cF_\Kh(z)\, \beta^\Kh\ .
\eeq
The K\"ahler manifold $\cM^{\rm cs}$ is furthermore special K\"ahler,
since $\cF_\Kh$ is the first derivative with respect to $Z^\Kh$ of 
a prepotential 
$\cF = \frac{1}{2} Z^\Kh \cF_\Kh$. This implies that $G_{K \bar L}$ 
is fully encoded in the holomorphic function $\cF$.
  
Note that $\Omega$ is only defined up to complex rescalings by a holomorphic function
$e^{-h(z)}$ which via \eqref{csmetric} 
also changes the K\"ahler potential by a K\"ahler transformation 
\beq\label{crescale}
\Omega\to\Omega\, e^{-h(z)}\ , \qquad \Kcs\to\Kcs + h +\bar h\ .
\eeq
This symmetry renders one of the periods (conventionally
denoted by $Z^0$) unphysical
in that one can always choose to fix a K\"ahler gauge and set $Z^0 = 1$. 
The complex structure
deformations can thus be identified with the remaining 
$h^{(1,2)}$ periods $Z^K$ by defining the special coordinates
$z^K = {Z^K}/{Z^0}$. A more  
detailed discussion of special geometry can be found in appendix \ref{specialGeom}.

Let us next turn to the second factor in \eqref{geom-mod} spanned by the complexified 
K\"ahler deformations $t^A$.  The metric on $\cM^{\rm ks}$ is given by \cite{Strominger,CdO}
\bea \label{Kmetric} 
  G_{A B} = \frac{3}{2\KK}
  \int_{Y}\omega_A \wedge *\omega_B = -\frac{3}{2}\left( \frac{\KK_{AB}}{\KK}-
  \frac{3}{2}\frac{\KK_A \KK_B}{\KK^2} \right) = \partial_{t^a} \partial_{\bar t^B} K^{\rm ks}\ ,
\eea
where $*$ is the six-dimensional Hodge-$*$ on $Y$ and the K\"ahler potential $K^{\rm ks}$
is given by 
\beq \label{Kpot_ks}
   K^{\rm ks} = - \ln \big[\tfrac{i}{6} \cK_{ABC}(t-\bar t)^A (t-\bar t)^B (t-\bar t)^C \big] = - \ln \tfrac{4}{3} \cK\ ,
\eeq  
where $\frac16 \cK$ is the volume of the Calabi-Yau manifold.
We abbreviated the intersection numbers as follows
\bea\label{int-numbers}
  \KK_{ABC} &=& \int_{Y}\omega_A \wedge \omega_B \wedge \omega_C\ , \qquad  \qquad 
  \KK_{AB}  = \int_{Y}\omega_A \wedge \omega_B \wedge J 
= \KK_{ABC}v^C\ ,  \\
  \KK_{A}   &=& \int_{Y}\omega_A \wedge J \wedge J
=\KK_{ABC}v^Bv^C \ , \qquad 
  \KK = \int_{Y}J \wedge J \wedge J
 =\KK_{ABC}v^Av^Bv^C \ ,\nonumber
\eea
with $J = v^A \omega_A $ being the K\"ahler form of $Y$. 
The manifold $\cM^{\rm ks}$ is once again special K\"ahler, since 
$K^{\rm ks}$ given in \eqref{Kpot_ks} can be derived from a single holomorphic 
function $f(t)=-\frac{1}{6} \cK_{ABC} t^A t^B t^C$ via \eqref{Kinz}.

\section{Type IIA on Calabi-Yau manifolds}
\label{revIIA}

Let us now apply these tools in Calabi-Yau compactifications of type IIA supergravity
following \cite{BCF,FS}. 
This theory is the maximally supersymmetric theory in ten spacetime dimensions,
which posses two gravitinos of opposite chirality. It is naturally obtained as the low energy limit of 
type IIA superstring theory. Thus the supergravity spectrum consists of the massless string modes. 
The bosonic fields are the dilaton $\hat \phi$, the ten-dimensional metric $\hat g$ and the two-form 
$\hat B_2$ in the NS-NS sector, 
while the one- and three-forms $\hat C_1,\hat C_3$ arise in 
the R-R sector.\footnote{We use a `hat'
to denote ten-dimensional quantities and omit it for 
four-dimensional fields.} Using form notation (our conventions are summarized in appendix~\ref{conventions})
the corresponding ten-dimensional type IIA supergravity action in the 
Einstein frame is given by \cite{JPbook}
\bea \label{10dact}
  S^{(10)}_{IIA} &=& \int -\tfrac{1}{2}\hat R*\mathbf{1} -\tfrac{1}{4} d\hat \phi\wedge * d\hat \phi
  -\tfrac{1}{4} e^{-\hat \phi}\hat H_3 \wedge *\hat H_3 
  -\tfrac{1}{2} e^{\frac{3}{2} \hat \phi}\hat F_2 \wedge *\hat F_2 \nn \\
  && -\tfrac{1}{2} e^{\frac{1}{2} \hat \phi}\hat F_4 \wedge *\hat F_4 
  -\tfrac{1}{2} \hat B_2 \wedge \hat F_4 \wedge \hat F_4\ ,
\eea
where the field strengths are defined as 
\bea \label{defHFF1}
  \hat H_3 = d \hat B_2\ , \quad \hat F_2 = d\hat C_1\ , \quad 
  \hat F_4 = d\hat C_3 - \hat C_1 \wedge \hat H_3\ .
\eea
In order to dimensionally reduce type IIA to a four-dimensional  
theory, we make the product Ansatz $\Mext \times \Mint$ and perform a Kaluza-Klein 
reduction. Since $Y$ is a Calabi-Yau manifold it posses one covariantly constant spinor $\eta$. 
Decomposing the two ten-dimensional gravitinos into $\eta$ times some four-dimensional spinor
leads to two gravitinos in $D=4$. Hence, compactifying type IIA supergravity on a Calabi-Yau
threefold $Y$ results in an $N=2$ theory in four space-time dimensions and the zero modes of 
$Y$ have to assemble into  massless  $N=2$ multiplets.
These zero modes are in one-to-one correspondence with harmonic forms
on $Y$ and thus their multiplicity is counted by the dimension
of the non-trivial cohomologies of the Calabi-Yau manifold.
For the dimensional reduction 
one chooses a block diagonal Kaluza-Klein Ansatz for the 
ten-dimensional background metric
\beq \label{lineel}
  ds^2\ = \ \eta_{\mu \nu}(x)\, dx^\mu dx^\nu + g_{i \jb}(y)\, dy^i dy^\jb\ ,
\eeq
where $\eta_{\mu \nu},\mu,\nu=0,\ldots,3$ 
is a four-dimensional Minkowski metric and 
$g_{i \jb},i,\jb=1 \ldots 3$ is a Calabi-Yau metric. Part of the four dimensional fields 
arise as variations around this background metric. They correspond to the four-dimensional 
graviton and the geometric deformations $v^A(x)$ and $z^K(x)$ defined in \eqref{def-v} and \eqref{cs}.  
Variations of the off-diagonal entries of this metric vanish due to the fact that $Y$ does 
not admit harmonic one-forms. Accordingly we expand 
the ten-dimensional gauge-potentials introduced in \eqref{defHFF1} in 
terms of harmonic 
forms on $Y$
\bea \label{fieldexp}
  \hat C_1 &=& A^0(x)\ ,\qquad 
\hat B_2\, = \, B_2(x) +  b^A(x) \, \omega_A\ ,\quad  
       A\ =\ 1,\ldots, h^{(1,1)}\ ,\\
  \hat C_3 &=& 
A^A(x) \wedge \omega_A + \,
              \xi^\Kh(x)\, \alpha_\Kh - \tilde \xi_\Kh(x)\, \beta^\Kh\ , \quad \Kh\ =\ 0,\ldots, h^{(2,1)}\ . \nn 
\eea  
Here $b^A,\xi^\Kh,\tilde \xi_\Kh$ are four-dimensional scalars, 
$A^0,A^A$ are one-forms and  $B_2$ is a two-form. 
The ten-dimensional one-form $\hat C_1$
only contains a four-dimensional one-form $A^0$ in the expansion
\eqref{fieldexp} since a Calabi-Yau 
threefold has no harmonic one-forms.

The geometric deformations $v^A,z^K$ together with the fields defined in 
the expansions \eqref{fieldexp} assemble into a gravity multiplet $(g_{\mu\nu},A^0)$,
$h^{(1,1)}$ vector multiplets $(A^A, v^A, b^A)$,
$h^{(2,1)}$ 
hypermultiplets $(z^K,\xi^K,\tilde \xi_K)$ and one tensor multiplet 
$(B_2,\phi,\xi^0,\tilde \xi_0)$ where we only give the bosonic 
components.
Dualizing the two-form $B_2$ to a scalar $a$ results in one
further hypermultiplet. We summarize the bosonic spectrum in 
table~\ref{tab-compIIAspec}.

\begin{table}[h]
\begin{center}
\begin{tabular}{| c | c | c |} \hline
   \rule[-0.3cm]{0cm}{0.8cm} gravity multiplet  &
   $1$ & {\small $(g_{\mu \nu},A^0)$} 
   \\ \hline
   \rule[-0.3cm]{0cm}{0.8cm} vector multiplets &
   $h^{(1,1)}$ & {\small $(A^{A}, v^A,b^A)$}\\ \hline
   \rule[-0.3cm]{0cm}{0.8cm} hypermultiplets  &
   $h^{(2,1)}$ & 
   {\small $(z^K,\xi^K,\tilde \xi_K)$}\\ \hline
\rule[-0.3cm]{0cm}{0.8cm} tensor multiplet  &
   1 &
   {\small $(B_2,\phi,\xi^0,\tilde \xi_0)$} \\ \hline

\end{tabular}
\caption{\small \label{tab-compIIAspec}
\textit{ $N=2$ multiplets for Type IIA supergravity compactified on a Calabi-Yau manifold.}}
\end{center}
\end{table}
In order to display the low energy  effective action in the standard 
$N=2$ form one needs to redefine the field variables slightly.
One combines the real scalars $v^A, b^A$ into  complex fields 
$t^A$ as done in \eqref{def-t_II} and defines a four-dimensional 
dilaton $D$ according to
\beq \label{4d-dilaton}
   e^{D} = e^{\phi} (\cK/6)^{-\frac{1}{2}}\ ,
\eeq
where $\cK$ is defined in \eqref{int-numbers}. Note that $v^A$, and hence the volume $\cK/6 = \text{Vol}_S(Y)$, 
are evaluated in string frame. In this frame the ten-dimensional 
Einstein-Hilbert term takes the form $\int \frac{1}{2} e^{-2\hat \phi} R * \mathbf{1}$
and $J=v^A \omega_A$ is obtained from the internal part of this string frame metric. 
The kinetic term for the ten-dimensional Einstein frame metric reads 
$\int \frac{1}{2} R * \mathbf{1}$ and hence $J$ is related to $J_E$ in the 
Einstein frame via $J = e^{\phi/2}J_E$.
Inserting the field expansions \eqref{fieldexp} into \eqref{defHFF1}, \eqref{10dact},
reducing the Riemann scalar $R$ by including the complex and K\"ahler 
deformations and performing a Weyl rescaling to the standard Einstein-Hilbert term,
one ends up with the four-dimensional $N=2$ effective action  
\cite{N=2review,BCF,FS}
\bea \label{IIA-4}
  S^{(4)}_{\rm IIA} & = &\int -\tfrac12 R * \mathbf{1} 
                     +  \tfrac12 \I \cN_{\Ah \Bh}\, F^{\Ah} \wedge *
                     F^{\Bh}
                   +  \tfrac12 \R \cN_{\Ah \Bh}\, F^{\Ah} \wedge  F^{\Bh} \\
                    & & - G_{A B}\, dt^A \wedge *  d\bar t^B 
                        - h_{uv}\, d\tilde q^u \wedge * d\tilde q^v \ , \nn
\eea
where $F^{\Ah} = dA^{\Ah}$. The couplings of the vector multiplets in the action
\eqref{IIA-4} are encoded by the metric $G_{A B}$ and the complex matrix 
$\cN_{\Ah \Bh}$. $G_{A B}$ only depends on the moduli $t^A$ (or rather
their imaginary parts) and is defined in \eqref{Kmetric} and \eqref{Kpot_ks}. 
The gauge-kinetic coupling matrix $\cN_{\Ah \Bh}$ also depends on 
the scalars $t^A$ and is given explicitly in \eqref{def-cN}.
It can be calculated from the same holomorphic prepotential like $G_{AB}$ as 
explained in appendix \ref{specialGeom}.

Next let us turn to the couplings of the hypermultiplet sector which are encoded in the
quaternionic metric $h_{uv}$. From the Kaluza-Klein reduction one obtains \cite{FS}
\bea \label{q-metr}
  h_{uv}\, d\tilde q^u\,  d\tilde q^v &=&  (dD)^2 + G_{K \bar L}\, dz^K d\bar z^L
                               +\tfrac{1}{4}e^{4D}\big(da -(\tilde \xi_\Kh d\xi^\Kh - \xi^\Kh d\tilde \xi_\Kh) \big)^2 \\
                              && - \tfrac{1}{2} e^{2D} (\text{Im}\; \cM)^{-1\ \Kh \Lh}
                                   \big(d\tilde \xi_\Kh - \cM_{\Kh \Nh} d\xi^\Nh \big)
                                   \big(d\tilde \xi_\Lh - \bar
                                   \cM_{\Lh \Mh} d\xi^\Mh \big)\ ,\nn 
\eea
where $G_{K \bar L}$ is the metric on the space of complex structure deformations given in 
\eqref{chi_barchi} and \eqref{csmetric}. The complex coupling matrix $\cM_{\Kh \Lh}$ 
appearing in \eqref{q-metr} depends on the complex structure deformations $z^K$ and is defined as
\cite{CDAF}
\bea \label{defM}
  \int \alpha_\Kh \wedge * \alpha_\Lh&=&-(\text{Im}\; \cM +(\text{Re}\; \cM)
  (\text{Im}\; \cM)^{-1}(\text{Re}\; \cM))_{\Kh \Lh}\ , \nn\\
  \int \beta^\Kh \wedge * \beta^\Lh &=&-(\text{Im}\; \cM)^{-1\ \Kh \Lh}\ ,  \\
  \int \alpha_\Kh\wedge * \beta^\Lh &=& 
  -((\text{Re}\; \cM)(\text{Im}\; \cM)^{-1})_{\Kh}^\Lh\ .  \nn
\eea
It can be calculated from the periods \eqref{pre-z} by using equation \eqref{gauge-c}.
Thus also in the hypermultiplet sector all couplings are determined
by a holomorphic prepotential and such metrics have been called dual or special
quaternionic \cite{CFGi,FS}.

As we have just reviewed the $N=2$ moduli space 
has the local product structure 
\beq \label{N=2modsp}
  \cM^{\rm SK} \times \cM^{\rm Q}\ ,
\eeq
where $\cM^{\rm SK}=\cM^{\rm ks}$ is the special K\"ahler manifold spanned
by the scalars in the vector multiplets or in other words
the (complexified) deformations of the Calabi-Yau K\"ahler form
and $\cM^{\rm Q}$ is a dual quaternionic manifold spanned by
the scalars in the hypermultiplets. 
$\cM^{\rm Q}$ has a special K\"ahler submanifold spanned by the 
complex structure deformations $\cM^{\rm cs}$. 

This ends our short review of Calabi-Yau compactifications of type IIA
supergravity. There is a second $N=2$ supersymmetric theory in 
ten dimensions which is the low energy effective theory of type IIB
string theory. Reviewing the Calabi-Yau reduction of this theory will be 
the task of the next section.

\section{Type IIB on Calabi-Yau manifolds}
\label{revIIB}

Now we turn to the review of type IIB compactifications 
on Calabi-Yau spaces \cite{BGHL}.
Type IIB supergravity is maximal supersymmetric in ten dimensions
and possesses two gravitinos of the same chirality. It 
consists of the same NS-NS fields as type IIA: the scalar  
dilaton $\hat \phi$, the metric $\hat g$ and a two-form $\hat B_2$. 
In the R-R sector type IIB consists of even forms, 
the axion $\hat C_0$, a two-form $\hat C_2$ and a 
four-form  $\hat C_4$.
The low energy effective action in the $D=10$ 
Einstein frame is given by 
\cite{JPbook} 
\begin{eqnarray}\label{10d-lagr}
  S^{(10)}_{IIB}&=&
  \int -\tfrac{1}{2} \hat R * \mathbf{1} - \tfrac{1}{4} d\hat \phi\wedge *d \hat \phi
  -\tfrac{1}{4} e^{-\hat \phi} \hat H_3 \wedge* \hat H_3   \\
  &&- \tfrac{1}{4} e^{2\hat \phi} \hat F_1 \wedge * \hat F_1 -
  \tfrac{1}{4} e^{\hat \phi} \hat F_3 \wedge * \hat F_3 -
  - \tfrac{1}{8}\hat F_{5} \wedge *\hat F_{5}
   -\tfrac{1}{4} \hat C_4 \wedge \hat H_3 \wedge \hat F_3\ ,
\nonumber  
\end{eqnarray}
with the field strengths defined as
\begin{eqnarray}
  \hat H_3 \ =\ d \hat B_2\ , \quad \hat F_1 = d\hat C_0\ ,\quad 
  \hat F_{q+1}\ =\ d \hat C_q - \hat C_{q-2} \wedge \hat H_3\ ,\quad q=2,4\ . \label{fieldstr}
\end{eqnarray}
The self-duality condition $\hat F_5=*\hat F_5$ is
imposed at the level of the equations of motion.

As in the type IIA compactifications discussed in the previous section 
we use the Ansatz \eqref{lineel} 
for the ten-dimensional background metric. Fluctuations around this background 
metric are parameterized by the four-dimensional graviton $g_{\mu \nu}$ and 
the geometric deformations of the Calabi-Yau metric. More precisely, we find 
$h^{(1,1)}$ real K\"ahler structure deformations $v^A$ 
introduced in \eqref{def-v} and $h^{(2,1)}$ complex structure deformations $z^K$ 
introduced in \eqref{cs}.
The type IIB gauge potentials appearing in the Lagrangian 
\eqref{10d-lagr} are similarly 
expanded in terms of harmonic forms on $Y$ according to 
\begin{eqnarray}\label{CYexpansion}
  \hat B_2 &=& B_2(x) + b^A(x)\, \omega_A\ , \qquad
\hat C_2\ =\ C_2(x) + c^A (x)\,\omega_A\ , \quad A=1,\ldots,h^{(1,1)}\ , \\
  \hat C_4 &=& D_2^A(x) \wedge \omega_A + V^{\hat K}(x) \wedge
               \alpha_{\hat K} - U_{\hat K}(x) \wedge \beta^{\hat K} +
               \rho_A(x)\, \tilde \omega^A\ , 
\quad \hat K=0,\ldots,h^{(1,2)}.\nonumber
  \label{full-exp}
\end{eqnarray}
The four-dimensional fields appearing in the expansion \eqref{CYexpansion}
are the scalars $b^A(x)$, $c^A(x)$ and $\rho_A(x)$, 
the one-forms $V^{\hat K}(x)$ and
$U_{\hat K}(x)$ and the two-forms $B_2(x),C_2(x)$ and $D_2^A(x)$.
The self-duality condition of $\hat F_5$ eliminates half of the 
degrees of freedom in $\hat C_4$ and in this section we choose to eliminate
$D^A_2$ and $U_{\hat K}$ in favor of $\rho_A$ and $V^{\hat K}$.
Finally, the two type IIB scalars $\hat \phi, \hat C_0$ also appear as
scalars in $D=4$ and therefore we drop the hats henceforth
and denote them by $\phi, C_0$.

In summary the massless $D=4$ spectrum consists of 
the gravity multiplet with bosonic components $(g_{\mu \nu}, V^0)$,
$h^{(2,1)}$ vector multiplets with bosonic components $(V^{K}, z^{K})$,
$h^{(1,1)}$ hypermultiplets with bosonic components
$(v^A, b^A, c^A, \rho_A)$ 
and one double-tensor multiplet \cite{BVT} with bosonic components
$(B_2, C_2, \phi, C_0)$ which can be dualized to an additional (universal) 
hypermultiplet. The four-dimensional spectrum is
summarized in Table \ref{tab-compIIBspec}.

\begin{table}[h]
\begin{center}
\begin{tabular}{| c | c | c |} \hline
   \rule[-0.3cm]{0cm}{0.8cm} gravity multiplet  &
   $1$ & {\small $(g_{\mu \nu},V^0)$} 
   \\ \hline
   \rule[-0.3cm]{0cm}{0.8cm} vector multiplets &
   $h^{(2,1)}$ & {\small $(V^{K}, z^{K})$}\\ \hline
   \rule[-0.3cm]{0cm}{0.8cm} hypermultiplets  &
   $h^{(1,1)}$ &
   {\small $(v^A, b^A, c^A, \rho_A)$
}\\ \hline
\rule[-0.3cm]{0cm}{0.8cm} double-tensor multiplet  &
   1 &
   {\small $(B_2, C_2,\phi,C_0)$ 
}\\ \hline

\end{tabular}
\caption{\small \label{tab-compIIBspec}
\textit{ $N=2$ multiplets for Type IIB supergravity compactified on a Calabi-Yau manifold.}}
\end{center}
\end{table}

The $N=2$ low energy effective action is computed by inserting
\eqref{fieldstr} and \eqref{CYexpansion} into the action \eqref{10d-lagr}
and integrating over the Calabi-Yau manifold.
For the details we refer the reader to the literature 
\cite{BGHL,Michelson,DallAgata,LM} and only recall the results here.
One finds 
\begin{eqnarray}\label{N=2}
  S_{IIB}^{(4)} &=& \int - \tfrac{1}{2} R *\! {\bf 1} 
+ \tfrac{1}{4} \RE\cM_{\hat K \hat L} {F}^{\hat K} \wedge {F}^{\hat L} + 
  \tfrac{1}{4} \IM_{\hat K\hat L} {F}^{\hat K} \wedge * {F}^{\hat L}\nonumber\\
&&- G_{K L} d z^K \wedge * d \bar{z}^{L} 
  - G_{AB} d t^A \wedge * d \bar t^B
  -  d D \wedge * d D -  \tfrac{1}{24} e^{2 D} \cK \dd l \wedge * \dd l \nonumber\\
&& 
  - \tfrac{1}{6} e^{2D} \cK G_{AB}
  \big(\dd c^A - l \dd b^A \big)\wedge * \big( \dd c^B - l \dd b^B
  \big)\\
  && - \tfrac{3}{8\cK} e^{2D} G^{AD} \big( \dd \rho_A - 
    \cK_{ABC} c^B \dd b^C \big) \wedge\! *\big( \dd \rho_D -
    \cK_{DEF} c^E \dd b^F \big) \nonumber \\
&& -\tfrac{1}{4}e^{-4D} \dd B_2 \wedge * \dd B_2 - \tfrac{1}{24}  
   e^{-2D} \cK 
  \big( \dd C_2 - l \dd B_2 \big) \wedge *\big( \dd C_2 - l \dd B_2 \big)
  \nn  \\
  && -  \tfrac{1}{2} dC_2 \wedge \big( \rho_A \dd b^A  - b^A d\rho_A \big)
     +\tfrac{1}{2} dB_2 \wedge c^A d \rho_A - \tfrac{1}{4}\cK_{ABC} c^A c^B dB_2 \wedge \dd b^C \nonumber \ ,
\end{eqnarray}
where $F^{\hat K}=dV^{\hat K}$. 
The gauge kinetic matrix $\cM_{\hat K\hat L}$ is related to the metric
on $H^3(Y)$ and given in \eqref{defM}. The metric $G_{K L}(z,\bar z)$ which appears in \eqref{N=2}
is the metric on the space of complex structure deformations given in \eqref{csmetric}.
It is a special K\"ahler metric in that it is entirely determined by the holomorphic prepotential $\mathcal{F}(z)$ 
\cite{Strominger2,CdO}. On the other hand, the metric $G_{AB}$ in \eqref{N=2} is the metric 
on the space of K\"ahler deformations defined in \eqref{Kmetric}.

In order to entirely express \eqref{N=2}  in terms
of vector- and hypermultiplets we dualize the 
$D=4$ two-forms $B_2,C_2$ to scalar fields. This can be done, since $B_2$ and $C_2$ are massless 
and posses the gauge symmetries $C_2  \rightarrow C_2 + d\Lambda_1$ and $B_2 \rightarrow B_2 + d\tilde \Lambda_1$. 
Let us first dualize $C_2$.
We replace $dC_2$ with $D_3$ and add the Lagrange multiplier $\frac12 h\, dD_3$ such 
that the differentiation with respect to $h$ yields $dD_3=0$. Locally this ensures 
that $D_3=dC_2$ for some two-form $C_2$.  The terms in \eqref{N=2} involving $D_3$ are simply
\beq
  \cL_{C_2} = - \tfrac{g}{4}   
   \big( D_3 - C_0\, \dd B_2 \big) \wedge * \big( D_3 - C_0\, \dd B_2 \big)
    -  \tfrac{1}{4} D_3 \wedge J_1 +\tfrac12 D_3 \wedge dh\ ,
\eeq
where we abbreviated $g = \frac{1}{6} e^{-2D} \cK$ and $J_1 = \rho_A \dd b^A  - b^A d\rho_A$.
Now we can consistently eliminate $D_3$ in favor of $h$ by its equation of motion. The 
dualized Lagrangian takes the form
\beq \label{dual_h}
  \cL_{h} = - \tfrac{1}{4 g} \big(dh - \tfrac12 J_1 \big) \wedge *
                                      \big(dh - \tfrac12 J_1 \big)
            + \tfrac12 C_0\, dB_2 \wedge \big(dh - \tfrac12 J_1 \big)\ . 
\eeq  
Similarly we can dualize the two-from $B_2$ into a scalar $\tilde h$. Having replaced 
$C_2,B_2$ by $h,\tilde h$ in \eqref{N=2} the effective action can be written in the standard 
$N=2$ form \cite{BaggerW,dWvP,N=2review}
\begin{eqnarray}
  S_{IIB}^{(4)} & = & \int -\tfrac{1}{2}R *{\mathbf 1} 
+ \tfrac{1}{4} \RE\cM_{\Kh\Lh} {F}^\Kh \wedge {F}^\Lh + \tfrac{1}{4} \I
  \cM_{\Kh\Lh} {F}^\Kh \wedge * {F}^\Lh\nonumber\\
&&\qquad - G_{KL} \dd z^K \wedge *\dd \bar{z}^{L} 
- h_{pq}\, \dd \tilde q^{p} \wedge * \dd \tilde q^{q} \ ,
  \label{action3}
\end{eqnarray}
where  $q^{p}$ denote the coordinates for all
$h^{(1,1)}+1$ hypermultiplets. The metric $h_{pq}$ is a quaternionic
metric  explicitly given by \cite{FS}
\begin{align} \label{q-metrB}
  h_{pq}\, d\tilde q^p\,  d\tilde q^q &=   (d D)^2 +  G_{AB} d t^A d\bar t^B
                                            + \tfrac{1}{24} e^{2 D} \cK (\dd C_0)^2 \nonumber\\
& + \tfrac{1}{6} e^{2D} \cK G_{AB} \big(\dd c^A - C_0\, \dd b^A \big)\big( \dd c^B - C_0\, \dd b^B \big) \\
  & + \tfrac{3}{8\cK} e^{2D} G^{AD} \big( \dd \rho_A - 
    \cK_{ABC} c^B \dd b^C \big) \big( \dd \rho_D -
    \cK_{DEF} c^E \dd b^F \big) \nonumber \\ 
   & + \tfrac{3}{2 \cK} e^{2D}\big(dh - \tfrac12 (\rho_A \dd b^A  - b^A d\rho_A) \big)^2 \nn\\
  & +  \tfrac12 e^{4D} \big(d\tilde h + C_0\, dh + c^A d\rho_A  +  \tfrac12 C_0\, (\rho_A \dd b^A  - b^A d\rho_A) 
                            - \tfrac{1}{4}\cK_{ABC} c^A c^B \dd b^C \big)^2  . \nn
\end{align}
In summary the scalar moduli space $\cM^{\rm SK}  \times  \cM^{\rm Q}$ of the $N=2$ theory is
the product of a quaternionic manifold $\cM^{\rm Q}$ spanned by the scalars
$q^{p}$ with metric \eqref{q-metrB} and a special
K\"ahler manifold $\cM^{\rm SK} = \cM^{\rm cs}$ spanned by the scalars $z^K$.
The complexified K\"ahler structure deformations span a special K\"ahler manifold $\cM^{\rm ks}$
inside $\cM^{\rm Q}$. In \cite{FS} it was shown that the quaternionic space 
can be constructed from the prepotential of $\cM^{\rm ks}$ such that 
$\cM^{\rm Q}$ is a special quaternionic manifold.
  
This ends our brief summary of type IIB compactified on
Calabi-Yau threefolds and its $N=2$ low energy effective action. 
We have seen that the effective actions of the type IIA and type
IIB indeed take the standard $N=2$ form. In both cases the metrics 
on the special K\"ahler and special quaternionic manifold are encoded by 
a corresponding prepotential. However, the role of the K\"ahler and complex 
structure deformations is exchanged in type IIA and type IIB compactifications.
As we will discuss momentarily this can be traced back to an underlying symmetry  
which finally enables us to identify both effective theories in the 
large volume -- large complex structure limit.

\section{N=2 Mirror symmetry}
\label{revMirror}

In this section we briefly discuss mirror symmetry for Calabi-Yau 
compactifications \cite{Mirror}. From a mathematical point of view, mirror 
symmetry is a duality in the moduli space of Calabi-Yau manifolds. 
It states that for a given Calabi-Yau manifold $Y$, there exists
a mirror Calabi-Yau $\tilde Y$ such that 
\beq \label{Hod_id}
  h^{(1,1)}(Y) = h^{(2,1)}(\tilde Y)\ , \qquad  h^{(2,1)}(Y) = h^{(1,1)}(\tilde Y)\ .
\eeq   
Applied to the Hodge diamond \eqref{hodge_diamond} this amounts to a reflection along the
diagonal. In other words, mirror symmetry identifies the odd and even cohomologies \eqref{odd_even_cohom}
of two topological distinct Calabi-Yau spaces
\beq \label{cohom_id}
  H^{ev}(Y)\  \cong \ H^{odd}(\tilde Y)\ ,\qquad H^{odd}( Y)\  \cong \ H^{ev}(\tilde Y)\ .
\eeq 
Moreover, it is much stronger than that, since
it also implies an identification of the moduli spaces of deformations of $Y$ and $\tilde Y$. 
As given in \eqref{geom-mod} the geometric moduli space of a Calabi-Yau manifold 
is a local product of two special K\"ahler manifolds $\cM^{\rm ks}$ and $\cM^{\rm cs}$.
Their complex dimensions are exactly given by $h^{(1,1)}$ and $h^{(2,1)}$. Motivated 
by \eqref{Hod_id} mirror symmetry conjectures the identifications
\beq \label{Modspace_id}
  \cM^{\rm ks}(Y)\  \equiv\  \cM^{\rm cs}(\tilde Y)\ , \qquad 
  \cM^{\rm cs}(Y)\ \equiv\ \cM^{\rm ks}(\tilde Y)\ ,
\eeq   
as special K\"ahler manifolds. Recall that the geometry of $\cM^{\rm cs}(Y)$ 
and $\cM^{\rm cs}(\tilde Y)$ are encoded in the variations of the holomorphic
three-forms $\Omega$ and $\tilde \Omega$ of the two Calabi-Yau manifolds 
$Y$ and $\tilde Y$. These can be expanded in a real symplectic basis of $H^{3}(Y)$ and 
$H^{3}(\tilde Y)$ respectively
\beq
  \Omega(z) = Z^\Kh \alpha_\Kh - \cF_\Kh \beta^\Kh\ , \qquad 
  \tilde \Omega(\tilde z) =\tilde Z^\Ah \alpha_\Ah - \tilde \cF_\Ah \beta^\Ah\ ,
\eeq
Under the large volume mirror map 
the coordinates on the two manifolds $\cM^{\rm ks}(Y)$ and $\cM^{\rm cs}(\tilde Y)$ as 
well as $\cM^{\rm cs}(Y)$ and $\cM^{\rm ks}(\tilde Y)$ 
are identified as 
\beq \label{mirror-map}
  t^A = {\tilde Z^A(\tilde z)}/{\tilde Z^0(\tilde z)} \ , \qquad {Z^K(z)}/{Z^0(z)} = \tilde t^K
\eeq
where $t^A$ and $\tilde t^K$ are the complexified K\"ahler deformations of $Y$ and $\tilde Y$. 
Equation \eqref{mirror-map} implies that $t^A,\tilde t^K$ are identified with special coordinates 
on $\cM^{\rm cs}$. Furthermore, recall that due to the special K\"ahler property the metric on both 
moduli spaces is encoded by a prepotential.  Applying \eqref{Modspace_id} it follows that 
these prepotentials $f_Y(t)$ and $f_{\tilde Y}(\tilde z)$ as well as 
$f_Y(z)$ and $f_{\tilde Y}(\tilde t)$ are identified under mirror symmetry. 
One immediately notices, that this can not be the full truth, since $\cM^{\rm ks}$ and $\cM^{\rm cs}$ have 
a different structure. $\cM^{\rm ks}$ is a cone and admits the
simple prepotential $f(t) = - \frac16 \cK_{ABC} t^A t^B t^C$, while the metric 
on $\cM^{\rm cs}$ is determined in terms of the (generically complicated)
periods of the holomorphic three-form. Hence, one expects corrections to
$f_Y(t)$ and $f_{\tilde Y}(\tilde t)$. These corrections get a physical interpretation
as soon as mirror symmetry is embedded in string theory.
They are due to strings wrapping two-cycles in $Y$ called world-sheet instantons.
Schematically one identifies 
\beq
   f_{Y}(t) = t^3 + \mathcal{O}(e^{-t})=f_{\tilde Y}(\tilde z)\ ,\qquad f_Y(z) 
            = \tilde t^3 + \mathcal{O}(e^{-\tilde t})=f_{\tilde Y}(\tilde t)\ .
\eeq
One can also turn the argument around and use mirror symmetry as a very powerful tool to calculate the world-sheet 
instanton corrections $\mathcal{O}(e^{-t})$ as done in the pioneering paper \cite{CdOGP}.
In most cases this is much simpler then a direct calculation of the world-sheet instanton
contributions.

The most prominent applications of mirror symmetry in string theory
is the identification of type IIA string theory compactified on $Y$ with 
type IIB string theory compactified on $\tilde Y$. It matches 
the full string theories including their low energy limits
and supersymmetric D-brane states. With the material presented in this chapter we can check
it on the level of the effective action by comparing \eqref{IIA-4} with 
\eqref{action3}. This amounts to matching the moduli spaces of the 
corresponding four-dimensional $N=2$ theories which take the standard $N=2$ 
form \eqref{N=2modsp}. Since we already discussed the special K\"ahler part in \eqref{N=2modsp},
let us now concentrate on the quaternionic manifolds $\cM^{\rm Q}(Y)$
and $\cM^{\rm Q}(\tilde Y)$. In accordance with \eqref{cohom_id} and \eqref{mirror-map}
one identifies the basis elements $(1, \omega_K,\tilde \omega^K, \vol(Y))$ of $H^{ev}(Y)$ with 
the basis $(\alpha_\Kh,\beta^\Kh)$ of $H^{odd}(\tilde Y)$ as
\beq \label{basis_id}
  1\leftrightarrow \alpha_0\ ,\quad \omega_K \leftrightarrow \alpha_K\ ,\quad    \vol(Y) \leftrightarrow \beta^0\ ,
  \quad \tilde \omega^K \leftrightarrow \beta^K\ .
\eeq
We will work in this basis in the following.
Let us now construct the explicit map for the quaternionic coordinates by using 
a slightly non-standard argument. We intend to apply the fact, that the fields of 
the quaternionic space describe the coupling to D-branes, which are extended 
non-perturbative objects present in both type II string theories. We will discuss the 
low energy dynamics and supersymmetry conditions of these objects more carefully in section 
\ref{D-branes}. All we need for constructing the mirror map for the quaternionic spaces 
is there coupling to the R-R forms in the supergravity theory and some information 
about supersymmetric branes in type IIA and type IIB string theory. 
It will become clear in section \ref{D-branes}, that the only supersymmetric Euclidean 
D-branes wrapping a cycle in a Calabi-Yau manifold are $D2$ 
branes in Type IIA and $D(-1)$, $D1$, $D3$ and $D5$ branes in type IIB.
The Chern-Simons action describes the coupling of the brane world-volume to the forms
\beq \label{CS_coupling}
   \text{IIA:}\quad (\sum_{p\ even} \hat C_{p} \wedge e^{-\hat B_2})_3\ , \qquad 
   \text{IIB:}\quad (\sum_{p\ odd} \hat C_{p} \wedge e^{-\hat B_2})_q\ ,\ 
   q = 0,2,4,6\ , 
\eeq 
where $\hat C_p$ and $\hat B_2$ are the ten-dimensional R-R and NS-NS forms introduced in 
section \ref{revIIA} and \ref{revIIB}. By $(\ldots)_q$ we indicate that we 
only consider the $q-$form appearing in the sum of forms inside the parenthesis.
Supersymmetry implies that the Euclidean D-branes, wrap cycles which are dual 
to harmonic forms. But the only odd harmonic forms are $(\alpha_{\hat K},\beta^{\hat K})$, while the even 
harmonic forms are 
$(1,\omega_K, \tilde \omega^K,\vol(Y))$. 
Next we match the Chern-Simons couplings \eqref{CS_coupling} for IIA and IIB Euclidean
D-branes. We decompose \eqref{CS_coupling} on the respective cohomology basis elements
by using the expansions \eqref{CYexpansion} of $\hat B_2$ and the R-R forms $\hat C_0,\hat C_2,\hat C_4$ 
as well as the expansion \eqref{fieldexp} of $\hat C_3$.
Applying the identification \eqref{basis_id} of the basis forms we find for the coefficients
of $\alpha_\Kh$ and $(1,\omega_K)$ that     
\beq
  \xi^0 = C_0\ ,\qquad \xi^K = c^K - C_0\ b^K\ . 
\eeq
Identifying the coefficients of $\beta^\Kh$ and $(\tilde \omega^K, \vol(Y))$
yields higher powers in $\hat B_2$ and we find \footnote{We have replaced $\int C_6$ by 
$h + \tfrac12 \rho_A b^A$.
This can be done since $C_6$ is dual to $C_2$, which was dualized to $h$ in \eqref{dual_h}.}
\bea
  \tilde \xi_K &=& \rho_K - \cK_{KLM} c^L b^M + \tfrac{1}{2} C_0\ \cK_{KLM} b^L b^M\ , \\
  \tilde \xi_0 &=& h - \tfrac{1}{2} \rho_K b^K + \tfrac{1}{2} \cK_{KLM} c^K b^L b^M 
                     - \tfrac{1}{6}C_0\  \cK_{KLM} b^K b^L b^M\ .\nn
\eea
It remains to identify the space-time two-forms from the 
NS-NS sectors. Since $B_2^A$ and $B^B_2$ are the only remaining two-forms in the spectrum, we 
are forced to set $B_2^A = B^B_2$. Dualized into scalars this amounts to
\beq
    a = 2\tilde h + C_0\, h + \rho_K(c^K - C_0\, b^K)  
\eeq
Thus, by using the explicit form of the Chern-Simons coupling to D-branes,
one can infer the mirror map for the coordinates on the quaternionic space.
Of course, that the established map indeed transforms $h^A_{uv}$ given in \eqref{q-metr} 
into $h^B_{uv}$ given in \eqref{q-metrB} can be checked by direct calculation as done in \cite{BGHL}.  

This ends our review section on Calabi-Yau compactifications of type IIA 
and type IIB supergravity. We now turn to their orientifold versions which 
break $N=2$ to $N=1$. The aim of the next chapter is 
to determine the characteristic data of the resulting supergravity theory.

%
%

\chapter{Effective actions of Type II Calabi-Yau orientifolds}
\label{effective_actO}

In this chapter we discuss the four-dimensional low energy effective supergravity theory 
obtained by compactifying type IIA and type IIB string theory on Calabi-Yau orientifolds.
Before entering the calculations we review some aspects of D-branes and orientifolds 
in section \ref{D-branes}. In particular, we introduce the low energy effective action
for D-branes. Later on this will allow us to comment on corrections due to wrapped 
Euclidean D-branes to the bulk supergravity theory.
As we already explained in section \ref{braneworlds} the inclusion of 
space-time filling D-branes is essential for consistency. However, we freeze their moduli 
and matter fields such that they do not appear in the low energy effective action.\footnote{%
This restriction was weakened e.g.~in \cite{GGJL,JL}, where the coupling to $D3$- and $D7$-bane moduli 
was determined by using the low energy effective action of the $D$ branes.}   
In a next step we turn to the main issue of this chapter and perform 
a Kaluza-Klein reduction by implementing the orientifold conditions and extract the 
resulting $N=1$ supergravity theory (sections \ref{oprojections} -- \ref{IIA_orientifolds}). 
Specifically we determine the K\"ahler potential and the gauge-kinetic 
coupling functions encoding the low energy effective theory. We end our analysis by checking 
mirror symmetry in the large complex structure and large volume limit in section \ref{Mirror_orientioflds}.
A derivation of a flux induced superpotential and possible gaugings will be presented in 
chapter \ref{fluxesAB}.

\section{D-branes and orientifolds}
\label{D-branes}

In this section we provide more details on D-branes and orientifolds 
as used in the construction of brane-world scenarios.
As already mentioned in section \ref{braneworlds} 
brane world scenarios are currently one of the promising approaches 
to construct phenomenologically interesting models from
string compactification \cite{reviewPP}. They consist of space-time filling 
D-branes serving as source for Abelian or non-Abelian gauge theories.
String theory implies a low energy effective action for this gauge theory
as well as the couplings to the bulk fields introduced in chapter \ref{TypeII}. 
More precisely, the gauge theory and the coupling to the NS-NS fields $\hat \phi$, $\hat g$ and $\hat B_2$ 
is captured by the Dirac-Born-Infeld action. The most simple example is provided by a single 
$Dp$-brane, which admits an $U(1)$ gauge theory on its $p+1$-dimensional world-volume. The corresponding 
bosonic part of the Dirac-Born-Infeld action reads in string frame \cite{D-branes,JPbook}
\begin{equation} \label{DBI}
   S_{\text{DBI}}^{\text{sf}}=-T_p\int_\WV d^{p+1}\xi\:e^{-\hat \phi}
         \sqrt{-\det\left(\Em^*(\hat g+\hat B_2)_{\hat \mu \hat \nu}+2\pi\alpha' \FD_{\hat \mu \hat \nu}\right)}\ ,
\end{equation}
where $T_p$ denotes the brane tension. The integral is taken over the $p+1$-dimensional 
world-volume $\WV$ of the $Dp$-brane, which is embedded in the ten-dimensional space-time 
manifold $\cM_{10}$ via the map $\Em:\WV\hookrightarrow \cM_{10}$. 
The Dirac-Born-Infeld action \eqref{DBI} contains an $U(1)$ field strength 
$\FD_{\hat \mu \hat \nu} = 2\partial_{[\hat \mu}A_{ \hat \nu]}$, which describes the 
$U(1)$ gauge theory to all orders in $\alpha' \FD$ \cite{Leigh:jq}. 
To leading order, the gauge theory reduces to an $U(1)$ gauge theory on the world-volume 
$\WV$ of the brane. The dynamics of the $Dp$-brane is encoded in the embedding map $\Em$. Fluctuations
around a given $\varphi$ are parameterized by charged scalar fields, which provide the matter
content of the low-energy effective theory.    
 
Since $Dp$-branes also carry R-R charges \cite{JP}, they couple as extended objects to appropriate
R-R forms of the bulk, namely the $p+1$-dimensional world-volume couples naturally to the R-R form $\hat C_{p+1}$. 
Moreover, generically $D$-branes contain lower dimensional $D$-brane charges, and hence interact also with 
lower degree R-R forms \cite{Douglas:1995bn}. 
All these couplings to the bulk are implemented in the Chern-Simons action 
\beq \label{CSaction}
   S_{\text{CS}}=\mu_p\int_\WV\Em^*\Big(\sum_q \hat C_{q} \wedge e^{-\hat B_2}\Big) \wedge e^{2\pi\alpha'\FD}\  ,
\eeq
where $\mu_p$ is the charge of the D-branes.
The lowest order terms in \eqref{CSaction} in the R-R fields are topological and represent the 
R-R tadpole contributions to the low energy effective action. Additionally, 
\eqref{CSaction} encodes the coupling of the gauged matter fields arising from 
perturbations of $\varphi$ to the R-R forms. 
The effective actions \eqref{DBI} and \eqref{CSaction} can be generalized to 
stacks of D-branes \cite{Myers}. This gives rise to non-Abelian gauge theories
and appropriate (intersecting) embeddings can yield Standard Model like gauge theories \cite{reviewPP}.

Generic brane world scenarios lead to non-supersymmetric low energy theories, which 
are plagued by various instabilities due to runaway potentials for the bulk moduli. 
In contrast, supersymmetric setups are under much better control and are therefore phenomenologically 
favored. 
However, the aim to preserve some supersymmetry poses strong conditions on the D-branes present 
in the setup. D-branes which preserve half of the original supersymmetries 
are called BPS branes and the corresponding supersymmetry conditions BPS conditions. 
In brane-world setups with a ten-dimensional background space-time
of the form $\Mext \times \Mint$ two types of branes will be of importance which 
preserve four dimensional Poncar\'e invariance. 
Firstly, one includes D-branes filling the space-time $\Mext$ and wrapping a cycle in 
the manifold $\Mint$. These provide the gauge theory and matter fields just discussed. 
Secondly, one might add Euclidean D-branes (called D-instantons) solely wrapping 
a cycle in $\Mint$. They induce corrections to the supergravity theory and 
their effects can be useful to stabilize bulk moduli. 
The BPS conditions for both types of 
branes demand that the brane tensions $T_p$ and charges $\mu_p$ are 
equal. This ensures stability since the net force between BPS branes 
vanishes \cite{JP}. Moreover, there are conditions 
on the cycles $\Lambda_{Dp}$ in $\Mint$ wrapped by the branes.
In \cite{BBS} it was shown that in a purely metric background with 
$\Mint$ being a Calabi-Yau manifold the only allowed cycles are 
special Lagrangian submanifolds of $\Mint$ in 
type IIA and holomorphic submanifolds in type 
IIB. More precisely special Lagrangian submanifolds are 
three-cycles $\Lambda^{(3)}$ in $Y$ for which 
\beq \label{spLagr-C}
  \vol(\Lambda^{(3)})=\tilde \varphi^*( \R \Omega)\ ,\qquad \tilde \varphi^*( \I \Omega) = 0\ , \qquad \tilde \varphi^* J = 0\ ,
\eeq
where $\vol(\Lambda^{(3)})=\det^{1/2}(\tilde \varphi^* g)\, d^3\lambda$ 
is the volume form on $\Lambda^{(3)}$, $J$ and $\Omega$ are the
K\"ahler form and holomorphic three-form of $Y$ as in chapter \ref{TypeII}
and $\tilde \varphi$ defines the embedding of the D-brane into $Y$.
On the other hand, holomorphic submanifolds are even-dimensional cycles $\Lambda^{(2)},\Lambda^{(4)}$ in 
$Y$ satisfying
\beq \label{holom-C}
  \vol(\Lambda^{(2)})=\tilde \varphi^*(J)\ ,\qquad \vol(\Lambda^{(4)})=\tfrac{1}{2}  \tilde \varphi^*(J \wedge J)\ ,
  \qquad   \varphi^*(\Omega) = 0\ .
\eeq
It can be shown that the conditions \eqref{spLagr-C} and \eqref{holom-C} ensure that such cycles 
minimizes their volume in their homology classes (see e.g. \cite{BBS}).

These conditions have to be adjusted as soon as one allows a non-trivial 
background of supergravity forms \cite{MMMS,CU}. As an example, the 
BPS conditions on the volume of the cycles in the presence of a non-trivial $\hat B_2$ field are given by \cite{MMMS}
\bea  \label{calcond}
   \text{IIA:} \qquad \vol_{DBI}(\Lambda^{(3)}_{Dp}) &=& e^{-i\theta_{Dp}}\ \tilde \varphi^* \big( \Omega \big)\ ,\\
   \text{IIB:} \qquad  \vol_{DBI}(\Lambda^{(q)}_{Dp}) &=& 
                       e^{-i\theta_{Dp}}\  \tilde\varphi^* \big(e^{-\hat B_2 + i J} \big)_q\ , \quad q=2,4,6\ , \nn
\eea
where $\vol_{DBI}(\Lambda^{(q)}_{Dp}) = \det^{1/2}(\tilde \varphi^*[g + \hat B_2])\, d^q\lambda$ is the Dirac-Born-Infeld 
volume form on $\Lambda^{(q)}_{Dp}$.
$e^{i\theta_{Dp}}$ denotes a constant phase which will be determined below.
The BPS conditions involving the volume elements split into real and imaginary 
parts, where the imaginary part has to vanish on $\Lambda^{(q)}_{Dp}$ by using reality 
of $\vol_{DBI}(\Lambda^{(q)}_{Dp})$. The cycles $\Lambda^{(q)}_{Dp}$ satisfying the conditions
\eqref{calcond} are called calibrated with respect to the form $e^{-i\theta_{Dp}}\,\Omega$ in type IIA
and calibrated with respect to $e^{-i\theta_{Dp}}\, e^{-\hat B_2 + i J}$ in type IIB.
In a setup with several D-branes some supersymmetry is preserved as 
soon as all D-branes are calibrated with respect to the same form.
However, as we already explained in section \ref{braneworlds} this 
is not the end of the story, since consisted supersymmetric theories 
have to include negative tension objects such as orientifold planes \cite{GKP}. 

Similar to D-branes, orientifold planes are hyper-planes 
of the ten-dimensional background. They arise in string theories 
which contain non-orientable world-sheets. Orientifold theories can 
be constructed by starting from a closed string theory such as type 
IIA or type IIB strings and dividing out a symmetry group \cite{AD,Ori} \footnote{%
As usual, dividing out a symmetry can be understood as a gauge fixing.} 
\beq \label{osym}
  G \cup S\Omega_p,
\eeq
where $G$ is a group of target space symmetries and $\Omega_p$ is the 
world-sheet parity, exchanging left and right movers. $S$ 
contains operations, which render $S\Omega_p$ to be a 
symmetry of the string theory. For orientifolds \eqref{osym} consists 
of evidently perturbative symmetries of the string theory, which can be imposed 
order by order in perturbation theory and are believed to be unbroken also 
non-perturbatively. Specifically this implies that the orientifold projection 
can be consistently imposed in a low energy description. 
The orientifold planes are the hyper-surfaces left invariant by $S$.
They naturally couple to the R-R forms and thus carry a charge. Moreover,
they can have negative tension.\footnote{%
Note that orientifold
planes are to lowest order non-dynamical in string theory. This is not anymore true 
to higher orders as can be inferred from their F-theory interpretation \cite{Sen}.} 
This allows to construct consisted D-brane setups with some fraction of supersymmetry preserved.
More precisely, in a background $\Mext \times \Mint$ orientifold planes 
wrap cycles in $\Mint$ arising as the fix-point set of $S$. If these
are calibrated with respect to the same form as the cycles wrapped 
by the D-branes in the setup, the brane-orientifold setup can preserves some 
supersymmetry. We will comment on these conditions later on in this chapter. 

Before we define the precise orientifold projections relevant for this work in section 
\ref{oprojections}, let us first collect some possible symmetry 
operations allowed in $S$. In the simplest example $S$ only consists 
of a target space symmetry $\sigma:\cM_{10}\rightarrow \cM_{10}$, 
such that $\Omega_p \sigma $ is a symmetry of the 
underlying string theory. This will be the case for IIB orientifolds with $O5$
or $O9$ planes. However, type IIB admits a second perturbative symmetry operation 
denoted by $(-1)^{F_L}$, where $F_L$ is the space-time fermion number in the left-moving 
sector.
Under the action of $(-1)^{F_L}$ R-NS and R-R states are odd 
while NS-R and NS-NS states are even. Orientifolds with $O3$ and/or $O7$ planes 
arise from projections of the form $(-1)^{F_L} \Omega_p \sigma$ as we will argue 
below. In summary let us display the transformation behavior of the massless bosonic 
states under these two operations \cite{JPbook,AD}
\beq \label{transf-AB} 
\begin{array}{cllll}  
  \Omega_p:&  \qquad \text{even:}\quad  &\hat \phi, \  \hat g,\ \hat C_1,\   \hat C_2 ,  
              \qquad& \text{odd:}\quad &\hat C_0, \ \hat B_2 ,\ \hat C_3, \ \hat C_4 \ ,\\
  (-1)^{F_L}:& \qquad \text{even:}\quad  &\hat \phi, \  \hat g,\  \hat B_2 ,  
              \qquad& \text{odd:}\quad &\hat C_0,\ \hat C_1,\ \hat C_2 ,\ \hat C_3, \ \hat C_4 \ , 
\end{array}
\eeq
where we have also displayed the transformation properties of the type IIA forms. 
With these transformations at hand one  easily checks that $\Omega_p$ as well as
$(-1)^{F_L}$ are symmetries of the ten-dimensional type IIB supergravity action.
This is in contrast to type IIA. By using \eqref{transf-AB} one immediately notices that
$\Omega_p$, \,$(-1)^{F_L}$ and $(-1)^{F_L}\Omega_p$ alone are no symmetries 
of the type IIA effective action \eqref{10dact}. However, orientifolds with 
$O6$ planes arise if $S$ includes $(-1)^{F_L}\Omega_p$ as well as some appropriatly 
chosen target space symmetry which ensures that $S\Omega_p$ leaves \eqref{10dact} invariant.  
Let us now make this more explicite by properly defining the Calabi-Yau orientifold 
projections.

\section{Orientifold projections} \label{oprojections}

After this brief introduction we are now in the position to specify the orientifolds 
under consideration and give an explicit definition 
of the orientifold symmetry group \eqref{osym}. 
We start from type II string theory and compactify 
on a Calabi-Yau threefold $Y$. In addition we 
mod out by orientation reversal of the string 
world-sheet $\Omega_p$ together with an  `internal'
symmetry $\sigma$ which acts solely on $Y$ 
but leaves the $D=4$ Minkowskian space-time untouched.
We will restrict ourselves to involutive symmetries ($\sigma^2 = 1$) of $Y$ 
and thus set $G$ in \eqref{osym} to be empty.\footnote{Calabi-Yau manifolds have only 
discrete isometries. For example in the case of the quintic, 
$\sigma$ could act 
by permuting the coordinates such that
the defining equation is left invariant. 
A classification of all possible involutions
of the quintic can be found in ref.\ \cite{BH}.} 
This avoids the appearance of further twisted 
sectors as they appear in general orbifold models \cite{DHVW}. In 
a next step we have to specify additional properties of $\sigma$ and the complete 
operation $S\Omega_p$ in order that it provides a symmetry of the 
string theory under consideration. To do that we discuss the type IIA and type IIB
case in turn.

\subsubsection{Type IIB orientifolds}

Let us start with type IIB Calabi-Yau orientifolds and define the orientifold projections following 
\cite{Sen,DP,AAHV,BH}. 
Later on, in section \ref{IIB_orientifolds} we show that gauge-fixing these symmetries indeed 
leads to an $N=1$ supergravity theory.  
In type IIB consistency requires 
$\sigma$ to be an isometric and holomorphic involution of $Y$ \cite{AAHV,BH}.
A holomorphic isometry leaves both the metric
and the complex structure of the Calabi-Yau manifold invariant. 
As a consequence also the K\"ahler form $J$
is invariant such that 
\beq
  \sigma^* J = J \ , 
\eeq
where $\sigma^*$ denotes the pull-back of the map $\sigma$.
Hence in our analysis we focus on the class of Calabi-Yau threefolds which 
admit such an involution but within this class we leave the
threefolds arbitrary. Since the involution is holomorphic 
it respects the Hodge decomposition \eqref{odd_even_cohom} and we find 
in particular $\sigma^* H^{(3,0)} = H^{(3,0)}$. Picking the
holomorphic three-form $\Omega$ as an representative of $H^{(3,0)}$
and using that $(\sigma^*)^2 =\text{id}$ one is left with two 
possible actions 
\beq \label{Omegatransf}
  (1)\quad O3/O7: \quad \sigma^* \Omega = - \Omega\ ,\qquad \qquad (2)\quad O5/O9: \quad  \sigma^* \Omega = + \Omega\ . 
\eeq
Correspondingly, depending on the transformation properties of  $\Omega$
two different symmetry operations $\mathcal{O}=S\Omega_p$
are possible \cite{Sen,DP,AAHV,BH} \footnote{%
The factor $(-1)^{F_L}$ is included in $\mathcal{O}_{(1)}$
to ensure that $\OO_{(1)}^2=1$ on all states.}
\beq \label{o3-projection}
\mathcal{O}_{(1)} = (-1)^{F_L} \Omega_p \, \sigma\ ,\qquad
\mathcal{O}_{(2)} = \Omega_p \, \sigma
\eeq 
where $\Omega_p$ is the world-sheet parity, $F_L$ is the space-time fermion number
in the left-moving sector introduced at the end of section \ref{D-branes}. 
This specifies the operation $S \Omega_p$ in \eqref{osym} and,
since $G$ is empty, the complete orientifold projection. We are now in the position to 
check if the orientifold projections are indeed a symmetry of the bosonic ten-dimensional 
type IIB supergravity action \eqref{10d-lagr}. We will do this check by concentrating only 
on some of the terms in \eqref{10d-lagr} keeping in mind that the analysis for the 
remaining terms is analoge. The background 
$\cM'=\Mext \times \sigma(Y)$ denotes the image of $\cM=\Mext\times Y$ under the geometric action 
$\sigma$. Also inserting the $\sigma$-transformed fields into \eqref{10d-lagr} one infers  
\footnote{Here we have used \eqref{wedge*comp}
in order to give the component expression of the kinetic terms in \eqref{10d-lagr}.}
\beq \label{tranfact}
 S^{(10)}_{IIB'}\ =\
  \int_{\cM'}\big( -\tfrac{1}{2} \hat R_{g'} *{'} \mathbf{1} 
                - \tfrac{1}{4}g'^{MN} (\partial_M \hat \phi')(\partial_N \hat \phi') *{'} \mathbf{1} - \ldots
     -\tfrac{1}{4} \hat C_4' \wedge \hat H_3' \wedge \hat F_3'\big)\ ,
\eeq
where $g'=\sigma^* g,\ \hat \phi'=\sigma^*\hat \phi$ etc.\ and the dots denote 
terms transforming similar to the kinetic term of $\phi'$. The Hodge star $*'$ is evaluated on
the manifold $\cM'$ with metric $g'$. Now we apply the properties of the involution. Since 
$\sigma$ is an isometry we find $g=g'$ and due to the holomorphicity of $\sigma$ we can deduce 
that the ten-dimensional volume element $*'\mathbf{1}$ does not change sign in going from $\cM'$ to $\cM$.\footnote{%
Holomorphic maps do not change the orientation of $M$.} This ensures that the 
Einstein-Hilbert term takes the from $\int_{\cM'} \sigma^* (- \frac{1}{2} R * \mathbf{1})$ and by applying 
\eqref{int-form1} and \eqref{int-form2} is invariant under the isometric map $\sigma$.
A similar reasoning applies to all other terms in \eqref{tranfact} and one concludes
that the effective action \eqref{10d-lagr} is indeed unchanged by $\sigma$. 
Combined with the invariance of \eqref{10d-lagr} under the world-sheet parity $\Omega_p$ and $(-1)^{F_L}$
one infers that the orientifold operations \eqref{o3-projection} are symmetries of the effective theory.

The fix-point set of the involutions $\sigma$ in \eqref{o3-projection} determines 
the location of the orientifold planes.
Modding out by $\mathcal{O}_{(1)}$
leads to the possibility of having $O3$- and $O7$-planes
while modding out by $\mathcal{O}_{(2)}$ allows 
$O5$- and $O9$-planes. To see this, recall that 
the four-dimensional Minkowski space is left invariant by
$\sigma$ such that the orientifold planes are necessarily space-time filling.
Using the fact that $\sigma$ is holomorphic they 
have to be even-dimensional (including the time direction) which 
selects $O3$-, $O5$-, $O7$- or $O9$-planes as the only possibilities.
The actual
dimensionality of the orientifold plane is then determined 
by the dimensionality of the fix-point set of $\sigma$ in $Y$.
In order to determine this dimensionality we need the induced
action of $\sigma$ on the tangent space at any point 
of the orientifold plane. 
Since one can always choose $\Omega \propto dy^1 \wedge dy^2 \wedge dy^3$
we see that for $\sigma^* \Omega  =  - \Omega$ 
the internal part of the orientifold plane  is 
either a point or a  surface of complex dimension two.
Together with the space-time filling part we thus can have 
$O3$- and/or $O7$-planes.
The same argument can be repeated  for $\sigma^* \Omega  =  \Omega$ 
which then leads to the possibility of 
$O5$- or $O9$-planes. There are no models with $O5$ and $O9$ planes, since
the appearance of a $O9$ plane implies that the complete background $\cM_{10}$
consist of fix-points of $\sigma=\text{id}$.
The case of $O9$ planes is special and coincides with type I if one 
introduces the appropriate number of $D9$-branes to cancel tadpoles.

Since the involution $\sigma$ is holomorphic the fix-point set of the involution
are holomorphic cycles $\Lambda_{Op}$. This implies that they are calibrated 
with respect to the forms $1$ and $J \wedge J$ in orientifolds with $O3/O7$ planes 
and with respect to $J$ or $J\wedge J\wedge J$ in orientifolds with $O5$ or $O9$ planes. 
More precicely, one finds that the volume forms on $\Lambda_{Op}$ equals the pull-back 
of $e^{iJ}$ to the cycle \footnote{%
Here we abbreviate the formal sum of $(q,q)$-forms
$ e^{iJ} = 1 + iJ + \frac{1}{2!}J \wedge J - \frac{i}{3!} J \wedge J\wedge J$.} 
\beq \label{cal_sOp}
 \vol(\Lambda_{Op}) = e^{-i\theta_{Op}}\, e^{iJ} \big|_{\Lambda_{Op}}\ , \qquad 
 \theta_{O3/7} = 0\ ,\quad \theta_{O5}=\tfrac{\pi}{2}\ ,\quad \theta_{O9}=-\tfrac{\pi}{2}\ ,
\eeq
where the phase depends on the type of orientifold planes in the setup. Furthermore
one has $\Omega|_{\Lambda_{Op}}=0$. Cycles fulfilling these conditions minimize their volume within
their homology class. Note that similar to \eqref{calcond} this condition has to be modified 
in the presence of a $\hat B_2$ field. In this case the form which 
calibrates the supersymmetric cycles is $e^{-\hat B_2+iJ}$. 
Let us check whether the fix-point sets $\Lambda_{Op}$ of $\sigma$ 
remain calibrated. 
In the two orientifold setups only fields are kept in
the spectrum which are invariant under the respective 
projection $\mathcal{O}_{(1/2)}$ given in \eqref{o3-projection}. Thus, by using \eqref{transf-AB} 
one infers that $\hat B_2$ has to transform
as $\sigma^* \hat B_2 = - \hat B_2$ for both orientifold projections. 
This implies that $\hat B_2$ restricted to the fix-point set of 
$\sigma$ vanishes. \footnote{%
Denoting $\rho^* \hat B_2 = \hat B_2|_{\Lambda_{Op}}$ the pull-back to the 
fix-point set $\Lambda_{Op}$ of $\sigma$ it follows 
$-\rho^*\hat B_2 = \rho^* (\sigma^*  \hat B_2) = (\sigma \circ \rho)^* \hat B_2 = \rho^* \hat B_2$
such that $\rho^*\hat B_2=0$.}
One concludes that the cycles $\Lambda_{Op}$ remain 
calibrated with respect to the generalized calibration form, i.e.\
\beq\label{cal-Op}
  \vol_{DBI}(\Lambda_{Op}) = e^{-i\theta_{Op}} e^{-\hat B_2+iJ} \big|_{\Lambda_{Op}}\ ,
\eeq
where $\theta_{Op}$ is as given in \eqref{cal_sOp} and $\vol_{DBI}(\Lambda_{Op})$ is defined as 
in \eqref{calcond}.
At this point, one can compare the calibration condition \eqref{cal-Op} for the 
orientifold planes with the one for the $Dp$-branes given in \eqref{calcond}. 
In order to preserve some supersymmetry all orientifold planes and 
D-branes, have to be calibrated with respect to the same form. This implies that 
the phases $\theta_{Dp}$ in \eqref{calcond} have to coincide with $\theta_{Op}$ given 
in \eqref{cal_sOp} (see also \cite{JL} for the case of $D3/D7$ branes). This is 
equivalently true for $Dq$-instantons wrapping $q+1$-cycles in $Y$. In supersymmetric setups 
with $O(q+3)$ planes one has to set $\theta_{Dq}=\theta_{O(q+3)}$, where $e^{i\theta_{Dq}}$
is the phase in the D-instanton calibration condition.

\subsubsection{Type IIA orientifolds}

Let us now turn to the type IIA Calabi-Yau orientifolds. 
In contrast to type IIB the orientifold projection has 
to include an anti-holomorphic and isometric involution $\sigma$ 
in order to preserve $N=1$ supersymmetry \cite{AAHV,BBKL,BH}. 
Hence, the K\"ahler form on $Y$ transforms as 
\beq \label{constrJ}
  \sigma^* J\ =\ -J\ , 
\eeq
since $\sigma$ preserves the metric but yields a minus sign when applied to 
the complex structure.
The complete projection takes the form 
\beq \label{oproj}
  \mathcal{O} = (-1)^{F_L} \Omega_p \sigma\ .
\eeq
In addition to the condition \eqref{constrJ} 
compatibility of $\sigma$ with the Calabi-Yau condition 
$\Omega \wedge \bar \Omega \propto J \wedge J \wedge J$ 
implies that $\sigma$ also acts non-trivially  on the three-form $\Omega$ as
\beq \label{constrO}
  \sigma^* \Omega\ =\ e^{2i\theta} \bar \Omega \ , 
\eeq
where $e^{2i\theta}$ is a constant phase and we included a factor 2 for later convenience.
Similar to the type IIB case we can check that the 
projection $\mathcal{O}$ is a symmetry of the type IIA supergravity 
action \eqref{10dact}. Note however, that $(-1)^{F_L} \Omega_p$ alone is not a symmetry 
of type IIA. Using \eqref{transf-AB} this can be already inferred from the fact that the kinetic and
topological terms in \eqref{10dact} transform with a different sign. 
On the other hand, under the action of the involution $\sigma$ the effective action changes as
\beq
 S^{(10)}_{IIA'}\ =\
  \int_{\cM'}\big( -\tfrac{1}{2} \hat R_{g'} *{'} \mathbf{1} 
                 - \tfrac{1}{4}g'^{MN} (\partial_M \hat \phi')(\partial_N \hat \phi') *{'} \mathbf{1} \ldots
     -\tfrac{1}{2} \hat B_2' \wedge \hat F_4' \wedge \hat F_4' \big)\ ,
\eeq
where as in \eqref{tranfact} we have set $g'=\sigma^* g,\ \hat \phi'=\sigma^*\hat \phi$ etc.\ and the Hodge star $*'$ is on
the manifold $\cM'=\Mext \times \sigma(Y)$ with metric $g'$. Using the fact that $\sigma$ is 
an anti-holomorpic isometric involution it changes the sign of the volume element 
$*\mathbf{1} \sim \vol(\Mext) \wedge J' \wedge J' \wedge J'$, such that $*'\mathbf{1}=-*\mathbf{1}$.
From equations \eqref{int-form1} and \eqref{int-form2} one finds that the topological term transforms with a minus sign
while the kinetic terms remain invariant. This extra sign cancels the minus from the action of
$(-1)^{F_L} \Omega_p$ such that $\mathcal{O}$ is indeed a symmetry of \eqref{10dact}.
In section \ref{IIA_orientifolds} we show that gauge-fixing this symmetry results in an $N=1$ supergravity
theory.

Type IIA orientifolds with anti-holomorphic involution generically contain $O6$ planes. This 
is due to the fact, that the fixed point set of $\sigma$ in $Y$ are  three-cycles 
$\Lambda_{O6}$ supporting the internal part of the orientifold planes. 
These cycles are special Lagrangian submanifolds of $Y$ as an 
immediate consequences of \eqref{constrJ} and \eqref{constrO}
which implies \cite{HitchinLec}
\beq \label{OLagr}
  J|_{\Lambda_{O6}} = 0\ , \qquad  \I(e^{-i\theta}\Omega)|_{\Lambda_{O6}} = 0\ .
\eeq
In other words, they are calibrated with respect to 
$\R(e^{-i\theta}\Omega)$
\beq \label{calibr-O6}
\rm{vol}(\Lambda_{O6})\sim \R(e^{-i\theta}\Omega)\ ,
\eeq
where the overall normalization of $\Omega$ will be determined
in \eqref{Omeganorm}. Once again this poses conditions on 
additional D-branes in the setup, if they are demanded to preserve the same
supersymmetry. More precicely, BPS branes have to be calibrated 
with respect to the same form as the orientifold planes. This implies 
by comparing \eqref{calcond} with \eqref{calibr-O6} that 
$\theta_{D6}=\theta$ for space-time filling $D6$-branes wrapping a 
three-cycle in $Y$. A similar condition $\theta_{D2}=\theta$ has 
to hold for supersymmetric $D2$-instantons wrapping a three-cycle in $Y$.

%
%

\section{Type IIB Calabi-Yau orientifolds \label{IIB_orientifolds}}

In this section we impose the projection \eqref{o3-projection}
on the type IIB theory and derive the massless spectrum 
(section~\ref{o3-spectrum}) and its 
low energy $N=1, D=4$ effective supergravity action 
(section~\ref{O37_effective_act}).
This generalizes similar derivations already performed in refs.\
\cite{GKP,BBHL}.
We restrict our analysis  to the bosonic fields of the compactification
keeping in mind that the couplings of the 
fermionic partners are fixed by  supersymmetry. Furthermore, we 
include space-time filling D-branes for consistency but fix their 
moduli, such that they do not appear in the low energy effective 
action. The compactification we perform is closely related to the compactification
of type IIB string theory on Calabi-Yau threefolds reviewed
in chapter \ref{TypeII}. The orientifold projection 
\eqref{o3-projection}
truncates the massless spectrum from $N=2$ to $N=1$
multiplets and also leads to a modification of the couplings
which render the low energy effective theory
compatible with $N=1$ supergravity.
Such truncation procedures from $N=2$ to $N=1$ supergravity has been carried
out from a purely supergravity point of view  
in refs.\ \cite{ADAF}.

\subsection{The $N=1$ spectrum \label{o3-spectrum}}
Before computing the effective action let us first
determine the massless spectrum when 
the orientifold projection is taken into account and see how 
the fields assemble in $N=1$ supermultiplets \cite{BH}.
In the 
four-dimensional compactified theory
only states invariant under the projection are kept.
Using equation \eqref{transf-AB} one immediately infers that 
the scalars $\hat \phi,\hat l$, the metric $\hat g$ and the
four-form $\hat C_4$ are even under $(-1)^{F_L} \Omega_p$
while both two forms $\hat B_2, \hat C_2$ are odd.
Using \eqref{o3-projection} this implies that the invariant
states have to obey
\begin{equation} \label{fieldtransfB}
\begin{array}{lcl}
\\
\sigma^*  \hat \phi &=& \  \hat \phi\ , \\
\sigma^*   \hat g &=& \ \hat g\ , \\
\sigma^*   \hat B_2 &=& -  \hat B_2\ ,
\end{array}
\hspace{1cm}
\begin{array}{lcl}
\multicolumn{3}{c}{ \underline{O3/O7}} \\[2ex]
\sigma^*  \hat C_0 &=& \ \ \hat C_0\ , \\
\sigma^*   \hat C_2 &=& - \hat C_2\ , \\
\sigma^*   \hat C_4 &=& \ \ \hat C_4\ , 
\end{array}
\hspace{1cm}
\begin{array}{lcl}
\multicolumn{3}{c}{ \underline{O5/O9}} \\[2ex]
\sigma^*   \hat C_0 &=& - \hat C_0\ , \\
\sigma^*   \hat C_2 &=& \ \ \hat C_2\ , \\
\sigma^*   \hat C_4 &=& - \hat C_4\ , 
\end{array}
\end{equation}
where the first column is identical for both involutions $\sigma$ 
in \eqref{o3-projection}. 
Since $\sigma$ is a holomorphic involution the cohomology groups $H^{(p,q)}$
(and thus the
harmonic $(p,q)$-forms) split into two eigenspaces 
under the action of $\sigma^*$ 
\beq\label{H3split}
H^{(p,q)} = 
H^{(p,q)}_+\oplus H^{(p,q)}_-\ .
\eeq
$H^{(p,q)}_+$ has dimension $h_+^{(p,q)}$ and denotes
the even eigenspace of $\sigma^*$ while
$H^{(p,q)}_-$ has  dimension $h_-^{(p,q)}$ and denotes
the odd eigenspace of $\sigma^*$. 
The Hodge $*$-operator commutes with $\sigma^*$ since $\sigma$ preserves the
orientation and the metric of the Calabi-Yau manifold and thus the Hodge
numbers obey $h^{(1,1)}_\pm=h^{(2,2)}_\pm$. Holomorphicity of $\sigma$ 
further implies $h^{(2,1)}_\pm = h^{(1,2)}_\pm$ while
\eqref{Omegatransf} leads to 
$h^{(3,0)}_+ = h^{(0,3)}_+=0, h^{(3,0)}_- = h^{(0,3)}_-=1$ for $O3/O7$ orientifolds 
and $h^{(3,0)}_+ = h^{(0,3)}_+=1, h^{(3,0)}_- = h^{(0,3)}_-=0$ for $O5/O9$ orientifolds.
Furthermore, the volume-form which is proportional
to $\Omega\wedge\bar\Omega$ is invariant under $\sigma^*$ and thus one has 
$h^{(0,0)}_+ = h^{(3,3)}_+=1, h^{(0,0)}_- = h^{(3,3)}_-=0$.
We summarize the non-trivial cohomology groups including
their basis elements in table~\ref{CYObasisB}.
\begin{table}[h]
\begin{center}
\begin{tabular}{|c || c | c || c| c || c | c |} \cline{1-7}
   setup &\multicolumn{2}{|c||}{\rule[-0.3cm]{0cm}{0.8cm} cohomology group} &
   \multicolumn{2}{|c||}{dimension} & \multicolumn{2}{|c|}{basis}
   \\ \cline{1-7}
   \multirow{3}{1.2cm}[-.3cm]{$O3/O7$ $\text{\ \ and}$ $O5/O9$}  &\rule[-0.3cm]{0cm}{0.8cm} $H^{(1,1)}_+$ & $H^{(1,1)}_-$  &
   $h^{(1,1)}_+$ & $h^{(1,1)}_- $ & $\omega_\alpha$ & $\omega_a$
   \\ \cline{2-7}
    &\rule[-0.3cm]{0cm}{0.8cm} $H^{(2,2)}_+$ & $H^{(2,2)}_-$  & $h^{(1,1)}_+$ & $h^{(1,1)}_-$ & 
   $\tilde \omega^\alpha$ & $\tilde \omega^a$
   \\ \cline{2-7}
    &\rule[-0.3cm]{0cm}{0.8cm} $H^{(2,1)}_+$  & $H^{(2,1)}_-$ 
   & $h^{(2,1)}_+$ & $h^{(2,1)}_-$ &
   $\chi_\kappa$ & $\chi_k$
   \\ \hline
   O3/O7&\rule[-0.3cm]{0cm}{0.8cm} $H^{(3)}_+$ & $H^{(3)}_-$  & $2h^{(2,1)}_+$ & $2h^{(2,1)}_-+2$ &
   $(\alpha_{\kappa},\beta^{\lambda})$ & $(\alpha_{\hat k},\beta^{\hat l})$ 
   \\ \hline 
   O5/O9&\rule[-0.3cm]{0cm}{0.8cm} $H^{(3)}_+$ & $H^{(3)}_-$  & $2h^{(2,1)}_+ +2$ & $2h^{(2,1)}_-$ &
   $(\alpha_{\kappa},\beta^{\lambda})$ & $(\alpha_{\hat k},\beta^{\hat l})$ \\ \hline
\end{tabular}
\caption{\small \label{CYObasisB}
\textit{Cohomology groups and their basis elements.}}
\end{center}
\end{table}

The four-dimensional invariant spectrum
is found by using the Kaluza-Klein expansion 
given in eqs.\ \eqref{def-v}, \eqref{cs} and \eqref{CYexpansion}
keeping only the fields which in addition obey \eqref{fieldtransfB}.
We see immediately that the $D=4$ scalar field arising from
$\hat\phi$ remains in the spectrum for both setups and as before we denote it by 
$\phi$. Since $\sigma^*$  leaves 
the K\"ahler form $J$ invariant  only the
$h_+^{(1,1)}$ even K\"ahler deformations $v^\alpha$ remain in the spectrum
and we expand 
\beq\label{transJo}
J =  v^{\alpha}\, \omega_{\alpha} \ ,\qquad
\alpha = 1,\ldots, h_+^{(1,1)}\ ,
\eeq 
where $\omega_\alpha$ denotes a basis of $H^{(1,1)}_+$.
{} Similarly, from eq.\ \eqref{cs}  we infer that the
invariance of the metric together with
\eqref{Omegatransf} implies that the complex structure deformations 
kept in the spectrum correspond to elements in $H^{(1,2)}_-$ for $O3/O7$
setups and to elements of $H^{(1,2)}_+$ for $O5/O9$. Hence, \eqref{cs} is 
replaced by
\bea\label{cso}
O3/O7:\quad \delta{g}_{ij} =  \frac{i}{||\Omega||^2}\, \bar z^{k}
(\bar \chi_{ k})_{i\ib\bj}\,
\Omega^{\ib\bj}{}_j \ ,\quad  k=1,\ldots,h_-^{(1,2)}\ , \\ \qquad 
O5/O9:\quad \delta{g}_{ij} =  \frac{i}{||\Omega||^2}\, \bar z^{\kappa}  
(\bar \chi_{\kappa})_{i\ib\bj}\,
\Omega^{\ib\bj}{}_j \ , \quad  \kappa =1,\ldots,h_+^{(1,2)}\ ,\nn
\eea
where $\bar\chi_{k}\ (\bar \chi_{\kappa})$ denotes a basis of $H^{(1,2)}_-\ (H^{(1,2)}_+)$.\footnote{%
In ref.\ \cite{BH} it is further shown that the
$h_\pm^{(1,2)}$ deformations form a smooth submanifold
of the Calabi-Yau  moduli space.}

{}From eqs.\ (\ref{fieldtransfB}) we learn that in the expansion of
$\hat B_2$ only odd elements are kept. Thus, for both orientifold setups we
have
\beq \label{exp-B}
  \hat B_2 = b^a\, \omega_a\ ,\qquad  a=1,\ldots, h_-^{(1,1)}\ ,
\eeq 
where $\omega_a$ is a basis of $H^{(1,1)}_-$. 
The orientifold projections differ in the R-R sector. For $O3/O7$ orientifolds
$\hat C_2$ is odd and $\hat C_4$ is even. Therefore the expansion \eqref{CYexpansion} 
is replaced by 
\beq\label{exp1}
  \hat C_2\ =\ c^a\, \omega_a\ , \qquad 
  \hat C_4\ =\  D_2^\alpha\wedge \omega_\alpha
+ V^{\kappa}\, \wedge \alpha_{\kappa} 
+ U_{\kappa}\wedge\beta^{\kappa}+
 \rho_\alpha\ \tilde \omega^\alpha\ ,
\eeq
where $\tilde\omega^\alpha$ is a basis
of $H^{(2,2)}_+$ which is dual to $\omega_\alpha$, and
$(\alpha_{\kappa}, \beta^{\kappa})$ is a real, symplectic
basis of $H^{(3)}_+ = H^{(1,2)}_+ \oplus H^{(2,1)}_+$ 
(c.f.\ table~\ref{CYObasisB}). From \eqref{fieldtransfB} we find that 
the axion $\hat C_0$ remains in the spectrum and we denote
the corresponding four-dimensional field by $C_0$.
Note that the two $D=4$ two-forms $B_2$ and $C_2$ present in the $N=2$
compactification (see \eqref{CYexpansion})
have been projected out and in the expansion of  $\hat B_2$ and $\hat C_2$
only the scalar fields $c^a, b^a$ appear.
The non-vanishing of $c^a,b^a$ and $V^\kappa$ is closely related to the 
appearance of $O7$-planes. To understand this in more detail
we recall, that $O3$-planes appear 
when the fix-point set of $\sigma$ is zero-dimensional in $Y$
or in other words all tangent vectors at this point are odd under
the action of $\sigma$.
This in turn implies that locally
two-forms  are even  under $\sigma^*$, while three-forms 
are odd. However, this is incompatible
with the expansions given in \eqref{exp1} for 
non-vanishing $b^a,c^a$ and $V^\kappa$. 
For a setup also including $O7$-planes we locally
get the correct transformation behavior, 
so that harmonic forms in $H^{(1,1)}_-$ and 
$H^{(2,1)}_+$ can be supported. 

For $O5/O9$ orientifolds the $\mathcal{O}_{(2)}$-invariant R-R forms transform exactly 
with the opposite sign under $\sigma$. Thus, 
the expansion \eqref{CYexpansion} reduces to 
\beq\label{expO5} 
\hat C_2\ =\ C_2+ c^\alpha\ \omega_\alpha\ , 
\qquad \hat{C}_4 \ =\ D_{2}^a \wedge \omega_a + V^{k} \wedge 
\alpha_{k} - U_{k} \wedge \beta^{k} + 
  \rho_a\, \tilde \omega^a\ .
\eeq
In this case the axion $\hat C_0$ is projected out and replaced by 
the $D=4$ antisymmetric tensor $C_2(x)$. As a consequence the $N=1$ 
spectrum contains a `universal' linear multiplet $(\phi,C_2)$ which in the massless 
case can be dualized to a chiral multiplet.
As for Calabi-Yau compactifications
imposing the self-duality on $\hat F_5$
eliminates half of the degrees of freedom in the expansions \eqref{exp1} and \eqref{expO5}
of $\hat C_4$. For the one-forms  $V^{\cdot},U_{\cdot}$ this corresponds to
the choice of electric versus magnetic gauge potentials.
On the other hand choosing the two forms $D_2^{\cdot}$ 
or the scalars $\rho_{\cdot}$ determines
the structure of the $N=1$ multiplets to be either a linear or a chiral 
multiplet and in chapter \ref{lin_geom_of_M} we discuss both cases.

Altogether the resulting $N=1$ fields for the two setups 
assembles into a gravitational
multiplet, $h_\pm^{(2,1)}$ vector multiplets and 
$(h_\mp^{(2,1)}+ h^{(1,1)}+1)$ chiral multiplets 
and are
summarized in  table~\ref{N=1spectrumtab} \cite{BH,TGL1}.

\begin{table}[h] 
\begin{center}
\begin{tabular}{|c|c|c||c|c|} \hline
 \rule[-0.3cm]{0cm}{0.8cm} & \multicolumn{2}{|c||}{$O3/O7$} & \multicolumn{2}{c|}{$O5/O9$}\\ \hline 
 \rule[-0.3cm]{0cm}{0.85cm} 
 gravity multiplet&1&$g_{\mu \nu} $ &1& $g_{\mu \nu}$ \\ \hline
 \rule[-0.3cm]{0cm}{0.85cm} 
 vector multiplets&   $\ h_+^{(2,1)}\ $&  $V^{\lambda} $& $\ h_-^{(2,1)}\ $ & $V^{k} $ \\ \hline
 \rule[-0.3cm]{0cm}{0.85cm} 
 \multirow{3}{30mm}[-3.5mm]{chiral multiplets} &   $h_-^{(2,1)}$& $z^{k} $ &   $h_+^{(2,1)}$& $z^{\lambda} $\\ \cline{2-5}
\rule[-0.3cm]{0cm}{0.85cm} 
 &  $ h^{(1,1)}_-$ &$( b^a, c^a)$ & $h^{(1,1)}_+$& $( v^\alpha, c^\alpha)$\\ \cline{2-5}
 \rule[-0.3cm]{0cm}{0.85cm} 
   & 1 & $(\phi,l)$ && \\ \hline
\rule[-0.3cm]{0cm}{0.85cm} 
\multirow{3}{44mm}[2mm]{chiral/linear multiplets } & $h^{(1,1)}_+$& $( v^\alpha, \rho_\alpha )$ & $h^{(1,1)}_-$& 
 $( b^a, \rho_a )$\\ \cline{2-5}
\rule[-0.3cm]{0cm}{0.85cm}  & && 1 & $(\phi,C_2)$ \\
\hline
\end{tabular}
\caption{\label{N=1spectrumtab} $N =1$ spectrum of Type IIB orientifold compactifications.}
\end{center}
\end{table} 

Compared to the $N=2$ spectrum of the Calabi-Yau compactification
given in table~\ref{tab-compIIBspec} we see that 
the graviphoton `left' the gravitational multiplet
while the $h^{(2,1)}$ $N=2$ vector multiplets decomposed
into $h_\pm^{(2,1)}$ $N=1$ vector multiplets plus $h_\mp^{(2,1)}$ 
chiral multiplets. Furthermore, the $h^{(1,1)}+1$ hypermultiplets
lost half of their physical degrees of freedom and are reduced 
into $h^{(1,1)}+1$ chiral multiplets. This is 
consistent with the theorem of \cite{AM,ADAF} where it was shown that 
any K\"ahler submanifold of a quaternionic manifold
can have at most half of its (real) dimension.

%
%

\subsection{The effective action \label{O37_effective_act}}

In following we derive the effective actions encoding the 
dynamics of the $N=1$ multiplets of the type IIB orientifold 
theories. However, before entering the actual computations 
a cautionary note is in order. 
In the presence of localized sources such as orientifold planes and 
D-branes as well as in the presence of non-trivial background fluxes 
the product Ansatz \eqref{lineel} for the metric is strictly speaking not anymore
suitable. This is due to the fact that the supergravity theory with source 
terms and fluxes does not have the background metric \eqref{lineel} as a solution 
\cite{BB1,Verlinde,GSS,GKP}. As deviation from the standard Calabi-Yau compactifications
a non-trivial warp factor $e^{-2A}$ has to be included into the
Ansatz for the metric \eqref{lineel} such that \cite{GKP,GP}
\beq 
ds^2=e^{2A(y)} {g_{\mu \nu}}(x) dx^{\mu} dx^{\nu}+ e^{-2A(y)} 
     {g_{i\bj}}(y) dy^i d\bar y^{\bj}\ .
\label{warpmetric}
\eeq
However, in this work we perform our analysis in the
unwarped Calabi-Yau manifold since in the large radius limit
the warp factor approaches one and the metrics 
of the two manifolds coincide  \cite{GKP,FP}. 
This in turn also implies that the metrics on the moduli space
of deformations agree and as a consequence the kinetic terms
in the low
energy effective actions are the same. The difference appears 
in the potential when some of the 
Calabi-Yau zero modes are rendered massive. However, 
the regime $e^{2A(y)} \approx 1$ should be understood as a 
special limit and it would be desirable to generalize compactifications
to warped backgrounds \eqref{warpmetric}.

Let us now turn to the derivation of the four-dimensional effective action 
by redoing the Kaluza-Klein reduction of the ten-dimensional
type IIB action given in \eqref{10d-lagr} for the truncated orientifold
spectrum. 

\subsubsection{The reduction of the $N=2$ vector sector} 

We first consider the reduction of the vector sector 
of the $N=2$ supergravity theory obtained by type IIB 
Calabi-Yau compactification. As discussed in section \ref{revIIB} the 
four-dimensional bosonic components of the vector multiplets
are $(z^K,V^K)$. The complex scalars $z^K$ parameterize the
complex structure deformations of $Y$. Under the 
orientifold projection these $N=2$ multiplets split into
chiral multiplets with bosonic components $(z^k)$ and vector
multiplets $(V^\lambda)$ for $O3/O7$ orientifolds and chiral multiplets $(z^\lambda)$
and vectors $(V^k)$ in $O5/O9$ orientifolds. Since the reduction of the vector 
sector is very similar for both the $O3/O7$ and $O5/O9$ case
we will first concentrate on the first case 
and later give a rule how to translate these results to 
$O5/O9$ orientifolds. 


Due to the split of the cohomology
$H^{(3)}= H^{(3)}_+\oplus H^{(3)}_-$ the real symplectic basis 
$(\alpha_\Kh , \beta^\Lh)$
of $H^{(3)}$ can be split into $(\alpha_{\kappa} , \beta^{\lambda})$
of $H^{(3)}_+$ and $(\alpha_{\hat k} , \beta^{\hat l})$
of $H^{(3)}_-$.  
Eqs.\ \eqref{int-numbers1} continue to hold which implies
that both basis are symplectic and obey
\begin{equation}\label{sbasiso}
 \int \alpha_{\kappa} \wedge \beta^{\lambda} 
= \delta^{\lambda}_{\kappa}\ ,
\qquad
\int \alpha_{\hat k} \wedge \beta^{\hat l} 
= \delta^{\hat l}_{\hat k}\ ,
\end{equation}
with all other intersections vanishing. Since $\hat C_4$ is even under $\sigma^*$
the expansion \eqref{exp1} led to $h^{(3)}_+ = h^{(2,1)}_+$ vectors 
$V^\kappa$. The three-form $\Omega$ is odd under $\sigma^*$ and thus has to be expanded
in a basis of $H^{(3)}_-$ according to 
\begin{equation}
\Omega(z^k) = Z^{\hat k}\alpha_{\hat k} - \mathcal{F}_{\hat k}
\beta^{\hat k}\ , 
\label{cond-1}
\end{equation} 
while the other periods $(Z^{\kappa},\mathcal{F}_{\kappa})$
vanish
\beq
 Z^{\kappa}|_{z^{\kappa}=0}= \int_Y \Omega\wedge\beta^\kappa = 0 \ , \qquad
\mathcal{F}_{\kappa}\big|_{z^{\kappa}=0} 
=  \int_Y \Omega\wedge\alpha_\kappa = 0 \ .
\eeq
As a consequence the metric on the space 
of complex structure deformations reduces to 
\beq\label{csmetrico} 
G_{ kl} = 
\frac{\partial}{\partial z^{k}}
\frac{\partial}{\partial\bar z^{l}}
\  K_{\rm cs}\ , \qquad
 K_{\rm cs} = -\ln\Big[ - i \int_Y \Ox \wedge \bar \Ox\Big] 
= -\ln i\Big[ Z^{\hat k} \bar \cF_{\hat k}    - \bar Z^{\hat k} {\mathcal{F}}_{\hat k} \Big]
\ ,
\eeq
replacing \eqref{csmetric}. The reduction of the kinetic terms for 
the $N=2$ vector sector thus yields \cite{TGL1}
\beq \label{red-vector}
S^{(4)\, vec}_{O3/O7}\ =\ \int - G_{k { l}} \; dz^{k} \wedge *d\bar z^{l}
    +\tfrac{1}{4}\text{Im}\; \cM_{\kappa \lambda}\; 
    F^{\kappa}\wedge *F^{\lambda}
     +\tfrac{1}{4}\text{Re}\; \cM_{\kappa \lambda}\;
     F^{\kappa}\wedge F^{\lambda}\ , 
\eeq
where $F^\lambda = dV^\lambda$. Recall that the vectors $V^k$ as well 
as the graviphoton are projected out by the orientifold projection \eqref{o3-projection}
and do not appear in \eqref{red-vector}. 
The coupling matrix $\cM_{\kappa \lambda}(z^k)$ in front of the remaining vectors
$V^\kappa$ is evaluated on the subspace where $z^\kappa=0$ and thus depends on 
$z^k$ only. The analysis for $O5/O9$ orientifolds is in complete anology to the
$O3/O7$ case, with the difference that the vectors $V^k$ and scalars $z^\lambda$
remain in the spectrum while $V^\lambda$ and $z^k$ is projected out. 
The equations \eqref{sbasiso} -- \eqref{red-vector} can be translated to this second case by replacing
the indices $k,l \rightarrow \kappa, \lambda$, $\kh \rightarrow \hat \kappa$ and 
$\kappa,\lambda \rightarrow k,l$. 
This is consistent with the fact that by 
\eqref{cso} the three-form $\Omega$ is in $H^{(3)}_-$ for $O3/O7$ setups and in $H^{(3)}_+$ for 
$O5/O9$ setups.
   
\subsubsection{The reduction of the $N=2$ quaternionic sector} 

Similar to the vector sector, we now perform the reduction 
of the hypermultiplet couplings \eqref{q-metrB}. One computes
the four-dimensional effective action by redoing the Kaluza-Klein reduction 
of the ten-dimensional type IIB action given in \eqref{10d-lagr} for the 
truncated orientifold spectrum. Equivalently, one can impose the 
orientifold constrains on the four-dimensional $N=2$ effective action 
\eqref{action3}. In type IIB the metric on the quaternionic manifold depends on the 
complexified K\"ahler deformations $t$ and the dilaton and is obtained from 
the intersection numbers in the even cohomologies. Hence, in order to perform the reduction 
to $N=1$ we first need to reconsider the structure of the metrics \eqref{Kmetric} and 
the intersection numbers \eqref{int-numbers} for the orientifold. 

Note that $\sigma^*J=J$ and  $\sigma^*\hat B_2=-\hat B_2$ holds for both IIB orientifold 
projections. This implies that the constraints on the space of 
K\"ahler structure deformations are the same for $O3/O7$ as well as $O5/O9$ setups.
Let us discuss them in the following.
Corresponding to the decomposition  
$H^{(1,1)}=H^{(1,1)}_+\oplus H^{(1,1)}_-$ also
the harmonic (1,1)-forms $\omega_A$
split into 
$\omega_A = (\omega_\alpha, \omega_a)$
such that 
$\omega_\alpha$ is a basis of 
$H^{(1,1)}_+$ and 
$\omega_a$ is a basis of 
$H^{(1,1)}_-$. This in turn results in a decomposition of the intersection 
numbers $\KK_{ABC}$ given in \eqref{int-numbers}.
Under the orientifold projection
only $\KK_{\alpha\beta\gamma}$ and $\KK_{\alpha bc}$ can be non-zero
while $\KK_{\alpha \beta c}= \KK_{abc} =0$ has to hold. 
Since the K\"ahler-form $J$ is invariant 
we also conclude from \eqref{int-numbers}
that $\KK_{\alpha b}=0=\KK_{a}$. To summarize,
keeping only the invariant intersection numbers results in
\begin{eqnarray}\label{constr}
  \KK_{\alpha \beta c}= \KK_{abc} =\KK_{\alpha b}=\KK_{a}=0\ ,
\end{eqnarray}
while all the other intersection numbers can be non-vanishing.\footnote{From
a supergravity point of view this 
has been also observed in refs.\ \cite{ADAF}.}
Inserting \eqref{constr} into \eqref{Kmetric} we derive
\begin{eqnarray} \label{splitmetr}
  G_{\alpha \beta}=
  -\frac{3}{2}\left( \frac{\KK_{\alpha \beta}}{\KK}-
  \frac{3}{2}\frac{\KK_\alpha \KK_\beta}{\KK^2} \right)\ , \qquad
  G_{a b}=-\frac{3}{2} \frac{\KK_{a b}}{\KK}\ , \qquad
G_{\alpha b}\ =\ G_{a \beta}\ =\ 0\ ,
\end{eqnarray}
where
\begin{equation}\label{intO3}
  \KK_{\alpha\beta}=\KK_{\alpha\beta\gamma}\; v^\gamma\ , 
\quad \ 
\KK_{ab}=\KK_{ab\gamma}\; v^\gamma\ ,\quad
\KK_{\alpha}=\KK_{\alpha \beta\gamma}\; v^\beta v^\gamma\ ,
  \quad \KK=\KK_{\alpha \beta \gamma} \; v^\alpha v^\beta v^\gamma
\ .
\end{equation}
We see that the metric $G_{AB}$ given in \eqref{Kmetric}
is block-diagonal with respect to the 
decomposition $H^{(1,1)}=H^{(1,1)}_+\oplus H^{(1,1)}_-$.
For later use let us also record the inverse metrics
\begin{eqnarray}\label{Ginvers}
  G^{\alpha \beta}
  =  -\frac{2}{3} \KK \KK^{\alpha \beta} + 2 v^\alpha v^\beta\ ,
\qquad
  G^{a b} =
   - \frac{2}{3} \KK \KK^{a b}\ ,
\end{eqnarray}
where $\KK^{\alpha \beta}$ and $\KK^{a b}$ are the inverse 
of $\KK_{\alpha \beta}$ and
$\KK_{a b}$, respectively. 

The $N=2$ hypermultiplet couplings are reduced by inserting \eqref{constr}  - \eqref{Ginvers}
and truncating to the orientifold spectrum as summarized in table \ref{N=1spectrumtab}. 
Since this the orientifold spectrum of $O3/O7$ setups differs from the one
of $O5/O9$ setups, one obtains two different effective actions. 
Together with the standard Einstein-Hilbert term and the contributions 
from the reduction of the $N=2$ vectors \eqref{red-vector} one finds after 
Weyl rescaling \cite{TGL1}
\begin{eqnarray}  \label{S_scalarO3}
S^{(4)}_{O3/O7} &=&\int -\tfrac{1}{2} R *\mathbf{1} 
  - G_{k \bar l} \; dz^{k} \wedge *d\bar z^{l}
  -G_{\alpha \beta} \; dv^\alpha \wedge *dv^\beta - G_{ab}\; db^a \wedge * db^b \nonumber \\
  && - dD \wedge * dD
    -\tfrac{1}{24}e^{2 D}\cK\, dl \wedge * dl -   \tfrac{1}{6}e^{2D} \cK
  G_{ab}\left(dc^a-l db^a \right) \wedge *\left(dc^b-l db^b \right)
  \nonumber \\
  &&-\tfrac{3}{8 \cK}e^{2D} G^{\alpha \beta}  \Big(d\rho_\alpha-
  \KK_{\alpha a b} c^a db^b \Big)
  \wedge
  *\Big(d\rho_\beta - \KK_{\beta cd} c^c db^d 
\Big) \nn \\
  &&+\tfrac{1}{4}\text{Im}\; \cM_{\kappa \lambda}\; 
    F^{\kappa}\wedge *F^{\lambda}
     +\tfrac{1}{4}\text{Re}\; \cM_{\kappa \lambda}\;
     F^{\kappa}\wedge F^{\lambda}\ ,  
\end{eqnarray}
and
\begin{eqnarray} \label{S_scalarO5} 
  S^{(4)}_{O5/O9}&=&\int -\tfrac{1}{2} R *\mathbf{1} 
  - G_{\kappa \bar \lambda} \; dz^{\kappa} \wedge *d\bar z^{\lambda}
  - G_{\alpha \beta} \; dv^\alpha \wedge *dv^\beta  
   \nn \\ 
  &&- G_{ab}\; db^a \wedge * db^b -d D \wedge * dD - \tfrac{1}{6}e^{2D} \cK G_{\alpha \beta}\; dc^\alpha \wedge * dc^\beta
   \nn \\ 
  &&- \tfrac{3}{2\KK}e^{2D}(dh+\tfrac{1}{2}(d\rho_a b^a -\rho_a db^a))
  \wedge *(dh+\tfrac{1}{2}(d\rho_a b^a - \rho_a db^a))\nonumber \\
  &&- \tfrac{3}{8 \cK}e^{2D}G^{ab}(d \rho_a - \KK_{ac\alpha} c^\alpha db^c)
  \wedge *(d \rho_b - \KK_{bd\beta} c^\beta db^d)\ . \nn\\
  &&+\tfrac{1}{4}\text{Im}\; \cM_{k l}\; 
    F^{k}\wedge *F^{l}
     +\tfrac{1}{4}\text{Re}\; \cM_{k l}\;
     F^{k}\wedge F^{l}\ ,  
\end{eqnarray} 
where we have expressed the result in a chiral basis and used the index conventions given in table \ref{CYObasisB}. In 
contrast to ref. \cite{TGL1} we have expressed the effective actions 
in terms of the string frame K\"ahler structure deformations $v^\alpha$ and
the four-dimensional dilaton 
\beq \label{def-D_B}
  e^D = e^\phi\ (\cK/6)^{-1/2}\ ,
\eeq
where $e^\phi$ is the ten-dimensional dilaton.  
This ends our computation of the orientifold bulk action. In remains to cast 
\eqref{S_scalarO3} and \eqref{S_scalarO5} into the standard $N=1$ form.

\subsection{The K\"ahler potentials and gauge-couplings}
\label{Kpo_gaugeIIB}

Our next task is to transform the actions \eqref{S_scalarO3} and \eqref{S_scalarO5} into the standard
$N=1$ supergravity form with chiral multiplets where it is 
expressed in terms of a K\"ahler potential $K$, 
a holomorphic superpotential $W$ and the holomorphic gauge-kinetic coupling 
functions $f$ as follows \cite{WB,GGRS}
\beq\label{N=1action}
  S^{(4)} = -\int \tfrac{1}{2}R * \mathbf{1} +
  K_{I \bar J} DM^I \wedge * D\bar M^{\bar J}  
  + \tfrac{1}{2}\text{Re}f_{\kappa \lambda}\ 
  F^{\kappa} \wedge * F^{\lambda}  
  + \tfrac{1}{2}\text{Im} f_{\kappa \lambda}\ 
  F^{\kappa} \wedge F^{\lambda} + V*\mathbf{1}\ ,
\eeq
where
\beq\label{N=1pot}
V=
e^K \big( K^{I\bar J} D_I W {D_{\bar J} \bar W}-3|W|^2 \big)
+\tfrac{1}{2}\, 
(\text{Re}\; f)^{-1\ \kappa\lambda} D_{\kappa} D_{\lambda}
\ .
\eeq
Here the $M^I$ collectively denote  all
complex scalars in chiral multiplets present in the theory  and 
$K_{I \bar J}$ is a K\"ahler metric satisfying
$  K_{I\bar J} = \partial_I \bar\partial_{\bar J} K(M,\bar M)$.
The scalar potential is expressed in terms of the 
K\"ahler-covariant derivative $D_I W= \partial_I W + 
(\partial_I K) W$. 

In the reduction we did not find any scalar potential, such that one immediately concludes
$W=0$ and $D_\kappa=0$. Next we need to find a complex structure on the space of
scalar fields such that the metrics computed in \eqref{S_scalarO3} and \eqref{S_scalarO5}
are manifestly K\"ahler. 

\subsubsection{The K\"ahler potential: $O3/O7$ setups}
 
As we saw in \eqref{csmetrico} the 
complex structure deformations $z^{k}$ are already good K\"ahler
coordinates with $G_{k\bar l}$ being the appropriate K\"ahler metric. 
For the remaining fields the definition of 
the K\"ahler coordinates is not so obvious. 
Guided by refs.\ \cite{HL,BBHL} we define
\beq \label{def-coordsO3}
 \fe - i\, \fa = i\tau + iG^a \omega_a - T_\alpha \tilde \omega^\alpha
\eeq
where 
\beq \label{def-A}
   \pev = \fe + i\, \feh =e^{-\phi} e^{-\hat B_2 + iJ}\ ,\qquad \fa = e^{-\hat B_2} \wedge \sum_{q=0,2,4,6} \hat C_q |_{scalar}\ , 
\eeq
are sums of even forms.
In \eqref{def-A} we have defined $\hat C_q |_{scalar}$ to be the part of $\hat C_q$ yielding scalars in $D=4$, 
e.g.~$\hat C_4|_{scalar} = \rho_\alpha\, \tilde \omega^\alpha$. Expanding all the forms in
\eqref{def-coordsO3} by using \eqref{def-A},\eqref{exp-B} and \eqref{exp1}
the coordinates take the form \cite{TGL1}
\bea \label{tau}
  \tau &=& C_0+ie^{-\phi} \ , \qquad  G^a =c^a-\tau b^a\ ,\nn \\
  T_\alpha &=&  i( \rho_\alpha - \tfrac{1}{2} \cK_{\alpha ab}c^a b^b) + \tfrac{1}{2} e^{-\phi} \cK_\alpha
               - \zeta_\alpha\ , 
\eea
where\footnote{The definition of $\zeta_\alpha$ is unique up to a constant
which does not enter into the metric. The possibility of a non-zero constant
is important for the formulation in terms of linear multiplets in 
section~\ref{IIB_lin}.}  
\beq\label{zetadef}
\KK_{\alpha} =  \KK_{\alpha\beta\gamma} v^\beta v^\gamma \ ,\qquad
\zeta_\alpha =  -\frac{i}{2(\tau-\bar \tau)}\ \KK_{\alpha b c}G^b (G- \bar G)^c\ .
\eeq
In ref.~\cite{TGL1} it was checked explicitly that in terms of these coordinates
the metric of \eqref{S_scalarO3} is K\"ahler with the K\"ahler potential \cite{TGL1}
\beq \label{kaehlerpot-O7-1}
   K = K_{\rm cs}(z,\bar z)   
   + K^{\rm Q}(\tau,T,G)\ , \qquad K_{\rm cs} = - \text{ln}\Big[-i\int_Y \Omega(z) \wedge \bar \Omega(\bar z) \Big]\ ,
\eeq
and 
\beq\label{kaehlerpot-Kk}
  K^{\rm Q} =  -\text{ln}\big[-i(\tau - \bar \tau)\big]
  - 2 \text{ln}\big[ \text{Vol}_E(\tau,T,G)\big]=-\ln\big[2 e^{-4D}\big] \ ,
\eeq
where we have used \eqref{def-D_B} in order to evaluate the last equality.
The Einstein frame volume 
$\text{Vol}_E (Y) = \frac{1}{6} e^{-\frac{3}{2} \phi} \KK_{\alpha\beta\gamma} v^\alpha v^\beta v^\gamma$
in \eqref{kaehlerpot-Kk}
should be understood 
as a function of the K\"ahler coordinates $(\tau,T,G)$ 
which enter by solving (\ref{tau}) for $e^{-\phi/2} v^\alpha$ in terms of $(\tau,T,G)$.
Unfortunately this solution cannot be given explicitly and therefore $\text{Vol}_E$ is known
only implicitly via $e^{-\phi/2} v^\alpha(\tau,T,G)$.\footnote{This is in complete analogy
to the situation encountered in compactifications
of M-theory on Calabi-Yau fourfolds studied in \cite{HL}. 
This is no coincidence and can be understood from the fact
that this theory can be lifted to F-theory on Calabi-Yau fourfolds
which in a specific limit  is related to  orientifold
compactifications of type IIB \cite{Sen}. In section \ref{F-theory}
we make this more explicit by checking this correspondence on the level of the 
effective actions.} 
In chapter \ref{lin_geom_of_M} we show that the definition of the
K\"ahler coordinates \eqref{tau} and the K\"ahler potential
\eqref{kaehlerpot-O7-1} can be understood somewhat
more conceptually in a dual formalism using linear multiplets 
$L^\alpha$ instead of the chiral multiplets $T_\alpha$.

Let us return to the K\"ahler potential (\ref{kaehlerpot-O7-1}).
$K_{cs}$ and the first term in \eqref{kaehlerpot-Kk} are the standard
K\"ahler potentials for the complex structure deformations
and the dilaton, respectively. 
$\text{Vol}_E(\tau,G,T)$ also depends on $\tau$ and therefore the metric mixes $\tau$
with $T_\alpha$ and $G^a$. It is block diagonal in the 
complex structure deformations which do not mix with the other scalars.
Hence, the moduli space locally has the form
\beq \label{modulispaceO3}
\cM_{N=1} = \tilde \cM^{\rm SK} \times\, \tilde \cM^{\rm Q}\ ,
\eeq
where each factor is a K\"ahler manifold. The manifold $ \tilde \cM^{\rm SK}$ 
has complex dimension $h_-^{(1,2)}$ and is a special K\"ahler manifold in 
that $K_{\rm cs}$ satisfies \eqref{csmetrico}. It parameterizes the complex 
structure deformations of $Y$ respecting the orientifold constraint \eqref{Omegatransf}.
On the other hand, $\tilde \cM^{\rm Q}$ is a $h^{(1,1)}+1$-dimensional K\"ahler manifold inside 
the quaternionic manifold $\cM^{\rm Q}$. Local coordinates are given by the 
fields $\tau, G^a,T_\alpha$ arising in the expansion \eqref{def-coordsO3}. 
Also the K\"ahler potential $K^{\rm Q}(\tau, G,T)$ fulfills
special properties inherited from the underlying 
special quaternionic manifold.   
To see this, let us bring $K^{\rm Q}$ in a slightly different form. 
Using the explicit expansion \eqref{def-coordsO3} of $\pev$ one checks 
that up to a trivial K\"ahler transformation the K\"ahler potential 
\eqref{kaehlerpot-Kk} can be rewritten as 
\beq \label{def-Phi}
   K^{\rm Q} = -2\ln\, \Phi_B(\fe)\ , \qquad \Phi_B(\fe)\equiv i\big<\pev, \pevb \big>\ , 
\eeq
where $\pev=\fe+i\feh$ is defined in \eqref{def-A} and $\feh(\fe)$ has 
to be evaluated. In \eqref{def-Phi} 
we abbreviated the skew-symmetric product $\big<\varphi, \psi \big>$ 
for two sums of even forms $\varphi=\varphi_0+\varphi_2 + \varphi_4 + \varphi_6$
and $\psi=\psi_0+\psi_2+\psi_4+\psi_6$ as \cite{HitchinGCM} 
\beq \label{symp-form}
  \big<\varphi, \psi \big> =  \int_Y\ \sum_{m} (-1)^{m} \varphi_{2m} \wedge \psi_{\,6-2m}\ .
\eeq
The function $\Phi_B$ can be identified with Hitchins functional on 
a generalized complex manifold \cite{HitchinGCM} evaluated for the simple form $\pev$ defined
in \eqref{def-A} (see \cite{GLprep} for more details).  
We discuss the geometry of $\tilde \cM^{\rm Q}$ in greater detail in section \ref{geom_of_modspace}. 

Although not immediately obvious from its definition
$K^{\rm Q}$ obeys a no-scale type condition in that it satisfies
\beq\label{NScond}
    \frac{\partial K}{\partial N^I}\,  (K^{-1})^{ I\bar J}\, 
    \frac{\partial K}{\partial \bar N^{\bar J}} = 4\ ,
\eeq
where $N^I = (\tau,G^a,T_\alpha)$.\footnote{For $G^a=0$
this has already been observed in \cite{GKP,BBHL,DWG,DAFT}.} 
This equality can be shown by direct computation as done in \cite{TGL1}.
Alternatively, it can be deduced from the fact that $\Phi_B$ defined in 
\eqref{def-Phi} is homogeneous of degree two, i.e.~$\Phi_B(a\, \fe) = a^2\, \Phi_B(\fe)$ for all 
$a\in \bbR $ \cite{HitchinGCM}. Using \eqref{def-coordsO3} a
simple calculation shows that $K^{\rm Q}=-2\ln \Phi_B$ satisfies \eqref{NScond}.
From \eqref{N=1pot} we see that 
\eqref{NScond} implies $V\ge 0$ which we also show 
in the linear multiplet formalism in section \ref{IIB_lin}.
For $\tau=\text{const.}$ the right hand side of \eqref{NScond} is found to
be equal to $3$ as it is the case in the standard no-scale K\"ahler potentials
of \cite{NS}. 

Let us relate \eqref{kaehlerpot-O7-1} to the known K\"ahler potentials 
in the literature. 
First of all, for $G^a=0$ and thus 
$T_\alpha=i \rho_\alpha + \frac{1}{2}\KK_\alpha $
 the K\"ahler potential 
\eqref{kaehlerpot-O7-1} reduce to the one given in \cite{BBHL}.
Secondly, for one overall K\"ahler modulus $v$ parameterizing the volume
(i.e. for $h^{(1,1)}_+=1$, $T_{\alpha}\equiv T$) 
the K\"ahler potential $K^{\rm Q}$ reduces to 
$K=-3 \text{ln}( T +\bar T )$
which coincides with the K\"ahler potential 
determined in \cite{GKP}.

Before we turn to the discussion of the $O5/O9$ case let us note that 
$K$ is invariant under the $SL(2,\bbR)$ transformations inherited from the
ten-dimensional type IIB theory. In the orientifold theory
this symmetry acts on $\tau$ by fractional linear transformations 
exactly as in $D=10$ and transforms $(b^a,c^a)$ as a doublet, such that
\beq \label{Sl2}
  \tau \to \frac{a\tau + b}{c\tau +d}\ ,\qquad G^a \to \frac{G^a}{c\tau +d}\ , \qquad ad-bc=1\ .
\eeq
Under the $SL(2,{\bf R})$ only the second term of $K$ 
given in \eqref{kaehlerpot-Kk}
transforms but this transformation is just a K\"ahler transformation.
This can be seen from \eqref{tau} and the fact that $e^{-\phi/2} v^\alpha$ and $z^k$ are invariant.
This symmetry reduces to $SL(2,\bbZ)$ in the full string theory, which is nothing
but the invariance group of a two-torus. This torus becomes part of the space-time 
in the formulation of `F-theory' \cite{Vafa}. We discuss in section \ref{F-theory} 
the embedding of $O3/O7$ orientifolds into this theory on the level of the effective action.

\subsubsection{The K\"ahler potential: $O5/O9$ setups}

In the action \eqref{S_scalarO5}  we immediately see that the complex
structure deformations $z^\kappa$ are again already good K\"ahler coordinates.
For the remaining fields we find the appropriate K\"ahler coordinates
to be
\beq \label{def-coordsO5} 
  \feh - i \fa\ =\  t^\alpha \omega_\alpha - A_b\, \tilde \omega^b - S\, \text{vol}(Y)\ ,
\eeq 
where $\feh=\I\, \pev$ and $\fa$ are defined in \eqref{def-A} and we have used that in 
$O5/O9$ setups the axion $C_0$ gets projected out. Furthermore, we denoted by 
$\text{vol}(Y)=\cK^{-1} J\wedge J \wedge J$ the to one normalized volume form of $Y$.
Using 
the expansions \eqref{transJo}, \eqref{exp-B}
and \eqref{expO5} we obtain the explicit expressions \cite{TGL1}
\begin{eqnarray}\label{Kcoord}
  t^\alpha &=& e^{-\phi} v^\alpha - i c^\alpha\ , \qquad A_a \ =\ \N_{ab} b^b + i \rho_a \ ,\\
  S & = & \tfrac{1}{6}  e^{-\phi}\, \cK + i h 
           - \tfrac{1}{4} (\text{Re}\N^{-1})^{ab} A_a (A+\bar A)_b\ , \nn
\end{eqnarray}
where we inserted 
\beq\label{Ndef}
   \N_{ab}(t) \equiv \KK_{ab \alpha} t^\alpha \ ,\qquad \int C_6 = h+\tfrac{1}{2} \rho_a b^a\ . 
\end{equation}
The matrix $\N_{ab}$ depends holomorphically on the coordinates $t^\alpha$ which
ensures that $\tilde \cM^{\rm Q}$ is K\"ahler \cite{FS,HL}.
In the variables given in \eqref{Kcoord} the K\"ahler potential reads \cite{TGL1}
\beq \label{O5-Kaehlerpot} 
  K \ =\ K_{cs}(z,\bar z) + K^{\rm Q}(S,t,A)\ , \qquad K_{cs} = -\text{ln}\Big[-i\int\Omega \wedge \bar \Omega \Big]
\eeq
with
\bea \label{kaehlerpot-KkO5}
 K^{\rm Q}& =& - \text{ln}\Big[\tfrac{1}{48}\KK_{\alpha \beta \gamma}(t+\bar t)^\alpha 
        (t+\bar t)^\beta (t+\bar t)^\gamma  \Big]\nn \\
     && - \text{ln}\Big[S + \bar S + \tfrac{1}{4} (A + \bar A)_a (\text{Re}\N^{-1})^{ab} 
        (A + \bar A)_b \Big]\\
     &=& -\ln\big[2 e^{-4D}\big]\ . \nn
\eea
where we used \eqref{def-D_B}.
The check that $K$ indeed reproduces \eqref{S_scalarO5} is straightforward, since 
\eqref{O5-Kaehlerpot} is closely related to the quaternionic
`K\"ahler potential' given in \cite{FS} and we can make use
of their results.\footnote{Note however,
that the complex structure changed non-trivially.
In \cite{FS} the standard $t \sim v + i b$ formed complex coordinates.}
The same reference  already observed 
that for a holomorphic matrix $\N$ the quaternionic geometry is also 
K\"ahler. This situation was also found in compactifications
of the heterotic string to $D=3$ on a circle \cite{HL}. 

{}From \eqref{O5-Kaehlerpot} we infer that the $N=1$ moduli space admits 
the local product structure $\tilde \cM^{\rm SK}\times \tilde \cM^{\rm Q}$
similar to \eqref{modulispaceO3}. However, in $O5/O9$ orientifolds $\tilde \cM^{\rm SK}$
is a special K\"ahler manifold spanned by the $h^{(2,1)}_+$ complex scalars $z^\kappa$, which 
are the ones projected out in $O3/O7$ orientifolds. $\tilde \cM^{\rm Q}$ is spanned by the complex scalars 
$S,t^\alpha,A_a$ and thus is of complex dimension $h^{1,1} + 1$ as in $O3/O7$ setups. 
Furthermore, also $K^{\rm Q}$ for orientifolds with $O5/O9$ planes can be rewritten in terms 
of the functional $\Phi_B(\feh)$ as
\beq \label{def-PhiO5}
  K^{\rm Q}\ =\ -2 \ln \Phi_B(\feh)\ ,\qquad  \Phi_B(\feh ) \equiv i\big<\pev,\pevb \big>\ ,
\eeq
where $\pev=\fe(\feh)+i\feh$ are defined in \eqref{def-A}.
The functional dependence of $K^{\rm Q}$ on $\pev$ is the same as in \eqref{def-Phi} for $O3/O7$ orientifolds.  
This can be understood from the fact that $\pev$ only depends on the NS-NS sector 
variables, which are the same in both types of orientifolds. 
Nevertheless, the local structure of $\tilde \cM^{\rm Q}$ is different for both orientifold theories.
This becomes appearent when one expresses $K^{\rm Q}$ in terms of
proper K\"ahler coordinates. In $O5/O9$ setups this corresponds to the fact that 
$\Phi_B$ is a function of $\feh$ as needed for \eqref{def-coordsO5}. Hence, in order 
to express $K^{\rm Q}$ in terms of the K\"ahler coordinates $S,t,A$ as in \eqref{kaehlerpot-KkO5} 
one evaluates $\fe(\feh)$. Let us end this discussion by remarking that
$\Phi_B$ is also homogeneous of degree two in $\feh$, such that 
by using \eqref{def-coordsO5} one extracts a no-scale type condition equivalent to \eqref{NScond}.

\subsubsection{The gauge-couplings: $O3/O7$ and $O5/O9$ setups}

Our next task is to determine
the gauge-kinetic coupling functions $f_{\kappa \lambda}$ 
and show that they are holomorphic in the moduli. We do 
this only for $O3/O7$ orientifolds, since the result easily 
translates to the $O5/O9$ case. As explained in section 
\ref{O37_effective_act} this is achieved by an appropriate replacement of the 
indices.
By comparing the actions \eqref{red-vector} and \eqref{N=1action} one finds
\beq \label{gauge-couplingsO3}
  f_{\kappa \lambda}
=-\tfrac{i}{2} \,
\bar{\cM}_{\kappa \lambda}\Big|_{z^{\kappa}=0=\bar z^{\kappa}}\ ,
\eeq 
where 
${\cM}_{\kappa \lambda}$ is the 
$N=2$ gauge kinetic matrix given in \eqref{defM}
evaluated at ${z^{\kappa}=\bar z^{\kappa}}=0$.
Its holomorphicity in the complex structure deformations $z^k$ is not
immediately obvious but can be shown by using
\eqref{defM} and  \eqref{gauge-c}.
More precisely, from \eqref{defM} together with 
the decomposition of $H^{(3)}$ expressed by \eqref{H3split}
and \eqref{sbasiso} we infer that ${\cM}_{\hat K \hat L}$
is block diagonal or in other words
${\cM}_{\kappa \hat l} = 0.$ Multiplying ${\cM}_{\kappa \hat l}$
with $X^{\hat l}$ and using $X^\lambda=0$ together with
\eqref{gauge-c} we further conclude 
\beq\label{Fdiag}
{\cF}_{\kappa \hat l}\Big|_{z^{\kappa}=0=\bar z^{\kappa}}=0\ .
\eeq
Finally inserting \eqref{cond-1} and \eqref{Fdiag}
into \eqref{gauge-c}
we arrive at \cite{TGL1}
\begin{eqnarray}\label{fholo}
  f_{\kappa \lambda} (z^k)
=-\tfrac{i}{2} 
\mathcal{F}_{\kappa \lambda}\Big|_{z^{\kappa}=0=\bar z^{\kappa}}\ ,
\end{eqnarray} 
which is manifestly holomorphic since $\mathcal{F}_{\kappa \lambda}(z^k)$
are holomorphic functions of the complex structure
deformations $z^k$.

%
%

\section{Type IIA Calabi-Yau orientifolds}
\label{IIA_orientifolds}

In this section we determine the $N=1$ supergravity action 
obtained by compactification of Type IIA string theory on 
a Calabi-Yau orientifold. The orientifold projection 
$\cO=(-1)^{F_L} \Omega_p \sigma$ was already defined in 
\eqref{oproj} and includes an anti-holomorphic isometric involution 
$\sigma$. In section \ref{spectrum-IIA} we extract the $N=1$ spectrum by 
identifying the fields invariant under $\cO$. The corresponding 
effective action is calculated in section \ref{eff_actIIA}.
It is shown to be compatible with $N=1$ supersymmetry in section \ref{Kpo_gaugeIIA},
where we determine the K\"ahler potential and gauge-kinetic coupling functions.

\subsection{The $N=1$ spectrum \label{spectrum-IIA}}

In order to determine the  $\mathcal{O}$-invariant states let us recall
that the ten-dimensional RR forms $\hat C_1$ and $\hat C_3$ 
are odd 
under $(-1)^{F_L}$ while all other fields are even. 
Under the world-sheet 
parity $\Omega_p$ on the other hand $\hat B_2, \hat C_3$ are odd
with all other fields being even.
As a consequence the 
$\mathcal{O}$-invariant states have to satisfy
\cite{BH}
\begin{equation} \label{fieldtransf}
\begin{array}{lcl}
\sigma^*  \hat \phi &=& \  \hat \phi\ , \\
\sigma^*   \hat g &=& \ \hat g\ , \\
\sigma^*   \hat B_2 &=& -  \hat B_2\ ,
\end{array}
\hspace{2cm}
\begin{array}{lcl}
\sigma^*   \hat C_1 &=&  -  \hat C_1\ , \\
\sigma^*   \hat C_3 &=& \  \hat C_3\ ,
\end{array}
\end{equation}
while the deformations of the Calabi-Yau metric are constrained
by \eqref{constrJ} and \eqref{constrO}.\footnote{%
Following the argument presented in 
\cite{BH} we note that the involution does not change
 under deformations of $Y$. 
This is due to its involutive property 
and the fact that we identify involutions which differ
by diffeomorphisms.
Therefore we fix an involution and restrict the deformation space by demanding 
\eqref{constrJ} and \eqref{constrO}. }

As we recalled in the previous section the massless modes are in one-to-one
correspondence with the harmonic forms on $Y$. The space of harmonic forms
splits under the involution $\sigma$ into even and odd eigenspaces
\beq \label{cohom-split}
   H^p(Y)\ =\ H^p_+ \oplus H^p_-\ \ .
\eeq
Depending on the transformation properties given in \eqref{fieldtransf}
the $\mathcal{O}$-invariant states reside either in $H^p_+$ or in $H^p_-$
and as a consequence the number of states is reduced.
We summarize all non-trivial cohomology groups including their basis elements 
in table \ref{CYObasis}.\\

\begin{table}[h]
\begin{center}
\begin{tabular}{| c || c | c| c | c | c | c |} \hline
   \rule[-0.3cm]{0cm}{0.9cm} cohomology group &  $\ H^{(1,1)}_+\ $ & 
   $\ H^{(1,1)}_-\ $ & $\ H^{(2,2)}_+\ $ & $\ H^{(2,2)}_-\ $ & $\ H^{(3)}_+\ $ & $\ H^{(3)}_-\ $
   \\ \hline
   \rule[-0.3cm]{0cm}{0.8cm} dimension &  $h^{(1,1)}_+$  & $h^{(1,1)}_- $  
                                       &  $h^{(1,1)}_-$  & $h^{(1,1)}_+$ 
                                       &  $h^{(2,1)}+1$  &  $h^{(2,1)}+1$ 
   \\ \hline
   \rule[-0.3cm]{0cm}{0.8cm} basis     & $\omega_\alpha$ & $\omega_a$
                                       & $\tilde \omega^a$ & $\tilde \omega^\alpha$
   & $ a_{\Kh}$ & $b^{\Kh}$ \\ \hline
\end{tabular}
\caption{\small \label{CYObasis}
\textit{Cohomology groups and their basis elements.}}
\end{center}
\end{table}

$\omega_\alpha, \omega_a$ denote
 even and odd $(1,1)$-forms while 
$\tilde\omega^\alpha, \tilde\omega^a$ denote odd and even
 $(2,2)$-forms. The number of even $(1,1)$-forms is equal to the number
of odd $(2,2)$-forms and vice versa since the 
volume form which is 
proportional to $J \wedge J \wedge J$ is odd and thus Hodge duality
demands $h^{(1,1)}_+ = h^{(2,2)}_- ,\ h^{(1,1)}_- = h^{(2,2)}_+$.
This can also be seen from the fact that the non-trivial
intersection numbers are 
\beq \label{basis-int}
  \int \omega_\alpha \wedge \tilde \omega^\beta =
  \delta^{\beta}_\alpha\ ,\quad \alpha,\beta = 1, \ldots, h^{(1,1)}_+\ ,
  \qquad 
  \int \omega_a \wedge \tilde \omega^b = \delta^{b}_a\ , \quad
  a,b=1,\ldots,h^{(1,1)}_-\ ,
\eeq
with all other pairings vanishing.
{}From the volume-form being odd 
one further infers $h^{(3,3)}_+=0,$ $h^{(3,3)}_-=1$ and 
$h^{(0,0)}_+=1,\ h^{(0,0)}_-=0$.

$H^{3}$ can be decomposed 
independently of the complex structure as
$H^{3}=H^3_+ \oplus H^3_-$ where 
the (real) dimensions of  both $H^3_+$ and $H^3_-$
is equal and given by $h^{3}_+ =h^{3}_-= h^{(2,1)}+1$.
Again this is a consequence of Hodge duality
together with the fact that the volume-form is odd.
It implies that for each element $a_\Kh \in H^3_+$ 
there is a dual element $b^\Lh \in H^3_-$
with the intersections
\beq \label{basis_ab}
 \int a_\Kh \wedge b^\Lh = \delta^\Lh_\Kh \ , \qquad 
 \Kh, \Lh = 0,\ldots, h^{(2,1)}\ .
\eeq
Compared to \eqref{int-numbers1} this amounts to a symplectic rotation
such that all $\alpha$-elements are chosen to be even and 
all $\beta$-elements are chosen to be odd but with the intersection
numbers unchanged.
The orientifold projection breaks this symplectic invariance 
or in other words fixes a particular symplectic gauge 
which groups all basis elements into even and odd. 
This in turn implies that the basis $(a_\Kh,b^\Kh)$ is only one possible choice. 
However, since the calculation simplifies considerably for this basis, we first restrict 
to this special case and later give the general results with calculations summarized in 
section \ref{IIA_lin}.

In the remainder of this subsection we determine the $N=1$ spectrum 
which survives the orientifold projections.
Let us first discuss the K\"ahler moduli. 
From the eqs. \eqref{constrJ} and \eqref{fieldtransf} we see that both
$J$ and $\hat B_2$ are odd and hence have to be expanded
in a basis $\omega_a$ of odd harmonic $(1,1)$-forms 
\beq \label{expJB}
  J\ =\ v^a(x)\, \omega_a\ ,\qquad  \hat B_2\ =\ b^a(x)\, \omega_a\ , \qquad a = 1,\ldots, h^{(1,1)}_-\ .
\eeq
In contrast to \eqref{fieldexp} the four-dimensional 
two-form $B_2$ gets projected out due to \eqref{fieldtransf} and the fact
that $\sigma$ acts trivially on the flat dimensions. 
$v^a$ and $b^a$ are space-time scalars and 
as in $N=2$ they can be combined into complex coordinates
\beq \label{def-t}
  t^a = b^a + i\,  v^a\ , \qquad \Jc = B_2 + i J\ ,
\eeq
where we have also introduced the complexified 
K\"ahler form $\Jc$.
We see that in terms of the field variables the same complex
structure is chosen as in $N=2$ but the dimension of the K\"ahler moduli
space is truncated from $h^{(1,1)}$ to $h^{(1,1)}_-$.

The number of complex structure deformations is similarly reduced since
\eqref{constrO} constrains the possible deformations.
To see this one performs a symplectic rotation on 
\eqref{Omegaexp} and expands $\Omega$ in the basis of
$H^p_+ \oplus H^p_-$, i.e.\ as\footnote{Let us stress that at this
point all $N=2$ relations are still intact since \eqref{Omegapm}
is just a specific choice of the standard $N=2$ basis \eqref{Omegaexp}.} 
\beq\label{Omegapm}
\Omega(z) = Z^\Kh(z)\, a_\Kh - \cF_\Lh(z)\, b^\Lh\ .
\eeq 
Inserted into \eqref{constrO} one finds 
\bea \label{Z=0}
   \I(e^{-i\theta} Z^\Kh)\ =\  0\ , \qquad
   \R(e^{-i\theta} \cF_\Kh )\ = 0\ .
\eea  
The first set of equations  are $h^{(2,1)}+1$ real conditions
for $h^{(2,1)}$ complex scalars $z^K$.
One of these equations is redundant due to the 
scale invariance \eqref{crescale} of $\Omega$.
More precisely, the phase of $e^{-h}$ can be used to 
trivially satisfy $\I(e^{-i\theta} Z^\Kh)= 0$ for one of the $Z^\Kh$.
Thus $\I(e^{-i\theta} Z^\Kh)=0$
projects out $h^{(2,1)}$ real scalars, i.e.\
half of the complex structure deformations. 
Furthermore, in section \ref{eff_actIIA} we will see 
the remaining real complex structure deformations 
span a Lagrangian submanifold $\cM^{\rm cs}_\bbR$
with respect to the K\"ahler form
inside $\cM^{\rm cs}$. 
Note that the second set of equations in \eqref{Z=0}
$\R (e^{-i\theta} \cF_\Kh ) = 0$ 
should not be read as equations determining
the $z^K$ but is a constraint on the periods (or equivalently
the Yukawa couplings) of the Calabi-Yau
which has to be fulfilled in order to admit an involutive symmetry 
with the property \eqref{constrO}.\footnote{This can also be seen 
as conditions arising in consistent truncations of 
$N=2$ to $N=1$ theories as discussed in ref.\ \cite{ADAF}.}

As we have just discussed 
the complex rescaling \eqref{crescale}
is reduced to the freedom of a real rescaling by \eqref{constrO}. 
Under these transformations $\Omega$ and the K\"ahler potential $\Kcs$ 
change as
\beq \label{real_K}
  \Omega\to\Omega\, e^{-\R(h)}\ , \qquad \Kcs\to\Kcs + 2 \R(h)\ ,
\eeq
when restricted to $\cM^{\rm cs}_\bbR$. This freedom can be used to set one of 
the $\R (e^{-i\theta}Z^{\Kh})$ equal to one and tells us
that $\Omega$ depends only on $h^{(2,1)}$ real deformation parameters. 
However, it will turn out to be more
convenient to leave this gauge freedom intact and define 
a complex `compensator' $C=re^{-i\theta}$ with the transformation property
$C\to C e^{\R (h)}$.\footnote{This is reminiscent of the situation
encountered in the computation of the entropy of $N=2$ black holes
where it is also convenient to leave this scale invariance intact \cite{OSV}.}
Later on we will relate $r$ to 
the inverse of the four-dimensional dilaton 
so that the scale invariant function $C\Omega$ depends on 
$h^{(2,1)}+1$ real parameters. 
Using \eqref{Omegapm} $C\Omega$ enjoys
the expansion 
\beq \label{decompO}
  C \Omega\ =\ \R (C Z^\Kh)\, a_\Kh - i\I (C \cF_\Lh)\,  b^\Lh\ .
\eeq

We are left with the expansion of the ten-dimensional fields
$\hat C_1$ and $\hat C_3$ into  harmonic forms. 
{}From \eqref{fieldtransf} we learn that $\hat C_1$ is odd
and so together with the fact that
$Y$ posses no harmonic one-forms 
and $\sigma$ acts trivially on the flat dimensions
the entire $\hat C_1$ is projected out. This 
corresponds to the fact that the $N=2$ graviphoton $A^0$ is removed
from the gravity multiplet, 
which in $N=1$ only consists of the metric $g_{\mu \nu}$ as 
bosonic component.
Finally, $\hat C_3$ is even and thus can be expanded according to
\beq \label{form-exp}
  \hat C_3 = \cc_3(x) 
+ A^\alpha(x) \wedge \omega_\alpha + \CC_3\ ,\qquad
\CC_3 \equiv \xi^\Kh(x)\,  a_\Kh \ ,
\eeq
where $\xi^\Kh$ are $h^{(2,1)}+1$ real 
scalars, $A^\alpha$ are $h^{(1,1)}_+$ one-forms
and $\cc_3$ is a three-form in four dimensions.
$\cc_3$ contains no physical degree of freedom but as we will see 
in section~\ref{O6sup} corresponds to a 
constant flux parameter in the superpotential.
The real scalars 
$\xi^\Kh$ have to combine with the 
$h^{(2,1)}$ real  complex structure deformations 
and the dilaton to form chiral multiplets.
In the next section we will find that the appropriate complex fields 
arise from the combination
\beq\label{Omegacdef}
  \Omegac\ =\ \CC_3 + 2i\R(C\Omega) \ .
\eeq
Expanding $\Omegac$  in a basis \eqref{basis_ab} of $H^3_+(Y)$ 
and using \eqref{decompO} and \eqref{form-exp} we have 
\beq\label{newO}
\Omegac\ =\ 2 N^\Kh a_\Kh \ ,\qquad
N^\Kh= \tfrac{1}{2} \int \Omega_c\wedge \beta^\Kh =
\tfrac{1}{2}\big(\xi^\Kh + 2i \R (C Z^\Kh)\big)\ . 
\eeq
Due to the orientifold projection the two three-forms 
$\Omega$ and $C_3$
each lost half of their degrees of freedom and combined
into a new complex three-form $\Omegac$. 
As we will show in more detail in the next section
the `good' chiral coordinates in the $N=1$ orientifold
are the periods of $C\Omega$ directly while in $N=2$
the periods agree with the proper field variables only 
in special coordinates.

Let us summarize the resulting $N=1$ spectrum.
It assembles into a gravitational multiplet,
$h^{(1,1)}_+$ vector multiplets and 
$(h^{(1,1)}_- + h^{(2,1)}+1)$ chiral multiplets.
We list the bosonic parts of the $N=1$ supermultiplets in table
\ref{N=1spectrumA} \cite{BH}. We see that the $h^{(1,1)}$ $N=2$
vector multiplets split into $h^{(1,1)}_+$ $N=1$ vector multiplets and 
$h^{(1,1)}_-$ chiral multiplets while the 
$h^{(2,1)}+1$ hypermultiplets are reduced to $h^{(2,1)}+1$ chiral multiplets.

\begin{table}[h]
\begin{center}
\begin{tabular}{|l|c|c|} \hline 
 \rule[-0.3cm]{0cm}{0.8cm} 
multiplets& multiplicity & bosonic components\\ \hline\hline
 \rule[-0.3cm]{0cm}{0.8cm} 
 gravity multiplet&1&$g_{\mu \nu} $ \\ \hline
 \rule[-0.3cm]{0cm}{0.8cm} 
 vector multiplets&   $h_+^{(1,1)}$&  $A^{\alpha} $\\ \hline
 \rule[-0.3cm]{0cm}{0.8cm} 
 {chiral multiplets} &   $h_-^{(1,1)}$& 
$t^a$ \\ \hline
 \rule[-0.3cm]{0cm}{0.8cm} 
{chiral multiplets} 
& $ h^{(2,1)}+1$ &$ N^\Kh$\\ 
\hline
\end{tabular} 
\caption{\label{N=1spectrumA} \textit{$N =1$ spectrum of $O6$ orientifold compactification.}}
\end{center}
\end{table}

\subsection{The effective action}
\label{eff_actIIA}

In this section we calculate the four-dimensional effective action of type 
IIA orientifolds by performing a Kaluza-Klein reduction of the 
ten-dimensional type IIA action \eqref{10dact} taking the 
orientifold constraints into account. Equivalently this amounts to
imposing the orientifold projections on the $N=2$ action of 
section~\ref{revIIA}.
Inserting \eqref{expJB}, \eqref{decompO}, \eqref{form-exp} into
the ten-dimensional type IIA action \eqref{10dact} and performing a Weyl 
rescaling of the four-dimensional metric 
we find \cite{TGL2}
\bea \label{act1}
  S^{(4)}_{O6} &=& \int -\tfrac{1}{2} R*\mathbf{1} 
- G_{a b}\, dt^a \wedge * d \bar t^b 
+ \tfrac{1}{2} \text{Im}\, \cN_{\alpha \beta}\ F^\alpha \wedge * F^\beta 
        + \tfrac{1}{2} \text{Re}\, \cN_{\alpha \beta}\ F^\alpha
        \wedge F^\beta   \nn\\
      && 
\quad      -\, d D \wedge * dD  -\, G_{K L}(q)\, dq^K \wedge * dq^L 
         +\tfrac{1}{2} e^{2D}\, \text{Im}\, \cM_{ \Kh  \Lh}\, 
         d\xi^{\Kh} \wedge * d\xi^{\Lh} \ ,
\eea
where $F^\alpha = dA^\alpha$.
Let us discuss  the different couplings appearing in \eqref{act1}
in turn. 
Apart from the standard Einstein-Hilbert term the first line arises
from the projection of the $N=2$ vector multiplets action.
As we already observed the orientifold projection reduces the number
of K\"ahler moduli from $h^{(1,1)}$  to $h^{(1,1)}_-$ ($t^A\to t^a$)
but leaves the complex structure on this component of the moduli space
intact.  Accordingly the metric $G_{ab}(t)$ is inherited from the
metric $G_{AB}$ of the $N=2$ moduli space 
$\cM^{SK}$ given in \eqref{Kmetric}.
Since the volume form is odd only intersection numbers with one or
three odd basis elements in 
\eqref{int-numbers} can be non-zero and consequently one has
\beq \label{van-int}
  \cK_{\alpha \beta \gamma} = \cK_{\alpha a b} = \cK_{\alpha a} = \cK_{\alpha} = 0\ ,
\eeq  
while all other intersection numbers can be non-vanishing.\footnote{From a supergravity 
point of view this has been discussed also in \cite{ADAF}.} 
This implies that the metric $G_{AB}(t^A)$ of \eqref{Kmetric} is block
diagonal and obeys
\begin{eqnarray} \label{splitmetrIIA}
  G_{a b}=
  -\frac{3}{2}\left( \frac{\KK_{a b}}{\KK}-
  \frac{3}{2}\frac{\KK_a \KK_b}{\KK^2} \right)\ , \qquad
  G_{\alpha \beta}=-\frac{3}{2} \frac{\KK_{\alpha \beta}}{\KK}\ , \qquad
G_{\alpha b}\ =\ 0\ ,
\end{eqnarray}
where
\begin{equation}\label{intO6}
  \KK_{ab}=\KK_{abc}\; v^c\ , 
\quad \ 
\KK_{\alpha \beta}=\KK_{\alpha \beta a}\; v^a\ ,\quad
\KK_{a}=\KK_{a b c}\; v^b v^c\ ,
  \quad \KK=\KK_{abc} \; v^a v^b v^c
\ .
\end{equation}
In comparison to type IIB orientifolds the opposite intersection 
numbers vanish as can be seen by comparing \eqref{van-int} with \eqref{constr}.
This is due to the fact that the K\"ahler form $J$ transforms in IIA and IIB 
orientifolds with a relative minus sign under the action of $\sigma$. 

The same consideration also truncates  the $N=2$ gauge-kinetic 
coupling matrix $\cN_{\Ah \Bh}$ explicitly given in \eqref{def-cN}. 
Inserting \eqref{van-int} and \eqref{intO6} one arrives at 
\beq \label{def-N_alph_bet} 
  \text{Re} \cN_{\alpha \beta} = - \cK_{\alpha \beta a} b^a\ , \qquad 
  \text{Im} \cN_{\alpha \beta} = \cK_{\alpha \beta}\ ,\qquad \cN_{a \alpha}=\cN_{0 \alpha}=0\ .
\eeq
(The other non-vanishing matrix elements $\cN_{\ah\bh}$ arise in the potential 
\eqref{U-pot} once fluxes are turned on.) 

Let us now discuss the 
terms in the second line of \eqref{act1} arising from the
reduction  of the $N=2$ hypermultiplet action which is 
determined by the quaternionic metric \eqref{q-metr}.
$D$ is the the four-dimensional dilaton defined in \eqref{4d-dilaton}.
The metric $G_{KL}$ is inherited from the $N=2$
K\"ahler metric $G_{K \bar L}(z,\bar z)$ given in \eqref{csmetric}
and thus is 
the induced metric on the submanifold $\cM^{\rm cs}_\bbR$ 
defined by the constraint \eqref{constrO}.
More precisely, the complex structure deformations respecting \eqref{constrO}
can be determined from \eqref{Kod-form}
by considering infinitesimal variations of 
$\Omega$
\beq
  \Omega(z + \delta z) \ =\ \Omega(z) + \delta z^K (\partial_{z^K} \Omega)_{z} \ 
              =\ \Omega(z) - \delta z^K( \Kcs_{z^K} \Omega - \chi_K)_z \ .
\eeq 
Now we impose the condition that both 
$\Omega(z+\delta z)$ and $\Omega(z)$ satisfy \eqref{constrO}. 
This implies locally 
\beq  \label{constr2}
  \delta z^K\, \partial_{z^K} \Kcs = \delta\bar z^K\, \partial_{\bar z^K} \Kcs\ , \qquad 
  \delta z^K\sigma^* \chi_K = e^{2i \theta} \delta \bar z^K\bar \chi_K \ ,
\eeq 
where $\partial_{z^K} \Kcs$ and $\chi_K$ are restricted to $\cM^{\rm cs}_\bbR$.
Using the fact that $\Kcs$ is a K\"ahler potential and therefore $\partial_{z^K}\Kcs\neq 0$, we conclude from
the first equation in \eqref{constr2} that for each $\delta z^K$ either the 
real or imaginary part has to be zero. This is consistent with the observation
of the previous section that coordinates of $\cM^{\rm cs}_\bbR$ can be 
identified with 
the real or imaginary part of the complex structure deformations $z^K$.
To simplify the notation we call these deformations collectively
$q^K$ and denote the embedding map by 
$\rho:\cM^{\rm cs}_\bbR \hookrightarrow \cM^{\rm cs}$.
Locally this corresponds to 
\beq \label{embmap1}
  \rho:\ q^K=(q^s,q^\sigma)\ \mapsto\ z^K=(q^s,iq^\sigma)\ ,
\eeq
for some splitting $z^K=(z^s,z^\sigma)$. In other words, 
the local coordinates on $\cM^{\rm cs}_\bbR$ 
are $\R z^s=q^s$ and $\I z^\sigma = q^\sigma$ while $\I z^s=0=\R z^\sigma$.
Using the second equation in \eqref{constr2}, the embedding
map \eqref{embmap1} and the expression \eqref{chi_barchi} for the $N=2$ metric $G_{K\bar L}$ we also deduce that
the K\"ahler form vanishes when pulled back to $\cM^{\rm cs}_\bbR$. 
In summary we 
have
\beq \label{def-G}
  \rho^*(G_{K \bar L}\, dz^K d \bar z^L)\, \equiv\, G_{KL}(q)\, dq^K dq^L\ , \quad
  \rho^*(iG_{K \bar L}\, dz^K \wedge d \bar z^L)\, =\, 0\ .
\eeq
The first equation defines the induced metric while the second equation
implies that $\cM^{\rm cs}_\bbR$
is a Lagrangian submanifold of $\cM^{\rm cs}$ with respect to the 
K\"ahler-form.

Finally, coming back to the action \eqref{act1}
the matrix $\cM_{\Kh \Lh}$ is defined in analogy with
\eqref{defM} as
\beq \label{defM2}
  \int a_\Kh \wedge * a_\Lh = -\text{Im}\; \cM_{\Kh \Lh} \ , \qquad 
 \int b^\Kh \wedge * b^\Lh\  = \ -(\text{Im}\; \cM)^{-1\ \Kh \Lh}\ , 
\eeq
where $\text{Im} \cM_{\Kh \Lh}$ can be given explicitly 
in terms of the periods by inserting \eqref{Z=0} into \eqref{gauge-c} \cite{TGL1}.
Similarly one obtains $\R\cM_{\Kh \Lh}=0$ 
consistent with the fact that \eqref{defM} implies that $\int a_\Kh\wedge * b^\Lh$ vanishes
for the special basis $(a_\Kh,b^\Kh)$.

This ends our discussion of
the effective action obtained by applying the orientifold projection. 
The next step is to rewrite the action \eqref{act1} 
in the standard $N=1$ supergravity form
which we turn to now.

%
%

\subsection{The K\"ahler potential and gauge-couplings}
\label{Kpo_gaugeIIA}

The standard $N=1$ supergravity the action is expressed 
in terms of a K\"ahler potential $K$, a holomorphic superpotential $W$ 
and the holomorphic gauge-kinetic coupling 
functions $f$ as given in \eqref{N=1action}. Hence, our task is to  
find $K,f$ and $W$ for the type IIA orientifolds. As an immediate 
observation one finds that \eqref{act1} includes no potential, such 
that $W=0$ and $D_\alpha=0$. It is also not difficult to read 
off the gauge-kinetic coupling function $f_{\alpha \beta}$.
Comparing \eqref{act1} with \eqref{N=1action} 
using \eqref{def-N_alph_bet} and \eqref{def-t}
one infers 
\beq \label{gauge-A}
  f_{\alpha \beta}\ =\ -i \bar \cN_{\alpha \beta}\ =\ i \cK_{\alpha \beta a}  t^a \ .
\eeq
As required by $N=1$ supersymmetry the $f_{\alpha \beta}$ 
are indeed holomorphic. Note that they are linear in the $t^a$ moduli
and do not depend on the complex structure and $\xi$-moduli.

{}From \eqref{act1} we also immediately observe that
the orientifold moduli space has the product structure  
\beq \label{direct-mod}
  \cM_{N=1}=\tilde\cM^{\rm SK} \times \tilde\cM^{\rm Q}\ .
\eeq
The first factor $\tilde\cM^{\rm SK}$ is a subspace of the 
$N=2$ moduli space $\cM^{\rm SK}$ with dimension $h^{(1,1)}_-$ 
spanned by  the complexified K\"ahler deformations $t^a$.
The second factor $\tilde\cM^{\rm Q}$ is a subspace of the quaternionic
manifold $\cM^{\rm Q}$ with
dimension $h^{(2,1)} +1$ 
spanned by  the complex structure deformations $q^K$, the dilaton $D$
and the scalars $\xi^{\hat K}$ arising from $C_3$.
Let us discuss both factors in turn.

As we already stressed earlier the metric $G_{ab}$ 
of \eqref{act1} defined in \eqref{splitmetr} is a trivial
truncation
of the $N=2$ special K\"ahler metric  \eqref{Kmetric} and therefore remains
special K\"ahler. The K\"ahler potential is given by
\beq \label{Kks}
K^{\rm K}\ =\ - \ln \Big[\tfrac{i}{6}\cK_{a b c} (t -\bar t)^a (t -\bar t)^b (t -\bar t)^c \Big]
  \ = \ - \ln \Big[\tfrac{4}{3}  \int_Y J \wedge J \wedge J\Big]\ ,
\eeq
where $J$ is the K\"ahler form in the string frame.
Moreover, $K^{\rm K}$ can be obtained from 
the prepotential $f(t)=-\tfrac{1}{6} \cK_{abc}t^a t^b t^c$ 
by using equation \eqref{Kinz}. 
It is well known that $K^{\rm K}$ obeys the standard no-scale condition
\cite{NS}
\beq \label{no-scale1}
  K_{t^a} K^{t^a \bar t^b} K_{ \bar t^b}\ =\ 3\ . 
\eeq

The geometry of the second component  $\tilde\cM^{\rm Q}$ in \eqref{direct-mod}
is considerably more complicated. This is due to the fact that 
\eqref{newO} defines a new complex structure on the field space. In the following
we sketch the calculation of the K\"ahler potential for the basis 
$(a_\Kh,b^\Kh)$ and only summarize the results for a generic symplectic basis.
The details of this more involved calculation will be presented in section \ref{IIA_lin}. 

To begin with, let us define the compensator $C$ introduced in section \ref{spectrum-IIA} as 
\beq \label{def-C}
   C\ =\ e^{-D-i\theta} e^{\Kcs(q)/2}\ , \qquad C \rightarrow C e^{\R\, h(q)}\ ,
\eeq 
where $\Kcs$ is the K\"ahler potential defined in \eqref{csmetric} restricted
to the real subspace $\cM^{\rm cs}_{\bbR}$. We also displayed the transformation 
behavior of $C$ under real K\"ahler transformations \eqref{real_K}. With this at hand one 
defines the scale invariant variable
\beq \label{l-def}
   l^\Kh \ =\  \R(C Z^\Kh(q))\ .
\eeq
Inserted into \eqref{act1} and using 
the Jacobian matrix encoding the change of variables $(e^D,q^K) \rightarrow l^\Kh$
the second line \eqref{act1} simplifies as\footnote{%
The calculation of this result can be found in section \ref{IIA_lin}.}
\beq \label{IIAQ}
  \cL^{(4)}_{\rm Q} =  2 e^{2D}\, \text{Im}\, \cM_{\Kh \Lh}\, 
(dl^\Kh \wedge * dl^\Lh  + \tfrac{1}{4} d\xi^{\Kh} \wedge * d\xi^{\Lh})\ .  
\eeq
We see that the scalars $l^\Kh$ and $\xi^\Kh$ nicely combine 
into complex coordinates 
\beq \label{Ncoords}
   N^\Kh\ =\ \tfrac{1}{2}\xi^\Kh +  i l^\Kh\ 
=\ \tfrac{1}{2}\xi^\Kh +  i \R(C Z^\Kh)
= \tfrac{1}{2} \int \Omega_c\wedge b^\Kh 
\ , 
\eeq
which we anticipated in equation \eqref{newO}.
The important fact to note here is that $\tilde\cM^{\rm Q}$
is equipped with a new complex structure and the corresponding
K\"ahler coordinates 
coincide with half of the periods of $\Omegac$.
This is in contrast to the situation in $N=2$ where one of the periods
($Z^0$) is a gauge degree of freedom and the K\"ahler
coordinates are the special coordinates $z^K = Z^K/Z^0$.

In order to show that the metric in \eqref{IIAQ} is K\"ahler we need the
explicit expression for the K\"ahler potential. Using \eqref{Z=0} in \eqref{gauge-c} 
one obtains straightforwardly
\beq
2 e^{2D} \text{Im}\, \cM_{\Kh \Lh} = \partial_{N^\Kh} \partial_{\bar N^\Lh} K^{\rm Q}\ ,
\eeq
where 
\beq \label{KQsimple}
 K^{\rm Q} = -2 \ln\big[4i\cF(CZ)\big]\ , \qquad
\cF\big(\R(CZ)\big) = \frac{i}2\,  \R(C Z^\Kh)\, \I(C\cF_\Kh)\ .
\eeq
Alternatively, using \eqref{decompO} and $*\Omega =- i \Omega$
one derives the integral representation
\beq \label{intKQ}
  K^{\rm Q}\ = - 2\ln\Big[2\int_Y \R(C \Omega)\wedge *\R(C\Omega)\Big]=\ - \ln\, e^{-4D} \ , 
\eeq
where in the second equation we used \eqref{def-C} and \eqref{csmetric}. In 
the form \eqref{intKQ} the dependence of $K^{\rm Q}$ on the coordinates $N^\Kh$
is only implicit and given by means of their definition \eqref{Ncoords}.  
Also $K^{\rm Q}$ obeys a no-scale type condition in that it 
satisfies
\bea \label{no-scale2}
  K_{N^\Kh} K^{N^\Kh \bar N^\Lh} K_{\bar N^\Lh} = 4\ ,
\eea
which can be checked by direct calculation.

The analysis so far started from the symplectic basis 
$(a_\Kh,b^\Kh)$ introduced in \eqref{basis_ab},
determined the K\"ahler coordinates in \eqref{Ncoords}
and derived  the K\"ahler potential $K^{\rm Q}$
in terms of the prepotential $\cF$ in \eqref{KQsimple} or as an 
integral representation in \eqref{intKQ}. Now we need to ask
to what extent  this result depends on the choice of 
the basis \eqref{basis_ab}. Or in other words let us redo
the calculation starting from an arbitrary symplectic basis
and determine the K\"ahler potential and the proper field variables
for the corresponding orientifold theory. 
Let us first recall the situation 
in the $N=2$ theory reviewed in section \ref{revIIA}. 
The periods $(Z^\Kh,\cF_\Kh)$ defined in \eqref{pre-z}
form a symplectic vector
of $Sp(2h^{(1,2)}+2,\bf Z)$
such that $\Omega$ given in \eqref{Omegaexp} and 
$\Kcs$ given in \eqref{csmetric} is manifestly invariant.
The prepotential  $\cF(Z) = \frac{1}{2} Z^\Kh \cF_\Kh$ on the other hand 
does depend
on the choice of the basis $(\alpha_\Kh,\beta^\Kh)$
and is not invariant.  

For $N=1$ orientifolds this situation is different 
since the orientifold projection \eqref{constrO} explicitly breaks the 
symplectic invariance.\footnote{A symplectic transformation $\cS$ preserve the
form $\big<\alpha,\beta\big> = \int \alpha \wedge \beta$, such that 
$\big<\cS \alpha,\cS \beta \big> = \big< \alpha,\beta \big>$.
On the other hand the anti-holomorphic involution satisfies
$\big<\sigma^* \alpha,\sigma^* \beta \big>  
= - \big< \alpha,\beta \big>$.}  
This can also be seen from the form
of the $N=1$ K\"ahler potential \eqref{KQsimple} which is expressed
in terms of the non-invariant prepotential.
One immediately concludes that the result \eqref{KQsimple} is 
basis dependent and $K^Q$ takes this simple form due to the special 
choice $a_\Kh \in H^{3}_+(Y)$ and $b^\Kh \in H^3_-(Y)$.\footnote{Note that this is in striking analogy to 
the background dependence of the B model partition function as discussed in \cite{BCOV,Witten2}.}
On the other hand, the integral representation \eqref{intKQ} only implicitly depends 
on the symplectic basis through the definition of the coordinates $N^\Kh$. 
This suggest, that it is possible to generalize our results by allowing for 
an arbitrary choice of symplectic basis in the definition of the $N=1$ coordinates. 
More precisely, let us consider the generic basis $(\alpha_\Kh,\beta^\Lh)$, 
where we assume that the $h^3_+=h^{2,1}+1$ basis elements $(\alpha_k,\beta^\lambda)$
span $H^3_+$ and the $h^3_-=h^{2,1}+1$ basis elements $(\alpha_\lambda,\beta^k)$ span $H^3_-$.
In this basis the intersections \eqref{int-numbers1} take the form
\beq \label{sp_alpha-beta}
  \int_Y \alpha_k \wedge \beta^l\ =\ \delta_k^l\ , \qquad 
  \int_Y \alpha_\kappa \wedge \beta^\lambda\ =\ \delta_\kappa^\lambda\ ,
\eeq
with all other combinations vanishing.
Applying the orientifold constraint \eqref{constrO} one concludes that 
the equations \eqref{Z=0} are replaced by 
\beq \label{Z=0gen}
  \I(C Z^k) =  \R (C \cF_k )\ =\ 0\ , \qquad 
   \R (C Z^\lambda) = \I(C \cF_\lambda)\ =\ 0\ .
\eeq 
Correspondingly, the expansions \eqref{decompO} and \eqref{form-exp}
take the form
\bea \label{decompO2}
  C \Omega &=& \R (C Z^k) \alpha_k + i\I (C Z^\lambda ) \alpha_\lambda -
             \R (C \cF_\lambda) \beta^\lambda - i\I (C \cF_k) \beta^k\ ,\nn\\
\CC_3 &=& \xi^k\, \alpha_k - \tilde \xi_\lambda\, \beta^\lambda\ ,
\eea
which implies that we also have to redefine the $N=1$ coordinates of 
$\tilde \cM^{\rm Q}$ in an appropriate way. 
In section~\ref{IIA_lin} we show that the 
new K\"ahler coordinates  $(N^k,T_\lambda)$ are again determined by the periods of $\Omegac$ and given  by 
\bea \label{Oexp}\label{def-NT}
  N^k &=&\tfrac{1}{2} \int \Omegac \wedge \beta^k \
 =\ \tfrac{1}{2}\xi^k + i \R(CZ^k)\ , \nn\\
  T_\lambda &=& i \int \Omegac \wedge \alpha_\lambda\ =\
i\tilde \xi_\lambda - 2 \R (C \cF_\lambda) \ ,
\eea
where we evaluated the integrals by using \eqref{Omegacdef} 
and \eqref{decompO2}.

The K\"ahler potential takes again the form \eqref{intKQ} but now 
depends on $N^k,T_\lambda$ and thus no longer simplifies to \eqref{KQsimple}.
Let us compare the situation to the original $N=2$ theory, which
was formulated in terms of the
$Z^\Kh$ or equivalently the special coordinates $z^K$. Holomorphicity
in these coordinates played a central role in defining the prepotential
encoding the special geometry of $\cM^{\rm cs}$ in $\cM^{\rm Q}$ (cf.~section 
\ref{revIIA}). In contrast, the $N=1$ orientifold constraints destroy this complex structure and force us 
to combine $\R(C\Omega)$ with the RR three-form $C_3$ into $\Omegac$. The 
K\"ahler coordinates are half of the periods of $\Omegac$ 
but now in this more general case also the
derivatives of $\cF$ can serve as coordinates as seen in \eqref{def-NT}. 
However, as it is shown in section 
\ref{IIA_lin}, $\R (C \cF_\lambda)$ and $e^{2D}\I (CZ^\lambda)$ are related
by a Legendre transformation of the K\"ahler potential. Working with this transformed
potential and the coordinates $\R(CZ^k)$ and $e^{2D}\I (CZ^\lambda)$ enables us 
to make contact to the underlying $N=2$ theory in its canonical formulation. 
From a supergravity point of 
view, this Legendre transformation corresponds to replacing the chiral multiplets 
$T_\lambda$ by linear multiplets as described in the next chapter.
This is possible due to the translational isometries of $K$, 
which arise as a consequence of the $C_3$ gauge invariance 
and which render $K$ independent of
the scalars $\xi$ and $\tilde\xi$. 
We show in section \ref{geom_of_modspace}
that this also enables us to construct $\tilde \cM^{\rm Q}$ from $\cM^{\rm cs}_\bbR$
similar to the moduli space of supersymmetric Lagrangian submanifolds in a
Calabi-Yau space as described by Hitchin \cite{Hitchin2}.  
This also allows us to interpret the no-scale condition \eqref{no-scale2}
geometrically.

Let us summarize the results obtained so far. We found that the moduli
space of $N=1$ orientifolds is indeed the product of two K\"ahler spaces
with the K\"ahler potential 
\beq \label{N=1Kpot}
  K\ =\ K^{\rm K} + K^{\rm Q} = - \ln \Big[\tfrac{4}{3}  \int_Y J \wedge J \wedge J\Big] 
        - 2\ln\Big[2\int_Y \R(C \Omega)\wedge *\R(C\Omega)\Big]\ .
\eeq 
The first term depends on the K\"ahler deformations of the orientifold
while the second term is a function of the real complex structure 
deformations and the dilaton.
The $N=1$ K\"ahler coordinates are obtained 
by expanding the complex combinations\footnote{This combination 
of forms has also appeared recently in ref.\ \cite{NOV}
in the discussion of $D$-instanton couplings in the A-model.
Here they appear as the proper chiral $N=1$ variables and as we will
see in the next section they linearize the D-instanton action.}
\beq \label{N=1coords}
  \Omegac\ =\ \CC_3 + 2i \R(C\Omega)\ ,\qquad 
  \Jc\ =\ \hat B_2 + iJ \ , 
\eeq
in a real harmonic basis of $H^{3}_+(Y)$ and $H^{(1,1)}_-(Y)$ respectively. 
Note that $K$ does not depend on the scalars arising in the expansion of 
$\hat B_2$ and $\hat C_3$, such that the K\"ahler manifold admits a set of 
$h^{(1,1)}_- + h^{(2,1)}+1$ translational isometries. In other words
$K$ consists of two functionals encoding the dynamics of the two-form $J$ 
and the real three-form $\R(C\Omega)$.
In type IIA orientifolds it is not difficult to rewrite $K^{\rm Q}$ in a form 
similar to \eqref{def-Phi}. Defining the odd form 
\beq \label{def-podd}
 \podd=\fu +i\, \fuh = C\Omega\ ,
\eeq
one finds 
\beq \label{symp-formodd}
  K^{\rm Q} = - 2 \ln \Phi_A(\fu) \ ,\qquad   \Phi_A(\fu) \equiv i\big<\podd,\poddb \big>=i\int_Y \podd \wedge \poddb\ .
\eeq
The function $\Phi_A(\fu)$ is 
known as Hitchins functional for the real 
three-form $\fu$ \cite{Hitchin1,HitchinGCM}. The orientifold constraint \eqref{constrO} 
restricts its domain to $\fu \in H^3_+(Y)$. Applying the fact that
$\Phi_A(\fu)$ is a homogeneous function of degree two $K^{\rm Q}$ obeys the no-scale type conditions
\eqref{no-scale2}, \eqref{no-scale4}. This is independent of the chosen basis 
and can be also shown directly as done in section \ref{IIA_lin}.

The no-scale conditions are violated when further stringy 
corrections are included. $K$ receives additional contributions due
to perturbative effects as well as world-sheet and $D2$ instantons.
It is well-known that the combination $\Jc=\hat B_2 + i J$ 
gives the proper coupling to the string world-sheet such that 
 world-sheet instantons correct the holomorphic prepotential as 
$f(t) = -\frac{1}{6}\cK_{abc}t^a t^b t^c + O(e^{-t})$. 
Since we divided out the world-sheet parity these corrections also
include non-orientable Riemann surfaces, such that the prepotential 
$f(t)$ consists of two parts $f(t) = f_{or}(t) + f_{unor}(t)$. 
The function $f_{or}$ counts holomorphic maps 
from orientable world-sheets to $Y$, while $f_{unor}$ counts holomorphic maps 
from non-orientable world-sheets to $Y$ \cite{BFM}. 
In the next section we show that $D2$ instantons naturally couple to the complex three-form  
$\Omegac$ and they are expected to correct 
$K^{\rm Q}$.

%
%

\section{Mirror symmetry \label{Mirror_orientioflds}}

In this section we discuss mirror symmetry 
for Calabi-Yau orientifolds from the point of view of the effective
action derived in the large volume limit. More precisely, we compare the $N=1$
data for type IIB orientifolds on $\tilde Y/\sigma_B$ (section \ref{Kpo_gaugeIIB}) 
with the data for type IIA orientifolds on $Y/\sigma_A$ (section \ref{Kpo_gaugeIIA}). 
Since we want to discuss mirror symmetry we choose $\tilde Y$ 
to be the mirror manifold of $Y$. This implies that the
non-trivial Hodge numbers $h^{(1,1)}$ and $h^{(2,1)}$ of $Y$ and $\tilde Y$ 
satisfy $h^{(1,1)}(Y)=h^{(2,1)}(\tilde Y)$  and $h^{(2,1)}(Y)=h^{(1,1)}(\tilde Y)$ as already 
given in section \ref{revMirror} where we briefly introduced $N=2$ mirror symmetry. 
In orientifolds we also have to specify the 
involutions $\sigma_A$ and $\sigma_B$ which are identified under mirror symmetry. Since the
discussion
in this article is quite generic and never specified any involution
$\sigma$ explicitly we also keep the discussion of mirror symmetry
generic. That is we assume that there exists a mirror pair
of manifolds $Y$ and $\tilde Y$ with a mirror pair of involutions
$\sigma_A, \sigma_B$. 
Matching the number of $N=1$ multiplets summarized in table \ref{numberM} 
implies an orientifold version of
\eqref{Hod_id},\footnote{For the sector of $\tilde \cM^{\rm Q}$ mirror
  symmetry
is a constraint on the couplings rather than the Hodge
numbers.}  i.e.\
\bea \label{matchchohm}
   O3/O7&: &\quad h^{1,1}_-(Y) = h^{2,1}_-(\tilde Y) \ , \qquad  h^{1,1}_+(Y) = h^{2,1}_+(\tilde Y) \ ,\nn\\
   O5/O9&: &\quad h^{1,1}_-(Y) = h^{2,1}_+(\tilde Y)   \ , \qquad h^{1,1}_+(Y) = h^{2,1}_-(\tilde Y) \ .
\eea
\begin{table}[h] 
\begin{center}
\begin{tabular}{|l|c|c|c|} \hline 
 \rule[-0.3cm]{0cm}{0.8cm} 
multiplets& IIA$_Y$ \  $O6$ & IIB$_{\tilde Y}$ \  $O3/O7$ & IIB$_{\tilde Y}$ \  $O5/O9$ \\ \hline\hline
 \rule[-0.3cm]{0cm}{0.9cm} 
 {vector multiplets} &   $h_+^{(1,1)}$ & $h_+^{(2,1)}$ & $h_-^{(2,1)}$ \\ \hline
\rule[-0.3cm]{0cm}{0.9cm} 
 chiral multiplets in $\tilde \cM^{\rm SK}$& $h_-^{(1,1)}$ & $h_-^{(2,1)}$ & 
                      $h_+^{(2,1)} $   \\ \hline
 \rule[-0.3cm]{0cm}{0.9cm} 
 chiral multiplets in $\tilde \cM^{\rm Q}$&$h^{(2,1)} + 1$&$h^{(1,1)} + 1$ & 
                      $h^{(1,1)} + 1$   \\ \hline
\end{tabular} 
\caption{ \textit{Number of $N=1$ multiplets of orientifold compactifications.}}\label{numberM}
\end{center}
\end{table}

Our next task will be to match the couplings of the mirror theories.
Since the effective actions on both sides 
 are only computed in the large volume limit
we can expect to find agreement only if we also take 
the large complex structure limit exactly as in the $N=2$ mirror
symmetry.
However, if one believes in mirror symmetry one can use the 
the geometrical results of the complex structure moduli space to
`predict' the corrections to its mirror symmetric component.
This is not quite as straightforward since the full $N=1$ moduli space is a
lot more complicated than the underlying $N=2$ space \cite{BH}.
Let us therefore start our analysis with the simpler situation of the 
special K\"ahler sectors $\tilde \cM^{\rm SK}_A,\, \tilde \cM_B^{\rm SK}$ in \eqref{direct-mod} 
and \eqref{modulispaceO3} and the vector multiplet couplings 
and postpone the analysis of $\tilde M^{\rm Q}_{A,B}$ 
to section \ref{O3O7mirror}.   

\subsection{Mirror symmetry  in $ \mathcal{M}^{\rm K}$}  \label{mirrorMK}
Recall that the manifold $\tilde \cM^{\rm SK}_A$ is spanned by the 
complexified K\"ahler deformations $t^a$ preserving the constraint 
\eqref{constrJ}.  Under mirror symmetry these moduli are mapped 
to the complex structure deformations which respect the constraint
\eqref{Omegatransf}.
In both cases the K\"ahler potential is merely a truncated
version of the $N=2$ K\"ahler potential and one has
\beq
  K^{\rm K}_A \ =\ - \ln \Big[\tfrac{4}{3}  \int_Y J \wedge J \wedge J\Big]
  \quad  \leftrightarrow \quad 
  K^{\rm cs}_B\ =\ -  \ln \Big[-i \int \Omega \wedge \bar \Omega \Big]\ .
\eeq
Both K\"ahler potentials can be expressed in terms of prepotentials
$f_A(t), f_B(z)$ and in the large complex structure limit
$f_B(z)$ becomes cubic and agrees with $f_A(t)$.
Mirror symmetry therefore equates these prepotentials 
and exchanges $J^3$ with $\Omega\wedge\bar\Omega$
exactly as in $N=2$ 
\beq\label{mirrorK}
f_A(t) =  f_B(z) \ , \qquad J^3 \leftrightarrow
\Omega\wedge\bar\Omega\ .
\eeq
In \cite{FMM} the $N=2$ version of this map was written into the form \footnote{The authors argued that
this should be true also for mirror symmetry of certain non-Calabi-Yau backgrounds. }
\beq \label{pure-spinor-map}
  e^{J_c}\ \leftrightarrow\ \Omega\ ,
\eeq 
where $J_c$ is given in \eqref{N=1coords}.  
Thus for $\tilde \cM^{\rm SK}$ mirror symmetry is a truncated
version of $N=2$ mirror symmetry. As we will see momentarily this also
holds for the gauge kinetic couplings 
which depend holomorphically on the moduli spanning $\tilde \cM^{\rm SK}$.

In type IIA the gauge-kinetic couplings
are given in \eqref{gauge-A} and read
$f_{\alpha \beta}(t) = i\cK_{\alpha\beta c}  t^c$.
The IIB couplings were determined in \eqref{fholo} to be 
\bea \label{gauge-B}
 f_{\alpha\beta}(z^a) = - {i} \bar \cM_{\alpha\beta}
   = - i\cF_{\alpha\beta}\ ,
\eea
where in order to not overload the notation we are using the same indices
for both cases.\footnote{We rescaled the type IIB gauge bosons
by $\sqrt 2$ in order to properly match the normalizations.} 
More precisely we are choosing
\bea
\alpha, \beta = 1, \ldots, h^{(2,1)}_+(\tilde Y)\ ,\qquad 
a, b = 1, \ldots, h^{(2,1)}_-(\tilde Y)\ , \qquad \textrm{for} \quad O3/O7\ ,\nn\\ 
\alpha, \beta = 1, \ldots, h^{(2,1)}_-(\tilde Y)\ ,\qquad 
a, b = 1, \ldots, h^{(2,1)}_+(\tilde Y)\ ,\qquad \textrm{for} \quad
O5/O9\ . 
\eea
The matrix $\cF_{\alpha\beta}(z^a)$ is  
holomorphic and the second derivatives of the prepotential restricted
to $\tilde \cM^{\rm  K}_B$. In the large complex structure limit 
$\cF_{\alpha\beta}$ is linear
in $z^a$ and therefore also agrees with the type IIA mirror
couplings. 
Thus mirror symmetry implies the map $\cN_{\alpha \beta}(\bar t^a) = \cM_{\alpha\beta}(\bar z^a)$
in both cases.

This concludes our discussions of mirror symmetry
for the chiral multiplets which span $\tilde\cM^{\rm SK}$.
We have shown that 
the K\"ahler potential and 
the gauge-kinetic coupling functions
agree in the large complex structure limit under mirror symmetry.
In this sector the geometrical quantities on the type IIB side include
corrections which are believed to 
compute world-sheet non-perturbative effects  
such as world-sheet instantons on the type
IIA side. 
This is analogous to the situation
in $N=2$ and  may be traced back to the 
fact, that it is still possible to formulate a topological 
A model counting 
world-sheet instantons for Calabi-Yau orientifolds \cite{AAHV,BFM}.

\subsection{Mirror symmetry in $ \cM^{\rm Q}$}
Let us now turn to the discussion of the K\"ahler manifolds $\tilde \cM^{\rm Q}_{A}$ and
$\tilde \cM^{\rm Q}_{B}$ arising in 
the reduction of the quaternionic spaces. 
On the IIA side the K\"ahler potential is given in \eqref{N=1Kpot}
which is expressed in terms of the $h^{(2,1)}+1$ coordinates
$(N^k,T_\lambda)$ defined in \eqref{Oexp}.
In this definition we did not fix the scale invariance \eqref{real_K}
$\Omega\to
\Omega e^{-\R (h)}$ or in other
words we defined the coordinates in terms of the scale invariant
combination $C\Omega$. Somewhat surprisingly there seem to be two 
physically inequivalent ways to fix this scale invariance.
In $N=2$ one uses the scale invariance to define special 
coordinates $z^K = Z^K/Z^0, z^0 = 1$ where  $Z^0$ is the coefficient
in front of the base element $\alpha_0$. The choice of $Z^0$
is convention and 
due to the symplectic invariance any other choice would be
equally good. 
However, as we already discussed in section 3.1 and 3.3 the 
constraint \eqref{constrO} breaks the symplectic invariance and  
$H^3$ decomposes into two eigenspaces $H^3_+\oplus H^3_-$.
Thus in \eqref{decompO2} we have the choice to scale one of the $Z^k$ 
equal to one or 
one of the $Z^\lambda$ equal to $i$.
Denoting the corresponding basis element by $\alpha_0$, 
these two choices are characterized by 
$\alpha_0 \in H^{3}_+$ or $\alpha_0 \in H^{3}_-$.
This choice identifies the dilaton direction inside the moduli space
and therefore is crucial in identifying the type IIB
mirror. This is related to the fact that in type IIB
the dilaton reside in a chiral multiplet for $O3/O7$ orientifolds and in a
linear multiplet for $O5/O9$ orientifolds as we make more explicit in section 
\ref{IIB_lin}. Let us discuss these two cases in turn.

%
%

\subsubsection{The Mirror of IIB orientifolds with $O3/O7$ planes}
\label{O3O7mirror}

We first want to show that in the large complex structure limit
$K^Q_A$ given in \eqref{intKQ} coincides with
$K^{\rm Q}_B$ given in \eqref{kaehlerpot-Kk} for 
orientifolds with $O3/O7$ planes.
It turns out that in order to do so we need to choose
$\alpha_0 \in H^3_{+}$ and the dual basis element
$\beta^0\in H^3_{-}$.
It is convenient to keep track of this choice and therefore
we mark the $\alpha$'s and $\beta$'s which contain $\alpha_0$
and $\beta^0$ by putting a hat on the corresponding index. 
Thus we work in the basis $(\alpha_\kh,\beta^\lambda)$ 
of $H^3_+$ and $(\alpha_\lambda,\beta^\kh)$
of $H^{3}_-$. Therefore, we rewrite the combination $C\Omega$ as 
\beq
  C\Omega = g_A^{-1}(\textbf{1}\, \alpha_0 + q^k \alpha_k + iq^\lambda \alpha_\lambda) + \ldots\ ,
\eeq
where we introduced $g_A$ and the real special coordinates 
\beq \label{realspC1}
  g_A =\frac{1}{\R(CZ^0)}\ ,\qquad q^k = \frac{\R(CZ^k)}{\R(CZ^0)}\ , \qquad q^\lambda = \frac{\I(CZ^\lambda)}{\R(CZ^0)}\ .
\eeq
We also need to express the prepotential $\cF(Z)$ 
in the special coordinates $q^k,q^\lambda$.
In analogy to \eqref{def-f} one defines a function $f(q)$ 
 such that 
\beq \label{def-h(q)}
  \cF\big(\R[CZ^\kh],i\I[CZ^\lambda] \big)\ =\ i\big(\R[ CZ^0]\big)^2\  f(q^k,q^\lambda) \ . 
\eeq
We are now in the position to rewrite the $N=1$ coordinates 
$N^\kh,T_\lambda$ given in 
\eqref{def-NT} in terms of $g_A$ and the special coordinates $q^K$. 
Inserting \eqref{realspC1} 
into \eqref{def-NT} one obtains
\beq \label{c-in-q37}
  N^0\ =\ \tfrac{1}{2} \xi^0 + i g_A^{-1}\ , \qquad
  N^k\ =\ \tfrac{1}{2} \xi^k + i g_A^{-1} q^k \ , \qquad
  T_\lambda\ =\ i \tilde \xi_\lambda - 2 g_A^{-1} f_\lambda(q)\ ,
\eeq
where $f_\lambda$ is the first derivative of $f(q)$ with respect to $q^\lambda$. 

The final step is to specify $f(q)$ in the large complex structure
limit. 
In this limit the $N=2$ prepotential is known to be
\beq \label{N=2pre}
 \cF(Z) = \tfrac{1}{6} (Z^0)^{-1}{\kappa_{KLM} Z^K Z^L Z^M}\ .
\eeq
Inserted into the orientifold constraints
\eqref{Z=0gen} one infers
\beq \label{vankappa37}
  \kappa_{klm} = \kappa_{\kappa \lambda l} = 0 \ , 
\eeq 
while $\kappa_{\kappa \lambda \mu}$ and $\kappa_{\kappa l m}$ can be non-zero.
Using \eqref{vankappa37}, \eqref{def-h(q)} and \eqref{realspC1} 
we arrive at
\beq\label{fori37}
  f(q)\ =\ - \tfrac{1}{6} \kappa_{\kappa \lambda \mu} q^\kappa q^\lambda q^\rho 
           + \tfrac{1}{2} \kappa_{\kappa kl} q^\kappa q^k q^l\ .
\eeq

In order to continue 
we also have to specify the range the indices $k$ and $\lambda$ 
take on the IIA side.
A priori it is not fixed and can be changed by a symplectic transformation.
Mirror symmetry demands 
\beq \label{na-nb}
 k = 1,\ldots, h^{(1,1)}_-(\tilde Y)\ , \qquad  
\lambda = 1,\ldots,h^{(1,1)}_+(\tilde Y)\ ,
\eeq
or in other words there have to be $h^{(1,1)}_-(\tilde Y)$ 
basis elements $\alpha_k$ and $h^{(1,1)}_+(\tilde Y)$ basis elements
$\beta^\lambda$ in $H^3_+(Y)$. In addition the 
non-vanishing couplings $\kappa_{\kappa \lambda \mu}$ and 
$\kappa_{\kappa l m}$
have to be identified with 
$\cK_{\kappa \lambda \mu}$ and $\cK_{\kappa l m}$ appearing
in the definition of the type IIB chiral coordinates \eqref{tau}.
With these conditions fulfilled
we can insert \eqref{fori37} into \eqref{c-in-q37} and compare with
\eqref{tau}. This leads to the identification
\beq
 N^{\kh} = (\tau, G^k) \qquad \textrm{and}\qquad
 T_{\lambda}^A = 2 T_{\lambda}^B\ ,
\eeq
which in terms of the Kaluza-Klein variables corresponds to 
\bea\label{phi=g}
 e^{\phi_B}&=& g_A \ ,\qquad  q^\lambda\ =\ v^\lambda\ ,\qquad  q^k\ =\ -b^k\ ,\nn\\
  \xi_0 &=& 2 C_0\ , \quad \xi^k=2(c^k-C_0 b^k)\ , \\
  \tilde \xi_\lambda &=& 2 \rho_\lambda - 2\cK_{\lambda kl}c^k b^l +
C_0 \cK_{\lambda kl}b^k b^l\ .\nn
\eea
With these identifications one immediately shows 
$e^{D_A} = e^{D_B}$, where $e^{D_A}$ and
$e^{D_B}$ are the four-dimensional dilatons of the type IIA and IIB theory.
This implies that the K\"ahler potentials \eqref{intKQ} and \eqref{kaehlerpot-Kk} 
of the two theories coincide in the large volume -- 
large complex structure limit. However, the corrections
away from this limit cannot be properly understood 
from a pure supergravity analysis. It is clear that 
$K^{\rm Q}_A$ includes corrections of the mirror IIB
theory but the precise nature of these corrections remains to be understood.

%
%

\subsubsection{The Mirror of IIB orientifolds with $O5/O9$ planes}
\label{O5O9mirror}

In this section we check mirror symmetry for type IIB orientifolds with 
$O5/O9$ planes with complex coordinates and K\"ahler potential determined 
in section \ref{Kpo_gaugeIIB}.
In order to find the same chiral data on the IIA side, we have to examine the 
case where $\alpha_0 \in H^3_{-}$. Therefore we choose a basis 
$(\alpha_k,\beta^{\hat \lambda})$ of $H^3_+$ and $(\alpha_{\hat \lambda},\beta^k)$
of $H^{3}_-$. We rewrite the combination $C\Omega$ in this basis as 
\beq
  C\Omega = g_A^{-1}(i\, \alpha_0 + i q^\lambda \alpha_\lambda  + q^k \alpha_k) + \ldots
\eeq
where we introduced the real special coordinates 
\beq \label{realspC2}
  g_A =\frac{1}{\I(CZ^0)}\ ,\qquad q^k = \frac{\R(CZ^k)}{\I(CZ^0)}\ , \qquad q^\lambda = \frac{\I(CZ^\lambda)}{\I(CZ^0)}\ .
\eeq
Let us also express the prepotential $\cF(Z)$ in terms of $q^k,q^\lambda$. As in $N=2$ one defines a 
function $f(q)$ such that 
\beq \label{def-h59}
  \cF\big(\R[CZ^k],i\I[CZ^{\hat\lambda}] \big) =- i\big(\I[ CZ^0]\big)^2\,  f(q^k,q^\lambda) \ . 
\eeq
We can now rewrite the $N=1$ coordinates $T_{\hat\lambda}, N^k$ 
given in \eqref{def-NT} in terms of 
$q^k,q^\lambda$ and $g_A$ as
\bea \label{c-in-q59}
  N^k &=& \tfrac{1}{2} \xi^k + i g^{-1}_A q^k \ , \qquad T_\lambda = i \tilde \xi_\lambda +2 g^{-1}_A f_\lambda(q)\ , \nn\\
  T_0 &=& i \tilde \xi_0 + 2 g^{-1}_A (2f(q)- f_\lambda q^\lambda - f_k q^k)\ , 
\eea
where $f_\lambda,f_k$ are the first derivatives of $f(q)$ with respect to $q^\lambda$ and $q^k$. 

Going to the large complex structure limit, the $N=2$ prepotential takes the form 
\eqref{N=2pre}. We split the indices as $K=(k,\hat \lambda)$ and apply the constraints 
\eqref{Z=0gen} to find that
\beq \label{vankappa59}
  \kappa_{\kappa \lambda \mu} = \kappa_{\kappa k l} = 0 \qquad \kappa_{klm} \neq 0\ ,\qquad 
  \kappa_{\kappa \lambda l} \neq 0\ .
\eeq 
Using \eqref{vankappa59} and \eqref{def-h59} we can calculate $f(q)$ as 
\beq
  f(q) = \tfrac{1}{6} \kappa_{ k l m} q^k q^l q^m - \tfrac{1}{2} \kappa_{\kappa \lambda k} q^\kappa q^\lambda q^k\ .
\eeq
In order to match the chiral coordinates $T_0,T_\lambda,N^k$ 
with the type IIB coordinates 
of \eqref{Kcoord} we need again to specify the range of the indices
on the type IIA side. Obviously we need
\beq \label{na-nb59}
k=1, \ldots, h^{(1,1)}_+(\tilde Y)\ , \qquad  \lambda= 1,\ldots,  h^{(1,1)}_-(\tilde Y)\ ,
\eeq
which is the equivalent of \eqref{na-nb} with the plus and minus sign interchanged. 
Thus the non-vanishing intersections can be identified with 
$\cK_{klm}$ and $\cK_{\kappa\lambda k}$ on the IIB side.
Inserting $f(q)$ back into the equations \eqref{c-in-q59} for the chiral 
coordinates $N^k,T_{\hat \lambda}$ and demanding \eqref{na-nb59} one can 
compare these to the type IIB coordinates \eqref{Kcoord}. 
One identifies 
\beq
 T_{\hat \lambda} = 2(S,A_\lambda)\ ,\qquad  N^{k} = it^k \ .
 \eeq
In terms of the Kaluza-Klein modes this amounts to the identification
\bea
 g_A &=& e^{\phi_B}\ , \qquad q^k = -v^k\ , \qquad 
q^\lambda = b^\lambda\ ,\qquad 
  \xi^k = -2 c^k\ , \nn \\ 
  \tilde \xi_\lambda &=& 2\rho_\lambda - 2 \cK_{\lambda \kappa l} c^l b^\kappa\ , \qquad 
  \tilde \xi_0 = 2h + \cK_{l\lambda \kappa} c^l b^\lambda b^\kappa - \rho_\lambda b^\lambda\ .
\eea  
With these identifications one shows again $e^{D_A} = e^{D_B}$ and as
a consequence the  K\"ahler potentials \eqref{intKQ} and \eqref{kaehlerpot-KkO5} agree
in the large volume -- large complex structure limit.

In summary, we found that it is indeed possible to obtain both type IIB 
setups as mirrors of the type IIA orientifolds. 
In analogy to \eqref{pure-spinor-map}
we found by comparing \eqref{N=1coords} with \eqref{def-coordsO3} and \eqref{def-coordsO5} the mirror relation
\bea\label{pure-spinor-map2}
O3/O7: & \quad \podd\ \leftrightarrow\ \pev\ , &\qquad C_3 \leftrightarrow \fa\ ,
\nn \\
O5/O9: & \quad \podd\ \leftrightarrow\ -i \pev \ ,& \qquad C_3 \leftrightarrow \fa\ ,
\eea
where $\podd,\pev$ and $\fa$ are defined in \eqref{def-podd} and \eqref{def-A}. 
Furthermore, we found that the functionals $\Phi_A$ and $\Phi_B$ have to identified
as 
\beq \label{mirror-hitchin}
  O3/O7: \quad \Phi_A(\fu)  \leftrightarrow \ \Phi_B(\fe)\ , \qquad  
  O5/O9: \quad \Phi_B(\fu)  \leftrightarrow \ \Phi_B(\feh)\ ,
\eeq 
such that the K\"ahler potentials are matched. However, the crucial role of the two 
definitions of special coordinates remains to be understood further. 

Let us close this chapter with a brief remark on the generalizations of this result. 
Formulated in this abstract fashion equations \eqref{pure-spinor-map2} and \eqref{mirror-hitchin}
are expected to hold even for orientifolds of generalized complex manifolds. This includes certain 
$SU(3)$ structure manifolds, such as half-flat manifolds. This looks very promising and 
deserves further investigation \cite{GLprep}.

%
%

\chapter{Linear multiplets and the geometry of the moduli space}
\label{lin_geom_of_M}

In this chapter we explore the geometry of the $N=1$ moduli 
space in more detail. Our attempt is to get some deeper understanding
of the properties of the K\"ahler manifolds obtained from the 
$N=2$ to $N=1$ reduction performed in the previous chapter. Recall 
that the orientifold moduli space is a direct product 
\beq \label{mod-spaceN=1}
  \tilde \cM^{\rm SK} \times \tilde \cM^{\rm Q}\ ,
\eeq
where $N=1$ supersymmetry demands each factor to be a K\"ahler manifold.
$\tilde \cM^{\rm SK}$ is a submanifold of the $N=2$ special K\"ahler 
manifold $\cM^{\rm SK}$ parameterizing complex structure deformations 
in type IIB and complexified K\"ahler structure deformations in type IIA.  
As we have shown also $\tilde \cM^{\rm SK}$ is special K\"ahler, since  
it inherits its complex structure from $\cM^{\rm SK}$ and admits
a K\"ahler metric obtained from a prepotential. 

The reduction of the hypermultiplet
sector is more `radical' since it defines a K\"ahler manifold $\tilde \cM^{\rm Q}$
inside of a quaternionic manifold $\cM^{\rm Q}$, which itself is not necessarily K\"ahler. 
This K\"ahler submanifold has half the dimension of the quaternionic space.
In general it is a difficult 
mathematical problem to characterize K\"ahler manifolds inside quaternionic
ones \cite{AM}. However, the quaternionic manifolds obtained by Calabi-Yau compactifications
of type IIA or type IIB supergravity posses special properties. As shown 
in \cite{CFGi,FS} they can be constructed from special K\"ahler manifold $\cM^{\text{SK}}$ via the 
local c-map,
\beq \label{c-map}
  \cM^{\rm SK}_{2n} \quad \xrightarrow{\text{c-map}}\quad  \cM^{\rm Q}_{4n+4}\ ,
\eeq
where $2n$ and $4n+4$ are the real dimensions of $\cM^{\rm SK}$ and $\cM^{\rm Q}$.  
These quaternionic manifolds are termed special or dual quaternionic.
One observes that their metric depends on only half of the bosonic fields in the 
hypermultiplets, or, in other words, on half of the quaternionic coordinates.
More precisely, the components of the metrics \eqref{q-metr} and \eqref{q-metrB}
on $\cM^{\rm Q}$ are functions of only NS-NS scalar fields $M^I_{\text{NS}}$. 
The second half are R-R scalar fields  denoted by $M_{I\, \text{RR}}$ which appear in the 
quaternionic metrics only as a differential and hence posses Peccei-Quinn shift symmetries
\beq
   M_{I\, \text{RR}} \rightarrow M_{I\, \text{RR}} + c_I\ ,
\eeq
for arbitrary constants $c_I$. 

The orientifold projection truncates half of the NS-NS fields and half of
the R-R fields. $N=1$ supersymmetry forces the remaining fields to span a K\"ahler 
manifold $\tilde \cM^{\rm Q}$. Furthermore, it can be seen in tables \ref{N=1spectrumtab} 
and \ref{N=1spectrumA} 
that supersymmetry combines each NS-NS field $M^I_{\text{NS}}$ together with a R-R field 
$M_{I\, \text{RR}}$ into a chiral multiplet with bosonic 
components $M^I = (M^I_{\text{NS}}, M_{I\, \text{RR}})$ spanning $\tilde \cM^{\rm Q}$.
The fact, that the R-R fields posses shift symmetries allows us to 
chose a set $M_{\alpha \, \text{RR}}$ and dualize them into two-tensors 
$D^{\alpha}_{2\, \text{RR}}$. This amounts to replacing 
the chiral multiplets $M^\alpha$ by linear multiples 
$L^\alpha=(M^\alpha_{\text{NS}},D^{\alpha}_{2\, \text{RR}})$, while keeping the
remaining fields $M^a$ chiral. The manifold $\tilde \cM^{\rm Q}_{L^\alpha}$
spanned by the real scalars $M^\alpha_{\text{NS}}$ and the complex scalars $M^a$
still contains all the information about the full K\"ahler space $\tilde \cM^{\rm Q}$. In 
that one can construct $\tilde \cM^{\rm Q}$ starting from $\tilde \cM^{\rm Q}_{L^\alpha}$,
\beq \label{dual-map}
   \tilde \cM^{\rm Q}_{L^\alpha} \quad \xrightarrow{\ \text{dualization of } D^\alpha_2\ }\quad \tilde \cM^{\rm Q}\ .
\eeq
This dualization procedure will be discussed in section \ref{linear_multiplets}. 
As we will explain there, the kinetic terms and couplings of the chiral and linear multiplets 
can be encoded by a single function $\tilde K$ being the Legendre transform of the K\"ahler potential. 
As an application we determine $\tilde K$ for all three orientifold setups. Firstly, in 
section \ref{IIB_lin} we apply the linear multiplet formalism to IIB orientifolds. Secondly, 
in section \ref{IIA_lin} we provide the missing calculation of the K\"ahler potential for $\tilde \cM^{\rm Q}$ for
general IIA orientifolds. In this derivation we apply the techniques connected with 
the map \eqref{dual-map}.

Finally, recall that the quaternionic space can be obtained from $\cM^{\rm SK}$ 
via the local c-map construction \eqref{c-map}. In section \ref{geom_of_modspace}
we construct the map 
\beq \label{N=1c-map}
   \cM^{\rm SK} \cap \tilde \cM^{\rm Q} \quad \xrightarrow{N=1 \text{ c-map} }\quad \tilde \cM^{\rm Q}\ ,
\eeq
which can be interpreted as the $N=1$ analog of the local c-map \eqref{c-map}. 
As we will show it is closely related to the dualization
in \eqref{dual-map}, when specifying the right chiral fields $M^\alpha$ for dualization.
This construction is inspired by the one presented in \cite{Hitchin2}, where 
the moduli space of Lagrangian submanifolds with $U(1)$ connection is discussed.
Furthermore, it provides the basis to extend the analysis to non-Calabi-Yau orientifolds.

\section{Linear multiplets and Calabi-Yau orientifolds\label{linear_multiplets}}

In this section we rewrite the bulk effective action of type IIB and type IIA orientifolds  
using the linear multiplet formalism of ref.\ \cite{BGG}. 
In this way we will be able to understand the definition of the K\"ahler
coordinates given in \eqref{tau}, \eqref{Kcoord} and \eqref{def-NT} as a superfield duality transformation
and furthermore discover the no-scale properties of $K^{\rm Q}$ 
somewhat more conceptually. In an analog three-dimensional situation this has 
also been observed in \cite{BHS}.

Let us first briefly review $N=1$ supergravity coupled to $n$ linear multiplets 
$L^\alpha, \alpha=1,\ldots, n$ and
$r$ chiral multiplets $N^A, A=1,\ldots,r$ following \cite{BGG}.
Linear multiplets are defined by the constraint 
\beq\label{linearc}
(D^2-8\bar R) L^\alpha = 0 = (\bar D^2-8R) L^\alpha\ ,
\eeq
where $D$ is the superspace covariant derivative and $R$ is the chiral 
superfield containing the curvature scalar.
As bosonic components $L$ contains a  real scalar field which we also
denote by $L$ and the field strength of a
two-form $D_2$.
The superspace Lagrangian (omitting the gauge interactions) is given by 
\beq\label{actionL}
S = - 3 \int   E\, F(N^A,\bar N^A, L^\alpha) 
+ \frac12 \int \frac{E}{R}\, e^{K/2}\ W(N)
+ \frac12 \int \frac{E}{R^\dagger}\, e^{K/2}\ \bar W(\bar N)
\ ,
\eeq
where $E$ is the super-vielbein and $W$ the superpotential.
The function $F$
 depends implicitly  on the K\"ahler potential 
$K(N^A,\bar N^A, L^\alpha)$
through the differential constraint\footnote{Strictly speaking
$K(N^A,\bar N^A, L^\alpha)$ is not a K\"ahler potential 
but as we will see it determines the kinetic terms in the action.}
\bea\label{Fcon}
 1- \frac{1}{3}L^\alpha K_{L^\alpha}  = F-L^\alpha F_{L^\alpha}\ ,
\eea
which ensures the correct normalization of the Einstein-Hilbert term.
The subscripts on $K$ and $F$ denotes differentiation, i.e.\
$K_{L^\alpha} = \frac{\partial K}{\partial L^\alpha},
F_{L^\alpha} = \frac{\partial F}{\partial L^\alpha}$, etc.\ .
Let us also define the kinetic potential $\tilde K$ and rewrite \eqref{Fcon} as
\beq \label{kinpo-def}
  \tilde K = K - 3 F\ ,\qquad F = 1-\tfrac{1}{3} \tilde K_{L^\alpha} L^\alpha\ .
\eeq
Expanding \eqref{actionL} into components one finds that $\tilde K$ 
determines the kinetic terms of the fields. More precisely,
the (bosonic) component Lagrangian derived from \eqref{actionL} 
is found to be\footnote{This is a straightforward generalization
of the Lagrangian for one linear multiplet given in \cite{BGG}.
The potential for this case has also been given in \cite{HL}.}
\bea\label{kinetic_lin}
\cL &=& -\tfrac{1}{2}R*\mathbf{1} - 
  \tilde K_{A\bar B}\, dN^A \wedge * d \bar N^{B}
  + \tfrac{1}{4} \tilde K_{L^\alpha L^\beta}\, 
  dL^\alpha \wedge * dL^\beta - V * \mathbf{1}\nn\\ 
  && + \tfrac{1}{4} \tilde K_{L^\alpha L^\beta}\, dD^\alpha_2 \wedge * dD^\beta_2
     - \tfrac{i}2\,  dD^\alpha_2 \wedge 
\big(\tilde K_{\alpha A}\, dN^A -\tilde K_{\alpha \bar A}\,d\bar N^A\big)
\ ,
\eea
where 
\beq\label{Lsc}
 V = e^K \Big(\tilde K^{A \bar B}D_AW D_{\bar B}\bar W - 
(3- L^\alpha K_{L^\alpha}) |W|^2  \Big)\ .
\eeq
We see that the function
$\tilde K(N,\bar N, L) = K - 3 F$ determines the kinetic terms of the fields 
$N^A$ and $L^\alpha$ as well as the couplings of the two-forms $D^\alpha_2$ to 
the chiral fields $N^I$. Note that only derivatives of  $F_{L^\alpha}$ appear leaving a 
constant piece in  $F_{L^\alpha}$ undetermined. This constant 
drops out from \eqref{Fcon}.

In a next step we like to recover the
standard $N=1$ effective action by dualizing the linear 
multiplets $L^\alpha$ into chiral multiplets $T_\alpha$. 
This establishes the map \eqref{dual-map}, which 
will be a useful tool in the remainder of this chapter.
From here we can proceed in two ways.
We can dualize the two-forms $D^\alpha_2$ 
in components and show that the resulting action is 
K\"ahler by determining the K\"ahler potential and 
complex coordinates.
This is done in appendix \ref{linm} and provides a simple,
but somehow more tedious dualization procedure.
However, performing the duality in superspace yields 
directly the proper K\"ahler coordinates $T_\alpha$ and 
K\"ahler potential $K(T,\bar T,N,\bar N)$.
  
The duality transformation in superfields is 
performed in detail in \cite{BGG} and here we only repeat the 
essential steps.
One first considers  the linear multiplets $L^\alpha$ to be 
unconstrained real superfields and modifies the action
\eqref{actionL} to read\footnote{We omit the superpotential
terms here since they only depend on $N$ and play no role
in the dualization.}
\beq\label{actionX}
S = - 3 \int E\, \Big(F(N^A,\bar N^A, L^\alpha) + 
    6 L^\alpha(T_\alpha + \bar T_\alpha) \Big) + \ldots\ ,
\eeq
where the $T_\alpha$ are chiral superfields and in order to be consistent
with our previous conventions we have included a factor $6$
in the second term. 
Variation with respect to $T_\alpha$ results in the constraint that $L^\alpha$ are linear multiplets
and one arrives back at the action \eqref{actionL}. 
Variation with respect to the (unconstrained) $L^\alpha$ yields the 
equations\footnote{Notice that there is a misprint
in the equivalent equation given in \cite{BGG}.}
\beq \label{bGl}
  6 (T_\alpha + \bar T_\alpha)  + F_{L^\alpha}
- \tfrac{1}{3} K_{L^\alpha} 
\big(F+ 6 L^\beta (T_\beta + \bar T_\beta)\big)  =0 \ ,
\eeq
where we have used 
$\delta_{L} E = -\tfrac{1}{3} E K_{L^\alpha} \delta L^\alpha$.
This equation determines  
$L^\alpha$ in terms of the chiral superfields $N^A,T_\alpha$ and is the looked
for duality relation.
However, depending on the specific form of $F$ and $K$ 
one might not be able to solve \eqref{bGl} explicitly
for $L^\alpha$ but instead only obtain an implicit
relation  $L^\alpha(N,\bar N, T+\bar T)$.
Nevertheless one should 
insert  $L^\alpha(N,\bar N, T+\bar T)$ back  into \eqref{actionX} 
which then expresses the Lagrangian (implicitly) in terms 
of $T_\alpha$ and therefore defines a Lagrangian in the chiral superfield
formalism. 
The unusual feature being that the explicit functional dependence is 
not known. A correctly normalized Einstein-Hilbert term is ensured by 
additionally imposing
\beq \label{normeq}
  F(N,\bar N,L) + 6 L^{\alpha}(T_\alpha + \bar T_\alpha) = 1\ .
\eeq 
Contracting \eqref{bGl} with $L^\alpha$ and using equation \eqref{normeq} one obtains
\eqref{Fcon}. Thus $F$
has to have the same functional dependence as before
and therefore eqn.~\eqref{kinpo-def} is unmodified, but 
one should insert $L(N,\bar N,T+\bar T)$ implicitly 
determined by \eqref{bGl}. Using \eqref{normeq} the duality 
condition \eqref{bGl} can be cast into the form 
\beq \label{dual_coords}
    T_\alpha + \bar T_\alpha = \tfrac{1}{2}\tilde K_{L^\alpha}\ ,
\eeq 
where $\tilde K$ is the kinetic potential defined in \eqref{kinpo-def}.
We also like to rewrite the K\"ahler potential $K\big( L(N,\bar N, T +\bar T),N,\bar N\big)$
in terms of $\tilde K$. Inserting \eqref{dual_coords} into \eqref{kinpo-def} one infers
\beq \label{Kpot_dual}
  K(N,\bar N, T+\bar T) = \tilde K(N,\bar N,L) - 2(T_\alpha + \bar T_\alpha) L^\alpha\ ,
\eeq  
where we removed a constant factor by means of a K\"ahler transformation. 
Equation \eqref{dual_coords} identifies $T_\alpha + \bar T_\alpha$ to be the canonical 
conjugate to $L^\alpha$ with respect to $\tilde K$, while by \eqref{Kpot_dual} the
K\"ahler potential $K$ is the Legendre transform of $\tilde K$. 
The equations \eqref{dual_coords} and \eqref{Kpot_dual} 
characterize the map \eqref{dual-map} and can 
be equivalently obtained by a component field dualization as shown in appendix \ref{linm}.
Before turning to the orientifold examples let us calculate the the bosonic 
effective action in terms of $\tilde K$ and the coordinates
\beq \label{coordinates}
 N^A\ ,\qquad  T_\alpha=i\tilde \xi_\alpha 
+ \tfrac{1}{4} \tilde K_{L^\alpha}\ ,
\eeq 
where $\tilde \xi_\alpha$ is the scalar dual to $D_2^\alpha$ and we have used \eqref{dual_coords}.
Using the K\"ahler potential \eqref{Kpot_dual} one obtains
\bea \label{dual_lagra}
\cL &=& -\tfrac{1}{2}R*\mathbf{1} - 
  \tilde K_{N^k\bar N^l}\, dN^k \wedge * d \bar N^{l}
  + \tfrac{1}{4} \tilde  K_{L^\kappa L^\lambda}\, 
  dL^\kappa \wedge * dL^\lambda  - V * \mathbf{1} \\ 
  && + 4 \tilde K^{L^\kappa L^\lambda} \Big(d\tilde \xi_\kappa - \tfrac{1}2
  \I \big(\tilde K_{L^\kappa N^l}\,dN^l\big)\Big)\wedge * 
  \Big(d\tilde \xi_\lambda - \tfrac{1}2
  \I \big(\tilde K_{L^\lambda N^k}\,dN^k\big)\Big) \ .\nn  
\eea
where $\tilde K$ is the kinetic potential appearing in \eqref{Kpot_dual}.
This is the dual Lagrangian to \eqref{kinetic_lin} as can be equivalently shown
by component field dualization (equation \eqref{eff_act1}). 
We now give some explicit examples for this dualization, by applying it
to the Calabi-Yau orientifolds studied in chapter \ref{effective_actO}.

\subsection{Two simple examples: Type IIB orientifolds \label{IIB_lin}}

\subsubsection{I.\quad $O3/O7$ orientifolds}

Let us now restrict to simple potentials $K(N,\bar N, L)$ and 
$F(N,\bar N, L)$, which describe the correct kinematics 
for $O3/O7$ orientifolds. Our aim is to rewrite the action \eqref{S_scalarO3}
in the linear multiplet formalism. As we are going to show this enables us to
circumvent the implicit definition of the K\"ahler potential \eqref{kaehlerpot-O7-1}.
In other words, replacing the chiral multiplets $T_\alpha$ with linear multiplets $L^\alpha$
as just described allows us to give an explicit expression for $K$ in terms of $\tau,z$ and 
$L^\alpha$ \cite{TGL1}. This is achieved by the K\"ahler potential
\beq\label{KL}
  K = K_0(N^A, \bar N^{A}) + \alpha\ln (\KK_{\alpha \beta \gamma} L^\alpha L^\beta L^\gamma)\ ,
\eeq
where we
leave $K_0(N^A, N^{\bar A})$ and the normalization constant
$\alpha$ arbitrary for the moment.
Inserting \eqref{KL} into \eqref{Fcon} shows that possible solutions $F$ have the form 
\beq\label{FL}
   F=1 - \alpha + \tfrac{1}{3}\, L^\alpha \zeta^R_\alpha(N^A, \bar N^{ A}) \ ,
\eeq
where  the real functions $\zeta^R_\alpha(N^A, \bar N^{A})$ are
not further determined by \eqref{Fcon}. In that sense the 
$\zeta^R_\alpha(N^A, \bar N^{A})$ are additional input functions
which determine the Lagrangian since they appear in the
kinetic potential \eqref{kinpo-def}.
Comparing \eqref{tau} with \eqref{dual_coords} by using \eqref{KL}
and \eqref{FL} we are led to identify\footnote{Strictly
speaking \eqref{dual_coords} only determines the real part
of $T_\alpha$. The imaginary part can be found by comparing 
the explicit effective actions \eqref{S_scalarO3} and \eqref{dual_lagra}.}
\beq\label{zetaid}
\alpha = 1\ , \qquad  L^\alpha =  \tfrac{3}{2} e^{\phi}\, \frac{ v^\alpha}{\KK} \ , \qquad 
 \zeta^R_\alpha = -\frac{i}{2(\tau-\bar \tau)} \KK_{\alpha b c} (G-\bar G)^b (G- \bar G)^c\ ,
\eeq
where $\zeta^R_\alpha = \zeta_\alpha+\bar \zeta_\alpha$ was already given in \eqref{zetadef}.
Hence, we have shown that the definition of the K\"ahler coordinates in \eqref{tau}
is nothing but the duality relation \eqref{dual_coords} obtained from the superfield
dualization of  the linear multiplets $L^\alpha$ to chiral multiplets $T_\alpha$.\footnote{%
The case $\alpha=1$ is a somewhat special situation 
in that the function $F$ does not have a constant piece but only the term
linear in $L^\alpha$.
This in turn requires that the $\zeta^R_\alpha$ cannot be chosen zero but that they
have at least a constant piece so that $F$ does not
vanish. This constant is otherwise irrelevant since it
drops out of all physical quantities.
(In a slightly different context 
the case $\alpha=1$ has also been discussed  in ref.\ \cite{Binetruy}.)}
It remains to determine $K_0$. Comparing \eqref{KL} by using \eqref{zetaid}
with \eqref{kaehlerpot-O7-1} one finds 
\beq\label{K0}
K_0 =  K_{\rm cs}(z,\bar z)  -\text{ln}\big[-i(\tau - \bar \tau)\big] \ .
\eeq
In summary, the low energy effective action for $O3/O7$ orientifolds 
can be rewritten by using chiral multiplets $(z^k,\tau,G^a)$ and linear
multiplets $L^\alpha$. This supergravity theory is determined (in the formalism of ref.\
\cite{BGG} and apart from $W$ and $f$ which we can neglect for this discussion)
by the independent functions $K$ and $F$ given in \eqref{KL} and \eqref{FL} together with
\eqref{zetaid} and \eqref{K0}. 
Inserted into \eqref{kinpo-def} we determine the kinetic potential
\beq\label{tildeK_O3}
 \tilde K(z, \tau, G,L) = K_{\rm cs}(z,\bar z) + 
      \ln\Big(\frac{1}{2}\frac{\cK_{\alpha \beta \gamma} L^\alpha L^\beta L^\gamma}{l^0} \Big)  
       - \frac{\cK_{\alpha a b} L^\alpha l^a l^b}{l^0}\ ,
\eeq
where we have defined $l^a = \I G^a$ and $l^0 = \I \tau$.
In the dual formulation where the linear multiplets $L^\alpha$ are dualized
to chiral multiplets $T_\alpha$ the Lagrangian is entirely determined
by the K\"ahler potential given in \eqref{kaehlerpot-O7-1}  with the `unusual'
feature that it is not given explicitly in terms of the chiral
multiplets but only implicitly via the constraint \eqref{dual_coords}.
In this sense the orientifold compactifications 
(and similarly the compactifications of F-theory on elliptic Calabi-Yau
fourfolds considered in \cite{HL} and section \ref{F-theory}) lead to 
a more general class of K\"ahler potentials
then usually considered in supergravity.
In fact the same feature holds for arbitrary $K_0$ and arbitrary $\zeta^R_\alpha$,
such that also $O3/O7$ orientifolds with space-time filling $D3$  and $D7$
branes fall into this class as shown in \cite{GGJL,JL}.

Furthermore, these `generalized' K\"ahler potentials are all of 
`no-scale type' in that they lead to a positive semi-definite potential $V$.
For $\alpha=1$ (and arbitrary $K_0$ and $\zeta_\alpha$) 
the K\"ahler potential \eqref{KL} obeys 
\beq
L^\alpha K_{L^\alpha} = 3\ ,
\eeq
and hence the
the second term in the potential \eqref{Lsc} vanishes leaving a positive semi-definite
potential with a supersymmetric Minkowskian ground state.
Since in the chiral formulation $K$ cannot even be given explicitly one can
consider such $K$s  as a `generalized' class of 
no-scale K\"ahler potentials.
The analogous property has also been observed in refs.\ \cite{HL,BBHL,DAFT}.
Finally note with what ease the no-scale property follows in the 
linear formulation compared to the somewhat involved computation
in the chiral formulation performed in \cite{TGL1}.

\subsubsection{II.\quad $O5/O9$ orientifolds}

As second simple example let us dualize the 
effective action \eqref{S_scalarO5} of orientifolds with $O5/O9$ planes.
In this case our motivation is slightly different, since in contrast
to $O3/O7$ orientifolds, the K\"ahler potential is already given explicitly in 
terms of the K\"ahler coordinates. Recall however, that type IIB compactified
on a Calabi-Yau naturally admits a double tensor multiplet $(\phi,C_0,B_2,C_2)$
which is truncated to the linear multiplet $L=(\phi,C_2)$ by the $O5/O9$ orientifold
projection. In section \ref{IIB_orientifolds} we  dualized $C_2$ to a scalar $h$ and extracted 
the K\"ahler potential in the chiral picture. However, 
with the techniques presented above, we are now in the position to 
formulate this $N=1$ theory by keeping the linear multiplet $L$ \cite{TGL1}.   

Let us determine $\tilde K=K-3F$ encoding the couplings of the 
chiral and linear multiplets in \eqref{kinetic_lin}. As we will show 
in a moment the potential $K(N,\bar N, L)$ and the function $F(N,\bar N,L)$ are given by
\beq\label{KpotL}
  K =  K_0 + \text{ln}\, L\ , \qquad F = \tfrac{2}{3} + \tfrac{1}{3}\, L \zeta^R\ ,
\eeq
which is readily checked to be a solution of the normalization condition \eqref{Fcon}.
Comparing equation \eqref{dual_coords} for $S$ by using the Ansatz \eqref{KpotL} with the 
definition \eqref{Kcoord} of $S$ one determines $L$ and $\zeta^R$ as 
\beq \label{Lzeta_O59}
  L = \tfrac{3}{2} e^{\phi}\cK^{-1}\ , \qquad 
  \zeta^R =  \tfrac{1}{4} (A+ \bar A)_a (\text{Re}\N^{-1})^{ab} (A+ \bar A)_b\ .
\eeq
Inserted back into \eqref{KpotL} indeed yields the K\"ahler potential \eqref{O5-Kaehlerpot}
if we identify 
\beq \label{KLdetail}
  K_0 = K_{\rm cs}(z,\bar z)
        - \text{ln}\Big[\tfrac{1}{48}\KK_{\alpha \beta \gamma}(t+\bar t)^\alpha 
        (t+\bar t)^\beta (t+\bar t)^\gamma  \Big]\ , 
\eeq
Thus we have shown that the kinetic terms can consistently
be described either in the chiral- or the linear multiplet formalism
and we have determined the appropriate coordinates. 

Let us supplement our analysis with another formulation of the $O5/O9$ setups. 
Namely we like to dualize the chiral multiplet $S$ 
as well as the chiral multiplets $A_a$ into a linear multiplets $L^0$ and $L^a$. 
As we will see, this will be a first case where $F(N,\bar N,L)$ is not linear
in the linear multiplets $L^0,L^a$ in contrast to \eqref{FL} and \eqref{KpotL}.
We will show  momentarily that the K\"ahler potential still has the form 
\beq \label{K_dualA}
  K(z,t,L) = K_0(z,t) + \ln L^0\ ,   
\eeq
where $K_0$ is the same as in \eqref{KLdetail}. $F$ can be deduced from 
equation \eqref{dual_coords}, which translates to 
\beq
  \tfrac{1}{2}\tilde K_{L^0} = S + \bar S\ , \qquad \tfrac{1}{2}\tilde K_{L^a} = A_a + \bar A_a
\eeq 
Inserting \eqref{K_dualA} and the coordinates $S,A_a$ given in \eqref{Kcoord} one easily concludes that
\beq \label{F_dualA}
  L^0=\tfrac{3}{2} e^\phi \frac{1}{\cK}\ ,\qquad L^a= \tfrac{3}{2} e^\phi \frac{b^a}{\cK}\ ,\qquad 
  F =  \tfrac{2}{3} - \tfrac{1}{3} (L^0)^{-1} \cK_{\alpha a b} (t^\alpha + \bar t^\alpha) L^a L^b\ .
\eeq
where $L^0$ is equal to $L$ in \eqref{Lzeta_O59}. Together with \eqref{K_dualA} 
this is consistent with the normalization equation \eqref{Fcon}.
Inserting \eqref{K_dualA} and \eqref{F_dualA} into \eqref{kinpo-def} the kinetic potential reads
\beq \label{tildeK_O5}
  \tilde K(z,t,A,L) = K_{\rm cs}(z,\bar z) 
      - \ln\Big(\frac{1}{6} \frac{\cK_{\alpha \beta \gamma} l^\alpha l^\beta l^\gamma}{L^0} \Big) 
      + 2 \frac{\cK_{\alpha a b} l^\alpha L^a L^b}{L^0}\ ,
\eeq
where we have defined $l^\alpha = \R\, t^\alpha$. 

Let us close this discussion by comparing this 
kinetic potential with the one obtained for $O3/O7$ orientifolds in \eqref{tildeK_O3}. They are identical 
under the identifications 
\beq
  \tilde K_{O3/O7}\ \rightarrow\ -\tilde K_{O5/O9}\ ,\qquad  L^\alpha \rightarrow\ l^\alpha \ ,\qquad 
  (l^a,l^0) \rightarrow\ (L^a,L^0)\ .
\eeq
Note however, that this is a rather drastic step, since we identify linear multiplets of the one 
theory with chiral multiplets of the other. It would be interesting to explore 
this duality in more detail. It corresponds in simple cases to two T-dualities and 
manifests itself by a rotation of the forms 
\beq
  \pev \ \rightarrow\ i\pev\ , \qquad (\fe, \feh)\ \rightarrow\ (-\feh, \fe)\ .
\eeq 
This ends our discussion of IIB orientifolds. As we have seen, much of the 
underlying K\"ahler geometry can be directly analyzed by simply switching to 
the linear multiplet picture.

\subsection{An involved example: Type IIA orientifolds \label{IIA_lin}}

Let us now turn to a more involved application of the linear multiplet 
formalism or rather the Legendre transform method behind \eqref{dual_coords} 
and \eqref{Kpot_dual}. Namely, we will present a more detailed analysis 
of the moduli space $\tilde \cM^{\rm Q}$ for type IIA orientifolds \cite{TGL2}.
Our aim is to show that the K\"ahler potential \eqref{intKQ} with coordinates 
$T_\lambda,N^k$ introduced in \eqref{Oexp} indeed encodes the correct 
low-energy dynamics of the theory obtained by Kaluza-Klein reduction.
Furthermore, we show that $K^{\rm Q}$ always obeys a no-scale
type condition equivalent to \eqref{no-scale2}.

Let us start by performing the reduction of the ten-dimensional 
theory by using the general basis $(\alpha_\Kh,\beta^\Kh)$ 
introduced in \eqref{decompO2}. It was chosen such that it splits on 
$H^3(Y)=H^{3}_+ \oplus H^3_-$ as 
\beq \label{basis1}
  (\alpha_k,\beta^\lambda) \in H^{3}_+(Y)\ , \qquad  (\alpha_\lambda,\beta^k) \in H^{3}_-(Y)\ ,
\eeq
where both eigenspaces are spanned by $h^{2,1}+1$ basis vectors.
As remarked above, we will only concentrate on the moduli space 
$\tilde \cM^{\rm Q}$, such that we can set $t^a=0$ and $A^\alpha=0$.
Due to \eqref{fieldtransf}, the ten-dimensional three-form $\hat C_3$ is expanded in 
elements of $H^{3}_+(Y)$ as 
\beq
  \CC_3 = \xi^k(x)\, \alpha_k - \tilde \xi_\lambda(x)\, \beta^\lambda \ ,
\eeq
where $\xi^k, \tilde \xi_\lambda$ are $h^{2,1}+1$ real space-time scalars in 
four-dimensions. Inserting this Ansatz into the ten-dimensional effective 
action one finds
\bea  \label{act2}
  S^{(4)}_{\tilde \cM^{\rm Q}} &=& \int -\, d D \wedge * dD  -\, G_{K L}(q)\, dq^K \wedge * dq^L 
         +\tfrac{1}{2} e^{2D}\, \text{Im}\, \cM_{ k  l}\, 
         d\xi^{ k} \wedge * d\xi^{l} \\
      &&  + \tfrac{1}{2} e^{2D}\, (\text{Im}\, \cM)^{-1\ \kappa \lambda}
      \big(d\tilde \xi_\kappa - \text{Re}\, \cM_{\kappa  l}\, 
      d\xi^{l} \big)
        \wedge * \big(d\tilde \xi_\lambda-\text{Re}\, \cM_{\lambda  k}\, d\xi^{ k} \big)\ , \nn 
\eea
where compared to \eqref{act1} only the terms involving $\xi^{k},\tilde \xi_\lambda$ have
changed. The metric $G_{K L}(q)$ was introduced in \eqref{def-G}
and is the induced metric on the space of real 
complex structure deformations $\cM^{\rm cs}_\bbR$ parameterized by $q^K$. 
It remains to comment on the kinetic and coupling terms of the 
scalars $\xi^k, \tilde \xi_\lambda$. In the quaternionic metric
\eqref{q-metr} of the $N=2$ theory they couple via the 
matrix $\cM_{\Kh \Lh}$ given in \eqref{defM}. Using the split of the symplectic basis 
$(\alpha_\Kh, \beta^\Kh)$ as given in \eqref{basis1} and the fact that by Hodge duality 
for a form $\gamma \in H^{3}_+$ one finds $ * \gamma \in H^{3}_-$ one concludes
\beq
  \text{Re} \cM_{\kappa \lambda}(q) = \text{Re} \cM_{k l}(q) = \text{Im} \cM_{\kappa k}(q) = 0\ , 
\eeq
whereas $\text{Re} \cM_{k \lambda}, \text{Im} \cM_{\kappa \lambda}, \text{Im} \cM_{k l}$ 
are generally non-zero on $\cM^{\text{cs}}_{\mathbb{R}}$. The explicit form of non-vanishing
components can be obtained by restricting \eqref{gauge-c} to $\cM^{\rm cs}_\bbR$ and
using the constraints \eqref{Z=0gen}.

In order to combine the scalars $e^D,q^K$ with $\xi^k, \tilde \xi_\kappa$ into 
complex variables, we have to redefine these fields and rewrite the first two 
terms in \eqref{act2}. Thus we define the $h^{2,1}+1$ real coordinates
\beq \label{lL-def}
   L^\lambda\ =\ - e^{2D}\, \I \big[C Z^\lambda(q) \big]\ ,\qquad 
   l^k \ =\  \R\big[C Z^k(q)\big]\ ,
\eeq
which is consistent with the orientifold constraint 
\eqref{Z=0gen}. The additional factor of $e^{2D}$ was included in order 
to match the dilaton factors later on.
Using \eqref{lL-def} one calculates the Jacobian matrix 
for the change of variables $(e^D,q^K)$ to $(l^k,L^\lambda)$ as 
explicitly done in \cite{TGL2}.
It is then straight forward to 
rewrite \eqref{act2} by using the 
identities \eqref{spconst} of special geometry as
\begin{align} \label{IIA1}
 S^{(4)}_{\tilde \cM^{\rm Q}} = & \int  2 e^{-2D} \text{Im} \cM_{\kappa \lambda}\, dL^\kappa \wedge * dL^\lambda
                        +  2 e^{2D} \text{Im} \cM_{k l}\, dl^k \wedge * dl^l
                        + \tfrac{e^{2D}}{2}  \text{Im} \cM_{ k  l}\, 
                          d\xi^{ k} \wedge * d\xi^{l}  \nn\\   
       &+ \tfrac{e^{2D}}{2} \, (\text{Im}\, \cM)^{-1\ \kappa \lambda}
      \big(d\tilde \xi_\kappa - \text{Re}\, \cM_{\kappa  k}\, 
      d\xi^{k} \big)
        \wedge * \big(d\tilde \xi_\lambda-\text{Re}\, \cM_{\lambda  k}\, d\xi^{ k} \big)\ . 
\end{align}
From \eqref{IIA1} one sees that the scalars $l^k$ and $\xi^k$ nicely combine 
into complex coordinates 
\beq
   N^k\ =\ \tfrac{1}{2}\xi^k +  i l^k\ =\ \tfrac{1}{2}\xi^k +  i \R(C Z^k)\ ,
\eeq
which corresponds to \eqref{def-NT}.
In contrast, one observes that 
the metric for the kinetic terms of the
scalars $\tilde \xi_\lambda$ is exactly the inverse of the one
appearing in the kinetic terms of the scalar fields $L^\lambda$. 
Hence, comparing \eqref{IIA1} with \eqref{dual_lagra} on concludes that this 
action is obtained by dualizing a set of linear multiplets 
$(L^\lambda, D^\lambda_2)$ into chiral multiplets 
$(L^\lambda,\tilde \xi_\lambda)$. To extract $\tilde K(L, N,\bar N)$ we 
compare \eqref{IIA1} with \eqref{dual_lagra} and read off the metric
\beq \label{lLmetric}
   \tilde K_{L^\kappa  L^\lambda}\ =\ 8\, e^{-2D} \IM_{\kappa \lambda}\ , \quad 
   \tilde K_{l^k l^l}\ =\ -8\, e^{2D} \IM_{k l} \ , \quad 
   \tilde K_{L^\kappa  l^l} \ =\  - 8\, \RM_{\kappa l}\ ,
\eeq
where we have used that the metric is independent of $\xi^k,\tilde \xi_\lambda$.
This metric can be obtained from a kinetic potential of the form
\beq \label{kinpo1}
  \tilde K(L,l)\ =\ - \ln\big[ e^{-4D} \big]+ 8e^{2D}\I \big[\rho^*\cF(CZ^k)\big]\ ,
\eeq
where $\cF$ is the prepotential of the special K\"ahler manifold $\cM^{\rm cs}$
restricted to the real subspace $\cM^{\rm cs}_\bbR$. The map $\rho$ was given 
in \eqref{embmap1} and enforces the constraints \eqref{Z=0gen}. To show that $\tilde K$
indeed yields the correct metric \eqref{lLmetric} one differentiates \eqref{kinpo1}
with respect to $e^{-D},q^K$ and uses the inverse of the Jacobian matrix 
for the change of variables $(e^D,q^K)$ to $(l^k,L^\lambda)$. Applying equations
\eqref{ML-hf} one finds its first derivatives  
\beq \label{first-der}
  \tilde K_{ L^\lambda} \ =\ - 8\, \R\big[C \cF_\lambda(q) \big] \qquad
   \tilde K_{l^k} \ = \ 8\, e^{2D}\, \I\big[C \cF_{k}(q) \big]\ .
\eeq
Repeating the procedure and differentiating \eqref{first-der}
with respect to $e^{-D},q^K$ and using once again the inverse Jacobian 
one applies \eqref{def-M} to show \eqref{lLmetric}. Knowing \eqref{kinpo1}
one can also extract the functions $F(L, N, \bar N)$ and $K(L,N,\bar N)$
by applying \eqref{kinpo-def}. As we will show momentarily 
$K$ and $F = \frac13 (K- \tilde K)$ are given by
\beq
  K(L,l) = - \ln \big[ e^{-4D} \big] \ , \qquad 
  F(L,l) = - \tfrac{8}{3} e^{2D}\I \big[\rho^*\cF(CZ)\big]+\tfrac{1}{3}\ . 
\eeq
It suffices to determine $K$ which expressed in the correct coordinates 
serves as the K\"ahler potential in the chiral description. 

As explained in the beginning of this section the actual K\"ahler potential of 
$\tilde \cM^{\rm Q}$ is the Legendre transform \eqref{Kpot_dual} of $\tilde K$ with 
respect to the variables $L^\lambda$. There we also found the explicit 
definition of the complex coordinates $T_\lambda$ combining $(L^\lambda,\tilde \xi_\lambda)$.  
Using \eqref{first-der} in \eqref{dual_coords} and fixing the normalization of the 
imaginary part of $T_\lambda$ by comparing \eqref{IIA1} with \eqref{dual_lagra}
one finds
\beq
  T_\lambda = i \tilde \xi_\lambda + \tfrac{1}{4}\tilde K_{L^\lambda} 
           = i \tilde \xi_\lambda - 2\, \R\big(C F_\lambda\big) \ ,
\eeq
which coincides with \eqref{def-NT} already quoted in section \ref{Kpo_gaugeIIA}.
To give an explicit expression for  $K^{\rm Q}$ we insert  
equation \eqref{kinpo1} into \eqref{Kpot_dual}. Applying the $N=2$ identity 
$\cF=\frac12 Z^\Kh \cF_\Kh$, the constraint equations \eqref{Z=0gen} 
and \eqref{lL-def},\eqref{first-der} we rewrite 
\beq  \label{K_lL}
    K^{\rm Q}= - \ln\big[e^{-4D} \big] + \tfrac{1}{2} (l^k \tilde K_{l^k} - L^\lambda \tilde K_{ L^\lambda})\ .
\eeq
It is possible to evaluate the terms appearing in the parentheses. In order to do that 
we combine the equations \eqref{lL-def} and \eqref{first-der} to the simple form 
\bea \label{lL}
  \R\big( C \Omega \big)\ =\ l^k \alpha_k + \tfrac{1}{8} \tilde K_{L^\lambda} \beta^\lambda\ ,\quad
   e^{2D} \I\big( C \Omega \big)\ =\ -L^\lambda \alpha_\lambda - \tfrac{1}{8} \tilde K_{l^k} \beta^k\ . 
\eea
We now use equation \eqref{csmetric} and the definition \eqref{def-C} of $C$
to calculate
\beq \label{skconstr}
 2 \int_Y \R( C\Omega) \wedge \I(C\Omega) = i \int_Y C\Omega \wedge \overline{C\Omega} = e^{-2D}\ .
\eeq
Inserting the equations \eqref{lL} into \eqref{skconstr} we find 
\bea \label{lL=4} 
  L^\lambda \tilde K_{L^\lambda} - l^k \tilde K_{l^k} = 4\ .
\eea
Inserted back into \eqref{K_lL} we have shown that the K\"ahler potential
has indeed the form \eqref{intKQ}.\footnote{By using the equation \eqref{skconstr} and $*\Omega=-i\Omega$ 
it is straight forward to show $e^{-2D}=2\int \R(C\Omega)\wedge * \R(C\Omega)$}
Moreover, \eqref{lL=4} directly translates into a no-scale type condition for $K^{\rm Q}$
\bea \label{no-scale4}
  K_{w^\Kh} K^{w^\Kh \bar w^\Lh} K_{\bar w^\Lh} = 4\ ,
\eea
where $w^\Kh=(T_\kappa, N^k)$.
In order to see this, one inserts the inverse K\"ahler metric \eqref{invKm1},
the K\"ahler derivatives \eqref{Kder} and the derivatives of \eqref{lL=4} back into
\eqref{lL=4}. In other words, we were able to translate one of 
the special K\"ahler conditions present in the underlying 
$N=2$ theory into a constraint on the geometry of 
$\tilde \cM^{\rm Q}$. Two non-trivial examples fulfilling 
\eqref{lL=4} are the $O3/O7$ and $O5/O9$ kinetic potentials 
\eqref{tildeK_O3} and \eqref{tildeK_O5}. They admit this 
simple form since instanton corrections are not taken into account.

\section{The geometry of the moduli space}
\label{geom_of_modspace}

In this section we give an alternative formulation of 
the geometric structures of the moduli space $\tilde \cM^{\rm Q}$ 
which is closely related the moduli space of 
supersymmetric Lagrangian submanifolds in a Calabi-Yau 
threefold \cite{Hitchin2}.\footnote{This 
analysis can equivalently be applied to the moduli space of 
$G_2$ compactifications of 
M-theory.}  In this set-up also
the no-scale conditions \eqref{no-scale2}, 
\eqref{lL=4} are interpreted geometrically. This provides 
a more elegant description of the $N=1$ moduli space and 
its special properties. Moreover, we construct the $N=1$
analog \eqref{N=1c-map} of the $N=2$ c-map \eqref{c-map}.
Our analysis can serve as a starting point for the analysis
of non-Calabi-Yau orientifolds by using the 
language of generalized complex manifolds invented by 
Hitchin \cite{HitchinGCM}.  

In section~\ref{IIA_orientifolds} we started from a $N=2$ quaternionic
manifold $\cM^{\rm Q}$ and determined the submanifold
$\tilde\cM^{\rm Q}$ by imposing the orientifold projection.
$N=1$ supersymmetry ensured that this submanifold is K\"ahler.
$\cM^{\rm Q}$ has a second but different K\"ahler submanifold
$\cM^{\rm cs}$ which intersects with $\tilde\cM^{\rm Q}$
 on the real manifold $\cM^{\rm cs}_\bbR$.
The c-map is in some sense the reverse operation where 
$\cM^{\rm Q}$ is constructed starting from $\cM^{\rm cs}$
and shown to be quaternionic \cite{CFGi,FS}.
In this section we analogously construct the K\"ahler manifold 
$\tilde\cM^{\rm Q}$ starting from $\cM^{\rm cs}_\bbR$.

\begin{figure}[h]
\begin{center}
\includegraphics[height=4cm]{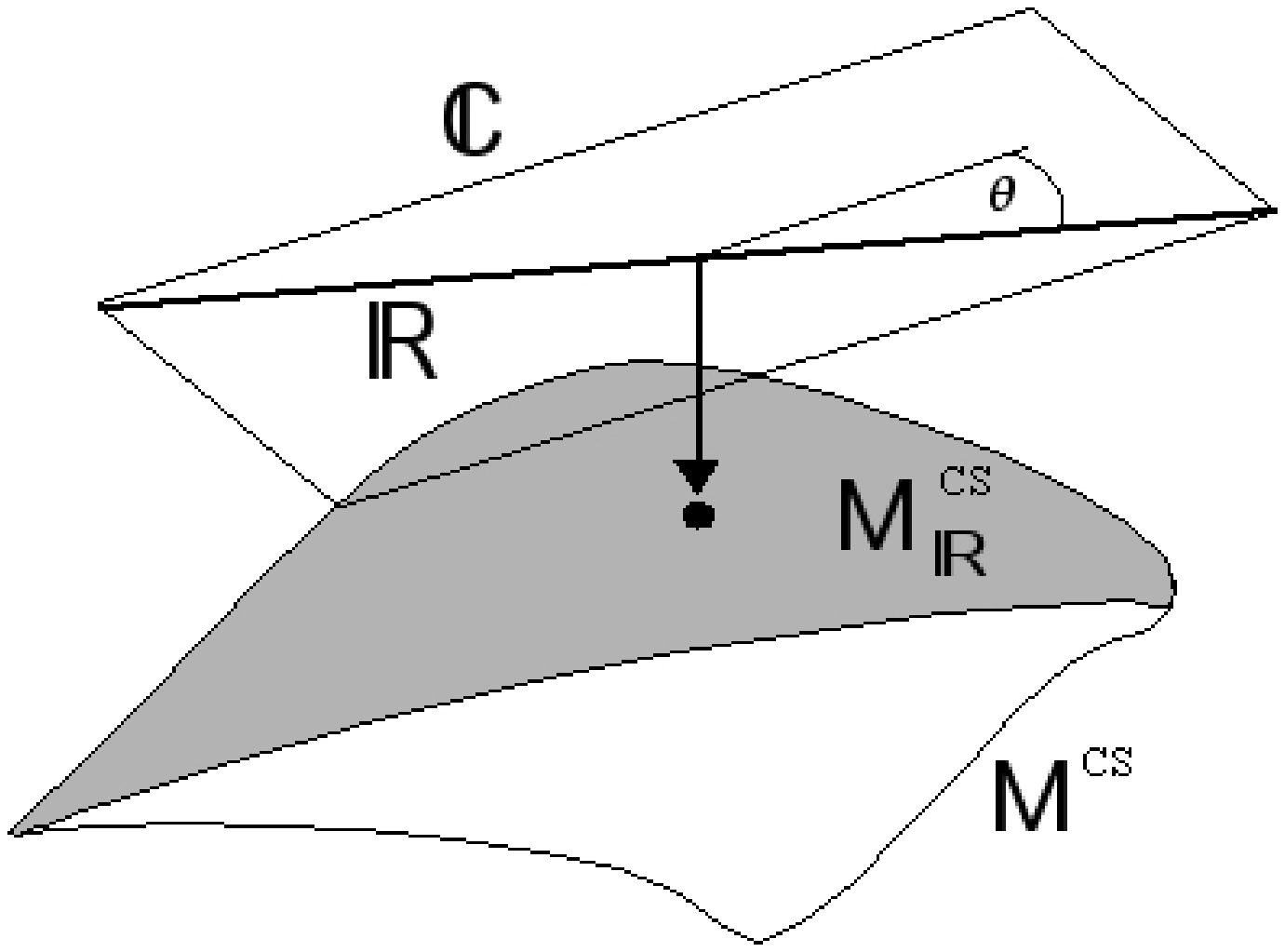}
\caption{\textit{The local moduli space $\cM_{\bbR} = \cM_{\bbR}^{\rm cs} \times \bbR$ in 
       $\cM^{\rm cs} \times \bbC \simeq \cM^{\rm cs} \times H^{(3,0)}$.}}
\label{loc_modspace}
\end{center}
\end{figure}

In fact the proper starting point is not $\cM^{\rm cs}_\bbR$ but rather
$\cM_\bbR=\cM_\bbR^{\rm cs} \times \bbR$ which is the local product of the
moduli space 
of real complex structure deformations of a Calabi-Yau orientifold 
times the real dilaton direction. The $N=2$ analog of 
$\cM_\bbR$ is the extended moduli space 
$\hat\cM^{\rm cs} = \cM^{\rm cs} \times \bbC$ where $\bbC$ 
is the complex line normalizing $\Omega$. The corresponding modulus
can be identified with the complex dilaton \cite{Witten2}.
The orientifold projection fixes the phase of the complex dilaton
(it projects out the four-dimensional $B_2$) to be $\theta$ and thus reduces
$\bbC$ to   $\bbR$ (figure \ref{loc_modspace}). 
The local geometry of $\cM_\bbR$ is encoded in the variations of the real and imaginary part of  the normalized holomorphic three-form $C\Omega$.
This form naturally defines an embedding 
\beq
  E: \cM_\bbR \rightarrow V \times V^*
= \ H^3_+(\mathbb{R}) \times H^3_-(\mathbb{R})\ .
\eeq
where $V =H^3_+(\mathbb{R})$ and we used the intersection 
form $\big<\alpha,\beta \big>=\int \alpha \wedge \beta $ on $H^{3}(Y)$ 
to identify $V^* \cong H^3_-(\mathbb{R})$. 
 $V \times V^*$ naturally admits a
symplectic form $\cW$ and an indefinite metric $\cG$ defined as
\bea
  \cW((\alpha_+,\alpha_-),(\beta_+,\beta_-)) = \big<\alpha_+, \beta_-\big> - \big<\beta_+ ,\alpha_-\big>\ , \nn\\
  \cG((\alpha_+,\alpha_-),(\beta_+,\beta_-)) = \big<\alpha_+, \beta_-\big> + \big<\beta_+ ,\alpha_-\big>\ ,
\eea
where $\alpha_\pm,\beta_\pm \in H^3_\pm(\mathbb{R})$.

Now we construct $E$ in such a way that 
$\cM_{\bbR}$ is a Lagrangian submanifold of $V \times V^*$ with respect to 
$\cW$ and its metric is induced from $\cG$, i.e.\ 
\beq\label{Lagr}
 E^*(\cW)=0 \ , \qquad E^*(\cG)=g
\eeq
where
\beq \label{metrQ}
  \tfrac{1}{2} g = dD \otimes dD +  G_{K L} d q^K \otimes d q^L\ 
\eeq 
 is the metric on $\cM_\bbR$ as determined in \eqref{act1}.
As we are going to show momentarily $E$ is given by
\bea  \label{embmap}
   E(q^\Kh) = 2\,\big(\fu ,  -e^{2D} \fuh \big)\ ,
\eea 
where $\fu+i\,\fuh = C\Omega$, $q^\Kh=(e^{-D},q^K)$ and 
$\Omega$ is evaluated at  $q^K \in \cM^{\rm cs}_\bbR$.
Additionally $E$ satisfies
\beq \label{no-scale3}
  \cG(E(q^\Kh),E(q^\Kh)) = 4\ ,
\eeq
for all $q^K $. This implies that the image of all points in 
$\cM_\bbR$ have the same distance from the origin. Later on we will show that
this translates into the no-scale condition \eqref{no-scale4}.

Let us first show that the $E$ given in \eqref{embmap} indeed satisfies
\eqref{Lagr} and \eqref{no-scale3}. The explicit calculation is straightforward 
and essentially included in the calculation presented in section \ref{IIA_lin}.\footnote{Formally one has to  
first evaluate $E_* (\partial_{Q^\Kh})$ and expresses the result in terms of the $(3,0)$-form $\Omega$ 
and the $(2,1)$-forms $\chi_K$. One then uses that by definition of the pull-back 
$E^* \omega(\partial_{q^\Kh},\cdot ) = \omega(E_* (\partial_{q^\Kh}),\cdot)$ for a form $\omega$ on $V \times V^*$. 
Applied to $\cG$ and $\cW$ one finds that the truncation of the special K\"ahler potential 
\eqref{csmetric} and \eqref{chi_barchi} indeed imply \eqref{Lagr}. This calculation does not
make use of any specific basis of $H^3_\pm$.} In order to connect with section 
\ref{IIA_lin} let us first recall how we applied the map \eqref{dual-map} to
extract the chiral data of the $N=1$ moduli space.
We started with a special K\"ahler manifold $\cM^{\rm sk}$ with metric determined 
in terms of a holomorphic prepotential $\cF(Z)$. 
Next we assumed that the $N=2$ theory with quaternionic space $\cM^{\rm Q}$ constructed 
via the local c-map \eqref{c-map} allows a reduction to $N=1$. Accordingly the section 
$\Omega(z)=Z^\Kh \alpha_\Kh - \cF_\Kh \beta^\Kh$ fulfills equation \eqref{Z=0gen} for some basis 
\beq \label{red-basis}
  (\alpha_k,\beta^\lambda) \in H^{3}_+ \ ,\qquad  (\alpha_\lambda,\beta^k) \in H^{3}_-\ .
\eeq
Using this basis we found the kinetic potential $\tilde K(L,l)$ given in \eqref{kinpo1}, which 
explicitly depends on the prepotential $\cF$. It encodes the metric on 
$\tilde \cM^{\rm Q} \subset \cM^{\rm Q}$ via the K\"ahler potential \eqref{Kpot_dual}. 
On the other hand, equation \eqref{dual_coords} defines the complex structure on 
$\tilde \cM^{\rm Q}$. 

These steps can be translated into the language of this section. Namely,
choosing the basis \eqref{red-basis} to expand the map $E$ defined in \eqref{embmap}
one finds   
\beq \label{Eincoords}
  E(q^\Kh)\ =\ \big(2l^k \alpha_k + \tfrac{1}{4} \tilde K_{L^\lambda} \beta^\lambda, 
                2L^\lambda \alpha_\lambda + \tfrac{1}{4} \tilde K_{l^k} \beta^k\big)\ ,
\eeq
where $l^k,L^\lambda$ and $\tilde K_{L^\lambda}, \tilde K_{l^k}$ are functions of $q^\Kh$ as given in 
\eqref{lL-def} and \eqref{first-der}.
We define coordinates $u^\Kh=(2l^k, \tfrac{1}{4}\tilde K_{L^\lambda})$ on $V$ and 
coordinates $v_\Kh=(\tfrac{1}{4}\tilde K_{l^k},-2L^\lambda)$ on $V^*$. 
In these coordinates the first two conditions in \eqref{Lagr} simply read
\beq \label{Lagrc}
    E^*(du^\Kh \wedge dv_\Kh)=0\ , \qquad E^*(du^\Kh \otimes dv_\Kh) = g\ .
\eeq
{}From section \ref{IIA_lin} we further know that 
$\tilde K_{L^\kappa}, \tilde K_{l^k}$ are derivatives 
 of a kinetic potential $\tilde K$ and thus we can evaluate $du^\Kh$ 
and $dv_\Kh$ in terms of $l^k,L^\kappa$.
Inserting the result into \eqref{Lagrc} 
the second equation can be rewritten as
\beq
  \tfrac{1}{2} g\ =\ \tfrac{1}{4} 
        \tilde K_{l^k l^l}\, dl^k \otimes dl^l - 
        \tfrac{1}{4} 
        \tilde K_{L^\kappa L^\lambda}\, dL^\kappa \otimes dL^\lambda\ ,
\eeq 
while the first equation is trivially fulfilled due to the symmetry of $\tilde K_{l^k l^l}$
and $\tilde K_{L^\kappa L^\lambda}$. This metric is exactly the one appearing in the action 
\eqref{IIA1} when using \eqref{lLmetric}. Expressing $g$ in coordinates 
$e^{D},q^K$ leads to \eqref{metrQ}, as we have already checked by going from 
\eqref{act2} to \eqref{IIA1} above.
Furthermore, inserting \eqref{Eincoords} into \eqref{no-scale3} 
it exactly translates 
into the no-scale condition \eqref{lL=4}, which was shown in section
\ref{IIA_lin} to be equivalent to \eqref{no-scale2}. 

We have just shown that $\cM_\bbR$ is a 
Lagrangian submanifold of  $V \times V^*$.
Identifying $T^*V \cong V \times V^*$ we conclude that  $\cM_\bbR$ 
can be obtained as the graph $(\alpha(u),u)$ 
of a closed one-form $\alpha$. This implies that we can locally find 
a generating function $K': V \rightarrow \mathbb{R}$ such that $\alpha = dK'$. 
In local coordinates $(v_\Kh,u^\Kh)$ this amounts to
\beq \label{v=K/u}
  v_\Kh(u) = \frac{\partial K'}{\partial u^{\Kh}}
\eeq 
such that 
\beq
  - L^\kappa(u) = 2\, \frac{\partial K'(u)}{\partial \tilde K_{L^\kappa}}\ , \quad 
  \tilde K_{l^k}(u) = 2\, \frac{\partial K'(u)}{\partial l^k}\ . 
\eeq
These equations are satisfied if we define $K'$ in terms of $\tilde K$ as
\bea \label{Legendre}
  2 K'\ =\ \tilde K(L(u),l) - \tilde K_{L^\kappa}(u)\, L^\kappa(u)\ , 
\eea  
which is nothing but the Legendre transform of $\tilde K$ with respect to $L^\kappa$.
Later on we show that the function ${2}K'$  is identified with the K\"ahler potential 
$K$ given in \eqref{intKQ}. 

In order to do that, we now extend our discussion to the 
full moduli space $\tilde \cM^{\rm Q}$ including the scalars 
$\zeta^\Kh=(\xi^k,\tilde \xi_\kappa)$ parameterizing the 
three-form $\hat C_3$ in $H^{3}_+(\bbR)$. Locally one has
\beq
   \tilde \cM^{\rm Q} = \cM_\bbR \times H^{3}_+(\bbR)\ .
\eeq
The tangent space at a point $p$ in $ \tilde \cM^{\rm Q}$ can be identified as
\beq
  T_p \tilde \cM^{\rm Q} \cong H^3_+(\bbR)\oplus H^3_+(\bbR) \cong H^3_+(\bbR) \otimes \bbC\ ,
\eeq
where the first isomorphism is induced by the embedding $E$ given in \eqref{embmap}.
This is a complex vector space and thus $\tilde \cM^{\rm Q}$ admits an 
almost complex structure $I$. In components it is given by
\beq \label{def-I}
  I(\partial_{q^\Kh}) = (\partial u^\Lh/\partial q^\Kh)\, \partial_{\zeta^\Lh}\ ,\qquad
  I((\partial u^\Lh/\partial q^\Kh)\, \partial_{\zeta^\Lh})=-\partial_{q^\Kh}\ ,
\eeq
where we have used that $I$ is induced by the embedding map $E$. One can show that
the almost complex structure $I$ is integrable, since  
\beq
  dw^\Kh = du^\Kh + i d \zeta^\Kh = (\partial u^\Lh/\partial q^\Kh) dq^\Kh + i d \zeta^\Kh\ ,
\eeq
are a basis of $(1,0)$ forms and $w^\Kh=u^\Kh+i\zeta^\Kh$ are complex coordinates on $\tilde \cM^{\rm Q}$. 
Using the definition of $u^\Kh$ one infers that as expected $w^\Kh = (N^k,T_\kappa)$. 
Moreover, one naturally extends the metric $g$ on $T \cM_\bbR$ to a hermitian metric 
on $T\tilde \cM^{\rm Q}$. The corresponding two-form is then given by
\beq
  \tilde \omega(\partial_{\zeta^\Lh}, \partial_{q^\Kh}) = g(I\partial_{\zeta^\Lh},\partial_{q^\Kh})\ ,
  \qquad \tilde \omega(\partial_{\zeta^\Kh},\partial_{\zeta^\Lh})= \tilde \omega(\partial_{q^\Kh},\partial_{q^\Lh})=0\ .
\eeq
Using the definition \eqref{def-I} of the almost complex structure and 
equation \eqref{Lagr}, one concludes that $\tilde \omega$ is given by
\beq \label{tildeo}
   \tilde \omega= dv_\Kh \wedge d\zeta^\Kh 
                = 2i \frac{\partial^2 K'}{\partial w^\Kh \partial \bar w^\Lh} dw^\Kh \wedge d\bar w^\Lh   \ ,
\eeq
where for the second equality we applied \eqref{v=K/u} and expressed the 
result in coordinates $w^\Kh=u^\Kh + i \zeta^\Kh$. Note that $K'$ is a function 
of $u^\Kh$ only, such that derivatives with respect to $w^\Kh$ translate to ones
with respect to $u^\Kh$. Equation \eqref{tildeo} implies that $K^{\rm Q}=2K'$ is indeed
the correct K\"ahler potential for the moduli space $\tilde \cM^{\rm Q}$. 

So far we restricted ourselves to type IIA orientifolds. However, by using the 
mirror map \eqref{pure-spinor-map2} one easily translates the above construction to IIB 
setups. In the IIB case the real manifold started with is simply the
local product $\cM^B_\bbR = \cM^{\rm ks}_\bbR \times \bbR$, where $\cM^{\rm ks}_\bbR$
is a real slice in the complexified K\"ahler cone $\cM^{\rm ks}$ and $\bbR$ parameterizes
the four-dimensional dilaton direction. $\cM^{\rm ks}_\bbR$ is locally spanned by 
the fields $v^\alpha$ and $b^a$ introduced in section \ref{IIB_orientifolds}. 
Once again we aim to find the embedding map $E$ 
\beq
  E: \cM^B_\bbR \rightarrow V \times V^*\ .
\eeq
In order to be more explicit we distinguish $O3/O7$ and $O5/O9$ setups and  
define 
\beq
  E_{O3/7}(q^\Kh)\ =\ 2\, (\fe,\ e^{2D_B} \feh)\ ,\qquad 
  E_{O5/9}(q^\Kh)\ =\ 2\, (\feh,\ e^{2D_B} \fe)\ ,
\eeq
where $\fe+i\feh = e^{-\phi} e^{-B+iJ}$ and
$q^{\Ah}=(e^{-D_B},v^\alpha,b^a)$. Correspondingly we need to set
\beq
  V_{O3/7}=H^{ev}_+\ ,\quad V_{O3/7}^*=H^{ev}_- \qquad
  V_{O5/9}=H^{ev}_-\ ,\quad V_{O5/9}^*=H^{ev}_+\ , 
\eeq
where we have abbreviated \footnote{Recall that $H^{(0,0)}_- = H^{(3,3)}_-=0$ as discussed in section 
\ref{IIB_orientifolds}.}
\beq \label{Heven}
  H^{ev}_+ = H^{(0,0)}_+ \oplus H^{(1,1)}_- \oplus H^{(2,2)}_+ \ ,\qquad 
  H^{ev}_- = H^{(1,1)}_+ \oplus H^{(2,2)}_- \oplus H^{(3,3)}_+\ .
\eeq
Given a vector space $V$ of even forms, the identification of $V^*$ with the 
respective cohomology groups is done by using the intersection form $\big<\cdot,\cdot\big>$ 
defined in \eqref{symp-form}.
To check that $E_{O3/7}$ and $E_{O5/9}$ are defined correctly, one 
proceeds in full analogy to the type IIA case. Once again, the calculation 
simplifies considerably by using the existence of the kinetic potentials \eqref{tildeK_O3} and 
\eqref{tildeK_O5}.  

Let us summarize our results. We constructed the metric and complex structure of 
the K\"ahler manifold $\tilde \cM^{\rm Q} \subset \cM^{\rm Q}$ by specifying a map 
\beq
   E: \cM_\bbR \rightarrow V \times V^*\ ,
\eeq
where $\cM_{\bbR}$ parameterizes the real four-dimensional dilaton direction times 
certain deformations of the Calabi-Yau orientifold. $V$ is an appropriately chosen 
vector space
\beq
  V_{IIA} = H^{odd}_+\ ,\qquad V_{IIB} = H^{ev}_\pm\ ,
\eeq
where $H^{odd}_+=H^3_+$ and $H^{ev}_\pm$ is given in \eqref{Heven}. 
More explicitly $E$ takes the form 
\beq
   E(q^\Kh)=2\, \big(\rho,- \hat \rho / \Phi_{A,B}\big)\ ,
\eeq
where $\Phi_{A,B}(\rho)$ is given in \eqref{def-Phi}, \eqref{def-PhiO5} and 
$\rho=(\fu,\fe,\feh)$ depending on the orientifold setup. In order 
to evaluate $\Phi_{A,B}(\rho)=e^{-2D}$ we use the definition of the four-dimensional dilaton 
\eqref{def-D_B}.
Since $\cM_{\bbR}$ is embedded as a Lagrange submanifold in $V\times V^*$ it can be locally given
by the graph of the one-form $dK'$. Moreover, since $E$ induces the metric on 
$\cM_{\bbR}$ and a complex structure on $\cM_{\bbR}\times V$ the function $2K'$ is nothing but the K\"ahler 
potential on the local moduli space $\tilde \cM^{\rm Q}=\cM_{\bbR}\times V$. Thus, the 
difficulty is to find the map $E$ or, by recalling \eqref{v=K/u}, the functional dependence 
$\hat \rho(\rho)$. This non-linear map 
\beq
  \rho\ \mapsto\ \hat \rho(\rho)\ ,   
\eeq 
lies at the heart of Hitchins approach to extract the geometry 
of even and odd forms on six-manifolds \cite{Hitchin1,HitchinGCM}. 
One may thus hope to generalize Calabi-Yau orientifolds to non-Calabi-Yau orientifolds \cite{GLprep}.

%
%

\chapter{Calabi-Yau orientifolds with NS-NS and R-R background fluxes}
\label{fluxesAB}

In this chapter we redo the reduction of type IIB and type IIA on Calabi-Yau 
orientifolds by additionally allowing for non-trivial R-R and 
NS-NS background fluxes. As we will show, these fluxes 
result in non-trivial potentials for the supergravity fields and 
can lead to charged scalars or massive tensors.

We first discuss the two type IIB setups.  In section \ref{O3_wflux} we 
show that in orientifolds with $O3/O7$ planes fluxes introduce a superpotential only. 
More intriguingly, we point out in section \ref{O5_wflux} that 
$O5/O9$ setups with background flux in general admit a superpotential as well as 
a massive linear multiplet. Thus, additionally to the kinetic terms studied in 
section \ref{IIB_lin} we find $D-$terms and a direct mass 
term for a linear multiplet \cite{TGL1,mass_tensors}. 
In both IIB orientifold cases the induced potentials 
depend only on some but not all bulk moduli fields in the theory. In order 
to find potentials for the remaining moduli one has to take non-perturbative
contributions into account. In \cite{Witten} it was argued that 
certain D-instantons induce corrections to the superpotential. 
To gain a better understanding of these corrections is subject
of various recent work \cite{non-pert,DSFGK}. 
In section \ref{non-pert_sup} we do only a very moderate step and check if the 
resulting leading order superpotentials are holomorphic in the bulk coordinates. 
Assuming a generic form of such a superpotential one might achieve that all bulk fields 
are stabilized in the vacuum \cite{KKLT,non-pert,DSFGK}. 

In type IIA orientifolds the situation is slightly different. As we 
show in section \ref{O6sup} generic NS-NS and R-R background fluxes 
induce a superpotential which depends on all bulk moduli of the theory.
Hence, appropriately chosen background fluxes could stabilize all geometric moduli 
in type IIA orientifolds. Additionally, one can attempt to include corrections due to 
non-perturbative effects. A brief discussion of superpotential contributions due
to world-sheet or D-instantons can be found in section \ref{non-pert_sup}.

\section{$O3/O7$ orientifolds: GVW superpotential}
\label{O3_wflux}

In this section we study $O3/O7$ orientifolds by also allowing
background three-form fluxes $H_3$ and $ F_3$ on the
Calabi-Yau manifold \cite{Michelson,TV,Mayr,GKP,TGL1}.
The Bianchi identities together with the equations of motion imply
that $H_3$ and $ F_3$ have to be harmonic three-forms. 
In orientifold compactifications they are further constrained 
by the orientifold
projection. {}From \eqref{fieldtransfB}
we see that for the projection given in \eqref{o3-projection}
they both have to be odd under $\sigma^*$
and hence are  parameterized by elements of $H^{(3)}_-(Y)$.\footnote{This
uses the fact that the exterior derivative on $Y$ commutes with $\sigma^*$.}
It is convenient to combine the two three-forms into a complex
$G_3$  according to 
\begin{equation}\label{fluxesB}
G_3 = F_3 -\tau H_3\ , \qquad \tau= C_0 + i e^{- \phi}\ .
\end{equation}
$G_3$ can  be explicitly expanded into a symplectic basis of $H^{(3)}_-$
as 
\beq\label{G3exp}
G_3 = m^{\hat k}\alpha_{\hat k} - e_{\hat k}\beta^{\hat k}\ , \qquad
\hat k = 0,\ldots, h^{(1,2)}\ ,
\eeq
with $2(h^{(1,2)}_-+1)$ complex  flux parameters 
\beq\label{mcomplex}
m^{\hat k} = m^{\hat k}_F -\tau  m^{\hat k}_H\ , \qquad 
e_{\hat k}  = e_{\hat k}^F -\tau  e_{\hat k}^H\ .
\eeq
However, in the following we do not need this explicit expansion and
express our results in terms of $G_3$. 

The reduction of the IIB theory is performed by replacing 
\beq
  d\hat B_2 \ \rightarrow \ d\hat B_2 + H_3\ , \qquad d\hat C_2 \ \rightarrow \ d\hat C_2 + F_3\ ,
\eeq
in the field-strengths \eqref{fieldstr}. $H_3$ and $F_3$ are the 
background value of the field strengths 
$\hat F_3$ and $\hat H_3$ but do not effect 
$\hat F_5$ since the only possible terms would be of
the form $H_3 \wedge C_2 $ or  $B_2 \wedge F_3$ 
but both $C_2$ and $B_2$ are projected 
out by the orientifold projection.\footnote{%
We neglect subtleties appearing when $\hat B_2,\hat C_2$ do not arise with 
a derivative. These can be approached along the lines of \cite{DeWGKT}.} 
The only effect of non-trivial background fluxes is the appearance of 
a potential $V$. It is manifestly positive semi-definite and found to be
\cite{TV,GKP,BBHL,DWG}
\beq\label{potential37}
 V= 
  e^{K}
  \Big( \int \Omega \wedge \bar G_3 \int \bar \Omega \wedge G_3 
  + G^{k { l}} \int \chi_{ k} \wedge G_3 
  \int \bar \chi_{ l} \wedge \bar G_3 \Big)\ ,
\eeq
where $K$ is given in \eqref{kaehlerpot-O7-1}, $\chi_{k}$ is a basis of $H^{(2,1)}_-$ defined in \eqref{cso} and 
the background flux $G_3$ is defined in \eqref{fluxesB}.
The details of the computation of $V$ can be found in \cite{GKP,TGL1}.

Strictly speaking the additional term 
$\cL^{(4)}_{\text{top}} \sim 
  \int_{Y}H_3 \wedge F_3 $
arises in the Kaluza-Klein reduction. 
However, consistency of the compactifications requires its cancellation 
against Wess-Zumino like couplings of the orientifold planes 
to the R-R flux \cite{GKP}.

Finally, one checks that the potential \eqref{potential37}
can be derived from a superpotential $W$ via the expression
given in \eqref{N=1pot} with vanishing $D$-term $D_\kappa=0$. 
For orientifolds with $c^a=b^a=0$
$W$  was  shown to be \cite{GVW,TV,GKP,BBHL,DWG}
\begin{equation}
  W(\tau,z^k) =  \int_{Y} \Omega \wedge G_3\ .
  \label{superpot}
\end{equation}
This continues to be the correct superpotential
also if  $c^a$ and $b^a$ are in the spectrum \cite{TGL1}, which is due to the 
fact that $K^{\rm Q}$ satisfies the no-scale condition \eqref{NScond}.
This ends our analysis for $O3/O7$ setups. Surprisingly, for $O5/O9$ orientifolds
the computation is more involved and forces us to once more apply and 
extend the linear multiplet techniques developed in chapter \ref{lin_geom_of_M}.

\section{$O5/O9$ orientifolds: Gaugings and the massive linear multiplet}
\label{O5_wflux}

We now turn to the effective action of $O5/O9$ orientifolds with 
background fluxes. In order to detect the 
changes due to this non-trivial background, we proceed as in the $O3/O7$ case and 
first evaluate the field strengths (\ref{fieldstr}) including 
the possibility of background three-form fluxes 
$H_3$ and $F_3$.
Since $\hat B_2$ and hence  $H_3$ is odd
it is again parameterized by $H^{(3)}_-$ while $\hat C_2$ and 
$F_3$ are even and therefore parameterized by $H^{(3)}_+$. 
As a consequence the 
explicit expansions of the background fluxes $H_3$ and $F_3$
are given by
\bea\label{mef}
H_3&=& m^k_H \alpha_k - e_k^H \beta^k\ , \qquad k = 1, \ldots, h^{(2,1)}_-\ ,
\nn\\
F_3 &=& m_F^{\hat \kappa} \, \alpha_{\hat \kappa} 
- e^F_{\hat \kappa}\, \beta^{\hat \kappa} \ ,
\qquad \hat\kappa = 0, \ldots, h^{(2,1)}_+ \ ,
\eea
where the $(m^k_H, e_k^H)$ are $2h^{(2,1)}_-$ constant flux 
parameters determining $H_3$ and 
$(m_F^{\hat \kappa}, e^F_{\hat \kappa})$ are $2h^{(2,1)}_++2$
constant flux 
parameters corresponding to $F_3$. 
Inserting \eqref{exp-B}, \eqref{expO5} and  \eqref{mef} into \eqref{fieldstr}
we obtain
\bea
 \hat H_3 &=& db^a \wedge \omega_a + m^k_H \alpha_{k} - 
             e_{k}^H \beta^{k}\ ,\qquad
 \hat F_3 \ = \ dC_2+dc^\alpha \wedge \omega_\alpha + F_3\ ,\nonumber \\
 \hat F_5 &=& dD_2^a \wedge \omega_a + \tilde F^{k} \wedge \alpha_{k}
               - \tilde G_{k}\wedge \beta^{k} 
               + d\rho_a \wedge \tilde \omega^a  \\ 
           && - db^a \wedge C_2 \wedge \omega_a - c^\alpha db^a \omega_a \wedge 
               \omega_\alpha\ ,\nn
\eea
where we defined
\beq\label{Fcech}
\tilde F^{k}= dV^{k} - m^{k}_H C_2\ , \qquad
\tilde G_{k}=dU_{k} - e_{k}^H C_2\ .
\eeq 
As in section \ref{IIB_orientifolds} the self-duality condition on $\hat F_5$
is imposed by a Lagrange multiplier \cite{DallAgata} and we eliminate 
$D^a_{2}$ and $U_{k}$ by inserting their equations
of motion into the action.
After Weyl rescaling the four-dimensional metric with a factor $\KK/6$ 
the ${N}=1$ effective action reads 
\bea\label{actiono5}
 S^{(4)}_{O5/O9} &=& \int -\tfrac{1}{2}R*\mathbf{1}-
  G_{\kappa\lambda} \; dz^{\kappa} \wedge *d\bar z^{\lambda}
  -G_{\alpha \beta} \; dv^\alpha \wedge *dv^\beta - G_{ab}\; db^a \wedge * db^b \nn \\
  &&   - \tfrac{e^{2D}}{6} \cK\, G_{\alpha \beta}\; dc^\alpha \wedge * dc^\beta
  - \tfrac{e^{-2D}}{24}  \cK\, dC_2 \wedge * dC_2 -\tfrac{1}{4} dC_2 \wedge (\rho_a db^a - b^a d\rho_a)  \nonumber \\ 
  && 
  - dD \wedge * dD 
  - \tfrac{3e^{2D}}{8\KK} G^{ab}(d \rho_a - \KK_{ac\alpha} c^\alpha db^c)
  \wedge *(d \rho_b - \KK_{bd\beta} c^\beta db^d)- V*\mathbf{1}\nn \\[.1cm]
&& 
 + \tfrac{1}{4} \text{Re}\; \mathcal{M}_{k l}\; \tilde 
  F^{k} \wedge \tilde F^{l} + 
  \tfrac{1}{4} \text{Im}\; \mathcal{M}_{k l}\; 
  \tilde 
  F^{k} \wedge * \tilde F^{l} +
  \tfrac{1}{4} e_{k} (dV^{k} +
  \tilde F^{k})\wedge C_2\ ,\nn \\
\eea
where 
\begin{align}\label{pot5}
V &=
  \frac{18i\ e^{4\phi}}{\cK^2 \int \Omega \wedge \bar \Omega }
  \left( \int \Omega \wedge F_3 \int \bar \Omega \wedge F_3 
  + G^{\kappa \lambda} \int \chi_{\kappa} \wedge F_3 
  \int \bar \chi_{\lambda} \wedge F_3 \right) \\
  &- 
   \tfrac{9\, e^{2\phi}}{\KK^2} \Big[
  m^{k}_H\, (\text{Im} \MM)_{ k l}\,   m^{l}_H +
  \big(e_{k}^H-(m_H \text{Re} \MM)_{k}\big)   \big(\text{Im} \MM\big)^{-1  k l} 
  \big(e_{l}^H-(m_H \text{Re} \MM)_{l}\big) \Big]\ . \nonumber
\end{align}
The derivation of this potential can be found in ref. \cite{TGL1}.\footnote{Note that in this class of 
orientifolds the topological term 
$\int_YH_3 \wedge F_3$ vanishes since there is no intersection between 
$H^{(3)}_+$ and $H^{(3)}_-$. Thus strictly speaking background
D-branes have to be included in order to satisfy the tadpole cancellation
condition.} 

The action \eqref{actiono5} has the standard one-form gauge invariance
$V^k\to V^k+d\Lambda^k_0$ 
but due to the modification in \eqref{Fcech} 
also a modified (St\"uckelberg) two-form gauge invariance given by
\beq\label{2gauge}
C_2\to C_2 +d\Lambda_1\ ,\qquad V^k\to V^k + m^k_H\Lambda_1\ .
\eeq
Thus for $m^k_H\neq 0$ one vector can be set to zero by an appropriate 
gauge transformation. 
This is directly related to the fact that  \eqref{actiono5}
includes mass terms  proportional to $m^k_H$ for $C_2$ arising from 
\eqref{Fcech}. In this case gauge invariance requires
the presence of Goldstone degrees of freedom which
can be `eaten' by $C_2$.\footnote{Exactly the same situation
occurs in Calabi-Yau compactifications of type IIB
with background fluxes where both $B_2$ and $C_2$
can become massive \cite{LM}.}
Finally note that the last term in \eqref{actiono5} also 
includes a standard $D=4$ Green-Schwarz term $F^k\wedge C_2$.

\subsection{Vanishing magnetic fluxes $m^k_H=0$}

The next step is to show that $S^{(4)}_{O5/O9}$ is consistent 
with the constraints of $N=1$ supergravity. However, due to the
possibility of $C_2$ mass terms this is not completely
straightforward. A massive $C_2$
is no longer dual to a scalar but rather to a vector.
We find it more convenient to keep the 
massive tensor in the spectrum and discuss 
the $N=1$ constraints in terms of a massive linear multiplet.
Before doing so, let us first discuss
the situation $m^k_H= 0$ where $\tilde F^{l}= F^{l}$ holds.
In this case the $C_2$ remains massless 
and  can be dualized to a scalar field $h$ which together with
the dilaton $ \phi$ combines to form a chiral multiplet 
$(\phi, h)$.
Using the standard dualization procedure (see section \ref{revIIB})
one obtains the effective action \eqref{S_scalarO5} plus the 
potential $V$ given in \eqref{pot5} evaluated at $m^k_H=0$.
Furthermore, due to electric NS-NS fluxes the scalar $h$ is gauged 
and we have to replace in \eqref{S_scalarO5}
\beq \label{hcov}
  dh\quad  \rightarrow \quad Dh=d h - e_k^H V^k\ .
\eeq
Hence, $h$ couples non-trivially to the gauge fields as a direct consequence
of the Green-Schwarz coupling $F^k\wedge C_2$
in \eqref{actiono5}. In the dualized
action the scalar $h$ then is charged 
under the $U(1)$ gauge transformation $h\to h + e_k^H \Lambda_0^k$ with $V^k \to V^k + d\Lambda_0^k$. 
Note that the gauge charges are set by the electric fluxes.

The K\"ahler potential \eqref{O5-Kaehlerpot} with chiral coordinates \eqref{Kcoord} 
and the gauge-couplings \eqref{fholo} remain unchanged for the theory with 
$m^k_H=0$. However, due to the non-trivial electric NS-NS fluxes the 
covariant derivative of $h$ given in \eqref{hcov} translates into the 
covariant derivative $DS = dS - i  e_{k} V^{k}$. It remains to cast 
the potential $V$ given in \eqref{pot5}, evaluated at $m^k_H=0$, into 
the standard $N=1$ supergravity form \eqref{N=1pot}.
{}From eq.\ \eqref{hcov}
we see that the axion is charged 
and as a consequence we
expect a non-vanishing $D$-term in the potential. Recall
 the general formula
for the $D$-term \cite{WB}
\beq
K_{I\bar J} \bar X^{\bar J}_k = i \partial_I D_k \ ,
\eeq
where $X^{I}$ is the Killing vector of the $U(1)$ gauge transformations 
defined as $\delta M^I = \Lambda^k_0 X_k^J \partial_J M^I$.
Inserting \eqref{O5-Kaehlerpot} 
and \eqref{Kcoord} we obtain 
\begin{equation}
  D_k = - e_{k}^H\, \frac{\partial K}{\partial \bar S} = 
  3\, e_{k}^H\, e^{\phi}\cK^{-1} \ .
\end{equation}
Using also \eqref{gauge-couplingsO3} we arrive at the $D$-term contribution 
to the potential 
\begin{eqnarray}
  \tfrac{1}{2}   (\text{Re}\; f)^{-1\ kl} D_k D_l = -
  \tfrac{9}{\cK^2} e^{2\phi} \; e_{k}^H\, (\text{Im}\; \MM)^{-1\ kl}\,
  e_{l}^H\ , \label{D-term}
\end{eqnarray}
which indeed reproduces the last term in \eqref{pot5} for 
$m^k_H=0$.

The first term in \eqref{pot5} arises from the  superpotential
\begin{equation}\label{W5}
  W= \int_{Y} \Omega \wedge F_3\ ,
\end{equation}
which follows from a calculation analog to the $O3/O7$ case \cite{TGL1}.
It is interesting  that for this class of orientifolds
the RR-flux $F_3$ results in a contribution to the superpotential while
the NS-flux $H_3$ contributes instead to a $D$-term.

%
\subsection{Non-vanishing magnetic fluxes $m^k_H\neq 0$}

Let us now turn to the case where both electric and magnetic fluxes are
non-zero and the two-form $C_2$ is massive.
In this case $C_2$ is dual to a massive vector or equivalently the massive
linear multiplet is dual to massive vector multiplet.
Here we do not discuss this duality but instead show how the couplings
of a massive linear multiplet is consistent with the action 
\eqref{actiono5} \cite{mass_tensors}.

In section \ref{IIB_lin} we already examined the kinetic terms and couplings
for the $O5/O9$ theory in the presence of one tensor multiplet $L=(\phi,C_2)$. 
We found that they are determined in terms of the generalized K\"ahler potential
and the function $F$ both given in \eqref{KpotL}.
Let us now briefly discuss the situation of a massive
linear multiplet coupled to $N=1$ vector- and chiral multiplets.
For simplicity we discuss the situation in flat space 
and do not couple the massive linear multiplet to supergravity.
However, we expect our results to generalize to the 
supergravity case. More details can be found in \cite{GGRS,mass_tensors}.

As we already said, a linear multiplet $L$ contains a real scalar (also denote by $L$)
and the field strength of a two-form $C_2$ as bosonic components. However, 
it does not contain the two-form itself
which instead is a member of the chiral `prepotential' $\Phi$ defined
as\footnote{We suppress the spinorial indices and use the convention
$D\Phi \equiv D^\alpha\Phi_\alpha$, 
$\bar D\bar\Phi \equiv \bar D^{\dot{\alpha}}\bar\Phi_{\dot{\alpha}}$.}
\beq
L= D\Phi +\bar D \bar \Phi\ , \qquad \bar D\Phi = 0\ .
\eeq
This definition solves the constraint \eqref{linearc} (in flat space).
The kinetic term for $L$ (or rather for $\Phi$) is given in 
\eqref{actionL} and a mass-term can be added via the 
chiral integral
\beq\label{Lmasst}
\cL_{m} = \tfrac14\int d^2\theta \Big[
f_{kl}(N) (W^k - 2i m^k_H\Phi)(W^l - 2i m^l_H\Phi)
+ 2 e_k^H (W^k - i m^k_H\Phi)\Phi\Big] + {\rm h.c.}\ ,
\eeq
where $W^k= -\tfrac14 \bar D^2 DV^k$ are the chiral field strengths supermultiplets
of the vector multiplets $V^k$ and $f_{kl}(N)$ are the gauge kinetic function
which can depend holomorphically on the chiral multiplets $N$.
$(m^k_H,e_k^H)$ are constant parameters which will turn out 
to correspond to the flux parameters defined in \eqref{mef}.
The Lagrangian \eqref{Lmasst} is invariant under the standard
one-form gauge invariance $V^k\to V^k +\Lambda_0^k + \bar \Lambda_0^k$
($\Lambda_0^k$ are chiral superfields)
which leaves both $W^k$ and $\Phi$ invariant.
In addition \eqref{Lmasst} has a two-form gauge invariance 
corresponding to \eqref{2gauge} given by
\beq\label{linearg}
\Phi \to \Phi +\tfrac{i}8 \bar D^2 D \Lambda_1\ ,\qquad
V^k\to V^k + m^k_H \Lambda_1\ ,
\eeq
where $\Lambda_1 $ now is a real superfield. 
{}From \eqref{linearg} we see that one entire vector multiplet
can be gauged away and thus plays the role of the Goldstone degrees
of freedom which are `eaten' by the massive linear multiplet.

In components one finds the bosonic action
\beq
\cL_{m} = -\tfrac{1}{2} \text{Re} f_{k l}\; \tilde 
  F^{k} \wedge  * \tilde F^{l} -
  \tfrac{1}{2} \text{Im} f_{k l}\; 
  \tilde 
  F^{k} \wedge \tilde F^{l} +
  \tfrac{1}{4} e_{k} (dV^{k} +
  \tilde F^{k})\wedge C_2 - V*\mathbf{1} \ ,
\eeq
where $\tilde F^{l}$ is defined exactly as in \eqref{Fcech}
and the potential $V$ receives
two distinct contributions
\beq
V=
\tfrac{1}{2}\, 
(\text{Re} f)^{-1 kl} D_{k} D_{l} + 2\, m^k_H\text{Re} f_{kl}\, m^l_H\, L^2\ ,
\qquad
D_{k} = \big(e_k^H + 2\,\text{Im}f_{kl}\, m^l_H \big)\, L \ .
\eeq
The first term arises from eliminating the $D$-terms in 
the $U(1)$ field strength $W^k$ while the second term is a 
`direct' mass term for the scalar $L$.\footnote{Note that this second term
is a contribution to the potential which is neither a $D$- nor an
$F$-term but instead a `direct' mass term whose presence is enforced
by the massive two-form.}
Inserting the $D$-term yields a second contribution to the mass term
and one obtains altogether
\bea
V&=&
\tfrac12 \big[\big(e_k^H +2\text{Im} f_{kp}\, m^p \big)
(\text{Re} f)^{-1 kl} \big( e_l^H +2 \text{Im} f_{lr}\, m^r\big)
+4 m^k_H\, \text{Re} f_{kl}\, m^l_H\big] L^2\ . \qquad 
\eea 
Using \eqref{Lzeta_O59}  and \eqref{gauge-couplingsO3} 
this precisely agrees with the second term in the potential \eqref{pot5}.

As before the first term in \eqref{pot5}  can be derived from  the superpotential
\eqref{W5}. This ends our discussion of 
type IIB orientifolds in a general NS-NS and R-R flux background. 
As we have seen, switching on fluxes yields a potential for only part of 
the moduli fields. This changes in IIA orientifolds to which we will 
turn now.

\section{$O6$ orientifolds: Flux superpotentials}
\label{O6sup}

In this section we derive the effective action of type IIA orientifolds
in the presence of background fluxes. 
For standard $N=2$ Calabi-Yau compactifications of type IIA a
similar analysis is carried out in refs.\ \cite{LM,KachruK}.
In order to do so
we need to start from the ten-dimensional action of massive 
type IIA supergravity which differs from the action \eqref{10dact} in
that the two-form $\hat B_2$ is massive. In the 
Einstein frame it is given by \cite{Romans}
\bea \label{10dactm}
  S^{(10)}_{MIIA} &=& \int -\tfrac{1}{2}\hat R*\mathbf{1} -\tfrac{1}{4} d\hat \phi\wedge * d\hat \phi
  -\tfrac{1}{4} e^{-\hat \phi}\hat H_3 \wedge *\hat H_3 
  -\tfrac{1}{2} e^{\frac{3}{2} \hat \phi}\hat F_2 \wedge *\hat F_2 \nn \\
  && -\tfrac{1}{2} e^{\frac{1}{2} \hat \phi}\hat F_4 \wedge *\hat F_4 
  -\tfrac{1}{2} e^{\frac{5}{2} \hat \phi}\, (m^0)^2 * \mathbf{1} + \cL_{\rm top}\ ,
\eea
where
\bea
  \cL_{\rm top}&=& -\tfrac{1}{2}\Big[ \hat B_2 \wedge d\hat C_3 \wedge d\hat C_3\  
                                   -(\hat B_2)^2 \wedge d\hat C_3 \wedge d\hat C_1
                                   + \tfrac{1}{3}(\hat B_2)^3 \wedge (d\hat C_1)^2 \nn \\
               & &                 - \tfrac{m^0}{3}(\hat B_2)^3 \wedge d\hat C_3
                                   + \tfrac{m^0}{4}(\hat B_2)^4 \wedge d\hat C_1
                                   + \tfrac{(m^0)^2}{20}(\hat B_2)^5 \Big]\ ,
\eea
and the field strengths are defined as
\bea \label{defHFF}
  \hat H_3 = d \hat B_2\ , \quad \hat F_2 = d\hat C_1+m^0 \hat B_2\ , \quad 
  \hat F_4 = d\hat C_3 - \hat C_1 \wedge \hat H_3-\tfrac{m^0}{2}(\hat B_2)^2\ .
\eea
Compared to the analysis of the previous section we now include
non-trivial background fluxes of the field strengths
$F_2$, $H_3$ and $F_4$ on the Calabi-Yau orientifold.
We keep the Bianchi identity and the equation of motion intact 
and therefore expand $F_2$, $H_3$ and $F_4$
in terms of  harmonic forms compatible with the orientifold
projection. From \eqref{fieldtransf} we infer that $F_2$ is expanded in
harmonic forms of $H^{2}_-(Y)$, 
$H_3$ in harmonic forms of $H^3_{-}(Y)$ and $F_4$ in harmonic forms
of $H^{4}_+(Y)$.\footnote{As we observed in the previous section
there is no $\hat C_1$
due to the absence of one-forms on the orientifold. 
Nevertheless its field strength $F_2$ 
can be non-trivial on the orientifold since $Y$ generically possesses
non-vanishing harmonic two-forms.}
 Explicitly the expansions read 
\bea \label{fluxesA}
 H_3\, =\, q^\lambda \alpha_\lambda - p_k\, \beta^k\ , \quad   F_2\, =\, -m^a \omega_a\ , \quad 
  F_4\, =\, e_a\, \tilde \omega^a\ ,
\eea
where $(q^\lambda,p_k)$ are $h^{(2,1)}+1$ real NS flux parameters 
while $(e_a,m^a)$ are $2h^{1,1}_-$ real RR flux parameters.
The harmonic forms $(\alpha_\lambda, \beta^k)$ are the elements of the real
symplectic basis of $H^3_-$ introduced in \eqref{sp_alpha-beta}. 
The basis $\tilde \omega^a$ of
$H^{(2,2)}_+$ is defined to be the dual basis of $\omega_a$ while the
basis $\tilde \omega^\alpha$ denotes a basis of $H^{(2,2)}_-$ dual to $\omega_\alpha$. 

Inserting \eqref{expJB}, \eqref{form-exp} and \eqref{fluxesA} into
\eqref{defHFF} we arrive at
\bea \label{fieldst}
  \hat H_3 &=& db^a\wedge \omega_a + q^\lambda \alpha_\lambda - p_k\, \beta^k\ ,  \qquad \qquad 
  \hat F_2 = (m^0 b^a + m^a)\, \omega_a\ ,\\
  \hat F_4 &=& dC_3 + dA^\alpha \wedge \omega_\alpha 
  + d\xi^k \wedge \alpha_k - 
            d\tilde \xi_\lambda \wedge \beta^\lambda +  
   \big(b^a m^b  -\tfrac12 m^0 b^a b^b\big)\, 
\cK_{abc}\tilde \omega^c + e_a\, \tilde \omega^a\ , \nn
\eea
where we have used $\omega_a \wedge \omega_b = \cK_{abc}\, \tilde \omega^c$.
Now we repeat the KK-reduction of the previous section using the
modified field strength 
\eqref{fieldst} and the action \eqref{10dactm} instead of \eqref{10dact}.
This results in%
\footnote{The action $S^{(4)}_{O6}$ is given in \eqref{act1}. However, due to the fact that 
we perform the Kaluza-Klein reduction in the generic basis introduced in \eqref{sp_alpha-beta} the kinetic 
terms for $\tilde \cM^{\rm Q}$ are replaced by \eqref{act2}. }
\beq\label{Sflux}
S^{(4)} = S^{(4)}_{O6} - \int  \tfrac{g}{2}\, d\cc_3 \wedge * d\cc_3 + {h}\, d\cc_3 +
 U * \mathbf{1}\ ,
\eeq
where $S^{(4)}_{O6}$ is given in \eqref{act1}.
$\cc_3$ 
is the four-dimensional part of the ten-dimensional 
three-form $\hat C_3$ defined in \eqref{form-exp} and 
its couplings to the scalar fields are given by
\beq
  g =  e^{-4 \phi} \left(\tfrac{\cK}6
       \right)^3\ , \qquad  h = e_a b^a + \tilde \xi_\lambda q^\lambda 
  - \xi^k p_k + \tfrac{1}{2}\R \cN_{0 \ah}\, m^\ah \ ,
\eeq
where we denoted $m^\ah=(m^0,m^a)$. The potential term $U$ of \eqref{Sflux}
is given by
\beq \label{U-pot}
  U = \tfrac{9}{\cK^2} e^{2\phi} \int_Y H_3 \wedge * H_3
        -  \tfrac{18}{\cK^2} e^{4\phi} \I \cN_{\ah \bh}\, m^\ah m^\bh
         +\tfrac{ 27 } {\cK^3} e^{4\phi} G^{ab}(e_a - \R \cN_{a\ah}\, m^\ah)(e_b - \R \cN_{b\bh}\, m^\bh)\ . 
\eeq
The matrix $\cN_{\ah \bh} (t,\bar t)$ is defined to be the corresponding 
part of the $N=2$ gauge-coupling matrix \eqref{def-cN} 
restricted to $\tilde \cM^{\rm SK}$ by applying \eqref{van-int} and \eqref{splitmetr}. 

In four space-time dimensions 
$\cc_3$ is dual to a constant which plays the role of
an additional electric flux $e_0$ in complete analogy with the
situation in $N=2$ discussed in \cite{LM}.
Eliminating $\cc_3$ in favor of $e_0$ by following \cite{LM} or \cite{BW}
the potential takes the form \cite{TGL2} 
\beq \label{V-pot1}
   V  = \tfrac{9}{\cK^2} e^{2\phi} \int H_3 \wedge * H_3 
      - \tfrac{18}{\cK^2} e^{4\phi} (\tilde e_\ah - \cN_{\ah \ch}\, m^\ch) (\I \cN)^{-1\, \ah \bh}
                                          (\tilde e_\bh -\bar \cN_{\bh
        \ch}\, m^\ch)\ ,
\eeq
where we introduced the shorthand notation 
$\tilde e_\ah=(e_0 + \xi_\lambda q^\lambda-\xi^\kh p_\kh,e_a)$ and $m^\ah=(m^0,m^a)$. 
Note that in the presence of NS flux 
one can absorb $e_0$ by shifting the fields 
$\xi,\tilde \xi$. This corresponds to adding an integral form to 
$\CC_3$ as carefully discussed in \cite{BW}. 
However, for the discussion of mirror symmetry it is more convenient to
keep the parameter $e_0$ explicitly in the action.

In order to establish the consistency with $N=1$ supergravity
one needs to rewrite $V$ given in \eqref{V-pot1} in terms of
\eqref{N=1pot} or in other words we need express $V$ in terms 
of a superpotential $W$ and appropriate $D$-terms. 
From \eqref{Sflux} we infer that turning on fluxes does not 
charge any of the fields and therefore all $D$-terms have to vanish. 
In \cite{TGL2} it was checked that the potential \eqref{V-pot1} can be entirely expressed in
terms of the superpotential 
\beq \label{superpot1}
  W\ =\  W^{\rm Q}(N,T) + W^{\rm K}(t)\ ,
\eeq 
where 
\bea \label{superpot2}
  W^{\rm Q}(N^k,T_\lambda)& =& \int_Y \Omegac \wedge H_3\ =\ 
        - 2N^k p_k - i T_\lambda q^\lambda\ , \\
  W^{\rm K}(t^a) &=& e_0 + \int_Y \Jc \wedge F_4 - \tfrac{1}{2} \int_Y \Jc \wedge \Jc \wedge F_2 
       - \tfrac{1}{6} m^0 \int_Y \Jc \wedge \Jc \wedge \Jc\ ,
\nn\\
&=& e_0 + e_a t^a + \tfrac{1}{2}\cK _{abc} m^a t^bt^c - \tfrac{1}{6} m^0  \cK _{abc} t^a t^bt^c\nn\ ,
\eea
and $\Omegac$ and $\Jc$ are defined in \eqref{N=1coords}. Using the 
definitions \eqref{symp-form} and \eqref{symp-formodd} of the skew-symmetric products $\big<\cdot,\cdot\big>$ 
for even and odd forms $W$ is rewritten as 
\beq
  W = \big<e^{\Jc}, F \big> + \big<\Omegac,H_3 \big>\ , \qquad F = m^0 - F_2 - F_4 + F_6\ , 
\eeq
where we have defined $F_6$ via $e_0= \int_Y F_6$. 
We see that the superpotential is the sum of two terms.
$W^{\rm Q}$  depends on the NS fluxes $(p_k,q^\lambda)$ of $H_3$ and the 
chiral fields $N^k,T_\lambda$ parameterizing the space $\tilde \cM^{\rm Q}$. 
$W^{\rm K}$ depends on the RR fluxes $(e_{\hat a}, m^{\hat b})$  
of $F_2$ and $F_4$ (together with $m^0$ and $e_0$) and
the complexified K\"ahler deformations $t^a$ parameterizing 
$\cM^{\rm SK}$. 
We see that contrary to the type IIB case both types of moduli, 
K\"ahler and complex structure deformations appear in the superpotential
suggesting the possibility that all moduli can be fixed in this set-up.
This was resently shown to be the case in refs.~\cite{VZ,DeWGKT}.

Let us end this section by comparing the R-R superpotentials of type IIA 
and type IIB orientifolds. Recall that for 
both IIB orientifold setups R-R fluxes induce superpotentials \eqref{superpot} and \eqref{W5} 
holomorphic in the complex structure deformations $z$. Hence, we compare 
\beq
  W_A(t) = \big<e^{\Jc}, F \big> \ , \qquad W_B(z) = \big<\Omega, F_3\big>\ ,
\eeq  
where the skew-products are defined in \eqref{symp-form} and \eqref{symp-formodd}.
As just discussed $F$ depends on $2h^{(1,1)}_- + 2$ RR fluxes $(e_\ah,m^\ah)$.
To count the flux parameters labeling $F_3$ recall that it transforms differently in the
two IIB orientifolds. $F_3$ sits in $H^{3}_-(\tilde Y)$ and is determined in terms of
$2h^{(2,1)}_-+2$ real flux parameters for the $O3/O7$ case and sits in $H^{3}_+(\tilde Y)$ 
depending on $2h^{(2,1)}_+ + 2$ real flux parameters
for the $O5/O9$ case. Therefore, the number of flux parameters matches when choosing mirror 
involutions satisfying \eqref{matchchohm}. Exchanging \cite{FMM}
\beq
  e^{\Jc}(t)\ \leftrightarrow\ \Omega(z) \ , \qquad F\  \leftrightarrow\ F_3\ ,
\eeq
as in equation \eqref{pure-spinor-map} the two superpotentials $W_A(t)$ and $W_B(z)$ get identified. 
In $N=2$ the mirror identification of the complex structure moduli space $\cM^{\rm cs}$ with the complexified K\"ahler 
moduli space $\cM^{\rm ks}$ can be used to calculate world-sheet instanton corrections to $\cM^{\rm ks}$.
It would be interesting to generalize this to $N=1$ orientifold theories which allow additionally for 
non-oriented world-sheets as discussed at the end of section \ref{IIA_orientifolds}.
In addition to world-sheet instantons also certain D-instantons induce correction terms 
to the superpotential. We will end this chapter by a few comments on their 
generic structure. 

\section{D-instanton corrections to the superpotentials}
\label{non-pert_sup}

Let us close this chapter by briefly discussing possible D-instanton
corrections to the superpotentials \eqref{superpot}, \eqref{W5} and \eqref{superpot1}. 
They can arise from wrapping $D(p-1)$-branes around $p$-cycles 
$\Sigma_p$ \cite{BBS}. In addition to corrections of the K\"ahler potential 
D-instantons induce extra superpotential terms \cite{Witten}. These 
depend on brane moduli as well as bulk fields and found recent phenomenological 
application in moduli stabilization \cite{KKLT, non-pert, DSFGK}. It would be interesting to fully 
incorporate these effects and to understand the additional contributions due to 
non-orientable world-volumes. First steps into this direction are done in the 
recent works \cite{non-pert, DSFGK}.
In this section we will take only a very moderate step and 
apply the calibration conditions to show that the D-brane action 
becomes linear in the bulk fields. This ensures holomorphicity of the induces 
superpotential terms when expressed in the proper K\"ahler variables of the 
respective orientifold setup. 

To make this more precise, recall that any correlation function
is weighted by the string-frame 
world-volume action of the wrapped Euclidean $D(p-1)$-branes
and thus includes a factor $e^{-S_{D(p-1)}}$ where
\beq \label{instact}
  S_{D(p-1)} = i\mu_{p}\, 
  \int_{\cW_{p}}\Big(d^{p} \lambda\ e^{- \hat \phi}  \sqrt{\det\big({\varphi^*(\hat g+ \hat B_2) +  \ell F}\big)} 
             - i \Em^*\Big(\sum_q \hat C_{q} \wedge e^{-\hat B_2}\Big) \wedge e^{\ell \FD}\Big)\ .
\eeq 
where $\ell=2\pi \alpha'$. This is the Euclidean analog of the 
Dirac-Born-Infeld action  \eqref{DBI} plus the Chern-Simons action \eqref{CSaction}.  
$\cW_p$ is the world-volume of the $D(p-1)$-brane and  $\varphi^*$ is the pull-back
of the map $\varphi$ which embeds $\cW_p$ into Calabi-Yau orientifold $Y$, 
$\varphi:\cW_p \hookrightarrow Y$.
We have chosen the R-R charge $\mu_p$ equal to the tension since 
the wrapped  $D(p-1)$-branes must be BPS in order to preserve $N=1$ supersymmetry.
In fact, as we already discussed in section \eqref{D-branes} there are 
additional condition arising from the requirement that the $Dp$-branes  preserves 
the same supersymmetry that is left intact
by the orientifold projections. This in turn implies 
that $O3/O7$ orientifolds can admit 
corrections from $D3$ instantons, $O5/O9$ setups from $D1$ and $D5$ instantons and 
$O6$ setups from $D2$ instantons. Moreover, these have to be calibrated 
with respect to the same forms as the internal parts of the orientifold planes.

The calibration conditions for Euclidean $D(p-1)$-branes
in a Calabi-Yau manifold have been derived in refs.\ \cite{BBS,MMMS}.
Let us first apply their results to type IIA orientifolds with $O6$ planes.
Recall that the unbroken supercharge has to be some linear combination  
$\epsilon=a^+ \epsilon_+ + a^- \epsilon_-$ of the two covariantly 
constant spinors $\epsilon_+$ and $\epsilon_-$ of the 
original  $N=2$ supersymmetry. Let us denote the relative phase 
of $a^+$ and $a^-$ by $a^-/a^+=-ie^{i\theta_{D2}}$ while the 
absolute magnitude can be fixed by the normalization of $\Omega$.
{} As forms $J$ and $\Omega$ have to obey the condition 
\beq
   J \wedge J \wedge J = \tfrac{3i}{2}e^{-2U} \Omega \wedge \bar \Omega\ 
\eeq 
at every point in the moduli space. Note however, that $J$ depends on 
K\"ahler structure deformations $v^a$ while $\Omega$ is a function of 
the complex structure deformations $q^K$. Hence, $e^U$ is a non-trivial function 
of $v^a$ and $q^K$ and from $\int J^3 =\frac{3i}{2}e^{-2U}\int \Omega\wedge\bar\Omega$
one infers 
\beq\label{Omeganorm}
e^{U}=\sqrt{2}\, e^{\frac{1}{2}(K^{\rm K}-\Kcs)}\ ,
\eeq
where K\"ahler potential $K^{\rm K}(t)$
is given in \eqref{Kks} while $\Kcs(q)$ is the restriction of the K\"ahler 
potential \eqref{csmetric} to the real slice $\cM^{\rm cs}_\bbR$.
The existence of $\epsilon$ imposes constraints
on the map $\varphi$. These BPS conditions read  \cite{BBS,MMMS}
\beq\label{sLagr-cond}
  \varphi^*(\Omega)\ =\ e^{U+i\theta_{D2}} \sqrt{\det\big({\varphi^*(\hat g+ \hat B_2) +  \ell F_2}\big)} 
                        d^3 \lambda\ , \qquad 
  \varphi^*\Jc + i 2\pi \alpha' F_2\ =\ 0\ ,
\eeq 
where $\Jc$ is given in \eqref{def-t}.
 The second condition in \eqref{sLagr-cond} enforces 
$\varphi^*(J)=0$ as well as $\varphi^*\hat B_2 + \ell F_2 =0$, such that the first equation 
simplifies to 
\beq \label{cal1}
  \varphi^*\R( e^{-i\theta_{D2}}\Omega)\ =\ e^U \sqrt{\det\big(\varphi^*\hat g\big)} d^3 \lambda\ , \qquad 
  \varphi^*\I( e^{-i\theta_{D2}}\Omega)\ =\ 0\ ,
\eeq
where we have used that the volume element on $\cW_3$ is real. For vanishing $F$ these conditions 
coincide with those displayed in equation \eqref{calcond}. Even in the general case 
\eqref{sLagr-cond} and \eqref{cal1} imply that the Euclidean $D2$ branes have to 
wrap special Lagrangian cycles in $Y$, which are calibrated with respect to 
$\R(e^{-U-i\theta_{D2}}\Omega)$. 
On the other hand, recall 
that the orientifold planes are located 
at the fixed points of the anti-holomorphic 
involution $\sigma$ in $Y$ which are
special Lagrangian cycles calibrated 
with respect to $\R(e^{-U-i\theta}\Omega)$
as was argued in  eqs.\
\eqref{OLagr} and \eqref{calibr-O6}.\footnote{$e^{-U}$ is the normalization factor which was left undetermined in \eqref{calibr-O6}.}
Thus, in order for the D-instantons to 
preserve the same linear combination of the supercharges as the orientifold, we have to 
demand  $\theta_{D2} =\theta$.
 Using this constraint and inserting the calibration conditions 
\eqref{cal1} back into \eqref{instact} one finds
\beq \label{instact2}
  S_{D2} = i\mu_3 \, 
  \int_{\cW_3} \big( \varphi^*\big[2\R( C\Omega) \big] - i \varphi^*(\hat C_3) \big)\ = \ 
  \, \int_{\cW_3} \varphi^*\Omegac \ ,
\eeq 
where $C=\frac{1}{2} e^{-\phi-i\theta} e^{-U}$ 
was defined in eqs.\ \eqref{def-C}, \eqref{4d-dilaton} and 
$\Omegac$ is given in \eqref{N=1coords}. The coefficients of $\Omegac$ 
expanded in a basis of $H^{3}_+(Y)$
are exactly the $N=1$ K\"ahler coordinates $(N^k,T_\lambda)$ introduced in \eqref{Oexp}. As a consequence the instanton action 
\eqref{instact2} is linear and thus holomorphic in these coordinates
which shows that $D2$-instantons 
can correct the superpotential.
Explicitly such corrections can be obtained by evaluating 
appropriate fermionic 2-point functions which are weighted 
by $e^{-S_{D2}}$ \cite{HM}. Applying \eqref{instact2}
and keeping only the lowest term in the fluctuations 
of the instanton one obtains corrections of the form 
\beq
   W_{D3} \propto  e^{-\int_{\Sigma_3}  \Omegac}\ , 
\eeq
where $\Sigma_3$ is the three-cycle wrapped by the $D2$ instanton.
 
This result can be lifted to M-theory by embedding Calabi-Yau orientifolds into 
compactifications on special $G_2$ manifolds.
In this case the $D2$ instantons correspond 
to membranes wrapping three-cycles in the $G_2$ space 
which do not extend in the 
dilaton direction \cite{HM,KMcG}. The embedding of IIA 
orientifolds into $G_2$ manifolds and the comparison of the
respective effective actions is the subject of section \ref{G2_embedding}.

Let us next extend this observation to IIB orientifolds. For 
simplicity we set $F=0$ for these cases, since brane fluxes would 
correct the K\"ahler coordinates as discussed e.g.\ in \cite{JL}. 
Hence, the calibration 
conditions for the respective D$(p-1)$-instantons read \cite{MMMS}
\beq \label{cal_IIB}
   \Em^*\big( e^{-B_2 + iJ} \big)_p = e^{i\theta_{D(p-1)}}  \sqrt{\det {\varphi^*(\hat g+ \hat B_2)}}\ d^p \lambda\ ,
   \qquad p=2,4,6\ ,
\eeq
where $\big( e^{-B_2 + iJ} \big)_p$ denotes the $p$-form in the sum over even forms.
In order that these instantons preserve the same supersymmetry as the orientifold planes 
we furthermore have to set $\theta_{D(p-1)} = \theta_{O(p+3)}$, where $\theta_{O(p+3)}$ is given in 
\eqref{cal_sOp}. Multiplying \eqref{cal_IIB} by $e^{-\phi}$ and comparing real and imaginary parts we find 
\beq \label{pull-e}
  \Em^*\fe_4 = e^{-\phi} \sqrt{\det {\varphi^*(\hat g+ \hat B_2)}}\ d^4 \lambda\ , 
\eeq 
where $\fe_4$ is the four-form in $\fe$ defined in \eqref{def-A} and 
we have only displayed the equation for $D3$ instantons. Furthermore, by comparing \eqref{instact} 
and \eqref{def-A} one finds that $\int_{\cW_{p}} \Em^* \fa$  exactly reproduces the Chern-Simons action,
since the vectors in the expansions of the R-R forms $C_p$ vanish when the pulled back to $\cW_p \subset Y$. 
Hence, together with \eqref{pull-e} we conclude that the instanton actions take the form 
\beq
    S_{D3}=i\mu_4 \int_{\cW_4} \Em^*\fe_4 - i \Em^* \fa = -i\mu_4\, T_\alpha\, \int_{\cW_4} \Em^* \tilde \omega^\alpha\ , 
\eeq 
where the definition of $T_\alpha$ is given in \eqref{def-coordsO3}. This shows that also in type IIB
orientifolds the $N=1$ K\"ahler coordinates defined in \eqref{def-coordsO3} and \eqref{def-coordsO5} linearize 
the instanton actions. By a similar reasoning as in the IIA case this ensures holomorphicity
of  instanton induced superpotentials in these coordinates. 

%
%

\chapter{Embedding into M- and F-theory}
\label{M-F-embedding}

In this chapter we discuss the embedding of type IIA and 
type IIB orientifolds into compactifications of M- and F-theory.
Let us first review the basic idea, by briefly introducing F- and
M-theory in the limit needed for our considerations. 

F-theory provides a geometrical interpretation of the non-perturbative 
$Sl(2,\mathbb{Z})$ symmetry \eqref{Sl2} of type IIB string theory.
Under this symmetry the complex dilaton $\tau$  
transforms in a non-trivial manner and  
can be interpreted as the complex structure modulus of 
a two-dimensional torus. In \cite{Vafa} this idea was 
put forward in arguing for a natural interpretation 
in terms of a twelve-dimensional F-theory. 
Compactifying this theory on a two-torus gives back type IIB in ten dimensions.
However, in going to lower dimensions, this torus can be fibered over the
internal manifold. Compactification of F-theory on such elliptically 
fibered manifolds $Y_{n+2} \rightarrow B_{n}$ is defined to be type IIB string
theory compactified on the base $B_n$, with a complex dilaton field $\tau$
varying over the internal manifold. One interesting case is when 
$Y_4$ is a elliptically fibered Calabi-Yau fourfold with base $B_3$. 
It was shown in \cite{Sen} that in a special limit which corresponds to 
a weak coupling limit of type IIB string theory the two-fold cover of $B_3$ 
is a Calabi-Yau manifold. Furthermore, the compactification on $B_3$ corresponds to an 
orientifold compactification with $O7$ planes and $D7$ branes, which are
located at points where the torus fibers become singular. This limit is called 
the orientifold limit 
\beq \label{orientifold_limit}
    \text{F-theory}\ /\ Y_4 \quad \xrightarrow[\text{limit}]{\text{\quad orientifold\quad }} 
    \quad \text{Type IIB}\ /\ \mathcal{O}Y_6 \ . 
\eeq
Section \ref{F-theory} is devoted to check this correspondence for the effective bulk 
actions of the two theories. However, since there is no known effective action
for F-theory we will take a detour over M-theory compactified on $Y_4$.
We compare the resulting three-dimensional effective action with the $D=3$ action 
obtained by compactifying the $O3/O7$ orientifold action on a circle. Later on 
we lift the correspondence to $D=4$ and compare it with \eqref{orientifold_limit}.   

In section \ref{G2_embedding} we discuss the embedding of Type IIA orientifolds into M-theory. 
Recall that type IIA supergravity can be obtained by compactifying 
11-dimensional supergravity (the low energy limit of M-theory) on a circle. 
Correspondingly the $D=4,N=2$ theories arising in Calabi-Yau compactifications 
are lifted as  
\beq
  \text{Type IIA}\ /\ Y_6 \quad \cong \quad  \text{M-theory}\ /\ S^1 \times Y_6 \ .
\eeq  
Hence, the immediate question is to find some analog for 
the orientifold compactifications. In order to do that, one
has to identify appropriate manifolds which upon compactification
of M-theory (understood as 11-dimensional supergravity) yield a 
four-dimensional $N=1$ theory. Recalling that the number of supersymmetries 
is related to the number of covariantly constant spinors, the only possible 
candidates are seven-manifolds with structure group or holonomy $G_2$. 
This implies that the reduction of the $SO(7)$ spinor 
representation yields one singlet, which in the case of $G_2$ holonomy is furthermore 
covariantly constant with respect to the Levi-Cevita connection.
It was argued in \cite{KMcG} that for a special class of 
$G_2$ manifolds $X$ the resulting four-dimensional theory coincides with 
the one of IIA Calabi-Yau orientifolds. Schematically one has
\beq
  \text{Type IIA}\ /\ \mathcal{O}Y_6  \quad \cong \quad  \text{M-theory}\ /\ X \ .
\eeq
In section \ref{G2_embedding} we verify this conjecture for a certain limit of the two
theories. This enables us to match the $N=1$ characteristic functions determined in 
section \ref{Kpo_gaugeIIA} for IIA orientifolds with the one obtained for $G_2$ compactifications
on $X$. As we will show, this includes the K\"ahler potential, the gauge-couplings as 
well as the flux superpotentials. In ref. \cite{TGL2} only part of the orientifold
superpotentials were found to have an origin in an M-theory compactification on a 
manifold with $G_2$ holonomy. As we will show, the remaining terms are due 
to a non-trivial fibration of a manifold with $G_2$ structure introduced in \cite{CS,CCDLM}.

%
%

\section{F-theory and $O3/O7$ orientifolds}
\label{F-theory}

In this section we discuss the embedding of $O3/O7$ orientifolds 
into a F-theory compactification, which corresponds to the limit \eqref{orientifold_limit}.
To analyze the two theories on the level of the effective bulk actions we start 
by compactifying M-theory on a Calabi-Yau four-fold. 
When shrinking the volume of the elliptic fiber the M-theory
compactification on $Y_4$ is equivalent to an F-theory 
compactification on $Y_4$. We only perform this limit at the 
very end and rather compare the two theories in three 
dimensions. In order to do that we first briefly review 
compactifications of eleven-dimensional supergravity on 
Calabi-Yau fourfolds following \cite{HL,BHS}. We determine the effective action
and characteristic functions encoding the supergravity theory.
Next we compactify the four-dimensional effective action 
of $O3/O7$ orientifolds to three dimensions on a circle. 
We are then in the position to show, that the characteristic 
data of the two three-dimensional theories coincide if we 
choose a Calabi-Yau fourfold of the form 
\beq \label{def-Z}
  Y_4 = (Y \times T^2)/\hat \sigma\ ,
\eeq
where $Y$ is a Calabi-Yau threefold and $\hat \sigma = (\sigma,-1,-1)$. 
The involution $\hat \sigma$ acts as a holomorphic isometric involution on $Y$ and
inverts both coordinates on $T^2$. Note that $Y_4$ generically admits singularities if
$\sigma$ has a non-trivial fix-point set. These have to be smoothed out which 
yields additional moduli in the theory. The analog on the orientifold are 
moduli corresponding to D-branes and orientifold planes. However, since 
we only restricted to the bulk fields we will also freeze moduli arising in 
the process of smoothing out $Y_4$ defined in \eqref{def-Z}. 
Having matched the three-dimensional theories we comment on the lift to 
$D=4$. Finally, we also include a brief discussion on the lift 
of orientifold three-form flux $G_3$ to four-form flux $G_4$.

\subsubsection{M-theory compactified on a Calabi-Yau fourfold}

Let us start by summarizing compactification of M-theory on a
Calabi-Yau fourfold by following the analysis of \cite{HL,BHS}. 
The low energy effective action of 11d supergravity is given by \cite{CJS}
\bea \label{11act}
 S^{(11)}=\int - \tfrac{1}{2} R *\mathbf{1} - \tfrac{1}{4} F_4 \wedge * F_4 
                   -\tfrac{1}{12} C_3 \wedge F_4 \wedge F_4\ ,
\eea
where $F_4=dC_3$ is the field strength of $C_3$. 
The three-form $C_3$ together with the eleven-dimensional metric are the 
only bosonic fields in the low energy description of M-theory.
Recall that the action \eqref{11act} is given to lowest order 
in $\kappa_{11}$. One-loop corrections associated to the sigma model anomaly of 
a $M5$-brane contribute additional terms to 
\eqref{11act} and induce a $C_3$ tadpole term 
$-\frac{\chi(Y_4)}{24}$ \cite{DLM,SVW}. 
This contribution can be canceled by considering setups 
with a certain number of background $M3$-branes or 
switched on background fluxes. 
However, for the moment we keep our analysis simple in sticking to
the action \eqref{11act} without extra source terms. 

The fields of the three-dimensional theory arise from the expansion 
of the eleven-dimensional supergravity fields into harmonic forms. 
For a Calabi-Yau fourfold $Y_4$, the only non-vanishing cohomologies are 
given by 
\bea \label{Z-cohom}
 H^{0}(Y_4)& =& H^{(0,0)}\ , \qquad H^{2}(Y_4)\ =\ H^{(1,1)}\ ,\qquad  H^{3}(Y_4)\ =\ H^{(2,1)} \oplus H^{(1,2)} \ ,\nn \\[.2cm]
 H^{4}(Y_4)&=&H^{(4,0)}  \oplus H^{(3,1)} \oplus H^{(2,2)} \oplus H^{(1,3)} \oplus H^{(0,4)}\ ,
\eea
with their Hodge duals $H^{5}$, $H^{6}$ and $H^{8}$.
Let us extract the spectrum obtained by expansion into harmonic basis forms of these
cohomologies. This is done in analogy to the case of type II compactifications discussed in 
chapter \ref{TypeII}.
The deformations of the metric of the fourfold respecting the 
Calabi-Yau condition split into two sets: $h^{(1,1)}(Y_4)$ real scalar K\"ahler
 structure deformations $M^\cA(x)$ and $h^{(3,1)}(Y_4)$ complex structure moduli $Z^\cK(x)$. 
Similar to \eqref{def-v} and \eqref{cs} for Calabi-Yau threefolds they parameterize  
the expansions
\beq \label{deform_4}
  J_F \ =\ M^\cA(x) e_\cA\ ,\qquad 
  \delta g_{\bi \bj} = -\frac{1}{3 ||\Omega ||^2} \bar \Omega_{F\, \bi}^{\ \ \ klm} Z^\cK(x) 
  \Phi_{\cK\, klm\bj}  
\eeq
where $J_F$ and $\Omega_F$ are the K\"ahler form and the holomorphic $(4,0)$-form on 
the Calabi-Yau fourfold. The harmonic forms $e_\cA, \cA=1,\ldots, h^{(1,1)}(Y_4)$
form a basis of  $H^{(1,1)}(Y_4)$, while $\Phi_\cK, \cK=1,\ldots, h^{(3,1)}(Y_4)$ 
form a basis of $H^{(3,1)}(Y_4)$. 
Also $C_3$ is expanded into harmonic forms via the Kaluza-Klein Ansatz
\beq \label{A3exp}
  C_3 = A^\cA(x)\wedge e_\cA + N^I(x)\, \Psi_I + \bar N^I(x)\, \bar \Psi_{ I}\ , 
\eeq
where $A^\cA(x)$ are vectors and $N^I(x)$ are complex scalars in three dimensions. 
The harmonic forms $\Psi_I,\bar \Psi_I,I=1,\ldots h^{(2,1)}$ 
define a basis of $H^{3}(Y_4)$, which can be chosen to obey
\footnote{This needs some words of justification. First, recall 
that for a complex manifold $Y_4$ the filtration $F^3(\MM) = H^{(3,0)}$,  
$F^2(\MM) = H^{(3,0)}  \oplus H^{(2,1)}$, 
etc.\ can be shown to consist of holomorphic bundles $F^i(\MM)$ over the space of complex 
structure deformations. Since $H^{(3,0)}$ is empty 
for Calabi-Yau fourfolds, $H^{(2,1)}$ is a holomorphic
bundle and one can locally choose a basis
$\psi_I(Z),\ \partial_{\bar Z^\cK}\psi_I=0 $. Hence, the holomorphic derivative is expanded as
$  \partial_{Z^\cK} \psi_I = (\sigma_\cK)^{J}_{I} \psi_J  + 
  (\lambda_\cK)^{\bar J}_{I} \bar \psi_{\bar J}, $
where $(\sigma_\cK)^{J}_{I},\ (\lambda_\cK)^{\bar J}_{I}$ are functions of $Z,\bar Z$.
One can now show, that there exists a basis $\Psi^I = M_I^{\bar J} \bar \psi$ (for some real 
$M_I^{\bar J}$) which obeys \eqref{der_Psi}.
In order that this is the case one has to demand: $\partial_{Z^\cK}\ln M^{\bar I}_J=A_{\cK J}^{\ \ \ I}$,
$B_{\bar \cK I}^{\ \ \ \bar J}=(M^{-1})^{\bar J}_K M^{\bar L}_I (\bar \lambda_{\bar \cK})^{K}_{\bar L}$
and $A_{\cK I}^{\ \ \ \bar J}=-(\sigma_\cK)^{J}_{I}$. A possible definition of $M^{\bar I}_J$ can
be found in \cite{HL}.}
\beq \label{der_Psi}
  \partial_{Z^{\cK}} \Psi_I = A_{\cK  I}^{\ \ \  J} \Psi_J \ , \qquad
  \partial_{\bar Z^{\cK}} \Psi_I = B_{\bar \cK I}^{\ \ \ \bar J} \bar  \Psi_J\ ,  
\eeq
where $A_{\cK  I}^{\ \ \ J}$ and $B_{\bar \cK I}^{\ \ \ \bar J}$ are model dependent 
functions of $Z$ and $\bar Z$. Differentiating these equations with respect 
to $Z^{\cK}$ and $\bar Z^{\cL}$ and comparing 
$\partial_{Z^\cK} \partial_{\bar Z^\cL} \Psi_I$ with $\partial_{\bar Z^\cL} \partial_{Z^\cK} \Psi_I$
we extract the consistency conditions
\beq \label{DE}
  \partial_{\bar Z^{\cK}}  A_{\cL  I}^{\ \ \  J} = B_{\bar \cK I}^{\ \ \ \bar L} \, \bar B_{\cL \bar L}^{\ \ \ J}\ ,
  \qquad
  \partial_{\bar Z^{\cK}} \bar B_{\cL \bar I}^{\ \ \ J} = A_{\bar \cK \bar I}^{\ \ \  \bar L} \bar B_{\cL \bar L}^{\ \ \ J}\ .
\eeq
In summary, the bosonic part of the $D=3,N=2$ supergravity spectrum obtained by compactification 
on a Calabi-Yau fourfold is displayed in table \ref{Mspectrum}.
   
\begin{table}[h] 
\begin{center}
\begin{tabular}{|l|c|c|c|} \hline 
 \rule[-0.3cm]{0cm}{0.9cm}
 {gravity multiplet} &   1  & $g^{(3)}_{pq}$\\ \hline
 \rule[-0.3cm]{0cm}{0.9cm}
 {vector multiplets} &    $h^{(1,1)}$ & $(M^\cA,A^\cA)$ \\ \hline
\rule[-0.3cm]{0cm}{0.9cm} 
 chiral multiplets & $h^{(3,1)} + h^{(2,1)}$ & $Z^\cK$, $N^I$ 
 \\ \hline
\end{tabular} 
\caption{\textit{$D=3,N=2$ spectrum for M-theory 
                 on a Calabi-Yau fourfold.}}\label{Mspectrum}
\end{center}
\end{table} 

Also the calculation of the three-dimensional low energy effective action is similar to
the analysis performed in chapter \ref{TypeII}. 
The field strength $F_4=dC_3$ is evaluated by using \eqref{A3exp} and \eqref{der_Psi}
as
\beq \label{fieldstr_4}
  F_4 = dA^\cA \wedge e_\cA + D N^I \Psi_I +  D \bar N^I \bar \Psi_I\ , 
\eeq
with
\beq
 DN^I = dN^I + (N^J  A_{\cK J}^{\ \ \ I} + \bar N^J B_{\cK \bar J}^{\ \ \ I}) dZ^\cK \ , \qquad D\bar N^I = \overline{DN^I}
\eeq
Inserting \eqref{deform_4}, \eqref{fieldstr_4} and \eqref{A3exp} and performing the standard Weyl rescaling
the effective action takes the form \cite{HL}
\bea \label{F-theory_act}
  S^{(3)}_{F}&= &\int -\tfrac{1}{2} R - G_{\cK \cL}\, dZ^\cK \wedge * dZ^\cL 
                       - \tfrac{1}{2} d \ln \cV \wedge * d \ln \cV - \tfrac{1}{2} G_{\cA \cB}\, dM^\cA \wedge * dM^\cB\nn\\ 
                &&       - \tfrac{1}{2} \cG_{I\bar J}\ D N^I \wedge * D\bar N^J 
                         - \tfrac{1}{2} \cV^2\, G_{\cA \cB}\ dA^\cA \wedge dA^\cB \nn \\
                &&       + \tfrac{i}{4} d_{\cA I \bar J}\ dA^\cA \wedge (N^I D\bar N^J - \bar N^I  D N^J)\ ,
\eea
where $G_{\cK \cL}$, $\cG_{I\bar J}$ and $G_{\cA \cB}$ are the metrics on $H^4$, $H^3$ and $H^2$ respectively
and will be discussed in turn.
Let us first comment on the complex structure and K\"ahler structure deformations. 
The higher-dimensional analog 
of \eqref{csmetric} is the metric $G_{\cK \cL}$ on the space of complex structure 
deformations of $Y_4$. It is K\"ahler and takes the form
\beq
  G_{\cK \bar \cL} = \partial_{Z^\cK} \partial_{\bar Z^\cL} K^{\rm cs}_F\ , \qquad  
  K^{\rm cs}_F=-\ln\big[\int_{Y_4} \Omega_F \wedge \bar \Omega_F \big]\ .
\eeq
In analogy to \eqref{Kmetric} and \eqref{int-numbers} 
we define on the space of 
$(1,1)$-forms intersection numbers $d_{\cA \cB \cC \cD}$ and 
a metric $G_{\cA \cB}$  via
\beq \label{d_ABCD}
 d_{\cA \cB \cC \cD} = \int_{Y_4} e_\cA \wedge e_\cB \wedge e_\cC \wedge e_\cD\ , \qquad 
 G_{\cA \cB} = \frac{1}{2 \cV} \int_{Y_4} e_\cA \wedge * e_\cB\ ,  
\eeq
where $\cV = \frac{1}{4!}\int J_F \wedge J_F \wedge J_F \wedge J_F$ is the volume of the 
Calabi-Yau four-fold. 

In contrast to a Calabi-Yau threefold the four-dimensional manifold $Y_4$ admits a third non-trivial 
cohomology $H^{3}(Y_4)$ with metric $G_{I \bar J}$. It has non-vanishing intersections 
$d_{\cA I \bar J}$ with $H^2$ such that
\beq \label{d_AIJ}
  d_{\cA I \bar J} = i\int_{Y_4} e_\cA \wedge \Psi_I \wedge \bar \Psi_{J}\ ,\qquad 
  \cG_{I\bar J} = \frac{1}{4 \cV} \int_{Y_4} \Psi_I \wedge * \bar \Psi_{J}= - \frac{\,M^\cA d_{\cA I \bar J}}{4 \cV}\ ,
\eeq 
where we have used $*\bar \Psi_I = -iJ_F \wedge \bar \Psi_I$ in order to evaluate 
the last equality.
However, in general $\cG_{I\bar J}$ as well as $d_{\cA I \bar J}$ depend on the complex structure
deformations $Z^\cK$, since their definition involves the forms $\Psi_I(Z,\bar Z)$.
Hence, by using \eqref{der_Psi} we obtain differential equations for $d_{\cA I \bar J}$ and $\cG_{I\bar J}$, 
which read 
\beq \label{Dd}
  \partial_{Z^\cK}  d_{\cA I \bar J}  = A_{\cK I}^{\ \ \ K}\,  d_{\cA K \bar J}\ ,\qquad 
  \partial_{Z^\cK} \cG_{I\bar J} =  A_{\cK I}^{\ \ \ K}\, \cG_{K\bar J}\ .
\eeq

Having determined the effective action \eqref{F-theory_act} we can now proceed in two ways. Either 
we dualize the vectors $A^{\cA}$ into scalars $P_\cA$ and combine them into chiral multiplets 
$T_\cA=(M^\cA,P_\cA)$. The K\"ahler potential of this $D=3, N=2$ theory was determined in \cite{HL}.
It takes the form
\beq
  K_F(Z,N,T) = - \ln\Big[ \int_{Y_4} \Omega_F \wedge \bar \Omega_F \Big] - 3 \ln \cV(T,N)\ , 
\eeq 
where $\cV(T,N)$ is the volume of $Y_4$, which depends implicitly on the K\"ahler coordinates.
This is indeed analog to the situation in type IIB orientifolds with $O3/O7$ planes.
However,  
in section \ref{IIB_lin} we explored a way around this implicit definition by 
changing to the dual picture. In $D=4$ this amounts to by keeping linear multiplets 
$(L^\alpha,D^\alpha_2)$ in the spectrum, which allows to give $K$ as an explicit function 
of $L^\alpha$. As we will review momentarily, this is equivalently 
true for the $D=3$ theory \eqref{F-theory_act} and amounts to keeping the vector multiplets $(M^\cA, A^\cA)$ in 
the spectrum \cite{BHS}. 

General $D=3,N=2$ supergravity theories with vector and chiral multiplets 
are discussed e.g.\ in \cite{BHS}. To avoid a detailed review of their results we make contact with 
section \ref{linear_multiplets} by observing that the effective action \eqref{kinetic_lin} for chiral and 
linear multiplets in $D=4$ can be translated to $D=3$ chiral-vector setups 
by replacing $dD_2^\cA$ with $dA^\cA$.\footnote{Furthermore, one has to replace in the potential 
\eqref{Lsc} the factor $3$ by a $4$ \cite{HL}.} Using these identifications, one compares 
\eqref{kinetic_lin} with \eqref{F-theory_act} to find 
\beq
  L^\cA = \frac{M^\cA}{\cV}\ ,\qquad  
  \tilde K_{L^\cA L^\cB} = - \tfrac{1}{2}\, \cV^2\, G_{\cA \cB}\ .
\eeq
The kinetic potential for the vector multiplet $(L^\cA, A^\cA)$ is found to be \cite{BHS}
\beq \label{kin_F}
 \tilde K(L,N,Z) = - \ln\Big[ \int_{Y_4} \Omega_F \wedge \bar \Omega_F \Big] 
                   + \ln\big(d_{\cA \cB \cC \cD} L^\cA L^\cB L^\cC L^\cD \big) + 
                    L^\cA \zeta_A 
\eeq
with
\beq
  \zeta^R_A = \tfrac12 d_{A I \bar J} \bar N^I N^J+ \omega_{A I J} N^I N^J + \omega_{A \bar I \bar J} \bar  N^I \bar N^J \ .
\eeq
The functions $\omega_{A \bar I \bar J}(Z,\bar Z)$ obey
\beq \label{Domega}
  \partial_{\bar Z^\cK} \omega_{\cA \bar I \bar J} = B_{\bar \cK I}^{\ \ \ \bar K} d_{\cA J \bar K}\ ,
\eeq
but are otherwise unconstraint. It is now straight forward to check, that 
the effective action determined in terms of $\tilde K(L,N,Z)$ is 
indeed equivalent to \eqref{F-theory_act} up to a total derivative \cite{BHS}.\footnote{More precisely 
one finds $\tfrac{i}{4} d_{\cA I \bar J}\ (\bar N^I DN^J - N^I  D \bar N^J) = \I(\tilde K_{L^\cA Q^m} d Q^m) +$ total derivative, where $Q^m=(N^I,Z^\cK)$.} This ends our review of the M-theory compactification. In order to compare 
\eqref{kin_O33} with the $O3/O7$ orientifold data, we first have to compactify the orientifold theory to 
three dimensions.

\subsubsection{The $O3/O7$ orientifolds in three-dimensions}

Let us now compactify the four-dimensional $O3/O7$ orientifold theory determined by 
\eqref{S_scalarO3} on a circle $S^1$. In order to do that we partly follow \cite{HL}, where general 
compactifications of $D=4,N=1$ theories are discussed. 
Due to the fact that $D=4$ chiral multiplets reduce to $D=3$ multiplets we turn our 
attention to the vectors $V^\kappa$ with kinetic terms \eqref{red-vector}. In three dimensions
vectors are dual to scalars and for four supercharges the dynamics can be encoded by a K\"ahler 
or kinetic potential.   
The Kaluza-Klein reduction is performed by choosing the Ansatz
\beq \label{3d-Ansatz}
  g^{(4)}_{\mu \nu} = \left(
  \begin{array}{cc}
   g^{(3)}_{pq} + r^2 A_p^0 A_q^0& r^2 A_q^0\\ r^2 A_p^0 & r^2
  \end{array} 
  \right)\ , \qquad V^\kappa_\mu = (A^\kappa_p+A^0_p\, n^\kappa, n^\kappa)\ ,
\eeq 
where $A^0_p,\ A^\kappa_p,\ p=1,2,3$ are vectors and $n^\kappa$ as well as $r$ (the radius of $S^1$) are
scalars in three dimensions. The resulting $D=3$ theory posses chiral multiplets $(z^k,\tau,G^a,T_\alpha)$ and 
vector multiplets $(A^0,r)$ and $(A^\kappa,n^\kappa)$. Next we dualize the vectors $A^\kappa$ 
into scalars $\tilde n_\kappa$
by the standard Lagrange multiplier method (see section \ref{revIIB}). However, we keep the vector multiplet 
$(A^0,r)$ and denote $L={r}^{-1}$.
The scalars $\tilde n_\kappa$ and $n^\kappa$ combine into complex scalars $D_\kappa$ via \cite{FS,HL}
\beq
  D_\kappa = -f_{\kappa \lambda}(z)\, n^\lambda  + i\, \tilde n_\kappa\ ,
\eeq 
where $f_{kl}(z)$ are the gauge-couplings of the $O3/O7$ theory given in \eqref{fholo}. One next inserts the 
Ansatz \eqref{3d-Ansatz} into the $D=4$ orientifolds action \eqref{S_scalarO3} and performs a Weyl rescaling
to obtain a $D=3$ effective action with standard Einstein-Hilbert term. Using the definition of $D_\kappa$ 
this action is encoded by a kinetic
potential 
\beq
  \tilde K_3 = - \ln\Big[ \int_Y \Omega \wedge \bar \Omega \Big]  + K^{k}(\tau, G,T) + \ln (L) + L \zeta^R\ ,
\eeq 
where $K^{k}(\tau, G,T)$ and $\zeta^R$ are given in \eqref{kaehlerpot-Kk} and \eqref{def-zR}.
Replacing the chiral multiplets $T_\alpha$ by vector multiplets $(A^\alpha,L^\alpha)$ we
apply \eqref{tildeK_O3} to rewrite the kinetic potential as 
\beq \label{kin_O33}
  \tilde K_3 = - \ln\Big[ \int_Y \Omega \wedge \bar \Omega \Big] 
               - \ln\big(-i(\tau -\bar \tau)\big) + 
               \ln( \cK_{\alpha \beta \gamma} L^\alpha L^\beta L^\gamma) +  \ln (L) + 
               L^\alpha \zeta^R_\alpha + L \zeta^R\ ,
\eeq
where 
\beq \label{def-zR}
  \zeta^R_\alpha = -\frac{i}{2(\tau-\bar \tau)}\ \KK_{\alpha b c}(G-\bar G)^b (G- \bar G)^c\ , \quad 
   \zeta^R = -\tfrac{1}{2} (D_k + \bar D_k) (\R f_{kl} )^{-1} (D_k + \bar D_k)\ .
\eeq
The function $\zeta^R_\alpha =\zeta_\alpha+\bar\zeta_\alpha$ was already given in \eqref{zetaid}.
$\tilde K_3$ fully encodes the dynamics of the chiral multiplets $z^k,\tau,G^a,D_k$
and the vector multiplets $(A^\alpha,L^\alpha)$ and $(A,L)$ in three-dimensions. This enables us 
to compare the orientifold theory with the M-theory compactification discussed at the 
beginning of this section.

\subsubsection{F-theory embedding of $O3/O7$ orientifolds}

In order to discuss the F-theory embedding of the $O3/O7$ bulk orientifold theory, we 
restrict to the simple fourfolds defined in \eqref{def-Z}. Working on these manifolds 
the $\hat \sigma$ invariant cohomologies split as  
\bea \label{cohom_splitF}
  H^2(Y_4) &=& H^{2}_+(Y) \oplus H^2_+(T^2) \ , \qquad H^3(Y_4)\ =\  H^{3}_+(Y) \oplus
                                              \big(H^{2}_-(Y)\wedge H^1_-(T^2)\big)\nn \\
  H^4(Y_4) &=& H^4_+ (Y) \oplus \big(H^3_-(Y)\wedge H^1_-(T^2)\big) \oplus  \big(H^{2}_+(Y) \wedge H^2_+(T^2)\big)\ ,
\eea
where $H^q_\pm(Y)$ are the cohomology groups of $Y$ introduced in \eqref{H3split} and we denote 
by $H^1_-(T^2),\ H^2_+(T^2)$ the cohomologies of $T^2$. We denote a basis of the $T^2$-cohomologies
by $\alpha^{(1,0)},\alpha^{(0,1)} \in H^1_-(T^2)$ and $\vol(T^2) \in  H^2_+(T^2)$.\footnote{ Recall, that for $T^2$ 
one finds $h^{(0,0)}_+=h^{(1,1)}_+=h^{(1,0)}_-=h^{(0,1)}_-=1$.}
We next analyze the spectrum and couplings of the three-dimensional 
theory \eqref{F-theory_act} on the manifolds \eqref{def-Z}. Let us start with the complex structure deformations $Z^\cK$.
{}From \eqref{cohom_splitF} one concludes, that the only $(3,1)$-forms in $H^4(Y_4)$ arise from 
the cohomology $H^{(2,1)}_-(Y)\wedge H^{(1,0)}_-(T^2)$ and $H^{(3,0)}_-(Y)\wedge H^{(0,1)}_-(T^2)$.
Hence we set 
\beq
  Z^\cK \equiv (\tau,\ z^k)\ ,  \qquad \cK=0,\ldots, h^{2,1}_-(Y) \ . 
\eeq 
This is consistent with the fact that in F-theory the complex dilaton 
$\tau$ becomes the complex structure modulus
of the torus fiber of the fourfold $Y_4$ given in \eqref{def-Z}. Hence, we 
will set $\alpha^{(1,0)}=dq - \tau dp$ and lift $\tau$
to one of the complex structure deformations of $Y_4$. 
Moreover, 
in the orientifold limit the complex structure deformations of the orientifold $z^k$ are
the complex structure deformations of the base of $Y_4$. On \eqref{def-Z} also 
the holomorphic four-form $\Omega_F$ splits as $\Omega_F=\Omega\wedge \alpha^{(1,0)}$,
such that 
\beq \label{split_OO}
  \ln\Big[ \int_{Y_4} \Omega_F \wedge \bar \Omega_F \Big] = \ln\Big[-i\int_Y \Omega \wedge \bar \Omega \Big] 
    + \ln\big[-i(\tau - \bar \tau) \big]\ ,
\eeq
where we have used $\int_{T^2} dq\wedge dp = 1$.

The K\"ahler structure deformations of $Y_4$ assembled into the vector multiplets 
$(M^\cA/\cV, A^\cA)=(L^\cA, A^\cA)$. These split under the decomposition 
\eqref{cohom_splitF} into one modulus parameterizing the torus volume and 
$h^{(1,1)}_+$ K\"ahler structure deformations of $Y/\sigma$. In three dimensions
this has an obvious counterpart in the orientifold theory, since an additional 
K\"ahler modulus $L=r^{-1}$ arose from the compactification on $S^1$. 
This leads us to identify 
\beq \label{id_L}
  L^\cA  \equiv (L,L^\alpha)\ ,\qquad  A^\cA \equiv (A^0,A^\alpha)\ ,\qquad \cA=0,\ldots, h^{1,1}_+(Y) \ .
\eeq
Note that this implies that one matches the volume modulus of $T^2$ with the inverse
radius $L=r^{-1}$ of the $S^1$. Also the corresponding intersection numbers 
\eqref{d_ABCD} split on the manifolds \eqref{def-Z} as
\beq
  d_{\cA \cB \cC \cD} \rightarrow d_{0 \alpha \beta \gamma}\ ,
\eeq
with all other intersections vanishing. Here we have chosen $e_0=\text{vol}(T^2)$ 
to be the invariant volume form of $T^2$. This implies that in the kinetic potential
\eqref{kin_F} the logarithm involving $L^\alpha$ splits as 
\beq \label{split_LLLL}
  \ln\big(d_{\cA \cB \cC \cD} L^\cA L^\cB L^\cC L^\cD \big) = \ln L + 
  \ln\big(\cK_{\alpha \beta \gamma} L^\alpha L^\beta L^\gamma \big)\ ,
\eeq 
where we have identified $d_{0\alpha \beta \gamma}=\cK_{\alpha \beta \gamma}$, being the
intersections of $H^2_+(Y)$.

Finally, the remaining chiral multiplets $N^I$  and the orientifold fields $G^a,D_\lambda$
have to be matched
\beq
  N^I \equiv (G^a,D_\lambda)\ , \qquad I = 1, \ldots, h^{(1,1)}_-(Y) + h^{(2,1)}_+(Y)\ . 
\eeq 
Once again, this is consistent with the split \eqref{cohom_splitF}
of $H^3(Y_4)$. The intersection numbers $d_{A I \bar J }$ given in \eqref{d_AIJ} 
decompose as 
\beq
  d_{\cA I \bar J} \rightarrow d_{0\kappa \lambda}\ , d_{\alpha a b}\ , 
\eeq
while all other intersections vanish. Note however, that in general 
$d_{\cA I \bar J}$ depends on the complex structure moduli $Z^\cK$ and
a naive identification $d_{\alpha a b} \cong \cK_{\alpha a b}$ can only 
be true up to a complex structure dependent part. To extract this dependence
we can proceed in two ways. Either we compare the two kinetic potentials
\eqref{kin_F} and \eqref{kin_O33} to determine 
$d_{\cA I\bar J}$ as well as $\omega_{\cA IJ}$ and check if the equations
\eqref{Dd}, \eqref{Domega} and \eqref{DE} are obeyed. However, we choose 
a different route and look for simple solutions of the consistency conditions 
\eqref{DE}. Having determined $A_{\cK I}^{\ \ \ J}$ and $B_{\bar \cK I}^{\ \ \ \bar J}$ 
we are in the position to solve \eqref{Dd}, \eqref{Domega} to determine $d_{\cA I \bar J}$
and $\omega_{\cA I \bar J}$. 

To construct a simple solution to \eqref{DE} we start with a 
holomorphic functions $f_{IJ}(Z)$, which can arise e.g.~as gauge couplings of a 
supersymmetric theory. In terms of $f_{IJ}$ the equations \eqref{DE} are solved by
\beq
  A_{\cK I}^{\ \ \ J} = - (\R f)^{-1\, JK}\, \partial_{Z^\cK} f_{KI} \ , \qquad 
  B_{\bar \cK I}^{\ \ \ \bar J} = (\R f)^{-1\, JK}\, \partial_{\bar Z^\cK} \bar f_{KI}\ .
\eeq
Relevant for the orientifold embedding are the two special cases 
\beq
  f_{\kappa \lambda}(Z^k) = f_{\kappa \lambda}(z^k)\ , \qquad f_{00}(Z^0) = -i\tau\ , 
\eeq
where $f_{\kappa \lambda}$ are the gauge-couplings of the orientifold given in \eqref{fholo}
and $-i\tau$ are the gauge-couplings of a gauge-theory on space-time filling $D3$ branes (see for example \cite{GGJL}).
Not to surprisingly, these are exactly the right functions to match the kinetic 
potentials \eqref{kin_F} and \eqref{kin_O33}. Namely, consistent with 
\eqref{Dd} and \eqref{Domega} we identify 
\beq \label{couplings}
  d_{0 \kappa \lambda} = \omega_{0 \kappa \lambda} = (\R f)^{\kappa \lambda}\ , \qquad 
  d_{\alpha a b} = \omega_{\alpha a b} = \frac{1}{\tau-\bar \tau} \cK_{\alpha a b}\ ,
\eeq
where $\cK_{\alpha a b}$  are the intersections on $Y$, which are independent of $\tau$ and $z^k$.
Equations \eqref{split_OO}, \eqref{split_LLLL} and \eqref{couplings} imply 
that the kinetic potential of the M-theory compactification 
reduces to the one for $O3/O7$ orientifolds on the Calabi-Yau fourfold \eqref{def-Z}.

The final step is to lift this correspondence to four dimensions. On the orientifold side 
this simply amounts to performing the decompactification limit $L_0=r_0^{-1} \rightarrow 0$,
where $r_0$ arises in $r_0+r(x)$ as the background radius. Of course, the resulting theory 
coincides with the $D=4$ orientifold theory, if identifying the correct four-vectors. 
More subtle is the lift of the M-theory compactification, which is known as the F-theory limit.
It amounts to shrinking the volume of the two-torus (identified in \eqref{id_L} with $L_0$)
on an elliptically fibered Calabi-Yau fourfold. However, for the simple manifold \eqref{def-Z}
this limes is rather straightforward and coincides with the decompactification limit for the 
orientifold.


In addition to the bulk theory one can allow for non-trivial four-form background flux
$G_4=\big<dC_3\big>$ on $Y_4$. The theory will be changed by a non-vanishing 
potential, which is obtained from the Gukov-Vafa-Witten superpotential $\int \Omega_F \wedge G_4$. 
In order to relate it to the 
$O3/O7$ orientifold three-form flux $G_3$ given in \eqref{fluxesB} 
one locally writes \cite{DRS,GSS,GKP} 
\beq
  G_{4} = - \frac{G_{3} \wedge \alpha^{(0,1)}}{\tau - \bar \tau}  + h.c. \ .
\eeq 
This implies that the Gukov-Vafa-Witten superpotential reduces as 
\beq
  \int_{Y_4} \Omega_F \wedge G_4 = \int_Y \Omega \wedge G_3\ ,
\eeq
which coincides with the orientifold superpotential found in \eqref{superpot}.

This ends our discussion of the F-theory embedding of $O3/O7$ orientifolds. 
Their are many directions for further research. It would be
desirable to include $D7$ branes into the setup, which correspond to certain 
singularities on the Calabi-Yau fourfold. The naive fourfold given 
\eqref{def-Z} is only valid in the regime were moduli for D-branes and orientifold
branes are frozen. F-theory compactifications provide  powerful tools to
approach regimes where these fields are included \cite{non-pert}. 
A second issue is to discuss moduli stabilization in those setups, 
resent results \cite{DSFGK} suggest that all moduli can be stabilized
in F-theory compactifications by including fluxes and non-perturbative corrections.


%
%

\section{Type IIA orientifolds and special $G_2$ manifolds}
\label{G2_embedding}

In this section we discuss the relationship between the type IIA 
Calabi-Yau orientifolds considered so far 
and $G_2$ compactifications of  M-theory. 
In refs.\ \cite{KMcG} it was argued that for a specific class
of $G_2$ compactifications $X$, type IIA orientifolds appear at special
loci in their moduli space. 
More precisely,  
these  $G_2$ manifolds have to be such that they admit the form 
\beq \label{spG_2}
  X\ =\ (Y \times S^1)/{\hat \sigma}\ ,
\eeq
where $Y$ is a Calabi-Yau threefold and $\hat \sigma = (\sigma,-1)$ 
is an involution which inverts the coordinates of the circle $S^1$ 
and acts as an anti-holomorphic isometric involution on $Y$. 
$\sigma$ and $\hat \sigma$ can have a non-trivial fix-point set
and as a consequence $X$ is a singular $G_2$ manifold. 
In terms of the 
type IIA orientifolds the fix-points of $\sigma$ are the locations 
of the $O6$ planes in $Y$ and as we already discussed earlier cancellation of 
the appearing  tadpoles require the presence 
of appropriate $D6$-branes. In this paper we froze all
excitation of the $D6$-branes and only discussed the effective action
of the orientifold bulk. In terms of $G_2$ compactification this 
corresponds to the limit where $X$ is smoothed out and all additional 
moduli arising in this process are frozen.

The purpose of this section is to check the embedding of type IIA
orientifolds into $G_2$ compactifications of M-theory 
at the level of the $N=1$ effective action.
For orientifolds the effective action was derived in sections
3 and 4 and so as a first step we need to 
recall the effective action of M-theory 
(or rather eleven-dimensional supergravity) on smooth $G_2$ manifolds
\cite{PT, HM, Hitchin1, GPap,BW}. 

The bosonic part of the eleven-dimensional 
supergravity theory was already given in equation \eqref{11act}. 
It encodes the dynamic of the bosonic components $g_{11}$ and $C_3$ of the supergravity multiplet.
As in the reduction on Calabi-Yau manifolds one chooses the background metric 
to admit a block-diagonal form
\beq \label{lin-el}
  ds^2 = ds^2_4(x) + ds^2_{G_2}(y)\ ,
\eeq
where $ds^2_4$ and $ds^2_{G_2}$ are the line elements of a Minkowski
and a $G_2$ metric, respectively.
The Kaluza-Klein Ansatz 
for the three-form $C_3$ reads 
\beq
  C_3 = c^i(x)\, \phi_i + A^\alpha(x) \wedge \omega_\alpha \ , \qquad i=1,\ldots,b^3(X)\ ,\quad \alpha = 1, \ldots, b^2(X)
\eeq
where $c^i$ are real scalars and $A^\alpha$ are one-forms in four space-time dimensions.
The harmonic forms $\phi_i$ and $\omega_\alpha$ span a basis of
$H^3(X)$ and $H^2(X)$, respectively.  
The $G_2$ holonomy allows for exactly one covariantly 
constant spinor which can be used to define a real, harmonic and 
covariantly constant 
three-form $\Phi$.\footnote{The covariantly constant
three-form is the analog of the holomorphic three-form $\Omega$ 
on Calabi-Yau manifolds.}
The deformation space of the $G_2$ metric has dimension $b^3(X)=\dim H^3(X,\bbR)$ and 
can be parameterized by expanding $\Phi$ 
into the basis $\phi_i$ \cite{Joyce} 
\beq
   \Phi = s^i(x)\, \phi_i \ .
\eeq
One combines the real scalars
$s^i$ and $c^i$ into complex coordinates according to
\beq
  S^i = c^i + i s^i\ ,
\eeq
which form the bosonic components of $b^{3}(X)$ chiral multiplets.
In addition the  effective four-dimensional supergravity features
$b^{2}(X)$ vector multiplets with the $A^\alpha$ as bosonic components. 
Due to the $N=1$ supersymmetry, 
the couplings of these multiplet are again expressed in terms of 
a K\"ahler potential $K_{G_2}$,  gauge-kinetic 
coupling functions $f_{G_2}$ and a (flux induced) superpotential $W_{G_2}$.
Let us discuss these functions in turn.

The K\"ahler potential was found to be \cite{HM,Hitchin1,GPap,BW}
\beq \label{G_2Kpo}
  K_{G_2}\ =\ - 3 \ln \big(  \tfrac{1}{ \kappa^2_{11}} \tfrac{1}{7} \int_X \Phi \wedge * \Phi \big)\ ,  
\eeq 
where $\frac{1}{7}\int \Phi \wedge * \Phi = \text{vol}(X)$ is the volume of the $G_2$ manifold $X$. 
The associated K\"ahler metric is given by 
\beq \label{G_2Kmetr}
  \partial_{i}\bar \partial_{\bj}  K_{G_2}\ =\ \tfrac{1}{4} \text{vol}(X)^{-1} \int_X \phi_i \wedge * \phi_j\ ,
  \qquad 
  \partial_{i}  K_{G_2}\ =\ \tfrac{i}{2} \text{vol}(X)^{-1} \int_X \phi_i \wedge * \Phi\ ,
\eeq
and obeys the no-scale type condition
\beq
  (\partial_{i}  K_{G_2})\,  K_{G_2}^{i \jb}\, (\partial_{\bj}  K_{G_2})  = 7\ . 
\eeq 

The holomorphic gauge coupling functions 
$f_{G_2}$ arise from the couplings of $C_3$ in 
\eqref{11act}. At the tree level they are linear in $S^i$ 
and read \cite{HM,GPap}
\beq \label{gauge-kinG}
  (f_{G_2})_{\alpha \beta} = \tfrac{i}{2 \kappa_{11}^2}\, 
S^i\int_X \phi_i \wedge \omega_\alpha \wedge \omega_\beta\ . 
\eeq

Finally, non-vanishing background flux $G_4$ of $F_4 =dC_3$ 
induces a scalar potential which via \eqref{N=1pot} 
can be expressed in terms of the superpotential 
\cite{Gukov,AS,BW}
\beq \label{G_2supo}
  W_{G_2}\ =\  \tfrac{1}{4 \kappa_{11}^2}\int_X \big(\tfrac{1}{2} C_3 +i\Phi) \wedge G_4\ .
\eeq 
(The factor $1/2$ ensures holomorphicity of $W_{G_2}$ in the
coordinates $S^i$ and compensates the quadratic dependence on $C_3$
\cite{BW}.)

In order to compare the low energy effective theory of $G_2$
 compactifications
with the one of the orientifold we first have to restrict to the special $G_2$ manifolds 
$X$ introduced in \eqref{spG_2}. 
This can be done by analyzing how the cohomologies of $X$ are related to the 
ones of $Y$. As in equation \eqref{cohom-split} we consider the splits $H^p(Y)=H^p_+ \oplus H^p_-$ 
of the cohomologies into eigenspaces
of the involution $\sigma$. Working on the $G_2$ manifold $X$ given in \eqref{spG_2} 
we thus find the $\hat \sigma$-invariant cohomologies
\beq \label{splcoho}
   \begin{array}{cclcrcl}
   H^2(X) &=& H^2_+(Y)\ , &\ &
   H^3(X) &=& H^3_+(Y) \oplus \big[H^{2}_-(Y)\wedge H_-^1(S^1)\big] \ ,\Big. \\
   H^5(X) &=& H^4_-(Y)\wedge H_-^1(S^1)\ , &&
   H^4(X) &=& H^4_+(Y) \oplus \big[H^{3}_-(Y)\wedge H_-^1(S^1)\big] \ ,   
  \end{array}
\eeq
where $H^2(X)$ and $H^5(X)$ as well as $H^3(X)$ and $H^4(X)$ are
Hodge duals. $H_-^1(S^1)$ is the one-dimensional space containing
the odd one-form of $S^1$. The split of $H^3(X)$ induces a split of the $G_2$-form
$\Phi$ which is most easily seen by introducing locally an orthonormal basis
$(e^1,\ldots,e^7) \in \Lambda^1(X)$ of one-forms. 
In terms of this basis one has \cite{Joyce,Hitchin1,CS}
\beq\label{Phidecomp}
  \Phi
      \ =\ J_M \wedge e^7 + \text{Re} \Omega_M \ ,
\qquad
*\Phi= \tfrac12 J_M\wedge J_M + \I \Omega_M\wedge e^7\ , 
\eeq
where 
\bea \label{defJO}
  J_M = e^1\wedge e^2 + e^3\wedge e^4 + e^5\wedge e^6\ , \quad 
\Omega_M = (e^1 + ie^2)\wedge(e^3+ie^4)\wedge(e^5+ie^6)\ .
\eea
Applied to the manifold \eqref{spG_2} 
we may interpret $e^7=dy^7$ as being the odd one-form
along $S^1$. Since  $\Phi$ is required to be invariant under 
$\hat\sigma$
and $\sigma$ is anti-holomorphic the decomposition 
\eqref{Phidecomp} implies
\beq \label{splitPhi}
  \hat \sigma^* J_M = - J_M\ , \qquad  
\hat \sigma^* \Omega_M = \bar \Omega_M\ .
\eeq  
In terms of the 
basis vectors $e^1,\ldots,e^6$ this is ensured by choosing
$e^4,e^5,e^6$ to be odd  and
$e^1,e^2,e^3$ to be even under $\sigma$.
We see that  $J_M$ and $\Omega_M$ satisfy
the exact same conditions as the corresponding forms of the
orientifold
(c.f.\ \eqref{constrJ}, \eqref{constrO}) and thus have to be proportional to
$J$ and $C\Omega$ used in section \ref{IIA_orientifolds}. In order to
determine the exact relation  
it is necessary to fix their relative normalization. 
The relation between
$J_M$ and the  K\"ahler form $J$ in the string frame 
can be determined from the relation of the respective metrics.
Reducing eleven-dimensional supergravity to type IIA supergravity in
the string frame 
requires the line element \eqref{lin-el} of the eleven-dimensional 
metric to take the form 
\beq \label{metransatz}
  ds^2 = e^{-{2 \hat \phi}/{3}} ds_4^2(x) + 
         e^{-{2 \hat \phi}/{3}} g_{(s)\, ab}\, dy^a dy^b + e^{{4 \hat \phi}/{3}} (dy^7)^2\ ,
\eeq   
where $a,b=1,\ldots,6$. 
The factors $e^{\hat \phi}$ of  the ten-dimensional dilaton are
chosen such that the type IIA
supergravity action takes the standard form with 
$g_{(s)}$ being the Calabi-Yau metric in string frame
(see e.g.~\cite{JPbook}). 
Consequently we have to identify 
\beq\label{JM}
J_M=e^{-{2 \hat \phi}/{3}} J\ .
\eeq

Similarly, using \eqref{defJO} we find that the normalization of $\Omega_M$ is given by
\bea \label{norm}
  J_M \wedge J_M \wedge J_M  = \frac{3i}{4}\, \Omega_M \wedge \bar \Omega_M\ .
\eea 
Integrating over $Y$ and using \eqref{JM}, \eqref{Kks} and \eqref{csmetric} we obtain
\bea \label{normO}
  \Omega_M =  e^{-\hat \phi-i\theta} 
  e^{\frac{1}{2}(\Kcs - K^{\rm K})}\, \Omega =  \sqrt{8} C\Omega\ ,
\eea
where $C$ is given in \eqref{def-C}.
The phase $e^{i\theta}$ drops out in \eqref{norm} such 
that we can choose it
as in \eqref{constrO} in order to fulfill \eqref{splitPhi}.
Inserting $J_M$ and $\Omega_M$ into equation \eqref{splitPhi} 
one arrives at
\beq \label{Phi-o}
  \Phi = J \wedge d\tilde y^7 + \sqrt{8} \text{Re}(C\Omega) \ ,
\eeq
where we defined $d\tilde y^7 = e^{-\frac{2 \hat \phi}{3}} dy^7$. The form 
$d\tilde y^7$ is normalized such that $\int_{S^1} d\tilde y^7=2\pi R$ where 
the metric \eqref{metransatz} was used and 
$R$ is the $\phi$-independent radius of the internal circle. 
We also set $\kappa^2_{10}=\kappa_{11}^2 / 2\pi R = 1$ henceforth.
Using \eqref{Phi-o}, \eqref{Phidecomp} 
and \eqref{def-C} we calculate 
\beq \label{voldec}
          \tfrac{1}{\kappa^2_{11}}\, \tfrac{1}{7} \int \Phi \wedge * \Phi 
           = e^{-\frac{4\hat \phi}{3}} \, \tfrac{1}{6} \int J \wedge J \wedge J \ , 
\eeq
which  equivalently  can be obtained by applying the split 
$\text{vol}(X)=\text{vol}(Y)\cdot\text{vol}(S^1)$ of the $G_2$ volume when evaluated in the metric \eqref{metransatz}.
Inserting \eqref{voldec} into \eqref{G_2Kpo} using \eqref{def-C}
we obtain
\beq\label{IIAori}
  K_{G_2}\ =\
       - \ln \Big[  \tfrac{1}{6} \int_Y J \wedge J \wedge J \Big]
       - 2\ln\Big[2\int_Y \R(C \Omega)\wedge *_6 \R(C\Omega)\Big]\ .
\eeq
 Thus we find exactly the K\"ahler potential 
$K$ of the type IIA orientifold as given in \eqref{N=1Kpot}.\footnote{%
In terms of the Hitchin functionals \cite{Hitchin1} recently discussed in 
\cite{DGNV,Nekrasov} the reduction of the 
$G_2$ K\"ahler potential \eqref{G_2Kpo} corresponds to
the split of the seven-dimensional Hitchin functional to the 
two six-dimensional ones \ref{IIAori}.}

In order to compare the gauge kinetic functions and the superpotential
we also need to identify the 
K\"ahler coordinates of the two theories. 
$C_3$ splits under the decomposition \eqref{splcoho} 
of the cohomologies as\footnote{We have introduced a factor of $\sqrt{2}$ for later convenience.} 
\beq \label{spl-C}
  C_3 = \hat B_2 \wedge d\tilde y^7 + \sqrt{2} \hat C_3 \ ,
\eeq
where $\hat B_2$ is an odd two-form on $Y$ 
and $\hat C_3$ an even three-form on $Y$.
Combining \eqref{Phi-o} and \eqref{spl-C} using \eqref{N=1coords}
one finds
\beq \label{spl-CPhi}
  S^i \phi_i\ =\ C_3 + i\Phi\ =\ \Jc \wedge d\tilde y^7 + \sqrt{2}\, \Omegac\ .
\eeq
As discussed after
\eqref{N=1coords} the coefficients arising in the expansions of $\Jc$ and $\Omegac$
into the basis $(\alpha_k,\beta^\lambda)$ of $H^{3}_+(Y)$
and $\omega_a$ of $H^2(Y)$ are exactly the orientifold coordinates and 
therefore we have to identify
$S^a \cong t^a$ and $S^K \cong (N^k,T_\lambda)$.
With this information at hand, it is not difficult to show that the 
gauge-kinetic couplings \eqref{gauge-kinG} coincide with \eqref{gauge-A}. 
One splits $\phi_a = \omega_a \wedge d\tilde y^7$ and obtains 
\beq
  (f_{G_2})_{\alpha \beta} 
= \tfrac{i}{2} S^a \int_Y \omega_a \wedge \omega_\alpha \wedge
\omega_\beta\ \sim i t^a \cK_{a\alpha\beta} = (f_{OY})_{\alpha \beta}\  
 ,
\eeq 
where the precise factor depends on the normalization of
the gauge fields.

It remains to compare the flux induced superpotentials \eqref{G_2supo} 
with \eqref{superpot1}. Using the 
cohomology splits \eqref{splcoho} and \eqref{spl-C} 
the background flux 
splits accordingly as $G_4 = H_3 \wedge d\tilde y^7 + \sqrt{2} F_4$. 
Inserted into \eqref{G_2supo}
using \eqref{spl-CPhi} we arrive at
\beq
  W_{G_2} = \tfrac{1}{\sqrt{8}}\int_Y \Jc \wedge F_4 + 
            \tfrac{1}{\sqrt{8}}\int_Y \Omegac \wedge H_3
\eeq
Compared to \eqref{superpot1} the superpotential $W_{G_2}$ only 
includes terms proportional to the fluxes $H_3$ and $F_4$.\footnote{%
The term proportional to $e_0$ in \eqref{superpot2} can be absorbed
into a redefinition of $\R t^a$ \cite{BW}.} 
An interesting question is to identify the remaining terms 
in \eqref{superpot1} which are likely to arise once manifolds with
$G_2$ structure (instead of $G_2$ holonomy) are considered. The term 
due to $F_2$ arises in compactifications on fibered 
$G_2$ manifolds $X \rightarrow Y$ \cite{CS,CCDLM}. In our case we 
restrict to circle fibrations over the quotient $Y/\sigma$, where $Y$
is a Calabi-Yau manifold. We introduce the projection $\pi:X \rightarrow Y$. The metric on such 
a manifold takes the form 
\beq
   g_{G_2} = \alpha \otimes \alpha + \pi^* g\ ,
\eeq 
where $g$ is the metric on $Y$ and $d\alpha=\pi^* F_2$. This implies 
that $X$ has not anymore $G_2$ holonomy but rather $G_2$ structure with 
$d\Phi = F_2\wedge J$ being not closed. Following \cite{BJ} this 
induces a superpotential term of the form 
\beq \label{non-G_2}
   W = \int_X (dC_3 + i d\Phi) \wedge (C_3+i\Phi) + \ldots = \int_Y F_2 \wedge \Jc \wedge \Jc + \ldots\ , 
\eeq
where $\Phi$ and $C_3$ are given in \eqref{Phi-o} and \eqref{spl-C} with $d\tilde y^7 = \alpha$
and $dC_3 = \hat B_2 \wedge F_2 + \ldots$. This reproduces exactly the $F_2$ superpotential term \eqref{superpot2}
in type IIA orientifolds. It remains to reveal the origin 
the superpotential term linear in $m^0$. Unfortunately, this is less straightforward and
is likely to involve more general $G_2$ manifolds \cite{Witt}.\footnote{We like to thank A. Micu for discussions on this point.}
It would be nice to make this more explicit and 
to point out the relation to the Scherk-Schwarz constructions of massive IIA supergravity.

\chapter{Conclusions}

In this work we determined the low energy effective action for type IIB and type IIA 
Calabi-Yau orientifolds in the presence of background fluxes. 
In our analysis we did not specify a particular Calabi-Yau manifold but
merely demanded that it admits an isometric involution $\sigma$.  
Furthermore, in order to preserve $N=1$ supersymmetry $\sigma$ was chosen to 
be a holomorphic map in type IIB and an anti-holomorphic map in type IIA. 
Depending on the explicit action of $\sigma$ on the holomorphic three-form $\Omega$, 
we analyzed three distinct cases: (1) orientifolds with $O3/O7$-planes, (2) 
orientifolds with $O5/O9$-planes and (3) orientifolds with $O6$-planes. 
For each case we calculated the characteristic functions of the 
corresponding $N=1$ supergravity theories and discussed their 
generic properties. 

In chapter \ref{effective_actO} we restricted to the case where background fluxes are absent
and no potential is generated. We computed the effective action by a Kaluza-Klein analysis
valid in the large volume limit and determined the
chiral variables, the K\"ahler potentials and the gauge kinetic functions for
all three setups. We found that the moduli space of the $N=1$ theory inherits 
a product structure $\tilde \cM^{\rm SK} \times \tilde \cM^Q$ from
the underlying $N=2$ theory obtained by ordinary Calabi-Yau
compactification of type II theories. $\tilde \cM^{\rm SK}$ is
a special K\"ahler manifold parameterized by the complex structure deformations 
in type IIB and by the complexified K\"ahler deformations in type IIA.
For type IIB orientifolds the second component $\tilde \cM^Q$ is parameterized
by the periods of the complex even form $\fe - i\, \fa$ for setups with 
$O3/O7$ planes and by the periods of $\feh - i\, \fa$ for setups with $O5/O9$. The 
form $\fe+ i\,\feh =e^{-\hat \phi }\, e^{-\hat B_2+iJ}$ comprises of the complexified 
K\"ahler deformations while $\fa$ is a sum of the even R-R forms defined in 
\eqref{def-A}. On the other hand, for type 
IIA orientifolds  with $O6$ planes $\tilde \cM^{\rm Q}$
is spanned by the periods of the complex three-form $\Omegac=C_3 + 2i \R C\Omega$ 
containing the complex structure deformations of the Calabi-Yau orientifold. 
$\tilde \cM^{\rm Q}$ is a K\"ahler submanifold inside the quaternionic manifold 
with a K\"ahler potential encoding the dynamics of the even/odd 
forms of the respective orientifold setup.
Finally we showed that in the large volume -- large complex structure limit
one finds mirror symmetric effective actions if one compares
type IIA and type IIB supergravity compactified on mirror manifolds
and in addition chooses a set of `mirror involutions'.
For $\tilde\cM^K$ mirror symmetry amounts to a truncated versions
of $N=2$ mirror symmetry in that it still relates
two holomorphic prepotentials. In this case the corrections computed
by mirror symmetry are likely to be analogous to the situation in $N=2$.
For $\tilde\cM^Q$
the situation is more involved since the geometry of the moduli
space changes drastically. Nevertheless we were able to show that
mirror symmetry holds in the large volume - large complex structure limit.
However, understanding 
the nature of the corrections computed by mirror symmetry
appear to be more involved and certainly deserves further study.
It is interesting to note that mirror symmetry can be understood as an exchange
of the odd form $\Omegac$ with the even forms $\fe+i\, \fa$ or $\feh+i\, \fa$ in accord with    
\cite{FMM}. Two choices of special coordinates in $\Omegac$ single out the corresponding orientifold 
setup on the mirror side. It would be desirable to reveal 
the origin of this mapping and finally to generalize it to non-Calabi-Yau compactifications. 

In chapter \ref{lin_geom_of_M} we presented a more detailed investigation of
the $N=1$ moduli space of Calabi-Yau orientifold compactifications.
The special K\"ahler manifold $\tilde \cM^{\rm SK}$ inherits its geometrical 
structure directly from $N=2$, such that we focused on
the K\"ahler manifold $\tilde \cM^Q$ inside the quaternionic space.  
It turned out that the definition of the K\"ahler coordinates as well as 
the no-scale type conditions on $\tilde \cM^{\rm Q}$ can be more easily understood 
in terms of the `dual' formulation where some chiral multiplets of the Calabi-Yau 
orientifold are replaced by linear multiplets. After a brief review of $N=1$ supergravity with 
several linear multiplets we reformulated all three orientifold setups by dualizing a certain set of
chiral multiplets. The transformation into 
linear multiplets corresponds to a Legendre transformation of 
the K\"ahler potential and coordinates. The new kinetic potential of $O3/O7$ and $O5/O9$ 
orientifolds takes a particularly simple form induced from a tree-level prepotential. 
In contrast for $O6$ orientifolds it is given in terms of a generic prepotential satisfying the 
orientifold constrains and generically includes correction corresponding to world-sheet instantons in 
type IIB. For orientifolds with $O6$ planes the Legendre transform was essential to make contact 
with the underlying $N=2$ special geometry. As a byproduct we determined an entire new class of no-scale
K\"ahler potentials which in the chiral formulation
can only be given implicitly as the solution of some constraint equation.
We closed this chapter by giving an explicit construction of the K\"ahler 
manifold $\tilde \cM^{\rm Q}$ replacing the $N=2$ c-map. The space  
$\tilde \cM^{\rm Q}$ was shown to admit a geometric structure similar to the one of the moduli
space of supersymmetric Lagrangian submanifolds \cite{Hitchin2}. This also 
provides the ground for a more general investigation of non-Calabi-Yau orientifolds. 
Namely, we found that the K\"ahler potential of $\tilde \cM^{\rm Q}$ is the 
logarithm of Hitchins functional for a generalized complex sixfold 
evaluated for the simple even and odd forms associated to the orientifold setup.

In chapter \ref{fluxesAB} we repeated the Kaluza-Klein compactification by additionally allowing
for non-trivial background fluxes. 
In the $O3/O7$ case the background fluxes induce a non-trivial scalar potential
which is determined in terms of a superpotential previously given in
\cite{GVW,TV,GKP,BBHL}. We also included the scalar fields $(b^a,c^a)$ arising from the two
type IIB two-forms $B_2$ and $C_2$. We showed that in this case the potential 
is unmodified which can be traced to the no-scale property of the K\"ahler potential.
For orientifolds with $O5/O9$ planes the influence of background fluxes is more
involved. This is due to the fact that the space-time two-form $C_2$ arising in the 
expansion of the RR field $\hat C_2$ remains in the spectrum. It combines with 
the dilaton into a linear multiplet, which only if it is massless can be dualized to 
a chiral multiplet. However, generic NS three-form background fluxes render this
form massive. We therefore first restricted our attention to the case were the mass
term vanishes which occurs if the magnetic fluxes arising from
the NS three-form $H_3$ are set to zero. In the resulting
chiral description the axion dual to $C_2$ is gauged with the gauge charges set by the 
electric fluxes. The scalar potential now consists of two distinct contributions. 
The term which depends on the RR fluxes arising from 
$F_3$ is obtained from a (truncated) superpotential of the previous case
whereas the second contribution depends on the electric fluxes of
$H_3$ and arises from $D$-terms which are present  due to the gauged isometry.    
Finally, we also analyzed non-vanishing magnetic fluxes 
in the NS sector which can be described by an $N=1$ theory including a massive linear
multiplet coupled to vector and chiral multiplets. 
In this case the scalar potential additionally 
includes a direct mass term for the scalar  
in the linear multiplet which is neither a $D$- nor an $F$-term.
For type IIA orientifolds all background fluxes induce a superpotential $W$ 
which depends on all geometrical moduli. It splits into the sum of two terms 
with one term depending on the RR fluxes and the complexified
K\"ahler form $J_c$ while the second term
features the NS fluxes and $\Omegac$.
Both terms are expected to receive non-perturbative corrections
from world-sheet- and D-brane instantons.
We showed that for supersymmetric type IIA and type IIB instantons the respective actions 
are linear in the chiral coordinates and thus can result in holomorphic
corrections to $W$.

In the last chapter \ref{M-F-embedding} we analyzed the embedding of type IIB and type IIA 
orientifolds into F- and M-theory compactifications.  
Orientifolds with $O3/O7$-planes can be obtained as a limit of 
F-theory compactified on elliptically fibered Calabi-Yau fourfolds \cite{Vafa,Sen}.
To check this correspondence on the level of the effective action we 
took a sideway by first compactifying M-theory on a Calabi-Yau fourfold.
This yields a three-dimensional $N=2$ supergravity theory determined in terms 
of the characteristic data of the Calabi-Yau fourfold. Restricting to a specific  
fourfold this effective theory can be compared to the one obtained by  
compactifying the effective action of $O3/O7$ orientifolds 
on a circle to $D=3$. We determined simple solutions to the 
fourfold consistency conditions for which we found perfect matching between the 
orientifold and M-theory compactifications. This correspondence can be lifted
to $D=4$ where the M-theory on the elliptically fibered fourfold descends to an F-theory compactification. 
In our analysis we neglected contributions due to singularities of 
the Calabi-Yau fourfold. Smoothed out they yield additional moduli, which 
are identified with $D7$ or $O7$ moduli in the orientifold limit. 
In a next step one can attempt to include these into the analysis and
later deform away from the orientifold limes. Non-trivial fibrations appear if
the orientifold charges are not canceled locally and the F-theory picture becomes
essential.  Finally we also discussed the embedding of type IIA
orientifolds into a specific class of 
$G_2$ compactification of M-theory. Neglecting the 
contributions arising from the singularities of the $G_2$ manifold
we were able to show agreement between the low energy effective
actions. Comparing the superpotentials we only discovered 
the terms which are due to four-form flux from in M-theory. 
However, relaxing the condition of $G_2$ holonomy we were able to 
identify one of the remaining terms as corresponding to a 
non-trivial fibration of a $G_2$ structure manifold. It remains to 
identify the counterpart of the orientifold  
superpotential term cubic in the complexified K\"ahler moduli. This 
term is propotional to the mass parameter of massive IIA supergravity
and plays the essential role in moduli stabilization.  

Let us end our conclusions with some directions for further research. 
Firstly, it would be desirable to include D-brane matter fields into 
the orientifold setups. For type IIB setups with $D3$ and $D7$ branes this was done, for example,
in refs. \cite{GGJL,JL}.  An important task is to extend these results to
type IIA orientifolds with space-time filling $D6$ branes. The knowledge 
of the full effective action enables to perform a calculation of soft supersymmetry 
breaking terms of semi-realistic D-brane scenarios. 

As already mentioned, a generalization to non-Calabi-Yau orientifolds is of 
particular interest \cite{GLprep}. Orientifolds allow for consistent $D=4$ 
Minkowski or Anti-de Sitter vacua for which the internal manifold possesses
non-trivial torsion. As we have argued, the orientifold projections specify
a K\"ahler submanifold in the quaternionic $N=2$ moduli space with geometry encoded 
by special even and odd forms. The K\"ahler potential is Hitchins functional truncated
by the projection. A similar analysis is likely to apply to orientifolds of 
generalized complex manifolds as introduced in \cite{HitchinGCM}. 
 
Brane worlds in orientifolds are a prominent arena for model building in
particle physics and cosmology. However, finding a particular vacuum featuring the
properties of our universe is a highly non-trivial task. One major step into this 
direction is to extract vacua with stabilized moduli fields. Assuming that this 
can be achieved, for example by background fluxes, one encounters a huge set of possible vacua
labeled by different flux quantum numbers.
In the pioneering paper \cite{Douglas} it was argued that a statistical analysis of this `landscape' 
could lead a deeper understanding of the vacuum structure of string theory. 
These considerations were mostly applied to type IIB orientifolds and
certain M-theory vacua. It is an interesting task to generalize this to type
IIA orientifolds. For early time cosmology a wave-function for flux vacua could 
yield an interesting attempt to approach quantum cosmological questions within
the framework of string theory \cite{OVV}. It would be nice to relate these 
new developments in topological string theories to the results of $N=1$
flux compactifications. Surprisingly various similarities appear, which hint 
to at least a formal relation.

\chapter{Appendix}

\appendix

\renewcommand{\thesection}{\Alph{section}}
\renewcommand{\theequation}{\Alph{section}.\arabic{equation}}
\section{Conventions}\label{conventions}
In this appendix we summarize our conventions.

\begin{itemize}
\item
The coordinates of the four-dimensional Minkowski space-time are 
denoted by $x^\mu, \mu=0,\ldots,3$.
The corresponding metric is chosen to have signature $(-,+,+,+)$.
The coordinates of the compact Calabi-Yau manifold  $Y$
are denoted by $y^i, \bar y^\bi,\ i,\bi=1,2,3$. 

\item
$p$-forms are expanded into a real basis according to 
\beq
  A_p\ =\ \frac{1}{p!}\, 
A_{\mu_1 \ldots \mu_p} dx^{\mu_1}\wedge \ldots \wedge dx^{\mu_p}\ .
\eeq
\item
$(p,q)$-forms are expanded into a complex basis as
\beq
  A_{p,q} = \frac{1}{p!q!} A_{i_1 \ldots i_p \bi_1 \ldots \bi_q} dy^{i_1}\wedge \ldots \wedge dy^{i_p}
            \wedge d\bar y^{\bi_1}\wedge \ldots \wedge d\bar y^{\bi_q}\ .
\eeq
\item
The exterior derivative is defined as 
\beq
  dA_p=\frac{1}{p!} \partial_\mu A_{\mu_1 \ldots \mu_p} dx^\mu\wedge dx^{\mu_1}\wedge \ldots \wedge dx^{\mu_p}\ .
\eeq
\item
The field strength of a $p$-form $F_{p+1}=dA_p$ 
is given by 
\beq
  F_{\mu_1 \ldots \mu_{p+1}} = (p+1)\, \partial_{[\mu_1}A_{\mu_2\ldots \mu_{p+1}]}\ .
\eeq
\item
The inner product for real forms is defined by using the Hodge-$*$ operator. In
components we have 
\beq \label{wedge*comp}
 \int F_p \wedge * F_p = \frac{1}{p!}\int F_{\mu_1 \ldots \mu_p} F^{\mu_1 \ldots \mu_p} *\mathbf{1}\ ,
\eeq
where $*\mathbf{1} = d^d x\, \sqrt{-g}$ is the $d$-dimensional measure. 

\item
The Hodge-$*$ satisfies 
$** F_p = (-1)^{p(d-p)+\kappa} F_p$, where $\kappa=1$ for Lorentzian signature 
and $\kappa=0$ for Euclidean signature. 

\item
Let $\sigma_1$ and $\sigma_2$ be an orientiation preserving and an orientation reversing map
$\sigma_{1,2}: M \rightarrow M$, where $M$ is an $n$-dimensional manifold. Then one finds
for a $n$-form $\omega$ on $M$ that
\beq \label{int-form1}
  \int_{\sigma_1(M)} \omega = \int_M \sigma_1^*(\omega)\ , \qquad  \int_{\sigma_2(M)} \omega = -\int_M \sigma_2^*( \omega)\ .
\eeq
However, if we choose $\omega_{M}=*\mathbf{1}$ to be the canonical volume form of $M$ then
$\omega_{\sigma_1(M)}= \sigma_1^* (\omega_{M})$ and $\omega_{\sigma_2(M)}= -\sigma_2^*( \omega_{M})$,
such that
\beq \label{int-form2}
  \int_{\sigma_{1,2}(M)} \omega_{\sigma_{1,2}(M)} = \int_M \sigma_{1,2}^*(\omega_M)\ .
\eeq

\end{itemize}

\section{N=2 supergravity and special geometry}
\label{specialGeom}

In this appendix we briefly summarize the $N=2$ special geometry
of the Calabi-Yau 
moduli space. A more detailed discussion can be found,
for example, in refs.\ \cite{CdO,Strominger2,Freed,N=2review,CRTV}.
A special K\"ahler manifold $\cM$ is a Hodge-K\"ahler manifold (with line bundle $\cL$)
of real dimension $2n$ with associated 
holomorphic flat $Sp(2n+2,\mathbb{R})$ vector bundle $\mathcal{H}$ over $\cM$. Furthermore 
there exists a holomorphic section $\Omega(z)$ of $\cL$ such that 
\beq\label{N=2KP}
  K(z,\bar z) = - \ln i \big<\Omega(z) , \bar \Omega(\bar z)  \big>\ , \qquad 
  \big<\Omega, \partial_{z^K} \Omega\big> = 0\ , \qquad K=1,\ldots n\ ,
\eeq
where $K$ is the K\"ahler potential of $\cM$ and 
$\big<\cdot,\cdot \big>$ is the 
symplectic product on the fibers. 
This is precisely what one encounters in the moduli space
of the complex structure deformations of a Calabi-Yau manifold
with $\Omega$ being the holomorphic three-form.
In this case 
one is lead to set $n=h^{(2,1)}$ and identify 
the fibers of the associated 
$Sp$-bundle with $H^3(Y,\mathbb{C})$. The symplectic product is given by 
the intersections on $H^3(Y,\bbC)$ as
\beq \label{sympl-f}
  \big<\alpha, \beta \big> = \int_Y \alpha \wedge \beta\ .
\eeq
The K\"ahler covariant derivatives of $\Omega$ are denoted by 
$\chi_K$ as explicitly given in \eqref{Kod-form}.
In terms of the symplectic basis $(\alpha_\Kh, \beta^\Kh)$
introduced in \eqref{int-numbers1} both
 $\Omega$ and $\chi_K$ enjoy the expansion
\beq \label{def-chi}
  \Omega = Z^\Kh\, \alpha_\Kh - \cF_\Kh\, \beta^\Kh\ , \qquad  
  \chi_K = \chi^\Lh_{K}\, \alpha_\Lh - \chi_{\Lh|K}\, \beta^\Lh\ .
\eeq 
The holomorphic functions $Z^\Kh(z)$ and $\cF_\Kh(z)$
are called the periods of $\Omega$, while $\chi^\Lh_{K}(z,\bar z)$ and $\chi_{\Lh|K}(z,\bar z)$ 
are the periods of $\chi_K$.  In terms of $Z^\Kh,\cF_\Kh$ the K\"ahler potential 
\eqref{N=2KP} can be rewritten as in \eqref{csmetric}.

For every special K\"ahler manifold there exists
a complex matrix $\cM_{\Kh \Lh}(z,\bar z)$ defined as 
\beq \label{def-M}
   \cM_{\Kh \Lh} = (\bar \chi_{\Kh|\bar M}\ \ \cF_\Kh)  (\bar \chi^\Lh_{\bar M}\ \ Z^\Lh)^{-1}\ ,
\eeq
where $\chi^{\Lh}_{K}$ and $\chi_{\Lh| K}$ are given in \eqref{def-chi}.
Furthermore, one extracts from \eqref{def-M} the  
identities
\bea \label{ML-hf}
  \cF_{\Kh} = \cM_{\Kh \Lh} Z^\Lh\ , \qquad   \chi_{\Lh| K} = \bar \cM_{\Lh \Mh} \chi^{\Mh}_{K}\ ,
\eea
which can be used to rewrite \eqref{N=2KP} as 
\bea \label{spconst}
   G_{M \bar N} & = & -2 e^{K} \chi^\Kh_M\, \I \cM_{\Kh \Lh}\, \bar \chi^\Lh_{\bar N}\ , \qquad
   1 \ = \ -2 e^{K} Z^\Kh\, \text{Im}\, \cM_{\Kh \Lh}\,\bar Z^\Lh \ , \\
   0 & = & -2 \bar \chi^\Kh_{\bar M}\, \text{Im}\, \cM_{\Kh \Lh}\,\bar Z^\Lh \ .\nn
\eea

If one assumes that  
the Jacobian matrix $\partial_{z^L}\big(Z^K/Z^0 \big)$ is invertible 
$\cF_\Kh$ is the derivative of a holomorphic prepotential $\cF$ with respect to the periods $Z^\Kh$. 
It is homogeneous of degree two and obeys
\beq
   \cF = \tfrac{1}{2}Z^\Kh \cF_{\Kh}\ , \qquad  \cF_\Kh =\partial_{Z^\Kh} \cF\ , \qquad  
   \cF_{\Kh \Lh} =\partial_{Z^\Kh} \cF_{\Lh}\ ,
\qquad \cF_{\Lh}= Z^\Kh \cF_{\Kh \Lh}\ ,
\eeq
which implies that $\cF_{\Kh \Lh}(Z)$ is invariant
under rescalings of $Z^\Kh$. 
Notice that $\cF$ is only invariant under a restricted 
class of symplectic transformations
and thus depends on the choice of symplectic basis. 

The complex matrix 
$\cM_{\Kh \Lh}$ defined in \eqref{def-M} can be rewritten in terms of the 
periods $Z^\Kh$ and the matrix $\cF_{\Kh\Lh}(Z)$ as
\bea \label{gauge-c}
   \cM_{\Kh \Lh}=\overline{ \mathcal{F}}_{\Kh \Lh}+2i \frac{(\text{Im}\; \mathcal{F})_{\Kh \Mh} Z^\Mh
   (\text{Im}\; \mathcal{F})_{\Lh \Nh}Z^\Nh }{Z^\Nh(\text{Im}\; \mathcal{F})_{\Nh\Mh} 
    Z^\Mh}\ .
\eea

Whenever the 
Jacobian matrix $\partial_{z^L}\big(Z^K/Z^0 \big)$ is invertible 
the $Z^\Kh$ can be viewed as projective coordinates of $\mathbb{P}_{h^{(2,1)}+1}$.
Going to a special gauge, i.e.~fixing the K\"ahler transformations
\eqref{crescale}, one introduces
special coordinates $z^K$ by setting $z^K=Z^K/Z^0$. 
Due to the homogeneity of $\cF$ it is possible to define
a holomorphic prepotential $f(z)$ which only depends on the special
coordinates as
\beq\label{def-f}
  \cF(Z) = (Z^0)^2 f(z)\ .
\eeq
In terms of $f$  the K\"ahler potential  given in \eqref{N=2KP} 
reads
\beq \label{Kinz}
  K\ =\ - \ln i|Z^0|^2 \big[2(f-\bar f)-(\partial_K\, f + \partial_{\bar K} \bar f)(z^K - \bar z^K) \big]\ . 
\eeq

A special example of the situation just discussed is the moduli
space spanned by the complexified K\"ahler deformations $t^A$ introduced 
in \eqref{4d-dilaton}. These fields can be interpreted as special coordinates on 
a special K\"ahler manifold $\cM^{\rm SK}(t,\bar t)$ \cite{CdO}. 
The K\"ahler potential of the metric $G_{AB}$ given 
in \eqref{Kmetric} is of the form \eqref{Kinz} with
\beq \label{pre-K}
  f(t)=-\tfrac{1}6 \cK_{ABC} t^A t^B t^C\ .
\eeq
Furthermore, inserting \eqref{pre-K}
into \eqref{gauge-c} using \eqref{def-f} 
one determine the gauge-couplings  $\cN_{\Ah \Bh}(t,\bar t)$ 
to be
\bea \label{def-cN}
  \text{Re} \cN &=& \ \
  \left(\ba{cc}-\frac13 \cK_{ABC}b^A b^B b^C &  \frac12 \cK_{ABC} b^B b^C \\
              \frac12 \cK_{ABC} b^B b^C & - \cK_{ABC}b^C  \ea \right)\ , \nn \\
  \text{Im} \cN &=& -\frac{\cK}{6}
  \left(\ba{cc}1 + 4 G_{AB}b^A b^B & -4 G_{AB}b^B  \\
             - 4 G_{AB}b^B &  4 G_{AB}  \ea \right)\ , \nn \\
  (\text{Im} \cN)^{-1} &=& - \frac{6}{\cK}
  \left(\ba{cc}1 & b^A  \\
         b^A &  \frac14 G^{AB} + b^A b^B \ea \right)\ ,
\eea
where $G_{AB}$ is given in \eqref{Kmetric}.

\section{Supergravity with several linear multiplets} \label{linm}

In this appendix we briefly discuss the dualization of several massless linear
multiplets to chiral multiplets. We only discuss the bosonic component fields and do not include possible couplings to vector 
multiplets. Our aim is to extract the K\"ahler potential 
for the $N=1,D=4$ supergravity theory with all linear multiplets replaced by chiral ones.
Let us begin by recalling the effective action for a set of 
linear multiplets $(L^\lambda, D^\lambda_2)$ couplet to  chiral multiplets
$N^k$. It takes the form\footnote{This action can be obtained by a straight forward 
generalization of the action for one linear multiplet given in \cite{BGG}.}
\bea\label{kinetic}
\cL &=& -\tfrac{1}{2}R*\mathbf{1} - 
  \tilde K_{N^k\bar N^l}\, dN^k \wedge * d \bar N^{l}
  + \tfrac{1}{4} \tilde K_{L^\kappa L^\lambda}\, 
  dL^\kappa \wedge * dL^\lambda \nn\\ 
  && + \tfrac{1}{4} \tilde K_{ L^\kappa L^\lambda}\, dD^\kappa_2 \wedge * dD^\lambda_2
     -  \tfrac{i}2\,  dD^\lambda_2 \wedge 
\big(\tilde K_{L^\lambda N^k}\,dN^k -\tilde K_{L^\lambda \bar N^k}\,d\bar N^k\big)
\ ,
\eea
where  $\tilde K(L,N,\bar N)$ is a function of the scalars $L^\lambda$ and
the chiral multiplets $N^k$. The kinetic potential $\tilde K$ is the 
analog of 
the K\"ahler potential in the sense that it encodes the dynamics of the linear
and chiral multiplets. In order to dualize the linear multiplets $(L^\lambda, D^\lambda_2)$ 
into chiral multiplets $(L^\lambda,\tilde \xi_\lambda)$ one replaces
$dD_2^\lambda$ by the form $D_3^\lambda$ and adds the term 
\beq
\cL \to \cL + \delta \cL\ , \qquad
  \delta \cL\ =\  
  - 2\tilde \xi_\lambda\, dD^\lambda_3\ =\ - 2 D_3^\lambda \wedge d\tilde \xi_\lambda\ ,\ 
\eeq
where $\tilde \xi_\lambda(x)$ is a Lagrange multiplier. Eliminating $\tilde \xi_\lambda$
one finds that $dD^\lambda_3=0$ such that locally $D_3^\lambda=dD_2^\lambda$ as required.
Alternatively one can consistently eliminate $D^\lambda_3$ by inserting 
its equations of motion
\beq
  *D_3^\kappa = 4 \tilde K^{L^\kappa L^\lambda}\Big(d\tilde \xi_\lambda + \tfrac{i}4 
  \big(\tilde K_{L^\lambda N^k}\,dN^k -\tilde K_{L^\lambda \bar N^k}\,d\bar N^k\big)\Big) 
\eeq
back into the Lagrangian \eqref{kinetic}. 
The resulting dual Lagrangian takes the form
\bea \label{eff_act1}
\cL &=& -\tfrac{1}{2}R*\mathbf{1} - 
  \tilde K_{N^k\bar N^l}\, dN^k \wedge * d \bar N^{l}
  + \tfrac{1}{4} \tilde  K_{L^\kappa L^\lambda}\, 
  dL^\kappa \wedge * dL^\lambda  \\ 
  && + 4 \tilde K^{L^\kappa L^\lambda} \Big(d\tilde \xi_\kappa - \tfrac{1}2
  \I \big(\tilde K_{L^\kappa N^l}\,dN^l\big)\Big)\wedge * 
  \Big(d\tilde \xi_\lambda - \tfrac{1}2
  \I \big(\tilde K_{L^\lambda N^k}\,dN^k\big)\Big) \ .\nn
\eea
Since we intend to use these results in the effective action for Calabi-Yau
orientifolds, we make a further simplification. We demand that the kinetic potential
$\tilde K$ is only a function of $L^\lambda$ and the imaginary part of $N^k$, which we 
denote by $l^k=\I N^k$. This implies that all chiral fields $N^k$ admit a Peccei-Quinn 
shift symmetry acting on the real parts of $N^k$ as it is indeed the case for the
orientifold setups. Thus the effective Lagrangian \eqref{eff_act1} simplifies to 
\bea \label{linaction}
\cL &=& -\tfrac{1}{2}R*\mathbf{1} - 
   \tfrac{1}{4}\tilde K_{l^k l^l}\, dN^k \wedge * d \bar N^l
  + \tfrac{1}{4} \tilde K_{L^\kappa L^\lambda}\, 
  dL^\kappa \wedge * dL^\lambda \\ 
  && + 4 \tilde K^{L^\kappa L^\lambda}
  \Big(d\tilde \xi_\kappa + \tfrac{1}{4}\tilde K_{L^\kappa l^l}\, d\, \text{Re}N^l\Big)\wedge * 
  \Big(d\tilde \xi_\lambda + \tfrac{1}{4}\tilde K_{L^\lambda l^k}\, d\, \text{Re}N^k\Big) \ .\nn
\eea
This $N=1$ Lagrangian is written completely in terms of chiral multiplets and therefore 
can be derived from a K\"ahler potential when choosing appropriate complex coordinates
$N^k$ and $T_\lambda=(L^\lambda, \tilde \xi_\lambda)$.
As we will see in a moment, a direct calculation yields that this K\"ahler potential 
is the Legendre transform of $\tilde K$ with respect to the scalars $L^\kappa$. 
It takes the 
form 
\beq \label{LegKP}
 K(T,N) = \tilde K(L, N - \bar N)  - 2 (T_\kappa +\bar T_\kappa) L^\kappa
\eeq
where $L^\kappa(N,T)$ is a function of the complex fields $N^k,T_\lambda$. This 
dependence is implicitly given via the definition of the coordinates $T_\lambda$
\beq\label{defT}
T_\lambda = i\tilde \xi_\lambda + \tfrac{1}{4}\tilde K_{L^\lambda}\ . 
\eeq
However, in order to calculate the K\"ahler metric, one only needs to determine 
the derivatives of $L^\kappa(N,T)$ with respect to 
$N^k,T_\lambda$. They are obtained by differentiating \eqref{defT} and simply read 
\beq \label{derL}
  {\partial L^\kappa}/{\partial T_\lambda} = 2 \tilde K^{L^\kappa L^\lambda}\ , \qquad 
  {\partial L^\kappa}/{\partial N^l} = - \tfrac{1}{2i} \tilde K^{L^\kappa L^\lambda} \tilde K_{L^\lambda l^l}\ .
\eeq
Using these identities one easily calculates the first derivatives of the K\"ahler 
potential \eqref{LegKP} as 
\beq \label{Kder}
  K_{T_\alpha} = -2 L^\alpha\ , \qquad K_{N^A} = \tfrac{1}{2i} \tilde K_{l^A}\ .
\eeq
Applying the equations \eqref{derL} once more when differentiating \eqref{Kder} 
one finds the K\"ahler metric
\bea \label{Km1}
  K_{T_\alpha \bar T_\beta} &=& -4 \tilde K^{L^\alpha L^\beta}\ , \quad 
  K_{T_\alpha \bar N^A}\ =\ i \tilde K^{L^\alpha L^\beta} \tilde K_{L^\beta l^A}\ , \nn \\ 
  K_{N^A \bar N^B} &=& \tfrac{1}{4} \tilde K_{l^A l^B} - \tfrac{1}{4} 
                     \tilde K_{ l^A L^\alpha }\, \tilde K^{L^\alpha L^\beta}\, \tilde K_{L^\beta l^B}\ ,
\eea
with inverse
\bea \label{invKm1}
  K^{T_\alpha \bar T_\beta} &=& - \tfrac{1}{4} \tilde K_{L^\alpha L^\beta}
                    + \tfrac{1}{4} \tilde K_{ l^A L^\alpha }\, \tilde K^{l^A l^B} \, \tilde K_{L^\beta l^B}\ , 
                    \nn \\
  K^{T_\alpha \bar N^B} & = & -i \tilde K^{l^A l^B}\, \tilde K_{ l^A L^\alpha }\ , \quad 
  K^{N^A \bar N^B} \ = \ 4 \tilde K^{l^A l^B}\ .
\eea
Finally, one checks that $K(T,N)$ is indeed the K\"ahler potential for the chiral part of the 
Lagrangian \eqref{linaction}. This is done by plugging in the definition 
of $T_\kappa$ given in \eqref{defT} and the K\"ahler metric \eqref{Km1} into 
\beq
 \cL =  -\tfrac{1}{2}R*\mathbf{1} - K_{M^I \bar M^J}\ dM^I \wedge * d\bar M^J\ , 
\eeq
where $M^I=(N^k, T_\lambda)$.

\newpage

\end{document}